\documentclass[aps,rmp,eqsecnum,twocolumn,groupedaddress,floats,showpacs]{revtex4-1}
\pdfoutput=1
\usepackage{latexsym}
\usepackage{dcolumn}
\usepackage{amssymb}
\usepackage{graphics}
\usepackage{amsmath}
\usepackage{bm}
\usepackage{subfigure}

\usepackage[dvips]{graphicx}
\usepackage{epsfig}

\newcommand{\vq}{\vec q}
\newcommand{\vk}{{\bf k}}

\renewcommand{\vec}[1]{\mathbf{#1}}

\newcommand{\beq}    {\begin{equation}}
\newcommand{\enq}    {\end{equation}}
\newcommand{\ceq}[1] {(\ref{#1})}
\newcommand{\rr}{{\bf r}}

\newcommand{\JJ}{{\bf J}}

\DeclareMathOperator{\sgn}{sgn}
\newcommand{\df}     {\equiv}
\newcommand{\smin}   {\sigma_{\rm min}}
\newcommand{\nimp}   {n_{\rm imp}}
\newcommand{\nrms}   {n_{\rm rms}}
\newcommand{\siot}   {{\rm SiO_2}}
\newcommand{\nav}    {\langle n \rangle}
\newcommand{\semt}   {\sigma_{\rm EMT}}
\newcommand{\limp}   {l_{\rm imp}}
\newcommand{\Vsc}    {V_{sc}}
\newcommand{\kk}     {{\bf k}}

\newcounter{sub3section}
\def\sub3section#1{\vspace*{3pt}
 \addtocounter{sub3section}{1}{\small
{\bf{\textsf{(\alph{sub3section}) #1:}}}}}

\begin{document}

\author{S. Das Sarma$^{1}$, 
Shaffique Adam$^{1,2}$, E. H. Hwang$^{1}$, 
and Enrico Rossi$^{1}$\footnote{Current address: Department of Physics, College of William and Mary, Williamsburg, VA 23187, USA}}
\affiliation{$^1$Condensed Matter Theory Center, Department of Physics,
University of Maryland, 
College Park, MD 20742-4111}
\affiliation{$^2$Center for Nanoscale Science and Technology, 
National Institute of Standards and Technology, 
Gaithersburg, Maryland 20899-6202, USA}

\title{Electronic transport in two dimensional graphene}

\date{\today}
\begin{abstract}
We provide a broad review of fundamental electronic properties of
two-dimensional graphene with the emphasis on density and temperature
dependent carrier transport in doped or gated graphene structures.  A
salient feature of our review is a critical comparison between carrier
transport in graphene and in two-dimensional semiconductor systems
(e.g. heterostructures, quantum wells, inversion layers) so that the
unique features of graphene electronic properties arising from its
gapless, massless, chiral Dirac spectrum are highlighted.  Experiment
and theory as well as quantum and semi-classical transport are
discussed in a synergistic manner in order to provide a unified and
comprehensive perspective. Although the emphasis of the review is on
those aspects of graphene transport where reasonable consensus exists
in the literature, open questions are discussed as well. Various
physical mechanisms controlling transport are described in depth
including long-range charged impurity scattering, screening,
short-range defect scattering, phonon scattering, many-body effects,
Klein tunneling, minimum conductivity at the Dirac point,
electron-hole puddle formation, p-n junctions, localization,
percolation, quantum-classical crossover, midgap states, quantum Hall
effects, and other phenomena.
\end{abstract}
\maketitle

\tableofcontents


\section{Introduction}  \label{sec_intro}

\subsection{Scope} \label{subsec_scope}

The experimental discovery of two dimensional (2D) gated graphene in 2004
by \citet{kn:novoselov2004} is a seminal event in electronic 
materials science, ushering in a tremendous outburst of scientific activity 
in the study of electronic properties of graphene which continues unabated upto
the end of 2009 (with the appearance of more than 5000 articles on graphene
during the 2005-2009 five-year period).  The subject has now reached a level
so vast that no single article can cover the whole topic in any 
reasonable manner, and most general reviews are likely to become obsolete
in a short time due to rapid advances in the graphene literature.

The scope of the current review (written in late 2009 and early 2010) is
transport in gated graphene with the emphasis on fundamental physics
and conceptual issues.  Device applications and related topics are not 
discussed \cite{avouris2007} nor are graphene's mechanical 
properties \cite{kn:bunch2007,kn:lee2008}.  The important subject of graphene materials
science, which deserves its own separate review, is not discussed
at all.
Details of the band structure properties and related 
phenomena are also not covered in any depth, except in the context 
of understanding transport phenomena.
What is covered in reasonable depth is the basic physics of carrier
transport in graphene, critically compared with the corresponding 
well-studied 2D semiconductor transport properties, with the
emphasis on scattering mechanisms and conceptual issues of 
fundamental importance.  In the context of 2D transport, it is
conceptually useful to compare and contrast graphene with the
much older and well established subject 
of carrier transport in 2D semiconductor structures (e.g. Si inversion
layers in MOSFETs, 2D GaAs heterostructures and quantum wells).  Transport
in 2D semiconductor systems has a number of similarities and key dissimilarities
with graphene.  One purpose of this review is to emphasize the key 
conceptual differences between 2D graphene and 2D semiconductors so 
as to bring out the new fundamental aspects of graphene transport 
which make it a truly novel electronic material qualitatively 
different from the large class of existing and well established
2D semiconductor materials.

Since graphene is a dynamically (and exponentially) evolving subject, with 
new important results appearing almost every week, the current 
review concentrates on only those features of graphene carrier 
transport where some qualitative understanding, if not a universal
consensus, has been achieved in the community.  As such, some 
active topics, where the subject is in flux, have
been left out.  Given the constraint of the size of this review,
depth and comprehension have been emphasized over breadth -- given
the huge graphene literature, no single review can attempt to 
provide a broad coverage of the subject at this stage.
There have already been several reviews of graphene 
physics in the recent literature.  We have made every effort
to minimize overlap between our article and these
recent reviews.  The closest in spirit to our review is
the one by \citet{kn:neto2009} which was written 2.5 years
ago (i.e. more than 3000 graphene publications have appeared
in the literature since the writing of that review).  
Our  review should be considered complimentary to \citet{kn:neto2009},
and we have tried avoiding too much repetition of the 
materials already covered by them, concentrating
instead on the new results arising in the literature following
the older review.  Although some repetition is necessary 
in order to make our review self-contained, we refer the 
reader to \citet{kn:neto2009} for details on the
history of graphene, its band structure considerations
and the early (2005-2007) experimental and theoretical
results.  Our material emphasizes the more mature phase
(2007-2009) of 2D graphene physics.

For further background and review of graphene physics
beyond the scope of our review, we mention in addition 
to the Rev. Mod. Phys. article by \citet{kn:neto2009}, the 
accessible reviews by Andrei Geim and his collaborators~\cite{kn:geim2007,geim-s-324-1530-2009}, 
the recent brief review by \cite{mucciolo2010},
as well as two 
edited volumes of Solid State Communications~\cite{kn:dassarma2007a,kn:falko2009}, where the 
active graphene researchers have contributed individual perspectives.

\subsection{Background}  
\label{subsec_background}

\begin{figure}
\hspace{0.1\hsize}
\begin{center}
\includegraphics[width=1.0\columnwidth]{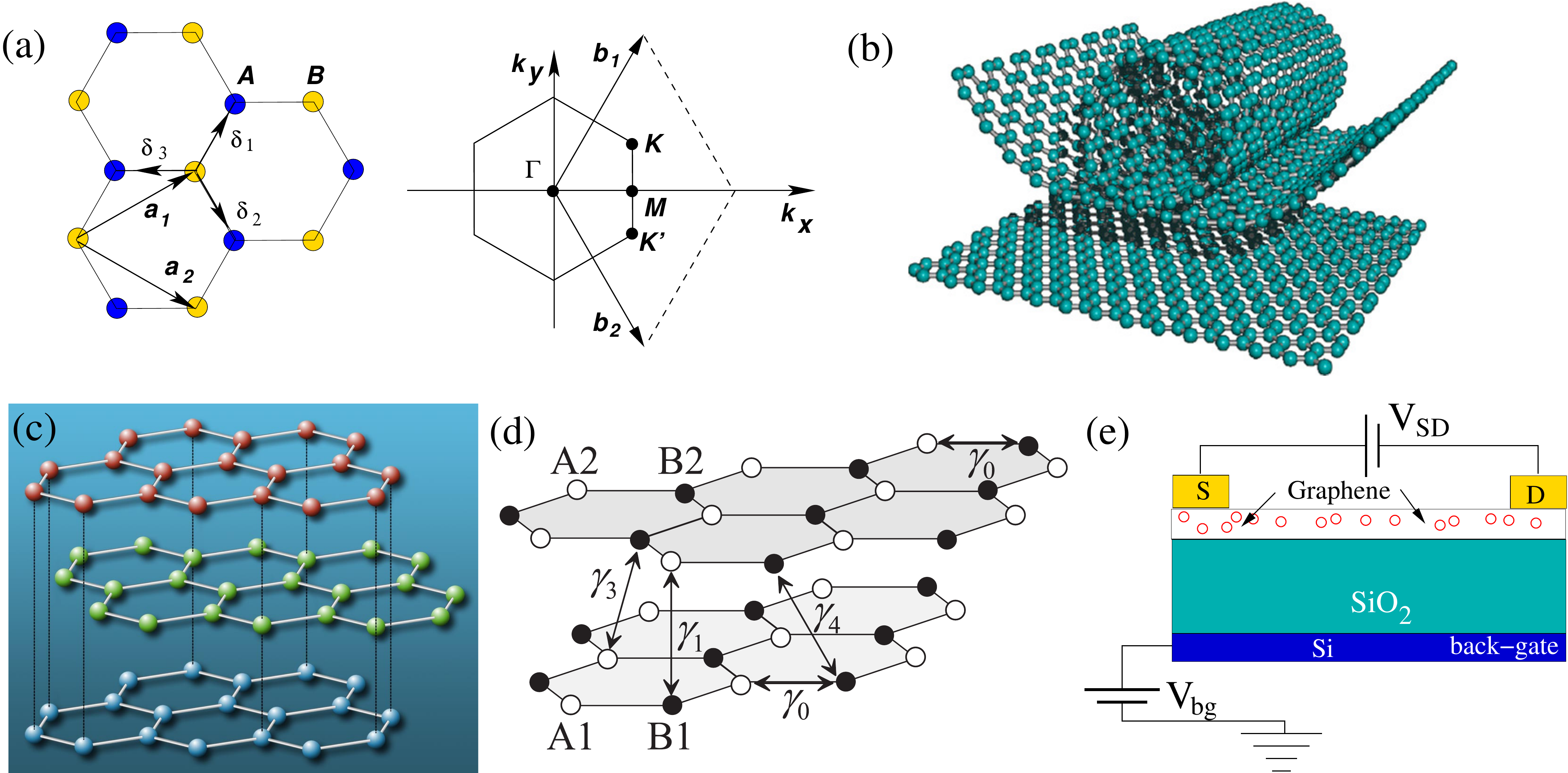}
\end{center}
\label{SDS_Fig_Nanotube}
\caption{(Color online)
(a) Graphene honeycomb lattice showing in different colors the
two triangular sublattices. Also shown is the graphene Brillouin
zone in momentum space. Adapted from \citet{kn:neto2009}.
(b) Carbon nanotube as a rolled up graphene layer. 
(c) Lattice structure of graphite, graphene multilayer.
    Adapted from \citet{neto-pw-2006}.
(d) Lattice structure of bilayer graphene. 
    $\gamma_0$ and $\gamma_1$ are respectively the intralayer and interlayer
    hopping parameters $t$, $t_\perp$ used in the text.
    The interlayer hopping parameters $\gamma_3$ and $\gamma_4$
    are much smaller than $\gamma_1\equiv t_\perp$ and are normally neglected.
    Adapted from \citet{mucha-kruczynski2010}
(e) Typical configuration for gated graphene.
}
\end{figure}          

\begin{figure}
\hspace{0.1\hsize}
\begin{center}
\includegraphics[width=1.0\columnwidth]{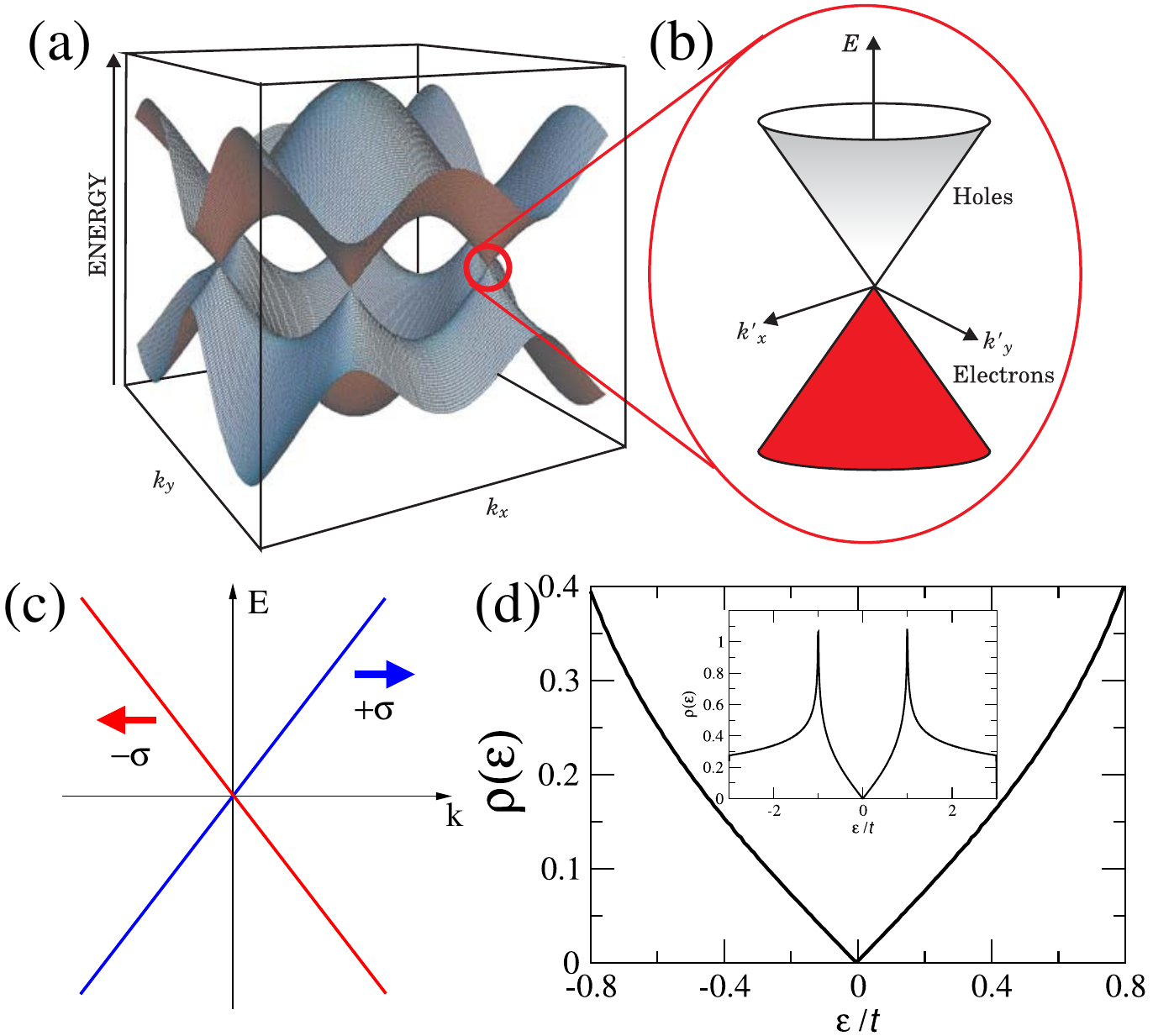}
\end{center}
\caption{(Color online)
(a) Graphene bandstructure. 
    Adpated from \citet{wilson2006}.
(b) Enlargment of the band structure close to the $K$ and $K'$
    points showing the Dirac cones.
    Adpated from \citet{wilson2006}.
(c) Model energy dispersion $E= \hbar v_{\rm F} |{\bf k}|$.
(d) Density of states of graphene close to the Dirac point.
    The inset shows the density of states over the full electron
    bandwidth. Adapted from \citet{kn:neto2009}.}
\label{SDS_Fig_BandStructure} 
\end{figure}

Graphene (or more precisely, monolayer graphene  -- in this 
review, we refer to monolayer graphene simply as ``graphene'') is
a single 2D sheet of carbon atoms in a honeycomb lattice.  As such
2D graphene rolled up in the plane is a carbon nanotube, 
and multilayer graphene with weak interlayer tunneling 
is graphite.  Given that graphene is simply a single 2D layer
of carbon atoms peeled off a graphite sample, there was 
early interest in the theory of graphene band structure
which was all worked out a long time ago.  In this review
we only consider graphene monolayers (MLG) and bilayers (BLG), which 
are both of great interest. 

\subsubsection{Monolayer graphene}
Graphene monolayers have been rightfully described as
the ``ultimate flatland'' \cite{geim2007}
i.e. the most perfect 2D electronic
material possible in nature, since the system is exactly
one atomic monolayer thick, and carrier dynamics
is necessarily confined in this strict 2D layer. The electron
hopping in 2D graphene honeycomb lattice is quite special 
since there are two equivalent lattice sites (A and B in 
Fig.~\ref{SDS_Fig_Nanotube}a) which give rise to the
``chirality'' in the graphene carrier dynamics.

The honeycomb structure can be thought of as a triangular 
lattice with a basis of two atoms per unit cell,
with the 2D lattice vectors ${\bf A_0} = (a/2) (3, \sqrt{3})$
and ${\bf B_0} = (a/2) (3, -\sqrt{3})$ 
($a \approx 0.142~{\rm nm}$ is the carbon-carbon distance).  
${\bf K} = (2 \pi /(3 a), 2 \pi/(3 \sqrt{3} a))$
and ${\bf K'} = (2 \pi /(3 a), -2 \pi/(3 \sqrt{3} a))$
are the inequivalent corners of the Brillouin zone and are 
called Dirac points.
These Dirac points are of great importance in the electronic 
transport of graphene, and play a role similar 
to the role of $\Gamma$ points in direct band-gap 
semiconductors such as GaAs.  Essentially all 
of the physics to be discussed in this review
is the physics of graphene carriers (electrons and/or holes)
close to the Dirac points (i.e. within a 2D wavevector
$q = |{\bf q}| \ll 2 \pi/a$ of the Dirac points) just as 
all the 2D semiconductor physics we discuss will 
occur around the $\Gamma$ point.

The electronic band dispersion of 2D monolayer 
graphene was calculated by \citet{kn:wallace1947}
and others~\cite{kn:mcclure1957,kn:slonczewski1958} a 
long time ago, within the tight-binding prescription,
obtaining upto the second-nearest-neighbor
hoping term in the calculation, the following
approximate analytic formula for the 
conduction (upper, $+$, $\pi^*$) band, and valence
(lower, $-$, $\pi$) band.
\begin{equation}
\label{SDS_Eq_Wallace}
E_{\pm}(q) \approx 3t' \pm \hbar v_{\rm F} |\vec{q}| - 
\left( \frac{9 t' a^2}{4} \pm \frac{3 t a^2}{8} \sin(3 \theta_q)  \right)|\vec{q}|^2,
\end{equation}
with $v_F=3ta/2$, $\theta_q = \arctan^{-1}[q_x/q_y]$, and where $t$, $t'$ are
respectively the nearest-neighbor (i.e. inter-sublattice, A-B) 
and next-nearest-neighbor (i.e intra-sublattice, A-A or B-B) hopping
amplitudes and $t (\approx 2.5~{\rm eV})  \gg t' (\approx 0.1~{\rm eV})$.

The almost 
universally used graphene band dispersion at long wavelength puts $t' = 0$, where
the band-structure for small $q$ relative to the Dirac point 
is given by 
\begin{equation}
\label{SDS_Eq_Linear}
E_{\pm}(q) = \pm \hbar v_{\rm F} q + {\mathcal O}(q/k)^2.
\end{equation}
\noindent 
%
Further details on the band structure of 2D graphene monolayers can be
found in the literature 
\cite{kn:wallace1947,kn:mcclure1957,kn:slonczewski1958,mcclure1964,reich2002,kn:neto2009}
and will not be discussed here.  Instead we
provide below, a thorough discussion of the implications of
Eq.~\ref{SDS_Eq_Linear} for graphene carrier transport.  Since much
of the fundamental interest is in understanding graphene transport in
the relatively low carrier density regime, complications arising from
the large $q~(\approx K)$ aspects of graphene band structure lead only
to small perturbative corrections to graphene transport properties and
are, as such, only of secondary importance to our review.

The most important aspect of graphene's energy dispersion (and the one
attracting most attention) is its linear energy-momentum relationship
with the conduction and valence bands intersecting at $q=0$, with no energy gap.
Graphene is thus a zero band-gap semiconductor with a linear rather
than quadratic long wavelength energy dispersion for both electrons (holes) in the
conduction (valence) bands.  The existence of two Dirac points at $K$
and $K'$ where the Dirac cones for electrons and holes touch 
(Fig.~\ref{SDS_Fig_BandStructure}b) each other in momentum space, gives
rise to a valley degeneracy $g_{\rm v}=2$ for graphene.  The presence
of any inter-valley scattering between $K$ and $K'$ points lifts this
valley degeneracy, but such effects require the presence of strong
lattice scale scattering.
Intervalley scattering seems to be weak and when they can
be ignored the presence of a second valley can be taken into
account symply via the degenercy factor $g_{\rm v}=2$. 
Throughout this introduction we neglect intervally scattering processes.
%

The graphene carrier dispersion $E_\pm(q) = \hbar v_{\rm F} q$, explicitly 
depends on the constant $v_{\rm F}$, sometimes called the graphene (Fermi) 
velocity.  In the literature different symbols ($v_{\rm F}, v_0, \gamma/\hbar$)
are used to denote this velocity.  The tight-binding prescription 
provides a formula for $v_{\rm F}$ in terms of the nearest-neighbor 
hopping $t$ and the lattice constant $a_2 = \sqrt{3} a$:
$\hbar v_{\rm F} = 3 t a /2$.
The best estimates of $t\approx 2.5~{\rm eV}$ and $a = 0.14~{\rm nm}$ 
give $v_{\rm F} \approx 10^{8}~{\rm cm/s}$
for the empty graphene
band i.e. in the absence of any carriers.  Presence of carriers
may lead to a many-body renormalization of the graphene velocity, 
which is, however, small for MLG, but could, in principle, be substantial 
for BLG. 

The linear long wavelength Dirac dispersion with Fermi velocity roughly 
$1/300$ of the corresponding velocity of light, is the most 
distinguishing feature of graphene in addition to its strict
2D nature.  It is therefore natural to ask about the precise
applicability of the linear energy dispersion, since it 
is obviously a long-wavelength continuum property of graphene carriers
valid only for $q \ll K \approx (0.1~{\rm nm})^{-1}$.

There are several ways to estimate the cut-off wavevector
(or momentum) $k_c$ above which the linear continuum Dirac dispersion    
 approximation breaks down for graphene.  The easiest is 
perhaps to estimate the carrier energy $E_c = \hbar v_{\rm F} k_c$,
and demand that $E_c < 0.4 t ~ 1.0~{\rm eV}$, so that
one can ignore the lattice effects (which lead to deviations from pure
 Dirac-like dispersion).  This leads
to a cutoff wavevector given by $k_c \approx 0.25~{\rm nm}^{-1}$.
%

The mapping of graphene electronic structure onto the massless 
Dirac theory is deeper than the linear graphene carrier
energy dispersion.  The existence of two equivalent, but independent,
sub-lattices A and B (corresponding to the two atoms per unit cell), 
leads to the existence of a novel chirality in graphene dynamics
where the two linear branches of graphene energy dispersion (intersecting
at Dirac points) become independent of each other,
indicating the existence of a pseudospin quantum number analogous 
to electron spin (but completely independent of real spin).  Thus graphene
carriers have a pseudospin index 
in addition to the spin and orbital index.  The 
existence of the chiral pseudospin quantum number is a natural byproduct
of the basic lattice structure of graphene comprising two independent sublattices.
The long-wavelength low energy effective 2D continuum Schr\"{o}dinger equation 
for spinless graphene carriers near the Dirac point therefore becomes
\begin{equation}
\label{SDS_Eq_Dirac}
-i \hbar v_F\boldsymbol\sigma \cdot \boldsymbol \nabla \Psi(\vec{r}) = E \Psi(\vec{r}),
\end{equation}
where $\boldsymbol\sigma = (\sigma_x, \sigma_y)$ is the usual vector of 
Pauli matrices (in 2D now), and $\Psi(\vec{r})$ is a 2D spinor wavefunction.
Eq.\ceq{SDS_Eq_Dirac} corresponds to the effective low energy Dirac Hamiltonian:
\begin{eqnarray}
{\mathcal H} = \hbar v_{\rm F}
\left( \begin{array}{cc} 
0 & q_x  - i q_y \\
q_x + i q_y & 0 \end{array} \right) = \hbar v_{\rm F} \boldsymbol \sigma \cdot \vec{q}.
\end{eqnarray} 

We note that Eq.~\ref{SDS_Eq_Dirac} is simply the equation for massless 
chiral Dirac Fermions in 2D (except that the spinor here refers to the graphene
pseudospin rather than real spin), although it is arrived at 
starting purely from the tight-binding Schr\"{o}dinger equation for
carbon in a honeycomb lattice with two atoms per unit cell.  This mapping
of the low energy, long wavelength electronic structure of graphene onto the
massless chiral Dirac equation, was discussed by \citet{kn:semenoff1984} 
more than 25 years ago.  It is a curious
historical fact that although the actual experimental discovery of 
gated graphene (and the beginning of the frenzy of activities leading 
to this review article) happened only in 2004, some of the key theoretical 
insights go back a long way in time and are as valid today for real graphene
as they were for theoretical graphene when they 
were 
introduced~\cite{kn:wallace1947,kn:semenoff1984,kn:haldane1988,kn:mcclure1957,kn:ludwig1994,kn:gonzalez1994}.

The momentum space pseudo-spinor eigenfunctions for Eq.~\ref{SDS_Eq_Dirac} 
can be written as
\begin{eqnarray}
\Psi(\vec{q}, K) = \frac{1}{\sqrt{2}} 
\left(  \begin{array}{c}
e^{-i \theta_q/2} \\ \pm e^{i \theta_q/2} \end{array} \right), \hspace{0.3cm}
\Psi(\vec{q}, K') = \frac{1}{\sqrt{2}} 
\left(  \begin{array}{c} 
e^{i \theta_q/2} \\ \pm e^{- \theta_q/2} \end{array} \right), \nonumber 
\end{eqnarray} 
%
where the $\pm$ signs correspond to the conduction/valence bands
with $E_\pm(q) = \pm \hbar v_{\rm F} q$.  It is easy to show using the
Dirac equation analogy that the conduction/valence bands come with 
positive/negative chirality, which is conserved, within the constraints
of the validity of Eq.~\ref{SDS_Eq_Dirac}.
We note that the presence of real spin,
ignored so far, would add an extra spinor structure to graphene's wavefunction
(this real spin part of the graphene wavefunction is similar to that of 
ordinary 2D semiconductors).  The origin of the massless Dirac description 
of graphene lies in the intrinsic coupling of its orbital motion to the
pseudospin degree of freedom due to the presence of A and B sublattices
in the underlying quantum mechanical description.  

%
 
\subsubsection{Bilayer graphene}

The case of bilayer 
graphene is interesting in its own right, since with two graphene monolayers
that are weakly coupled by interlayer carbon hopping, it is intermediate between 
graphene monolayers and bulk graphite. 

The tight-binding description can be adapted to study the bilayer 
electronic structure assuming specific stacking of the two layers
with respect to each other (which controls the interlayer hopping 
terms).  Considering the so-called A-B stacking of the two layers (which 
is the 3D graphitic stacking), the low energy, long wavelength electronic structure of bilayer 
graphene is described by the following energy dispersion relation
\cite{dresselhaus2002,brandt1988,kn:mccann2006b,mccann2006z}
\begin{eqnarray} \label{SDS_Eq_Bilayer}
E_{\pm}(q) &=& \left[ V^2 + \hbar^2 v_{\rm F}^2 q^2 + t_{\perp}^2/2 \right. \nonumber \\
 &&\left.  \mbox{} \pm \left( 4V^2\hbar^2 v_{\rm F}^2 q^2 + t_{\perp}^2 \hbar^2 v_{\rm F}^2 q^2 + 
t_{\perp}^4/4 \right)^{1/2} \right]^{1/2},
\end{eqnarray}
where $t_\perp$ is the effective interlayer hopping energy (and $t$,
$v_{\rm F}$ are the intra-layer hopping energy and graphene 
Fermi velocity for the monolayer case).  We note that $t_\perp (\approx
0.4~{\rm eV}) < t (\approx 2.5~{\rm eV})$, and we have neglected 
several additional interlayer hopping terms since they are 
much smaller than $t_\perp$.  The quantity $V$ with dimensions
of energy appearing in Eq.~\ref{SDS_Eq_Bilayer} for bilayer dispersion
corresponds to the possibility of a real shift (e.g. by an applied
external electric field perpendicular to the layers, $\hat{z}$-direction) in the electrochemical
potential between the two layers, which would translate into 
an effective band-gap opening near the Dirac point
\cite{ohta2006,castro2007,kn:oostinga2007,zhang-n-2009}  

Expanding Eq.~\ref{SDS_Eq_Bilayer} to leading order  in momentum,
and assuming ($V \ll t$), we get
\begin{equation} \label{SDS_Eq_Hyperbolic}
E_{\pm}(q) = \pm [V - 2 \hbar^2 v_{\rm F}^2 V q^2/t^2_\perp + \hbar^4 v_{\rm F}^4 q^4/(2t_\perp^2 V)].
\end{equation}
\noindent We conclude that (i) For $V\neq 0$, bilayer graphene 
has a minimum bandgap of $\Delta = 2V-4V^3/t^2_\perp$ at $q= \sqrt{2}V/\hbar v_{\rm F}$, and (ii)
for $V=0$, bilayer graphene is a gapless semiconductor with 
a parabolic dispersion relation $E_{\pm} (q) \approx \hbar^2 v_{\rm F}^2 q^2/t_\perp
= \hbar^2 q^2/(2 m)$, where $m = t_\perp/(2 v_{\rm F}^2)$ for small $q$.  The 
parabolic dispersion (for $V=0$) applies only for small values of 
$q$ satisfying $\hbar v_{\rm F} q \ll t_\perp$, whereas in the opposite 
limit $\hbar v_{\rm F} q \gg t_\perp$, we get a linear band dispersion
$E_\pm(q) \approx \pm \hbar v_{\rm F} q$, just like the monolayer case.
We note that using 
the best estimated values for $v_{\rm F}$ and $t_\perp$, the bilayer
effective mass is $m \approx (0.03~\mbox{to}~0.05)~m_{\rm e}$, which corresponds 
to a very small effective mass.

To better understand the quadratic to linear crossover in the
effective BLG band dispersion, it is convenient to rewrite the BLG
band dispersion (for $V=0$) in the following hyperbolic form
\begin{equation}
 E_{\rm BLG} = \mp m v_{\rm F}^2 \pm m v_{\rm F}^2 \left[ 1 + (k/k_0)^2 \right]^{1/2},
\end{equation}
where $k_0 = t_{\perp}/(2 \hbar v_{\rm F})$ is a characteristic wavevector.  In this form
it is easy to see that $E_{\rm BLG} \rightarrow k^2 (k)$ for $ k \rightarrow 0
(\infty)$ for the effective BLG band dispersion with $k \ll k_0$ ($k
\gg k_0$) being the parabolic (linear) band dispersion regimes,  $k_0
\approx 0.3~{\rm nm}^{-1}$ for $m \approx 0.03~m_e$.  Using the best
available estimates from band structure calculations, we conclude that
for carrier densities smaller (larger) than $5 \times 10^{12}~{\rm
  cm}^{-2}$, the BLG system should
have parabolic (linear) dispersion at the Fermi level.

What about chirality for bilayer graphene?  Although the bilayer
energy dispersion is non-Dirac like and parabolic, the system is
still chiral due to the A/B sublattice symmetry giving rise to the
conserved pseudospin quantum index.  The detailed chiral 4-component
wavefunction for the bilayer case including both layer and sublattice
degrees of freedom can be found in the original literature
\cite{kn:mccann2006b,mccann2006z,kn:nilsson2006,kn:nilsson2006b,kn:nilsson2008}.

The possible existence of an external bias-induced band-gap and the 
parabolic dispersion at long wavelength distinguish bilayer graphene
from monolayer graphene, with both possessing chiral carrier dynamics.
We note that bilayer graphene should be considered a single 2D system,
quite distinct from ``double-layer'' graphene, \citet{kn:hwang2009b} 
which is a composite 
system consisting of two parallel single layers of graphene, 
separated by a distance in the $\hat{z}$-direction.  The 2D energy dispersion
in double-layer graphene is massless Dirac-like (as in the monolayer case),
and the interlayer separation is arbitrary, whereas bilayer graphene
has the quadratic band dispersion with a fixed inter-layer separation
of $0.3~{\rm nm}$ similar to graphite.

\begin{figure}
\hspace{0.1\hsize}
\begin{center}
\includegraphics[width=0.95\columnwidth]{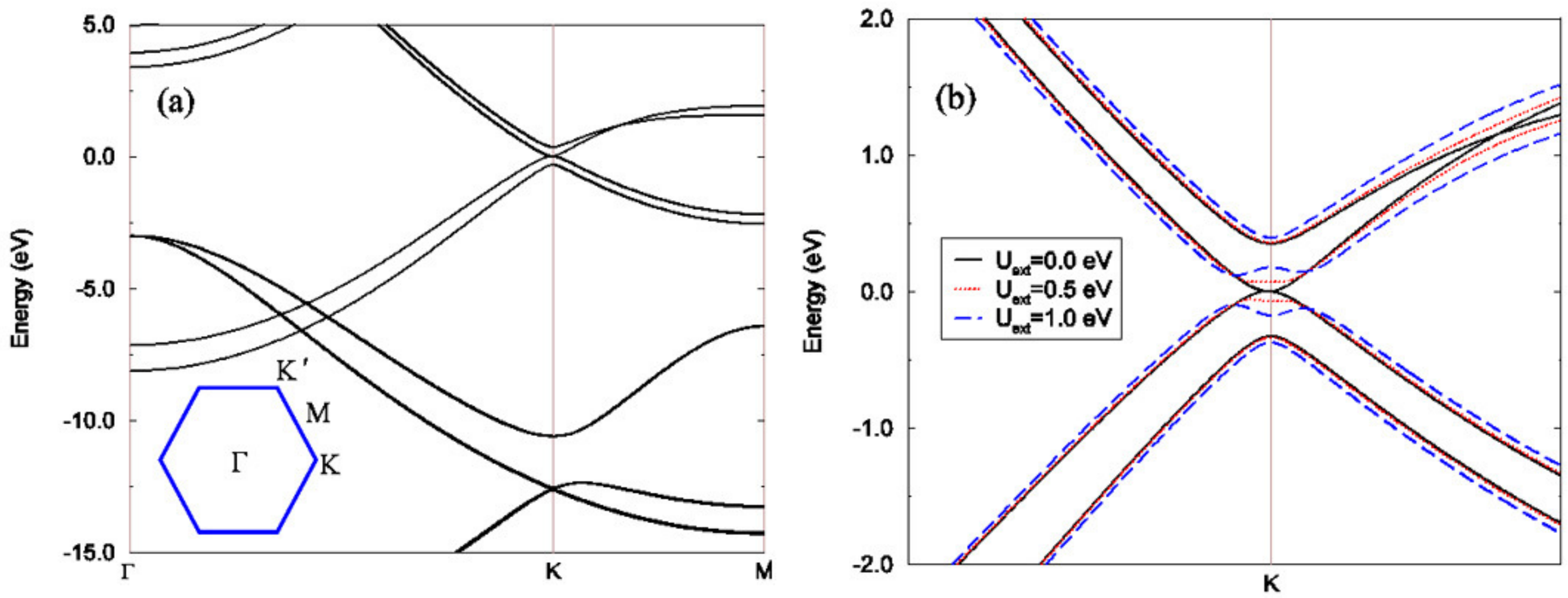}
\end{center}
\caption{\label{SDS_Fig_BilayerGraphene} (Color online)
 (a) Energy band of bilayer graphene for $V=0$.
 (b) Enlargment of the energy band close to the neutrality point $K$
     for different values of $V$.
     Adapted from \citet{kn:min2007}
     }
\end{figure}

\subsubsection{2D Semiconductor structures}

Since one goal of this review is to understand graphene electronic 
properties in the context of extensively studied (for more than 40 years)
2D semiconductor systems (e.g. Si inversion layers in MOSFETs, GaAs-AlGaAs
heterostructures, quantum wells, etc.), we summarize in this section 
the basic electronic structure of 2D semiconductor systems which 
are of relevance in the context of graphene physics, without giving 
much details, which can be found in the 
literature~\cite{kn:ando1982,kn:bastard1991,kn:davies1998}.  

There are, broadly speaking, four qualitative differences between
2D graphene and 2D semiconductor systems.  (We note that there 
are significant quantitative and some qualitative differences
between different 2D semiconductor systems themselves).  
These differences are sufficiently 
important so as to be emphasized right at the outset. \\ 

{\em (i)} First, 2D semiconductor systems typically have very large
($>1~{\rm eV}$) bandgaps so that 2D electrons and 2D holes must be
studied using completely different electron-doped or hole-doped
structures.  By contrast, graphene (except biased graphene bilayers
that have small band-gaps) is a gapless semiconductor with the nature
of the carrier system changing at the Dirac point from electrons to
holes (or vice versa) in a single structure.  A direct corollary of
this gapless (or small gap) nature of graphene, is of course, the
``always metallic'' nature of 2D graphene, where the chemical
potential (``Fermi level'') is always in the conduction or the 
valence band.  By contrast, the 2D semiconductor 
becomes insulating below a threshold
voltage, as the Fermi level enters the bandgap. \\

{\em (ii)} Graphene systems are chiral, whereas 2D semiconductors
are non-chiral.  Chirality of graphene leads to some important consequences
for transport behavior, as we discuss later in this review. (For example,
$2 k_{\rm F}$-backscattering is suppressed in MLG at low temperature.)   \\

{\em (iii)} Monolayer graphene dispersion is linear, whereas 2D semiconductors
have quadratic energy dispersion. This leads to substantial quantitative
differences in the transport properties of the two systems.  \\ 

{\em (iv)} Finally, the carrier confinement in 2D graphene is ideally 
two-dimensional, since the graphene layer is precisely one atomic 
monolayer thick.  For 2D semiconductor structures, the quantum dynamics
is two dimensional by virtue of confinement induced by an external 
electric field, and as such, 2D semiconductors are quasi-2D systems, 
and always have an average width or thickness 
$\langle z \rangle$ ($\approx 5~{\rm nm}$ to $50~{\rm nm}$) in the
third direction with $\langle z \rangle \lesssim \lambda_{\rm F}$, 
where $\lambda_{\rm F}$ is the 2D Fermi wavelength (or equivalently 
the carrier de Broglie wavelength).  The condition $\langle z \rangle < \lambda_{\rm F}$ defines a 2D electron system. \\

The carrier dispersion of 2D semiconductors is given by 
$ E(q) = E_0 + \hbar^2 q^2/(2 m^*)$,
%
%
where $E_0$ is the quantum confinement energy of the lowest quantum 
confined 2D state and $\vec{q} = (q_x, q_y)$ is the 2D wavevector.  If
more than one quantum 2D level is occupied by carriers -- usually 
called ``sub-bands'' -- the system is no longer, strictly speaking, two
dimensional, and therefore a 2D semiconductor is no longer two dimensional 
at high enough carrier density when higher subbands get populated.  

The 2D effective mass entering  $m^*$ is known from
bandstructure calculations, and within the effective mass approximation
$m^* = 0.07~m_{\rm e}$ (electrons in GaAs), $m^* = 0.19~m_{\rm
  e}$ (electrons in Si 100 inversion layers), $m^* = 0.38~m_{\rm e}$
(holes in GaAs), and $m^* = 0.92~m_{\rm e}$ (electrons in Si 111
inversion layers).  In some situations, e.g. Si 111, the 2D effective
mass entering the dispersion relation may have anisotropy in the $xy$
plane and a suitably averaged $m^* = \sqrt{m_x m_y}$ is usually used.

The 2D semiconductor wavefunction is non-chiral, and is derived
from the effective mass approximation to be 
\begin{equation}
\Phi(\vec{r}, z) \sim e^{i \vec{q} \cdot \vec{r}} \xi(z),
\end{equation}
where $\vec{q}$ and $\vec{r}$ are 2D wavevector and position,
and $\xi(z)$ is the quantum confinement wavefunction in the 
${\hat z}$-direction for the lowest sub-band.  The confinement
wavefunction defines the width/thickness of the 2D semiconductor
state with $\langle z \rangle = |\langle \xi | z^2 | \xi \rangle|^{1/2}$.  The
detailed form for $\xi(z)$ usually requires a quantum mechanical 
self-consistent local density approximation calculation 
using the confinement potential, and we refer the reader to the 
extensive existing literature for the details on the confined 
quasi-2D subband structure 
calculations~\cite{kn:bastard1991,kn:davies1998,kn:ando1982,kn:stern1984}.

Finally, we note that 2D semiconductors may also in some situations
carry an additional valley quantum number similar to graphene.  But the
valley degeneracy in semiconductor structures e.g. Si MOSFET 2D electron
systems, have nothing whatsoever to do with a pseudospin chiral index.
For Si inversion layers, the valley degeneracy ($g_v = 2,4$ and $6$
respectively for Si 100, 110 and 111 surfaces) arises from the bulk 
indirect band structure of Si which has 6 equivalent ellipsoidal conduction 
band minima along the 100, 110 and 111 directions about 85~\% to 
the Brillouin zone edge.  The valley degeneracy 
in Si MOSFETs, which is invariably slightly lifted ($\approx 0.1~{\rm meV}$), 
is a well established experimental fact.   
    
\begin{figure}
\hspace{0.1\hsize}
\begin{center}
\includegraphics[width=1.0\columnwidth]{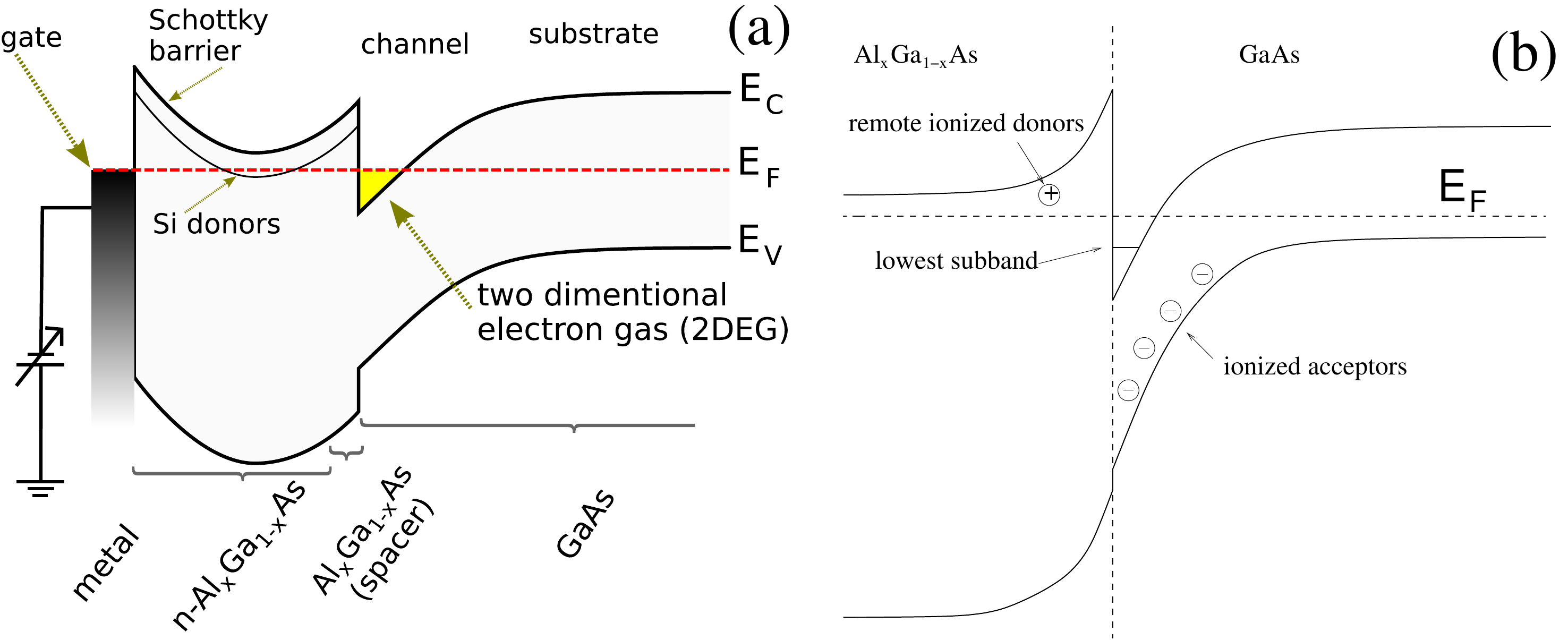}
\end{center}
\caption{\label{SDS_Fig_2DEG} (Color online)
(a) Heterostructure inversion layer quantum well. 
(b) Band diagram showing the bending of the bands at the
    interface of the semiconductors and the 2 dimensional subband.
  }
\end{figure}

\begin{center}
\begin{table*}[t!!!!]
\caption{\label{tab:table2}
Elementary electronic quantities. Here $E_F$, $D(E)$, $r_s$, and
$q_{TF}$ represent the Fermi energy, the density of states, the
interaction parameter, and Thomas Fermi wave vector,
respectively. $D_0=D(E_F)$ is the density of states at the Fermi
energy and $q_s = q_{TF}/k_F$. } 
\begin{ruledtabular}
\begin{tabular}{ccccccc}
 & $E_F$ & $D(E)$ & $D_0=D(E_F)$ & $r_s$ & $q_{TF}$ & $q_s$ \\ 
\hline 
MLG & $\hbar v_F \sqrt{\frac{4\pi n}{g_sg_v}}$ & $\frac{g_sg_vE}{2\pi(\hbar v_F)^2}$
& $\frac{\sqrt{g_s g_v n}}{\sqrt{\pi} \hbar v_F}$ & $\frac{e^2}{\kappa \hbar v_F}
\frac{\sqrt{g_s g_v}}{2}$ & $\frac{\sqrt{4\pi g_s g_v n}e^2}{\kappa \hbar v_F}$ &
$\frac{g_s g_v e^2}{\kappa \hbar v_F}$ \\
BLG/2DEG & $\frac{2\pi \hbar^2 n}{mg_s g_v}$ & $\frac{g_s g_v m }{2 \pi \hbar^2}$
& $ \frac{g_s g_v m }{2\pi \hbar^2}$ & $ \frac{me^2}{2\kappa \hbar^2}\frac{ g_s g_v}
{\sqrt{\pi n}}$ & $ \frac{g_s g_v m e^2 }{\kappa \hbar^2} $ &
$\frac{(g_s g_v)^{3/2} me^2}{\kappa \hbar^2 \sqrt{4\pi n}}$ \\
\end{tabular}
\end{ruledtabular}
\hspace{0.1\hsize}
\end{table*}
\end{center}

\subsection{Elementary electronic properties} \label{subsec_elementary}

We describe, summarize, and critically contrast the elementary electronic properties of 
graphene and 2D semiconductor
based electron gas systems based on their long wave length effective
2D energy dispersion discussed in the earlier sections.  Except 
where the context is obvious, we use the following abbreviations 
from now on: MLG (monolayer graphene), BLG (bilayer graphene), 
and 2DEG (semiconductor based 2D electron gas systems).  The 
valley degeneracy factors are typically $g_{\rm v} = 2$ for 
graphene and Si 100 based 2DEGs, whereas $g_{\rm v} = 1~(6)$ for 2DEGs
in GaAs (Si 111). The spin degeneracy is always $g_{\rm s} = 2$,
except at high magnetic fields.  
%
%
The Fermi wavevector for all 2D systems is given simply by filling 
up the non-interacting momentum eigenstates upto $q = k_{\rm F}$.  
\begin{equation}
n = g_{\rm s} g_{\rm v} \int_{|\vec{q}|\le k_F} \frac{d\vec{q}}{(2 \pi)^2} \rightarrow
k_{\rm F} = \sqrt{\frac{4 \pi n}{g_{\rm s} g_{\rm v}}},
\end{equation} 
where $n$ is the 2D carrier density in the system.  Unless otherwise
stated, we will mostly consider electron systems (or the conduction 
band side of MLG and BLG).
Typical experimental values
of $n \approx 10^{9}~{\rm cm}^{-2}$ to $5 \times 10^{12}~{\rm cm}^{-2}$ 
are achievable in graphene and Si MOSFETs, wheras in GaAs-based
2DEG systems $n \approx 10^{9}~{\rm cm}^{-2}$ to 
$5 \times 10^{11}~{\rm cm}^{-2}$.



\subsubsection{Interaction parameter $r_{\rm s}$}
The interaction parameter, also known as the Wigner-Seitz radius or 
the coupling constant or the effective fine-structure constant, is 
denoted here by $r_s$, which in this context is the ratio of 
the average inter-electron Coulomb interaction energy to the Fermi 
energy.  Noting that the average Coulomb energy is simply
$\langle V \rangle = e^2/(\kappa \langle r \rangle )$, where
$\langle r \rangle = (\pi n )^{-1/2}$ is the average inter-particle
separation in a 2D system with $n$ particles per unit area, and 
$\kappa$ is the background dielectric constant, we obtain $r_{\rm s}$
to be $r_s \sim n^0$ for MLG and $r_s \sim n^{1/2}$ for BLG and 2DEG.

A note of caution about the nomenclature is in order here,
particularly since we have kept the degeneracy factors $g_{\rm s}
g_{\rm v}$ in the definition of the interaction parameter.  Putting
$g_{\rm s} g_{\rm v}=4$, the usual case for MLG, BLG and Si 100 2DEG,
and $g_{\rm s} g_{\rm v}=2$ for GaAs 2DEG, we get $r_{\rm s} =
e^2/(\kappa \hbar v_{\rm F})$ (MLG), $r_{\rm s} = 2 m e^2/(\kappa
\hbar^2 \sqrt{\pi n})$ (BLG and Si 100 2DEG), and $r_{\rm s} = m
e^2/(\kappa \hbar^2 \sqrt{\pi n})$ (GaAs 2DEG).  The traditional
definition of the Wigner-Seitz radius for a metallic Fermi liquid is
the dimensionless ratio of the average inter-particle separation to
the effective Bohr radius $a_{\rm B} = \kappa \hbar^2 /(m e^2)$.  This
gives for the Wigner-Seitz radius $r_{\rm s}^{\rm WS} = m e^2/(\kappa
\hbar^2 \sqrt{\pi n})$ (2DEG and BLG), which differs from the
definition of the interaction parameter $r_{\rm s}$ by the degeneracy
factor $g_{\rm s} g_{\rm v}/2$.  We emphasize that the Wigner-Seitz
radius from the above definition is meaningless for MLG because the
low-energy linear dispersion implies a zero effective mass (or more
correctly, the concept of an effective mass for MLG does not apply).
For MLG, therefore, an alternative definition, widely used in the
literature defines an ``effective fine structure constant ($\alpha$)''
as the coupling constant $\alpha = e^2/(\kappa \hbar v_{\rm F})$, which
differs from the definition of $r_{\rm s}$ by the factor $\sqrt{g_{\rm
    s} g_{\rm v}}/2$.  Putting $\sqrt{g_{\rm s} g_{\rm v}} = 2$ for MLG
  gives the interaction parameter $r_{\rm s}$ equal to the effective
  fine structure constant $\alpha$, just as setting $g_{\rm s}
    g_{\rm v} = 2$ for GaAs 2DEG gave the interaction parameter equal
    to the Wigner-Seitz radius.  Whether the definition 
of the interaction parameter should or should not contain the degeneracy
factor is a matter of taste, and has been discussed in the literature
in the context of 2D semiconductor 
systems~\cite{kn:dassarma2009b}.

A truly significant aspect of the monolayer graphene interaction 
parameter, which follows directly from its equivalence with the fine
structure constant definition, is that it is a carrier density 
independent constant, unlike $r_{\rm s}$ parameter for the 2DEG (or
BLG), which increases with decreasing carrier density as $n^{-1/2}$.  In 
particular the interaction parameter for MLG is bounded, i.e.,
$0 \leq r_{\rm s} \lesssim 2.2$,
since $1 \leq \kappa \leq \infty$, and as discussed earlier $v_{\rm F}
\approx 10^{8}~{\rm cm/s}$ is set by the carbon hopping parameters and
lattice spacing.  This is in sharp contrast to 2DEG systems where 
$r_{\rm s} \approx 13$ (for electrons in GaAs 
with $n \approx 10^{9}~{\rm cm}^{-2}$) and  $r_{\rm s} \approx 50$ 
(for holes in  GaAs with $n \approx 2 \times~10^{9}~{\rm cm}^{-2}$) have been 
reported \cite{huang2006z,kn:dassarma2005b,kn:manfra2007}.

Monolayer graphene is thus, by comparison, always a 
fairly weakly interacting system, while bilayer graphene could
become a strongly interacting system at low carrier density.  We point out,
however, that the real low density regime in graphene (both MLG and BLG) 
is dominated entirely by disorder in currently available samples, 
and therefore a homogeneous carrier density of 
$n \lesssim 10^{10}~{\rm cm}^{-2}$ ($10^{9}~{\rm cm}^{-2}$) 
is unlikely to be accessible for gated (suspended) samples 
in the near future.   Using the
BLG effective mass $m = 0.03~m_{\rm e}$, we get 
the interaction parameter for
$\mbox{BLG:} \ \ r_{\rm s} \approx 68.5/(\kappa {\sqrt{\tilde n}})$,
where ${\tilde n} = n/10^{10}~{\rm cm}^{-2}$. 
For a comparison, 
the $r_{\rm s}$ parameters for GaAs 2DEG ($\kappa = 13, 
m^* = 0.67~m_{\rm e}$), Si 100 on SiO$_2$ ($\kappa = 7.7, 
m^* = 0.19~m_{\rm e}, g_{\rm v} = 2$) are 
$r_{\rm s} \approx \frac{4}{\sqrt{\tilde n}}$, and
$r_{\rm s} \approx \frac{13}{\sqrt{\tilde n}}$
respectively.

For the case when the substrate is SiO$_2$, $\kappa = (\kappa_{{\rm
    SiO}_2} + 1)/2 \approx 2.5$ 
for MLG and BLG we have $r_{\rm s} \approx 0.8$ and 
$r_{\rm s} \approx 27.4/(\sqrt{\tilde n})$ respecitvely.
In vacuum, $\kappa = 1$ and
$r_{\rm s} \approx 2.2$ for MLG and $r_{\rm s} \approx 68.5/(\sqrt{\tilde n})$
for BLG.

\subsubsection{Thomas-Fermi screening wavevector $q_{\rm TF}$}
Screening properties of an electron gas depend on the density 
of states, $D_0$ at the Fermi level.  The simple Thomas-Fermi 
theory leads to the long wavelength Thomas-Fermi screening
wavevector
\begin{equation}
 q_{\rm TF} = \frac{2 \pi e^2}{\kappa} D_0.
\end{equation}
%
%
%
\noindent The density independence of long wavelength screening in BLG and 
2DEG is the well known consequence of the density of states being 
a constant (independent of energy), whereas the property that 
$q_{\rm TF} \sim k_{\rm F} \sim n^{1/2}$ in MLG, is a direct consequence 
of the MLG density of states being linear in energy.

A key dimensionless quantity determining the charged impurity 
scattering limited transport in electronic materials is 
$q_{\rm s} = q_{\rm TF}/k_{\rm F}$ which controls the dimensionless 
strength of quantum screening.  From Table~\ref{tab:table2} 
we have $q_s \sim n^0$ for MLG and $q_s \sim n^{-1/2}$ for BLG and 2DEG.
%
%
 Using the usual substitutions $g_{\rm s} g_{\rm v}= 4 (2) $ for Si
 100 (GaAs) based 2DEG system, and taking the standard values of $m$
 and $\kappa$ for graphene-SiO$_2$, GaAs-AlGaAs and Si-SiO$_2$
 structures, we get (for ${\tilde n} = n/10^{10}~{\rm cm}^{-2}$)
  \begin{subequations} \label{SDS_Eq_qs}
\begin{eqnarray}
\mbox{MLG:} \  q_{\rm s} &\approx& 3.2, \;\;\;\;\; \mbox{BLG:}  \ q_{\rm s}
\approx {54.8}/{\sqrt{{\tilde n}}}, \\ \mbox{n-GaAs:}  \ q_{\rm s} 
&\approx& {8}/{\sqrt{{\tilde n}}}, \;\; \mbox{p-GaAs:} \  q_{\rm s} 
\approx {43}/{\sqrt{{\tilde n}}}.
\end{eqnarray}
 \end{subequations}  
%
 
 We point out two important features of the simple screening
 considerations described above: {\it (i)} In
 MLG, $q_{\rm s}$ being a constant implies that
 the screened Coulomb interaction has exactly the same behavior as the
 unscreened bare Coulomb interaction.  The bare 2D Coulomb interaction
 in a background with dielectric constant $\kappa$ is given by
 $v(q) = 2 \pi e^2/(\kappa q)$
and the corresponding long-wavelength screened interaction is given by
$u(q) = 2 \pi e^2/(\kappa (q + q_{\rm TF}))$.
Putting $q = k_{\rm F}$ in the above equation, we get, $u(q)
\sim (k_{\rm F} + q_{\rm TF})^{-1} \sim k_{\rm F}^{-1} (1 + q_{\rm
  TF}/k_{\rm F})^{-1} \sim k_{\rm F}^{-1}$ for MLG.  Thus, in MLG, the
functional dependence of the screened Coulomb scattering on the
carrier density is exactly the same as unscreened Coulomb scattering,
a most peculiar phenomenon arising from the Dirac linear dispersion.
{\it (ii)} In BLG (but not MLG, see above) and in 2DEG, the effective
screening becomes stronger as the carrier density decreases since
$q_{\rm s} = q_{\rm TF}/k_{\rm F} \sim n^{-1/2} \rightarrow
\infty~(0)$ as $n \rightarrow 0~(\infty)$.  This counter-intuitive
behavior of 2D screening, which is true for BLG systems also, means
that in 2D systems, effects of Coulomb scattering on transport
properties increases with increasing carrier density, and at very high
density, the system behaves as an unscreened system.  This is
in sharp contrast to 3D metals where the screening effect increases 
monotonically with increasing electron density.

Finally in the context of graphene, it is useful to give a direct
comparison between screening in MLG versus screening in BLG:
$ q_{\rm TF}^{\rm BLG}/q_{\rm TF}^{\rm MLG} \approx 16/\sqrt{{\tilde n}} $,
%
showing that as carrier density decreases, BLG screening becomes much
stronger than SLG screening.

\subsubsection{Plasmons}
Plasmons are self-sustaining normal mode oscillations of a carrier
system, arising from the long-range nature of the inter-particle
Coulomb interaction.  The plasmon modes are defined by the zeros of
the corresponding frequency and wavevector dependent dynamical
dielectric function.  The long wavelength plasma oscillations are
essentially fixed by the particle number (or current) conservation,
and can be obtained from elementary considerations.  We write down the
long-wavelength plasmon dispersion $\omega_{\rm p}$
\begin{subequations} \label{SDS_Eq_plasmons}  
\begin{eqnarray}
 \mbox{MLG:} \ \ \omega_{\rm p} (q \rightarrow 0) &=& \left( \frac{e^2
  v_{\rm F} q}{\kappa \hbar} \sqrt{\pi n g_{\rm s} g_{\rm v}}
\right)^{1/2}, \\ \mbox{BLG \& 2DEG:} \ \ \omega_{\rm p} (q
\rightarrow 0) &=& \left( \frac{2 \pi n e^2}{\kappa m} q
\right)^{1/2}.
\end{eqnarray}
\end{subequations}
A rather intriguing aspect of MLG plasmon dispersion is that it is
non-classical (i.e. $\hbar$ appears explicitly in
Eq.~\ref{SDS_Eq_plasmons}, even in the long wavelength limit).  This
explicit quantum nature of long wavelength MLG plasmon is a direct
manifestation of its linear Dirac like energy-momentum dispersion,
which has no classical analogy~\cite{kn:dassarma2009}.

\subsubsection{Magnetic field effects}
Although magnetic field induced phenomena in graphene and 2D
semiconductors (e.g. quantum Hall effect and fractional quantum Hall
effect) 
are briefly covered in section~\ref{sec:qhe},
we want to mention at this point a few elementary electronic
properties in the presence of an external magnetic field perpendicular
to the 2D plane leading to the Landau orbital quantization of the
system.

\sub3section{Landau Level Energetics}
The application of a strong perpendicular external magnetic field (B)
leads to a complete quantization of the orbital carrier dynamics of
all 2D systems leading to the following quantized energy levels
$E_{\rm n}$, the so-called Landau levels
%
\begin{subequations} \label{SDS_Eq_LandauLevels}  
 \begin{eqnarray} 
 \mbox{MLG:} \ \ E_{\rm n} &=& {\rm sgn}(n) v_{\rm F} \sqrt{2 e \hbar
   B |n|}, \nonumber \\ 
   &&\ \ \mbox{with $n=0, \pm 1, \pm 2, \cdots$}, \\ 
   \mbox{BLG:}
   \ \ E_{\rm n} &=& \frac{{\rm sgn}(n)}{\sqrt{2}} \left[ \frac{}{} (2
   |n| + 1)(2 e B v_{\rm F}^2 \hbar) + 4 m^2 v_{\rm F}^4 \right. \nonumber \\ 
   && \left. \hspace{-2.5cm} - \sqrt{(2 m v_{\rm F} )^4 + 2 (2 |n| + 1)(2 e
   B v_{\rm F}^2 \hbar)( 2 m v_{\rm F}^2 )^2 + (2 e B v_{\rm F}^2
   \hbar)^2} \right], \nonumber \\ 
   && \mbox{with $n=0, \pm 1, \pm 2, \cdots$},\\ 
   \mbox{2DEG:} \ \ E_{\rm n} &=& (n+1/2 )\left( \frac{e B
   \hbar}{mc} \right), \nonumber\\
   && \mbox{with $n=0, 1, 2, \cdots$}.
 \end{eqnarray}
\end{subequations} 
%
 \noindent The hallmark of the Dirac nature of graphene is the
 existence of a true zero-energy ($n=0$ in
 Eq.~\ref{SDS_Eq_LandauLevels}) Landau level, which is equally shared
 by electrons and holes.  The experimental verification of this zero
 energy Landu level in graphene is definitive evidence for
 the long wavelength Dirac nature of the system~\cite{kn:novoselov2005,kn:zhang2005,kn:miller2009}.

\sub3section{Cyclotron Resonance}

External radiation induced transitions between Landau levels gives
rise to the cyclotron resonance in a Landau quantized system, which
has been extensively studied in 2D semiconductor~\cite{kn:ando1982} 
and graphene systems~\cite{henriksen2008,kn:jiang2007,kn:henriksen2009}.
The cyclotron resonance frequency in MLG and 2DEG is given by
\begin{subequations}
\begin{eqnarray}
 \mbox{MLG:} \ \ \omega_{\rm c} &=& v_{\rm F} \sqrt{2 e \hbar B}
 \left( \sqrt{n+1} - \sqrt{n} \right), \\ \mbox{2DEG:} \ \ \omega_{\rm c} &=& \frac{eB}{mc}.
\end{eqnarray}
\end{subequations}
\noindent For BLG, the cyclotron frequency should smoothly interpolate
between the formula for MLG for very large $n$, so that $E_{\rm n}$ in
Eq.~\ref{SDS_Eq_LandauLevels} is much larger than $2 m v_{\rm F}^2$, to
that of the 2DEG for small $n$ so that $E_{\rm n} \ll 2 m v_{\rm
  F}^2$ (where $m \approx 0.033$ is the approximate $B=0$ effective
mass of the bilayer parabolic band dispersion).  Experimental 
BLG cyclotron resonance studies~\cite{kn:henriksen2009} indicate
the crossover from the quadratic band dispersion (i.e. 2DEG-like) 
for smaller $q$ to the linear band dispersion (i.e. MLG-like) 
at larger $q$ seems to happen at lower values of $q$ than that implied
by simple band theory considerations.

A particularly interesting and important feature of cyclotron
resonance in graphene is that it is affected by electron-electron
interaction effects unlike the usual parabolic 2DEG, where the
existence of Kohn's theorem prevents the long wavelength cyclotron
frequency from being renormalized by electron-electron 
interactions~\cite{kn:kohn1961,kn:ando1982}.
For further discussion of this important topic, we refer the reader to
the recent literature on the subject~\cite{kn:henriksen2009,shizuya2009}.

\sub3section{Zeeman splitting}

In graphene, the spin splitting can be large since the
Lande g-factor in graphene is the same ($g=2$) as in vacuum.  The
Zeeman splitting in an external magnetic field is given by ($\mu_{\rm B}$: Bohr magneton)
$E_{\rm z} = g \mu_{\rm B} B = 0.12~B [T]~{\rm meV}$,
for $g=2$ (MLG, BLG, Si 2DEG) and
$E_{\rm z} = - 0.03~B [T]~{\rm meV}$
 for $g = -0.44$ (GaAs 2DEG).
We note that the relative value of $E_{\rm z}/E_{\rm F}$ is 
rather small in graphene, $E_{\rm z}/E_{\rm F} \approx 0.01
(B[T]/\sqrt{\tilde n}) \rightarrow 0.01$ for $B= 10~{\rm T}$ and $n=
10^{12}~{\rm cm}^{-2}$.  Thus the spin splitting is only $1~{\rm
  percent}$ even at high fields.  Of course, the polarization 
effect is stronger at
low carrier densities, since $E_{\rm F}$ is smaller.

%
%

\begin{center}
\begin{table*}[t]
\caption{\label{tab:table1}
Electronic quantities for monolayer graphene.
Note: The graphene Fermi velocity ($v_F=10^8$ cm/s) and the degeneracy factor
$g=g_sg_v=4$, i.e. the usual spin degeneracy ($g_s=2$) and a valley degeneracy
($g_v=2$), are used in this table.  Here 
$\tilde{n} =n/(10^{10}~{\rm cm}^{-2})$, and $B$,
$q$, and $\sigma$ are measured in T, cm$^{-1}$, and
$e^2/h = 38.74$ $\mu S$ (or $h/e^2=25.8$ k$\Omega$) respectively.} 
\hspace{0.1\hsize}
\begin{ruledtabular}
\begin{tabular}{cc}
Quantity & Scale values \\ \hline Fermi wave vector ($k_F$) &
$1.77\times10^5 \sqrt{\tilde{n}}$ \; [cm$^{-1}$] \\ Thomas Fermi wave vector
($q_{TF}$) & $ 1.55\times 10^6 \sqrt{\tilde{n}}/\kappa $ \;
[cm$^{-1}$]\\ Interaction parameter ($r_s$) & 2.19/$\kappa$ \\ DOS at
$E_F$ ($D_0\equiv D(E_F)$) & $1.71\times 10^9 \sqrt{\tilde{n}}$ \;
[meV$^{-1}$cm$^{-2}$] \\ Fermi energy ($E_F$) & $11.65\sqrt{\tilde{n}}
$ \; [meV] \\ Zeeman splitting ($E_z$) & $ 0.12 B$ \; [meV]
\\ Cyclotron frequency ($\omega_c$) & $ 5.51\times 10^{13}\sqrt{B}$ \;
   [$s^{-1}$]\\ Landau level energy ($E_n$) & sgn($l$)
   $36.29\sqrt{B|l|}$ \; [meV], $l=0$, $\pm 1$, $\pm 2$ ...\\ Plasma
   frequency ($\omega_p(q)$) & $5.80\times 10^{-2}
   \sqrt{\sqrt{\tilde{n}}q/\kappa}$ \; [meV] \\ Mobility ($\mu$) &
   $2.42 \times 10^4 \; \sigma/\tilde{n}$ \; [cm$^2$/Vs] \\ Scattering
   time ($\tau$) & $2.83 \times 10^{-14} \; \sigma/\sqrt{\tilde{n}}$
   \; [s] \\ Level broadening ($\Gamma$) & $11.63
   \sqrt{\tilde{n}}/\sigma$ \; [meV]
\\
\end{tabular}
\end{ruledtabular}
\hspace{0.1\hsize}
\end{table*}
\end{center}

\subsection{Intrinsic and extrinsic graphene} \label{subsec_intrinsic}

It is important to distinguish between intrinsic and
extrinsic graphene because gapless graphene (either MLG or BLG) has a
charge neutrality point (CNP), i.e. the Dirac point, where its
character changes from being electron-like to being hole-like.  Such a
distinction is not meaningful for a 2DEG (or BLG with a
large gap) since the intrinsic system is simply an undoped system with
no carriers (and as such is uninteresting from the
electronic transport properties perspective).

In monolayer and bilayer graphene, the ability to gate (or dope) the
system by putting carriers into the conduction or valence band by
tuning an external gate voltage enables one to pass through the CNP
where the chemical potential ($E_{\rm F}$) resides precisely at the
Dirac point.  This system, with no free carriers at $T=0$, and $E_{\rm F}$
precisely at the Dirac point is called intrinsic graphene with a
completely filled (empty) valence (conduction) band.  Any
infinitesimal doping (or, for that matter, any finite temperature)
makes the system ``extrinsic'' with electrons (holes) present in the
conduction (valence) band.  \cite{kn:muller2009}. 
Although the intrinsic system is a set of
measure zero (since $E_{\rm F}$ has to be precisely at the Dirac
point), the routine experimental ability to tune the system from being
electron-like to to being hole-like by changing the external gate
voltage, manifestly establishes that one must be going through the
intrinsic system at the CNP.
If there is an insulating
regime in between, as there would be for a gapped system, then 
intrinsic graphene is not being accessed.

Although it is not often emphasized, the great achievement of
\citet{kn:novoselov2004} in producing 2D graphene 
in the laboratory is not just fabricating 
\cite{kn:novoselov2005b}
and identifying
\cite{kn:ferrari2006,ferrari2007z}
stable monolayers of graphene flakes on substrates, but also
establishing its transport properties by gating the graphene device
using an external gate which allows one to simply tune an
external gate voltage and thereby continuously controlling the 2D graphene carrier
density as well as their nature (electron or hole).  If all that could
be done in the laboratory was to produce beautiful 2D graphene flakes,
with no hope of doping or gating them with carriers, the subject of
graphene would be many orders of magnitude smaller and less
interesting.  What led to the exponential growth in graphene
literature is the discovery of gatable and density tunable 2D graphene
in 2004.

Taking into account the quantum capacitance in graphene the doping
induced by the external gate voltage $V_g$ is given by the relation \cite{fang2007,fernandez-rossier07}:
\beq
 n = \frac{CV_g}{e} + n_Q\left[1 - \sqrt{1 + \frac{CV_g}{e n_Q}}\right];
 \label{eq:ln:qcap}
\enq 
where $C$ is the gate capacitance, 
 $e$ the absolute value of the electron charge and
$n_Q \df \frac{\pi}{2}\left(\frac{C\hbar v_F}{e^2}\right)^2$.
The second term on the r.h.s. 
of \ceq{eq:ln:qcap} is analogous to the term due to the so-called quantum
capacitance in regular 2DEG. Notice that in graphene, due to the linear
dispersion, contrary to parabolic 2D electron liquids, 
the {\em quantum capacitance} depends on $V_g$.
For a background dielectric constant $\kappa \approx 4$ 
and gate voltages larger than few millivolts
the second term on the r.h.s.
of \ceq{eq:ln:qcap} 
can be neglected for thicknesses of the dielectric larger than
few angstroms. In current experiments on exfoliated graphene
on $\siot$ the oxide is 300~nm thick and therefore quantum-capacitance
effects are completely negligible. 
In this case, 
a simple capacitance model connects the 2D carrier density ($n$) with
the applied external gate voltage $V_{\rm g}$
$n \approx C V_{\rm g}$,
where $C \approx 7.2\times 10^{10}~{\rm cm}^{-2}/{\rm V}$ for graphene 
on SiO$_2$ with roughly $300~{\rm nm}$ thickness.  This approximate 
value of the constant $C$ seems to be pretty accurate, and the following
scaling should provide $n$ 
for different dielectrics 
\begin{equation}
n~[ 10^{10}~{\rm cm}^{-2}] = 7.2 \times \frac{ t~[{\rm nm}]}{300} \frac{\kappa}{3.9}
V_{\rm g}~[V],
\end{equation}    
where $t$ is the thickness of the dielectric (i.e. the distance from the gate
to the graphene layer) and $\kappa$ is the dielectric constant of 
the insulating substrate.

It is best, therefore, to think of 2D graphene on SiO$_2$ (see
Fig. 1(e)) as a
metal-oxide-graphene-field-effect-transistor (MOGFET) similar to the
well known Si MOSFET structure, with Si replaced by graphene where the
carriers reside.  In fact, this analogy between graphene and Si 100
inversion layer is operationally quite effective: Both have the
degeneracy factor $g_{\rm s} g_{\rm v} = 4$ and both typically have
SiO$_2$ as the gate oxide layer.  The qualitative and crucial difference
is, of course, that graphene carriers are chiral, massless, with
linear dispersion and with no band gap, so that the gate allows one to
go directly from being $n$ type to a $p$ type carrier system through
the charge neutral Dirac point.  Thus a graphene MOGFET is not a
transistor at all (at least for MLG), since the system never becomes
insulating at any gate voltage \cite{avouris2007}.

We will distinguish between extrinsic
(i.e. doped) graphene with free carriers and intrinsic (i.e. undoped)
graphene with the chemical potential precisely at the Dirac point.
All experimental systems (since they are always at $T\neq 0$) are
necessarily extrinsic, but intrinsic graphene is of theoretical
importance since it is a critical point. In particular, intrinsic
graphene is a non-Fermi liquid in the presence of electron-electron
interactions~\cite{kn:dassarma2007b}, whereas 
extrinsic graphene is a Fermi liquid.  Since the
non-Fermi liquid fixed point for intrinsic graphene is unstable to the
presence of any finite carrier density, the non-Fermi liquid nature of
this fixed point is unlikely to have any experimental
implication. But it is important to keep this non-Fermi liquid nature
of intrinsic graphene in mind when discussing graphene's electronic
properties.  We also mention (see Sec.~\ref{sec_low}) that disorder,
particularly long-ranged disorder induced by random charged 
impurities present in the environment, is a relevant strong
perturbation affecting the critical Dirac point, since the
system breaks up into spatially random electron-hole puddles, thus
masking its zero-density intrinsic nature.

\subsection{``Other topics''}  \label{subsec_other}

There are several topics that are of active current research that we
could not cover in any depth in this review article.  Some of these
remain controversial and others are still poorly understood.  Yet
these subjects are of importance, both in terms of
fundamental physics and for the application of graphene for useful
devices.  Here we sketch the status of these important subjects at
the time of writing this article.  For example, there have recently emerged several novel 
methods of fabricating graphene including chemical vapor deposition on 
nickel~\cite{ISI:000263064700037} and 
copper~\cite{XuesongLi06052009}, as well as directly unzipping carbon
nanotubes~\citep{ISI:000265182500039,sinitskii-apl-95-253108-2009} 
and other chemical 
methods~\citep{jiao-n-458-877-2009}.  As of
early 2010, all of these other fabrication processes are just in their
infancy.  The notable exception is ``epitaxial graphene'' manufactured
by heating SiC wafers, causing the Si atoms to desorb, resulting in
several graphene layers at the surface~\citep{ISI:000225925100006,kn:berger2006,kn:first2010,kn:deheer2010,kn:emtsev2008}
that are believed to be very weakly coupled and of very 
good quality ~\citep{kn:hass2008,kn:rutter2007,kn:miller2009,PhysRevLett.101.267601}.  We note that graphene can be used as a component of more complicated structures by exploiting
its spin~\citet{TombrosNature,1704408,cho:123105,PhysRevB.74.155426,PhysRevLett.103.146801,PhysRevLett.102.137205,PhysRevB.80.241403}  or valley~\citet{rycerz2007b} degeneracy
or by patterning gates with a periodic super-potential~\cite{ParkNaturePhysics,PhysRevLett.103.046809}.  Graphene can also be made to superconduct by
coupling it to superconducting leads and through the 
proximity effect~\cite{PhysRevLett.97.067007,heersche-n-446-56-2007,PhysRevB.77.184507,RevModPhys.80.1337} or other novel proposals~\cite{PhysRevLett.101.106402,ISI:000263786700010}.  This review could not cover any of these topics in any reasonable depth.

 \subsubsection{Optical conductivity}

It was pointed out as early as 1994 by \citet{kn:ludwig1994} that if one examined the conductivity of Dirac Fermions in linear response theory, keeping a finite frequency, i.e. $\sigma(\omega)$  while
taking the limit of zero temperature ($T \rightarrow 0$) and vanishing disorder ($\Gamma \rightarrow 0$),
then one obtained a universal and frequency independent optical conductivity 
(i.e. electrical conductivity at finite frequency)
\begin{subequations}
\label{Eq:Intro_cond}
\begin{equation}
\label{Eq:Intro_optcond}
\sigma(\omega) = g_s g_v \frac{\pi e^2}{8 h}. 
\end{equation}
\noindent \citet{kn:ludwig1994} also noted that this result did not commute with the d.c. conductivity where one first took the limit $\omega \rightarrow 0$ and then $\Gamma \rightarrow 0$, in which case one obtained
\begin{equation}
\sigma_{\rm min} = g_s g_v \frac{e^2}{\pi h}.
\end{equation}
\end{subequations}
\noindent  These $T=0$ results apply to intrinsic graphene 
where $E_{\rm F}$ is precisely at the Dirac point. The crossover between 
these two theoretical intrinsic limits remains an open problem~\cite{kn:ostrovsky2006,kn:katsnelson2006}.

The optical conductivity (Eq.~\ref{Eq:Intro_optcond}) has been
measured experimentally both by infrared
spectroscopy~\cite{kn:li2008b} and by measuring the absorption of
suspended graphene sheets~\cite{kn:nair2008}.  In the IR measurements,
$\sigma(\omega)$ is close to the predicted universal
value for a range of frequencies $ 4000~{\rm cm}^{-1}< \omega <
6500~{\rm cm}^{-1}$.  While in the absorption experiment, the
attenuation of visible light through multilayer graphene scales as $\pi \alpha$ 
per layer.  The authors claimed that this was an accurate measurement
of the fine structure constant $\alpha$, and is a direct consequence
of having $\sigma(\omega)$ being a universal and frequency independent
constant.  In some sense, it is quite remarkable that disorder and
electron-electron interactions do not significantly alter the value of
the optical conductivity.  This has attracted considerable theoretical
interest~\cite{gusynin-prb-73-245411-2006, mishchenko2007,
  kn:kuzmenko2008, peres-e-84-38002-2008,
  peres-ijmpb-22-2529-2008, kn:stauber2008b, kn:katsnelson2008b, kn:min2009,
  kn:mishchenko2009, kn:herbut2008, kn:sheehy2009},
where it has been argued that it is a fortuitous cancellation of
higher order terms that explains the insensitivity of $\sigma(\omega)$
to interaction effects.  We refer the reader to these works for
detailed discussion of how interaction effects and disorder change
$\sigma(\omega)$ from the universal value, although, a
consensus is yet to emerge on whether these effects could be observed
experimentally or how accurate $\sigma(\omega)$ is for a measure of
the fine structure constant~\cite{kn:mak2008,kn:gusynin2009}.

\subsubsection{Graphene nanoribbons}

It was realized in the very first graphene transport experiments that
the finite minimum conductivity (Eq.~\ref{Eq:Intro_cond}) would be an
obstacle for making a 
useful transistor since there is no ``off" state.  One way to
circumvent this problem is to have a quasi-1D geometry that
confines the graphene electrons in a strip of (large) length $L$ and a
finite (small) width $W$.  The confinement gap typically scales as
$1/W$~\cite{kn:wakabayashi1999}, however this depends on the imposed
boundary conditions.  This is quite similar to carbon nanotubes 
(since a nanotube is just a nanoribbon with
periodic boundary conditions).  The nomenclature in graphene is
slightly different from carbon nanotubes, where a zigzag-edge nanoribbon is
similar to an armchair nanotube in that it is always metallic within
the tight-binding approximation.  Similarly, an armchair nanoribbon is
similar to a zigzag nanotube in that it can be either metallic or
semiconducting depending on the width.  Early theoretical
calculations~\cite{kn:son2006,son2006,kn:yang2007} used a density functional
theory to calculate the bandgap of armchair graphene nanoribbons and
found that just like carbon nanotubes, the energy gaps come in three
families that are all semiconducting (unlike the tight-binding
calculation, which gives one of the families as metallic).
\citet{kn:brey2006} showed that simply quantizing the Dirac
Hamiltonian (the low energy effective theory) gave quantitatively
similar results for the energy gaps as the tightbinding calculation, while
\citet{kn:son2006} showed that the density functional results
could be obtained from the tightbinding model with some added edge disorder.
By considering arbitrary boundary conditions,
\citet{kn:akhmerov2008} demonstrated that the behavior of the
zigzag edge is the most generic for graphene nanoribbons.  These
theoretical works gave a simple way to understand the gap in graphene
nanoribbons.

\begin{figure}
\hspace{0.1\hsize}
\begin{center}
 \includegraphics[width=1.0\columnwidth]{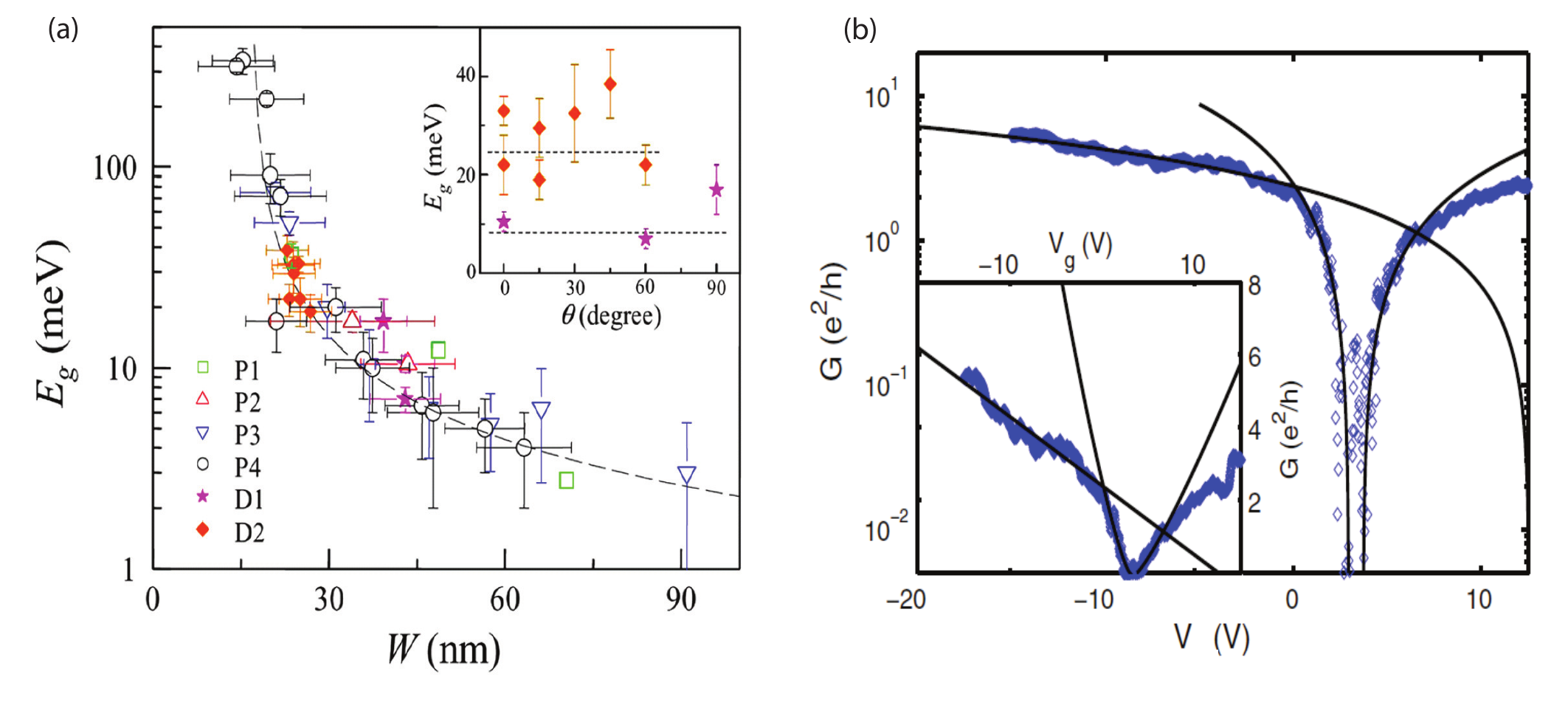} 
\end{center}
\caption{\label{SA_Fig:GNR} (Color online)
(a) Graphene nanoribbon energy gaps as a
function of width, adapted from \citet{kn:han2007}.  Four devices
(P1-P4) were orientated parallel to each other with varying width,
while two devices (D1-D2) were oriented along different
crystallographic directions with uniform width.  The dashed line is a
fit to a phenomenological model with $E_g = A/(W-W*)$ where $A$ and $W*$ are
fit parameters.  The inset shows that contrary to predictions, the
energy gaps have no dependence on crystallographic direction.  The
dashed lines are the same fits as in the main panel.  (b) Evidence for
a percolation metal-insulator transition in graphene nanoribbons,
adapted from \citet{kn:adam2008c}.  Main panel shows graphene ribbon
conductance as a function of gate voltage.  Solid lines are a fit to
percolation theory, where electrons and holes have different
percolation thresholds  (seen as separate critical gate voltages $V_c$).
The inset shows the same data in a linear scale, where even by eye the
transition from high-density Boltzmann behavior to the low-density
percolation transport is visible. }
\end{figure}

However, the first experiments on graphene
nanoribbons~\cite{kn:han2007} presented quite unexpected results.  As
shown in Fig.~\ref{SA_Fig:GNR} the transport gap for narrow ribbons is much
larger than that predicted by theory (with the gap diverging
at widths $\approx 15~{\rm nm}$), while wider ribbons have a much
smaller gap than expected.  Surprisingly, the gap showed no dependence
on the orientation (i.e. zig-zag or armchair direction) as required by
the theory.  These discrepancies 
have prompted several 
studies~\cite{kn:sols2007,kn:martin2007b,kn:adam2008c,abanin-prb-78-035416-2008,areshkin-nl-7-204-2007,basu-apl-92-042114-2008,biel-nl-9-2725-2009,biel-prl-102-096803-2009,chen-pesn-40-228-2007,stampfer-prl-102-056403-2009,todd-nl-9-416-2009,dietl2009}.      
In particular, \citet{kn:sols2007}
argued that fabrication of the nanoribbons gave rise to very rough
edges breaking the nanoribbon into a series of quantum dots. Coulomb
blockade of charge transfer between the dots
\cite{ponomarenko2008z}
explains the larger gaps
for smaller ribbon widths.  In a similar spirit,
\citet{kn:martin2007b} showed that edge disorder qualitatively
changed the picture from that of the disorder-free picture presented
earlier, giving a localization length comparable to the sample width.
For larger ribbons, \citet{kn:adam2008c} argued that charged
impurities in the vicinity of the graphene would give rise to
inhomogeneous puddles so that the transport would be governed by
percolation (as shown in Fig.~\ref{SA_Fig:GNR}, points are
experimental data, and the solid lines, for both electrons and holes,
show fits to $\sigma \sim (V-V_c)^\nu$, where $\nu$ is close to $4/3$,
the theoretically expected value for percolation in 2D systems).  The
large gap for small ribbon widths would then be explained by a
dimensional crossover as the ribbon width became comparable to the
puddle size.  A numerical study including the effect of quantum
localization and edge disorder was done by
\citet{kn:mucciolo2008} who found that a few atomic layers of
edge roughness was sufficient to induce transport gaps appear that are
approximately inversely proportional to the nanoribbon width.  Two
very recent and detailed
experiments~\cite{kn:han2010,kn:gallagher2010} seem to suggest that a
combination of these pictures might be at play 
(e.g. transport trough quantum dots that
are created by the charged impurity potential), although as yet, a
complete theoretical understanding remains elusive.  The phenomenon
of the measured transport gap being much smaller 
than the theoretical bandgap seems to be a generic feature
in graphene, occuring not only in nanoribbons but also in biased
bilayer graphene where the gap measured in transport experiments
appears to be substantially smaller than the theoretically 
calculated bandgap~\cite{kn:oostinga2007} or even the measured 
optical gap~\cite{zhang-n-2009,kn:mak2009}.

\subsubsection{Suspended graphene}

Since the substrate affects both the morphology of
graphene~\cite{kn:meyer2007,kn:ishigami2007,kn:stolyarova2007} as well
as provides a source of impurities, it became clear that one needed to
find a way to have electrically contacted graphene without the
presence of the underlying substrate.  The making of ``suspended
graphene" or ``substrate-free" graphene was an important experimental
milestone~\cite{kn:bolotin2008, kn:du2008,kn:bolotin2008b} where after
exfoliating graphene and making electrical contact, one then etches
away the substrate underneath the graphene so that the graphene is
suspended over a trench that is approximately $100~{\rm nm}$ deep.  
As a historical note,
we mention that suspended graphene without electrical contacts was
made earlier by \citet{kn:meyer2007}.  Quite surprisingly, the
suspended samples as prepared did not show much difference from
unsuspended graphene, until after current
annealing~\cite{moser-apl-91-163513-2007,barreiro-prl-103-076601-2009}.
This suggested that most of impurities limiting the transport
properties of graphene were stuck to the graphene sheet and not buried
in the substrate.  After removing these impurities by driving a large
current through the sheet, the suspended graphene samples showed both
ballistic and diffusive carrier transport properties. 
Away from the charge
neutrality point, suspended graphene showed near-ballistic transport
over hundreds of ${\rm nm}$, which prompted much theoretical
interest~\cite{kn:adam2008b,kn:stauber2008,kn:fogler2008c,kn:muller2009}.
One problem with suspended graphene is that only a small gate voltage
($V_g \approx 5~{\rm V}$) could be applied before the graphene buckles
due to the electrostatic attraction between the charges in the gate
and on the graphene sheet, and binds to the bottom of the trench that
was etched out of the substrate. This is in contrast to graphene on a
substrate that can support as much as $V_g \approx 100~{\rm V}$ and a
corresponding carrier density of $\approx 10^{13}~{\rm cm}^{-2}$.  To
avoid the warping, it was proposed that one should use a top gate with
the opposite polarity, but at the time of writing, this has yet to be
demonstrated experimentally.  Despite the limited variation in carrier
density, suspended graphene has achieved a carrier mobility of more
than $200,000~{\rm cm}^2 /{\rm
  Vs}$~\cite{kn:bolotin2008,kn:bolotin2008b,kn:du2008}.  Very
recently, suspended graphene bilayers were demonstrated
experimentally~\cite{kn:feldman2009}.

\subsubsection{Many-body effects in graphene}

The topic of many-body effects in graphene is itself a large subject,
and one that we could not cover in this transport review.  As already
discussed earlier, for intrinsic graphene, the many-body ground state
is not even a Fermi liquid~\cite{kn:dassarma2007b}, an indication of
the strong role played by interaction effects.  Experimentally, one
can observe the signature of many body effects in the
compressibility~\cite{kn:martin2008} and using
ARPES~\cite{kn:bostwick2007,zhou2007}.  Away from the Dirac point,
where graphene behaves as a normal Fermi liquid, the calculation of
the electron-electron and electron-phonon contribution to the
quasiparticle self-energy was studied by several
groups~\cite{PhysRevB.77.081412,PhysRevLett.102.076803,PhysRevB.76.205411,PhysRevLett.99.086804,PhysRevLett.99.236802,kn:hwang2007b,kn:hwang2007,kn:barlas2007,kn:polini2007,PhysRevB.77.081411,PhysRevB.81.045419},
and show reasonable agreement with
experiments~\cite{kn:bostwick2007,kn:brar2010}.  For both bilayer
graphene~\cite{kn:min2008b} and for double layer
graphene~\cite{PhysRevB.78.121401} an instability towards an exitonic
condensate has been proposed.  In general, monolayer graphene is a
weakly interacting system since the coupling constant ($r_s \leq 2$)
is never large \cite{muller2009z}. In principle, bilayer graphene
could have arbitrarily large coupling at low carrier density where
disorder effects are also important.  We refer the reader to these
original works for details on this vast and interesting subject.

\subsubsection{Topological insulators}

There is a deep connection between graphene and topological 
insulators \cite{kane2005,sinitsyn2006}.  
Graphene has a Dirac cone where the
``spin" degree of freedom is actually related to the sublattices in
real space, whereas it is the real electron spin that provides the Dirac
structure in the topological insulators~\cite{kn:hasan2010} on
the surface of BiSb and BiTe~\cite{ISI:000255208600034,Chen07102009}.  
Graphene is a weak topological insulator because it has
two Dirac cones (by contrast, a strong topological insulator is
characterized by a single Dirac cone on each surface); but in
practice,the two cones in graphene are mostly decoupled and it behaves like
two copies of a single Dirac cone. Therefore many of the results
presented in this review, although intended for graphene, should also
be relevant for the single Dirac cone on the surface of
a topological insulator.  In particular, we expect the interface 
transport properties of topological insulators to be similar to the physics
described in this review as long as the bulk is a true gapped insulator.

\subsection{2D nature of graphene}  \label{subsec_2D}

As the concluding section of the Introduction,
we ask: what
precisely is meant when an electronic system is categorized as 2D and
how can one ensure that a specific sample/system is 2D from the
perspective of electronic transport phenomena.

The question is not simply academic, since 2D does not necessarily
mean a thin film (unless the film is literally one atomic monolayer
thick as in graphene, and even then, one must consider the possibility
of the electronic wavefunction extending somewhat into the third
direction).  Also the definition of what constitutes a 2D may depend
on the physical properties or phenomena that one is considering.  For
example, for the purpose of quantum localization phenomena, the system
dimensionality is determined by the width of the system being smaller
than the phase coherence length $L_\phi$ (or the Thouless length).
Since $L_\phi$ could be very large at low temperature,
metal films and wires can respectively be considered 2D and 1D for
localization studies at ultra-low temperature.
For our purpose, however,
dimensionality is defined by the 3D electronic wavefunction being
``free'' plane-wave like (i.e. carrying a conserved 2D wavevector) in
a 2D plane, while 
being a quantized bound state in the third dimension.  This ensures
that the system is quantum mechanically 2D.

Considering a thin film
of infinite (i.e. very large) dimension in the $xy$-plane, and a finite
thickness $w$ in the $z$-direction, where $w$ could be the typical
 confinement width of a potential well creating the film,
the system  is considered 2D if  $\lambda_{\rm F} = \frac{2 \pi}{k_{\rm F}} > w$,
For graphene, we have
 ${\lambda_{\rm F}} \approx (350/\sqrt{{\tilde  n}})~{\rm nm}$,
 where ${\tilde  n} =n/( 10^{10}~{\rm cm}^{-2})$, and since
 $w \approx 0.1~{\rm nm}$ to $0.2~{\rm nm}$ (the monolayer atomic
 thickness), the condition $\lambda_{\rm F} \gg w$ is always
 satisfied, even for unphysically large
 $n = 10^{14}~{\rm cm}^{-2}$.

Conversely, it is essentially impossible to create 2D electronic
systems from thin metal films since the very high electron density of
metals, provides $\lambda_{\rm F}\approx 0.1~{\rm nm}$, so that even
for a thickness $w \approx 1~{\rm nm}$ (the thinnest metal film that
one can make), $\lambda_{\rm F} < w$, making them effectively 3D.  By
virtue of the much lower carrier densities in semiconductors, the
condition $\lambda_{\rm F} > w$ can be easily satisfied for $w =
5~{\rm nm}$ to $50~{\rm nm}$ for $n=10^{9}~{\rm cm}^{-2}$ to
$10^{12}~{\rm cm}^{-2}$, making it possible for 2D semiconductor
systems to be readily available since confinement potentials with
width $\approx 10~{\rm nm}$ can be implemented by external gate
voltage or band structure engineering.

We now briefly address the question of the experimental verification
of the 2D nature of a particular system or sample.  The classic technique
is to show that the orbital electronic dynamics is sensitive only
to a magnetic field perpendicular to the 2D plane (i.e. $B_z$)
 (Practically, there could be complications
if the spin properties of the system affect the relevant dynamics, since
the Zeeman splitting is proportional to the total magnetic field).  Therefore,
if either the magnetoresistance oscillations
(Shubnikov-de Hass effect) or cyclotron resonance properties
depend only on $B_z$, then the 2D nature is established.

 Both of these are true in graphene.  The most definitive evidence for
 2D nature, however, is the observation of the quantum Hall effect,
 which is a quintessentially 2D phenomenon.  Any system manifesting an
 unambiguous quantized Hall plateau is 2D in nature, and therefore the
 observation of the quantum Hall effect in graphene in 2005 by
 \citet{kn:novoselov2005} and \citet{kn:zhang2005}
absolutely clinched its 2D nature.  In fact, the quantum Hall effect
in graphene persists to room temperature~\cite{kn:novoselov2007},
indicating that graphene remains a strict 2D electronic material even
at room temperature.
                 
Finally, we remark on the strict 2D nature of graphene from a
structural viewpoint.  The existence of finite 2D flakes of graphene with
crystalline order at finite temperature does not in any way violate
the Hohenberg-Mermin-Wagner-Coleman theorem which rules out the
breaking of a continuous symmetry in two dimensions.  This is because
the theorem only asserts a slow power law decay of the crystalline
(i.e. positional order) correlation with distance, and hence, very
large flat 2D crystalline flakes of graphene (or for that matter, of
any material) are manifestly allowed by this theorem. In fact, a 2D
Wigner crystal, i.e. a 2D hexagonal classical crystal of electrons in
a very low-density limit, was experimentally observed more than thirty
years ago \cite{grimes1979} on the surface of liquid He4
(where the electrons were bound
by their image force). A simple back of the envelope calculation shows
that the size of the graphene flake has to be unphysically large
for this theorem to have any effect on its
crystalline nature \cite{kn:thompson2009}.
There is nothing mysterious or
remarkable about having finite 2D crystals with quasi-long-range
positional order at finite temperatures, which is what we have in 2D
graphene flakes.

\setcounter{sub3section}{0}


\section{Quantum transport}
\subsection{Introduction}
\label{SA_Sec_Intro}
The phrase ``quantum transport'' usually refers to the charge current
induced in an electron gas in response to a vanishing external
electric field in the regime where quantum interference effects are
important~\cite{kn:rammer1998,kn:akkermans2007}.  This is
relevant at low temperatures where the electrons are coherent
and interference effects are not washed out by dephasing.  Theoretically, this
corresponds to the systematic application of diagrammatic perturbation
theory or field theoretic techniques to study how quantum interference changes
the conductivity.  For diffusive transport in two dimensions
(including graphene), to lowest order in this perturbation theory,
interference can be neglected, and one recovers the Einstein
relation $\sigma_0 = e^2 D(E_{\rm F}) {\mathcal D}$, where $D(E_{\rm
  F})$ is the density of states at $E_F$ and ${\mathcal D} = v_{\rm F}^2
\tau/2$ is the diffusion constant.  This corresponds to the classical
motion of electrons in a diffusive random walk scattering
independently off the different impurities.

Since the impurity potential is typically calculated using the 
quantum-mechanical Born approximation, this leading order contribution to the
electrical conductivity is known as the semi-classical transport
theory and is the main subject of Sec.~\ref{subsec_boltzmann} below.
Higher orders in perturbation theory give ``quantum corrections'' to
this semi-classical result i.e. $\sigma = \sigma_0 + \delta \sigma$,
where $\delta \sigma << \sigma$.  In some cases these corrections can
be divergent; a result that simultaneously implies a formal breakdown
of the perturbation theory itself, while suggesting a phase transition
to a non-perturbatively accessible ground state.

For example, it is widely accepted that quantum interference between
forward and backward electron trajectories is the microscopic
mechanism responsible for the Anderson metal-insulator
transition~\cite{kn:abrahams1979}.  For this reason the leading
quantum correction to the conductivity is called ``weak localization''
and is interpreted as the precursor to Anderson localization.

Weak localization is measured experimentally by using a magnetic field
to break the symmetry between the forward and backward trajectories
causing a change in the resistance.  In this case the zero
field conductivity $\sigma(B=0)=\sigma_0 + \delta \sigma$ includes the
quantum corrections while $\sigma(B>B^*)=\sigma_0$ has only the
semiclassical contribution; ($B^*$ is the approximately the magnetic
field necessary to thread the area of the sample with one flux
quantum.)

The second hallmark of quantum transport is mesoscopic conductance
fluctuations.  If one performed the low temperature magnetotransport
measurement discussed above, one would notice fluctuations in the
magnetoresistance that would look like random noise.  However, unlike
noise, these traces are reproducible and are called
magneto-fingerprints.  These magneto-fingerprints depend on the
positions of the random impurities as seen by the electrons. Annealing
the sample relocates the impurities and changes the fingerprint.  The
remarkable feature of these conductance fluctuations is that their
magnitude is universal (depending only on the global symmetry of the
system) and notwithstanding the caveats discussed below, are
completely independent of any microscopic parameters such as material
properties or type of disorder.

While the general theory for weak localization and universal
conductance fluctuations is now well established~\cite{kn:lee1985a},
in Section~\ref{Sec:WL_UCF} we discuss its application to graphene.

The discussion so far has concerned diffusive transport, in 
what follows, we also
consider the ballistic properties of non-interacting electrons in
graphene.  Early studies on the quantum mechanical properties of the
Dirac Hamiltonian revealed a peculiar feature -- Dirac carriers could
not be confined by electrostatic potentials~\cite{kn:klein1929}.

An electron facing such a barrier would transmute into a hole and
propagate through the barrier.  In Section~\ref{Sec:KT_LC} we study
this Klein tunneling of Dirac carriers and discuss how this formalism
can be used to obtain graphene's ballistic universal minimum
conductivity.  There is no analog of this type of quantum-limited
transport regime in two dimensional semiconductors.  The ``metallic
nature'' of graphene gives rise to several interesting and unique
properties that we explore in this section including the absence of
Anderson localization for Dirac electrons and a metal-insulator
transition induced by atomically sharp disorder (such as
dislocations). We note that many of the results in this section can be
also obtained using field-theoretic
methods~\cite{kn:fradkin1986,kn:ludwig1994,kn:ostrovsky2006,kn:schuessler2008,kn:altland2006,kn:fritz2008,ryu-prl-99-116601-2007}.

\subsection{Ballistic transport}
\label{Sec:KT_LC}

\subsubsection{Klein tunneling}


In classical mechanics, a potential barrier whose height is greater
than the energy of a particle will confine that particle.  In quantum
mechanics, the notion of quantum tunneling describes the process
whereby the wavefunction of a non-relativistic particle can leak out
into the classically forbidden region.  However, the transmission
through such a potential barrier decreases exponentially with the
height and width of the barrier.  For Dirac particles, the
transmission probability depends only weakly on the barrier height,
approaching unity with increasing barrier
height~\cite{kn:katsnelson2006b}.  One can understand this effect by
realizing that the Dirac Hamiltonian allows for both positive energy
states (called electrons) and negative energy states (called holes).
Whereas a positive potential barrier is repulsive for electrons, it is
attractive for holes (and vice versa).  For any potential barrier one
needs to match the electron states outside the barrier with the hole
states inside the barrier.  And since the larger the barrier, the greater
the mode matching between electron and hole states, the greater the 
transmission.  For an infinite barrier, the transmission becomes perfect.
This is called Klein tunneling~\cite{kn:klein1929}.

By solving the transmission and reflection coefficients for both the
graphene p-n junction~\cite{kn:cheianov2006b,low-prb-80-155406-2009}
and the p-n-p junction~\cite{kn:katsnelson2006b}, it was found that
for graphene the transmission at an angle normal to the barrier 
was always perfect (although there could be
some reflection at other angles).  This can be understood in terms of
pseudospin conservation. At normal incidence, the incoming electron
state and the reflected electron state are of opposite chirality
resulting in vanishing probability for reflection.

At finite angles of incidence, the transmission depends on how sharp
the barrier is.  In the limit of a perfectly sharp step, the
transmission probability is determined only by pseudospin conservation
and given by $T_{\rm step}(\theta) = \cos^2\theta$.  For a smooth
variation in the electrostatic potential that defines the p-n junction
(characterized by a length-scale $\xi$), the transmission probability
was shown by \citet{kn:cheianov2006b} to be $T_{\xi}(\theta) =
\exp[-\pi (k_{\rm F} \xi) \sin^2\theta)]$.  This implies that for
both sharp and smooth potential barriers, a wavepacket of Dirac Fermions 
will collimate in a direction
perpendicular to the p-n junction.  One can estimate the 
conductance of a single p-n junction (of width $W$) to be
\begin{equation}
G_{p-n} = \frac{4 e^2}{h} (k_{\rm F} W) \int \frac{ d \theta}{2 \pi} 
T_\xi(\theta)
\stackrel{k_{\rm F} \xi \gg 1}{\longrightarrow} \frac{2 e^2}{\pi h} \sqrt{\frac{k_{\rm F}}{\xi}} W.
\end{equation}
\noindent Although the conductance of smooth p-n junctions are smaller 
by a factor of $\sqrt{k_{\rm F} \xi}$ compared to sharp ones, this 
result suggests that the presence of p-n junctions would make a 
small contribution to overall resistivity of a graphene sample (see
also Sec.~\ref{subsec:emt} below), i.e. graphene p-n junctions
are essentially transparent.  

The experimental realization of p-n junctions came shortly after
the theoretical
predictions~\cite{kn:huard2007,kn:lemme2007,kn:oezyilmaz2007,kn:williams2007}.
At zero magnetic field the effect of creating a p-n junction was to modestly
change the device resistance.  More dramatic was the change at high magnetic 
field in the quantum Hall regime (see Sec.~\ref{sec:qhe} below).
  
More detailed calculations of the zero field conductance of 
the p-n junction were performed taking into account the
effect of non-linear electronic screening.  This tends to make the p-n junction
sharper, and for $r_s \ll 1$, increasing the conductance by a factor
$r_s^{1/6}$~\cite{kn:zhang2008b}, and thereby further reducing the overall
contribution of p-n junctions to the total resistance.  The effect of
disorder was examined by \citet{kn:fogler2008} who studied how the p-n
junction resistance changed from its ballistic value in the absence of
disorder to the diffusive limit with strong disorder.  More recently,
\citet{kn:rossi2009} used a microscopic model of charged
impurities to calculate the screened disorder potential and solved for
the conductance of such a disordered n-p-n junction numerically.   
\begin{figure}[htb]
 \begin{center}
\includegraphics[width=1.0\columnwidth]{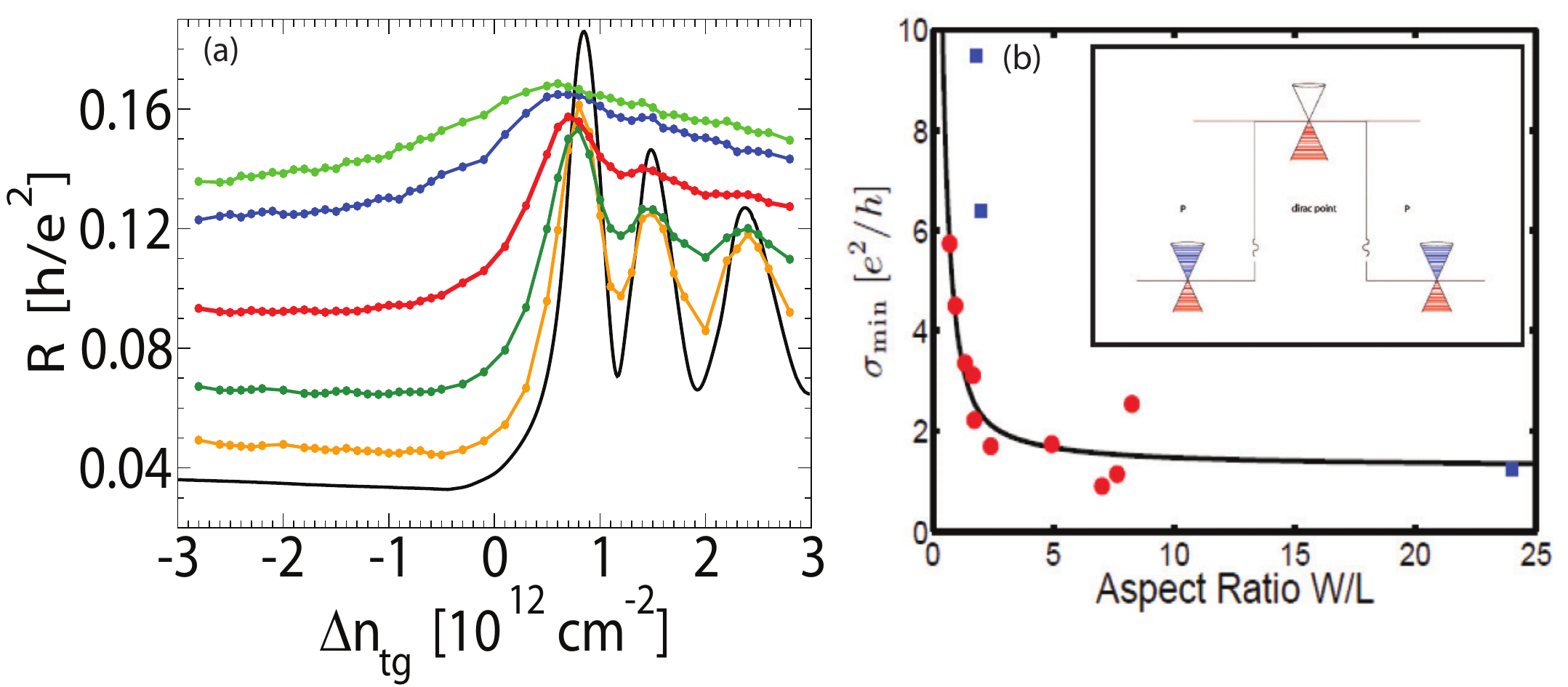}
  \caption{
           (Color online).
           (a) Disorder averaged resistance as a function of top gate voltage
           for a fixed
           back gate density $n_{\rm bg}=5\times 10^{11}{\rm cm}^{-2}$ 
           and several values of the impurity density (from bottom to top 
           $n_{\rm imp} = 0, 1, 2.5, 5, 10,$ and $15 \times 10^{11}{\rm cm}^{-2}$).
           Results were obtained using $10^3$ disorder realizations for a square samples of size $W=L=160$~nm
in presence of a top gate placed in the middle of the sample 10~nm 
above the graphene layer, 30~nm long and of width W.  The charge impurities
were assumed at a distance $d=1$~nm and the uniform dielectric constant $\kappa$
was taken equal to 2.5.  Adapted from \cite{kn:rossi2009}.
(b)  Solid line is Eq.~\ref{SA_Eq:UniMinCond} for armchair boundary
  conditions showing the aspect ratio dependence of the Dirac point
  ballistic conductivity~\cite{kn:tworzydlo2006}.  For $W\gg L$ the
  theory approaches the universal value $4e^2/\pi h$.  Circles show
  experimental data taken from \citet{kn:miao2007} and squares
  show the data from \citet{kn:danneau2008}.
Inset: Illustration of the configuration used to calculate 
graphene's universal minimum conductivity.  For
$V_g > 0$, one has a p-p-p junction, while for $V_g < 0$,
one has a p-n-p junction.  The figure illustrates 
the ballistic universal conductivity that occurs at the transition 
between the p-p-p and p-n-p junctions when  $V_g = 0$
 } 
  \label{fig:ln:qt1}
 \end{center}
\end{figure} 
The broad oscillations visible in  Fig.~\ref{fig:ln:qt1} arise
from resonant tunneling of the few modes with smallest transverse momentum.  
These results demonstrate that the signatures of the Klein tunneling are observable for impurity densities as high as $10^{12}\;{\rm cm}^{-2}$ and would not be washed away by disorder as long as the impurity limited mean-free-path is greater than length of
the middle region of the opposite polarity.  This implies that at
zero magnetic field the effects of Klein tunneling are best seen with 
a very narrow top-gate.  Indeed, recent experiments have succeeded in 
using an ``air-bridge''~\cite{kn:gorbachev2008, kn:liu2008b} or 
 very narrow top gates~\cite{kn:young2009, kn:stander2009}.  The observed
oscillations in the conductivity about the semi-classical value are in good agreement 
with theory of \citet{kn:rossi2009}.   

There is a strong similarity between the physics of 
phase-coherent ballistic trajectories of electrons and that of light waves 
that is often exploited~\cite{kn:ji2003b,kn:cheianov2007b,kn:shytov2008}.  In 
particular, \citet{kn:liang2001} demonstrated that one could construct
a Fabry-Perot resonator of electrons in a carbon nanotube.  This relies 
on the interference between electron paths in the different valleys $K$ 
and $K'$.  The same physics has been seen in ``ballistic'' graphene,
where the device geometry is constructed such that the source and
drain electrodes are closer than the typical electronic
mean-free-path~\cite{kn:miao2007,kn:cho2009}.

\subsubsection{Universal quantum limited conductivity}
\label{SA_Sec_UBMC}

  
An important development in the understanding of graphene transport 
is to use the formalism of Klein tunneling
to address the question of graphene's minimum
conductivity~\cite{kn:katsnelson2006,kn:tworzydlo2006}.  This of course
considers non-interacting electrons at zero temperature, and in the limit
of no disorder.  As shown in inset of Fig.~\ref{fig:ln:qt1}~(b), the insight is
to consider the source-graphene-drain configuration as the n-p-n or
n-n-n junction i.e. the leads are heavily electron doped, while the
graphene sheet in the middle could be electron doped, hole doped or be
pinned at the Dirac point with zero doping.  Since there is no
disorder, the electronic mean free path is much longer than the
distance between the source and drain ($\ell \gg L$).  It is this
situation we have in mind when we talk of graphene's ``ballistic
conductivity''.

For a non-Dirac metal, at finite carrier density, the absence of
scattering would imply that the semi-classical electrical 
conductivity is infinite,
since there is nothing to impede the electron motion.  However, the
conductance would then vanish as the carrier density is tuned to
zero.  This metal-insulator transition will be discussed in more
details later in the context of two dimensional semiconductors.

The situation is quite different for graphene.  From studying the
Klein tunneling problem, we already know that both the n-p-n junction
and the n-n-n junction have finite transmission coefficients.  The
interesting question is: what is the tunnelling at the precise point
where the junction changes from a n-p-n junction to the n-n-n
junction?  The conductivity at this transition point would then be the
quantum limited (ballistic) conductivity of graphene at the Dirac
point.

The solution is obtained by finding the transmission probabilities and 
obtaining the corresponding ballistic conductivity.  This is analogous 
to the quantum mechanics exercise of computing the transmission through 
a potential barrier, but now for relativistic electrons.  Using the 
non-interacting Dirac equation 
\begin{equation}
\label{SA_Eq:Dirac}
\left[\hbar v_{\rm F} \boldsymbol\sigma \cdot \vec{k} + e V(x)
\right] \Psi(\vec{r}) = \varepsilon \Psi(\vec{r}),
\end{equation}
with the boundary conditions corresponding to: $V(x<0) = V(x>L)= V_{\infty}$ to
represent the heavily doped leads and $V(x) = V_g$ for $0<x<L$.  For the case of $V_{\infty}
\rightarrow \infty$ and at the Dirac point ($V_g = \varepsilon =0$),
the transmission probability (i.e. the square of the transmission
amplitude) is given by purely evanescent modes~\cite{kn:tworzydlo2006}
\begin{equation}
\label{SA_Eq:Transmission} 
  T_{n} = \left| \frac{1}{\cosh(q_n L)} \right|^2.
\end{equation}
\noindent This is in contrast to the non-relativistic electrons (i.e.
with the usual parabolic dispersion), where for fixed $q_n$, the
analogous calculation gives vanishing transmission probability $T_n
\sim 1/V_{\infty}$.  The remaining subtle point is determining the form
transverse wave-vector $q_n$.  While it is clear that $q_n \sim nW^{-1}$
for large $n$, the choice of the boundary condition changes the
precise relation e.g. $q_n = n \pi/W$ for metallic armchair edges
and $q_n = (n \pm 1/3)\pi/W$ for semiconducting armchair edges.
Following \citet{kn:tworzydlo2006}, we use twisted boundary
conditions $\Psi(y=0) = \sigma_x \Psi(y=0)$ and $\Psi(y=W) = -
\sigma_x \Psi(y=W)$ which gives $q_n = (n+1/2) \pi/W$.  This boundary
condition is equivalent to having massless Dirac Fermions inside the
strip of width $W$, but infinitely massive Dirac Fermions outside of
the strip, thereby confining the electrons~\cite{kn:ryu2007}.

The Landauer conductivity is then given by 
\begin{eqnarray}
\label{SA_Eq:UniMinCond}
\sigma &=& \frac{L}{W} \times \frac{g_s g_v e^2}{h} \sum_{n=0}^{\infty} T_n  \\
&=& \frac{4e^2}{h} \sum_{n=0}^{\infty} \frac{L}{W \cosh^2\left[\pi(n+1/2) L/W \right]} \stackrel{W \gg L}{\longrightarrow} \frac{4 e^2}{\pi h}. \nonumber
\end{eqnarray}
Since at the Dirac point (zero energy) there is no energy scale in the
problem, the conductivity (if finite) can only depend on the aspect
ratio $L/W$.  The remarkable fact is that for $W\gg L$, the sum in
Eq.~\ref{SA_Eq:UniMinCond} converges to a finite and universal value
-- giving for ballistic minimum conductivity $\sigma_{\rm min} = 4 e^2/\pi h$.  This result also agrees with that obtained using
linear response theory in the limit of vanishing disorder 
suggesting that the quantum mechanical
transport through evanescent modes between source and drain (or
equivalently, the transport across two p-n junctions with heavily
doped leads), is at the heart of the physics behind the universal
minimum conductivity in graphene.

\citet{kn:miao2007} and \citet{kn:danneau2008} have probed
this ballistic limit experimentally using the two-probe geometry.  Their results, shown in
Fig.~\ref{fig:ln:qt1}~(b), are in good agreement with the theoretical
predictions.  Although, it is not clear what role contact
resistance~\cite{kn:golizadeh2009,kn:giovannetti2008,kn:huard2008,blake-ssc-149-1068-2009,blanter-prb-76-155433-2007,cayssol-prb-79-075428-2009,lee-nn-3-486-2008} played in these 2-probe measurements.

\subsubsection{Shot noise}

Shot noise is a type of fluctuation in electrical current caused by
the discreteness of charge carriers and from the randomness in their
arrival times at the detector or drain electrode.  It probes any
temporal correlation of the electrons carrying the current, quite
distinct from ``thermal noise'' (or Johnson-Nyquist noise) which
probes their fluctuation in energy.  Shot noise is quantified by the
dimensionless Fano factor~${\mathcal F}$, defined as the ratio between
noise power spectrum and the average conductance.  Scattering theory
gives~\cite{kn:buettiker1990}
\begin{equation}
{\mathcal F} = \frac{\sum_n T_n(1-T_n)}{\sum_n T_n}.
\end{equation}
Some well known limits include ${\mathcal F} = 1$ for ``Poisson
noise'' when $T_n \ll 1$ (e.g. in a tunnel junction), and ${\mathcal
  F} = 1/3$ for disordered metals~\cite{kn:beenakker1992}.  For
graphene at the Dirac point, we can use Eq.~\ref{SA_Eq:Transmission} to
get ${\mathcal F} \rightarrow 1/3$ for $W\gg
L$~\cite{kn:tworzydlo2006}.  One should emphasize that obtaining the
same numerical value for the Fano factor ${\mathcal F} = 1/3$ for
``ballistic'' quantum transport in graphene as that of diffusive
transport in disordered metals could be nothing more than a
coincidence~\cite{kn:dragomirova2009}. \citet{kn:cheianov2006b} found that 
the shot noise
of a single p-n junction was ${\mathcal F} = 1 - \sqrt{1/2}$ that is
numerically quite close to $1/3$.

Since several different mechanisms all give ${\mathcal F} \approx
1/3$, this makes shot noise a complicated probe of the underlying
physical mechanism.  Recent numerical studies by
\citet{kn:lewenkopf2008,kn:san-jose2007,sonin-prb-79-195438-2009,sonin-prb-77-233408-2008} treated the role of
disorder to examine the crossover from the ${\mathcal F} = 1/3$ in
ballistic graphene to the diffusive regime (see also
Sec.~\ref{SA_SEC_QC_crossover}).  Within the crossover, or away from
the Dirac point, the Fano factor is no longer universal and shows
disorder dependent deviations.  The experimental situation is less
clear.  \citet{kn:danneau2008} measured the Fano factor decrease
from ${\mathcal F} \approx 1/3$ with increasing carrier density  to 
claim agreement with the ballistic theory.
While \citet{kn:dicarlo2008} found that ${\mathcal F}$ was mostly
insensitive to carrier type and density, temperature, aspect ratio and
the presence of a p-n junction, suggesting diffusive transport in the
dirty limit.

Since shot noise is, in principle, an independent probe of the nature
of the carrier dynamics, it could be used as a separate test of the
quantum-limited transport regime.  However, in practice, the
coincidence in the numerical value of the Fano factor with that of
diffusive transport regime makes this prospect far more challenging.

\subsection{Quantum interference effects}
\subsubsection{Weak antilocalization}
\label{Sec:WAL}
Over the past 50 years, there has been much progress towards
understanding the physics of Anderson localization (for a recent
review, see \citet{kn:evers2008}).  Single particle Hamiltonians
are classified according to their global symmetry.  Since the Dirac
Hamiltonian (for a single valley) ${\mathcal H} = \hbar v_{\rm F}
\boldsymbol\sigma \cdot {\bf k}$ is invariant under the transformation
${\mathcal H} = \sigma_y {\mathcal H^*} \sigma_y$ (analogous to
spin-rotation symmetry (SRS) in pseudospin space) it is in the AII class
(also called the symplectic Wigner-Dyson class).  The more familiar
physical realization of the symplectic class is the usual disordered
electron gas
with strong spin-orbit coupling.
\begin{eqnarray}
\label{SA_Eq:SO}
{\mathcal H}_{\alpha \beta} &=& \frac{\hbar^2 k^2}{2 m} \delta_{\alpha \beta}  
+ {\mathcal V}_{\alpha \beta}, \nonumber \\
 {\mathcal V}_{\alpha \vec{k}, \beta \vec{k}'} &=& 
V_{\vec{k}-\vec{k'}} - i V^{\rm so}_{\vec{k}-\vec{k'}} (\vec{\hat{k}'}
\times \vec{\hat{k}})\cdot \boldsymbol\sigma_{\alpha \beta},   
\end{eqnarray}
\noindent where $\alpha$ and $\beta$ are (real) spin indices, $\boldsymbol\sigma_{\alpha \beta}$ a vector of Pauli matrices.  Notice that this
Hamiltonian is also invariant under SRS, ${\mathcal H} = \sigma_y {\mathcal H^*} \sigma_y$. We have 
\begin{equation}
\langle V_\vec{q} V_\vec{q'} \rangle = \frac{\delta(\vec{q} - \vec{q'})}{2 \pi \nu \tau} \ \ \ \langle V^{\rm so}_\vec{q} V^{\rm so}_\vec{q'} \rangle = \frac{\delta(\vec{q} - \vec{q'})}{2 \pi \nu \tau_{\rm so}}.
\end{equation}   
\noindent It was shown by \citet{kn:hikami1980} that when
the classical conductivity is large ($\sigma_0 \gg e^2/h$), the
quantum correction to the conductivity is positive 
\begin{equation}
\delta \sigma = \frac{e^2}{\pi h} \ln(L/\ell).
\label{SA_Eq:WAL}
\end{equation}          

Equivalently, one can define a one parameter scaling 
function~\cite{kn:abrahams1979} 
\begin{equation}
\beta(\sigma) = \frac{d \ln \sigma}{d \ln L},
\end{equation}
where for the symplectic class, it follows from Eq.~\ref{SA_Eq:WAL}
that $\beta(\sigma) = 1/(\pi \sigma)$ for
large $\sigma$.  To have $\beta>0$ means that the conductivity 
increases as one goes to larger system sizes or adds more disorder.
This is quite different from the usual case of an Anderson transition
where a negative $\beta$-function means that for those same changes, 
the system becomes more insulating.  

Since perturbation theory only gives the result for $\beta(\sigma \gg
e^2/h)$, the real question becomes what happens to the $\beta$
function at small $\sigma$.  If the $\beta$ function crosses zero and
becomes negative as $\sigma \rightarrow 0$, the system exhibits the 
usual Anderson metal-insulator transition.  Numerical studies
of the Hamiltonian (Eq.~\ref{SA_Eq:SO}), show that for the spin-orbit
system, the $\beta$-function vanishes at $\sigma^* \approx 1.4$ and
below this value, the quantum correction to the classical
conductivity is negative resulting in an insulator at zero
temperature.  $\sigma^*$ is an unstable fixed point for the symplectic
symmetry class. 

However, as we have seen in Sec.~\ref{SA_Sec_UBMC}, graphene has a
minimum ballistic conductivity $\sigma_{\rm min} = 4e^2/\pi h$ and
does not become insulating in the limit of vanishing disorder.  This 
makes graphene different from the spin-orbit Hamiltonian discussed
above, and the question of what happens with increasing disorder
becomes interesting.

\begin{figure}
\hspace{0.1\hsize}
\begin{center}
\includegraphics[width=1.0\columnwidth]{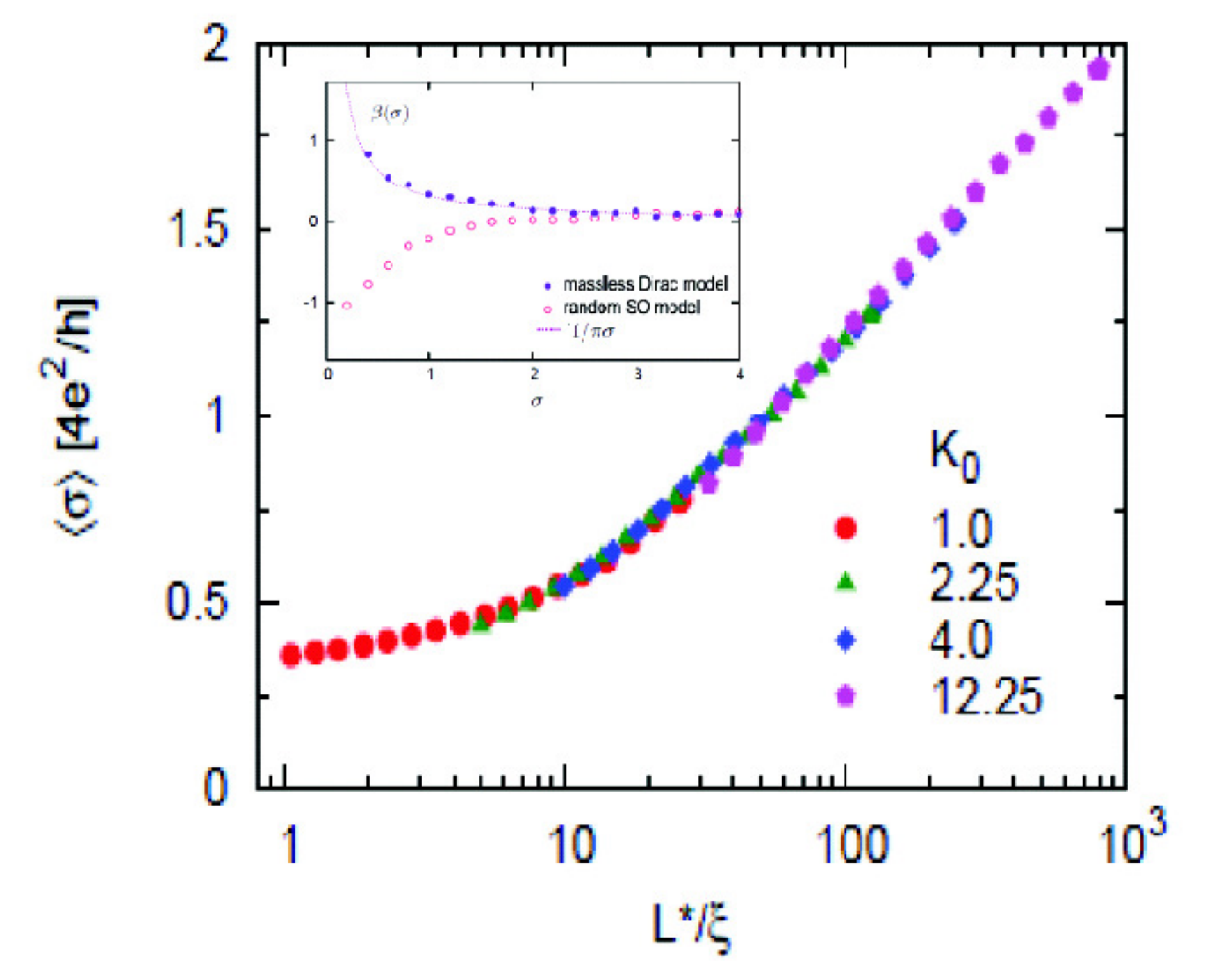}
\end{center}
\caption{\label{SA_Fig_bardarson} (Color online)
Demonstration of one parameter scaling at the Dirac
point~\protect{\cite{kn:bardarson2007}}.  Main panel shows the
conductivity as a function of $L^*/\xi$, where $L^* = f(K_0) L$ is the
scaled length and $\xi$ is the correlation length of the disorder
potential.  Notice that $\beta = d \ln \sigma/d \ln L >0$ for any
disorder strength.  The inset shows explicit
comparison~\protect{\cite{kn:nomura2007c}} of the $\beta$ function for
the Dirac Fermion model and for the symplectic (AII) symmetry class.}
\end{figure}

\citet{kn:bardarson2007} studied the Dirac Hamiltonian 
(Eq.~\ref{SA_Eq:Dirac}) with the addition of a Gaussian correlated
disorder term $U(\vec{r})$, where
\begin{equation}
\label{SA_Eq:GaussDis}
\langle U(\vec{r}) U(\vec{r'}) \rangle = K_0 \frac{(\hbar v_{\rm F})^2}{2 \pi \xi^2} \exp\left[\frac{-|\vec{r} - \vec{r'}|^2}{2 \xi^2}\right].
\end{equation}
\noindent  One should think of $K_0$ as parameterizing the strength of
the disorder and $\xi$ as its correlation length.  If the theory of
one-parameter scaling holds for graphene, then it should be possible
to rescale the length $L^* = f_0(K_0) L$, where $f_0$ is a scaling
function inversely proportional to the effective electronic
mean free path. Their 
numerical results are shown in Fig.~\ref{SA_Fig_bardarson} and 
demonstrate that (i) graphene does exhibit one-parameter scaling
(i.e. there exists a $\beta$-function) and (ii) the $\beta$-function is 
always positive unlike the spin-orbit case.  Therefore, Dirac Fermions 
evade Anderson localization and are always
metallic.  Similar conclusions were obtained by 
\citet{kn:nomura2007c,kn:san-jose2007,titov-e-79-17004-2007,tworzydlo-prb-78-235438-2008}.  

\begin{figure}
\hspace{0.1\hsize}
\begin{center}
\includegraphics[width=.7\columnwidth]{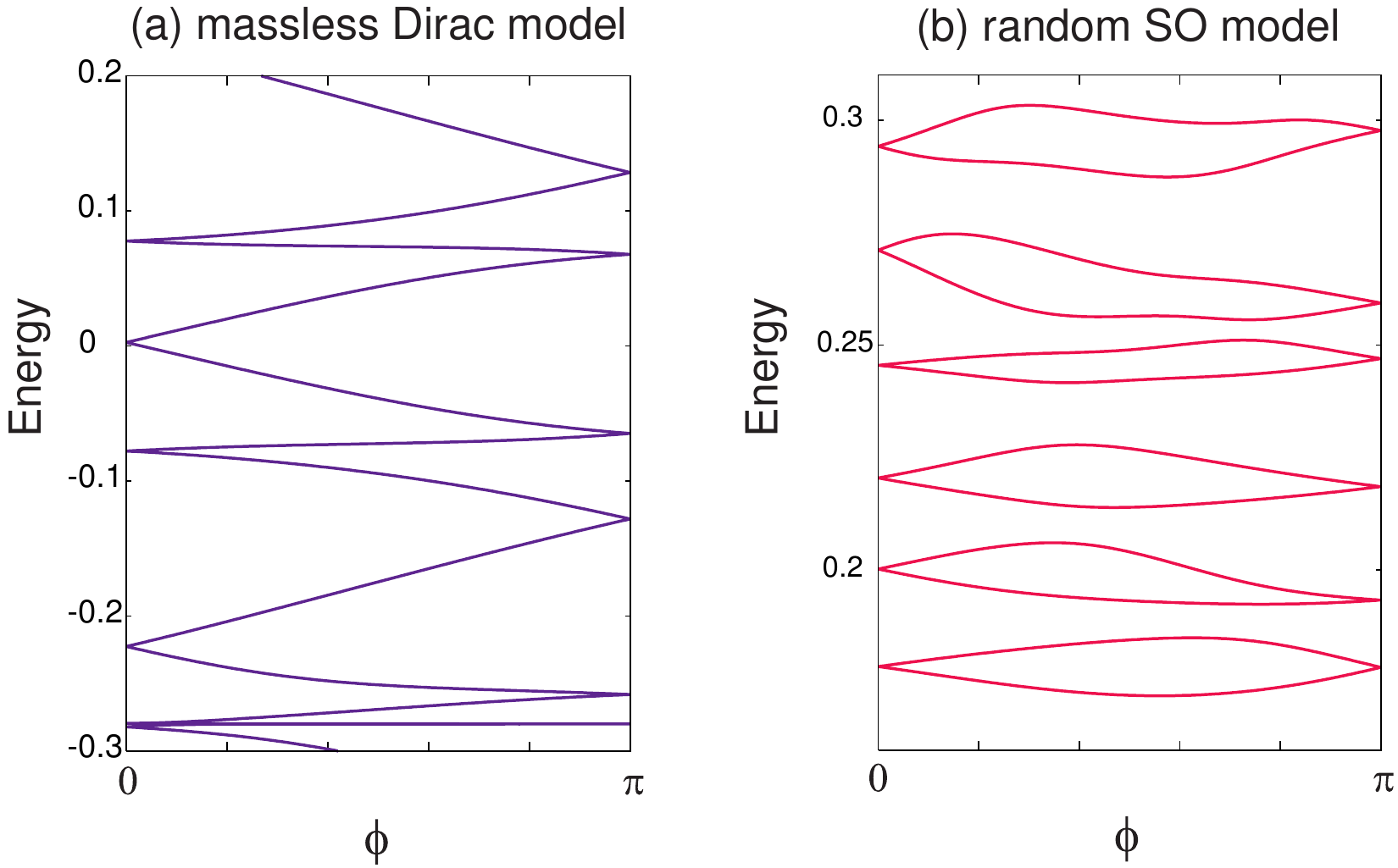}
\end{center}
\caption{\label{SA_Fig_Nomura2} (Color online)
Picture proposed by~\citet{kn:nomura2007c} 
to understand the difference in topological structure between 
the massless Dirac model and the random spin-orbit symmetry class.}
\end{figure}

The inset of Fig.~\ref{SA_Fig_bardarson} shows an explicit computation of the
$\beta$-function comparing Dirac Fermions with the spin-orbit model.
The difference between these two classes of the AII symmetry class has
been attributed to a topological term (i.e. two possible choices for
the action of the field theory describing these
Hamiltonians).
Since it allows for only two possibilities, it has been
called a $Z_2$ topological symmetry~\cite{kn:kane2005,kn:evers2008}.  The
topological term has no effect at $\sigma \gg e^2/h$, but is
responsible for the differences at $\sigma \approx e^2/h$ and
determines the presence or absence of a metal-insulator transition.
\citet{kn:nomura2007c} present an illustrative visualization of
the differences between Dirac Fermions and the spin-orbit symplectic
class shown in Fig.~\ref{SA_Fig_Nomura2}.  By imposing a twist
boundary condition in the wave-functions such that $\Psi(x=0) = \exp[i
  \phi] \Psi(x=L)$ and $\Psi(y=0) = \Psi(y=W)$, one can examine the
single particle spectrum as a function of the twist angle $\phi$.  For
$\phi = 0$ and $\phi = \pi$, the phase difference is real and
eigenvalues come in Krammers degenerate pairs.  For other values of
$\phi$, this degeneracy is lifted.  As seen in the figure, for
massless Dirac Fermions all the energy states are connected by a
continuous variation in the boundary conditions.  This precludes
creating a localized state, which would require the energy variation
with boundary condition (also called Thouless Energy) be smaller than
the level spacing.  Since this is a topological effect,
\citet{kn:nomura2007c} argued that this line of reasoning should
be robust to disorder.

The situation for the spin-orbit case is very different. The same
Krammer's pairs that are degenerate at $\phi = 0$, reconnect at
$\phi=\pi$.  In this case, there is nothing to prevent localization if
the disorder would push the Krammer's pairs past the mobility edge.
Similar considerations regarding the $Z_2$-symmetry hold also 
for topological insulators where the metallic surface state should
remain robust against localization in the presence of disorder.

\subsubsection{Crossover from the symplectic universality class}
\label{SA_Sec_symplectic_crossover}

As is already apparent in the preceding discussion, each Dirac cone
is described by the Dirac Hamiltonian ${\mathcal H} =
\mathbf{\sigma}\cdot {\bf p}$.  The effective SRS ${\mathcal H} =
\sigma_y {\mathcal H}^* \sigma_y$ is preserved in each cone, and for
most purposes graphene can be viewed as two degenerate copies of the
AII symplectic symmetry class.  However, as was first pointed out by
\citet{kn:suzuura2002}, a material defect such as a missing atom
would couple the two Dirac cones (and since each cone is located in a
different ``valley'', this type of interaction is called intervalley
scattering).  One can appreciate intuitively why such scattering is
expected to be small.  The two valleys at points K and K' in the
Brillouin zone are separated by a large momentum vector that is inversely
proportional to the spacing between two neighboring carbon atoms.  This 
means that the potential responsible for such intervalley coupling would
have to vary appreciably on the scale of 0.12~{\rm nm} in order to couple
the $K$ and $K'$ points.  

We note that while such defects and the corresponding coupling between 
the valleys are commonly observed in STM studies on epitaxial
graphene~\cite{kn:rutter2007}, they are virtually absent in all similar
studies in exfoliated graphene~\cite{kn:ishigami2007,kn:stolyarova2007, 
kn:zhang2008}.  In the presence of such atomically sharp disorder, 
\citet{kn:suzuura2002} proposed a model for the two-valley 
Hamiltonian that captures the effects of intervalley scattering.
The particular form of the scattering potential 
is not important, and below in Sec.~\ref{Sec:WL_UCF} we will 
discuss a generalized Hamiltonian that includes all non-magnetic 
(static) impurities consistent with the honeycomb symmetry and 
is characterized by five
independent parameters~\cite{kn:aleiner2006,kn:mccann2006}.  Here, the
purpose is simply to emphasize the qualitative
difference between two types of disorder: long-range
(i.e. diagonal) disorder ${\mathcal U}^{\rm LR}$ that preserves the
effective SRS, and a short-range potential ${\mathcal U}^{\rm
  SR}$ that breaks this symmetry.  

We note that with the intervalley term ${\mathcal U}^{\rm SR}$, the
Hamiltonian 
belongs to the Wigner-Dyson orthogonal symmetry class, while as was 
discussed in
Sec.~\ref{Sec:WAL} including only diagonal disorder ${\mathcal U}^{\rm
  LR}$, 
one is in the Wigner-Dyson symplectic class.

A peculiar feature of this crossover is that it is
governed by the concentration of short-range impurities 
thus questioning the notion that 
the ``universality class'' is determined only by the
global symmetries of the Hamiltonian and not by microscopic details.
However, a similar crossover was observed in
\citet{kn:miller2003} where the strength of spin-orbit
interaction was tuned by carrier density, moving from weak
localization at low density and weak spin-orbit interaction, to weak
antilocalization at high density and strong spin-orbit interaction.  

\begin{figure}
\hspace{0.1\hsize}
\begin{center}
\includegraphics[width=.8\columnwidth]{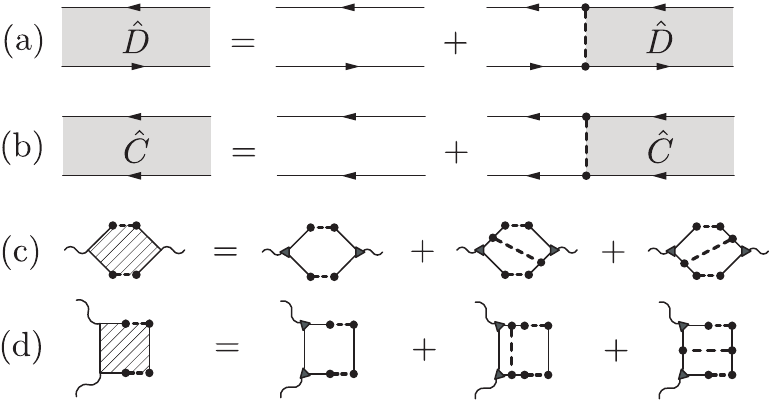}
\end{center}
\caption{\label{SA_Fig_DressedHikami} (Color online)
Diagrammatic representation 
for (a) diffuson, (b) Cooperon, and dressed Hikami 
boxes [(c) and (d)]. Figure adapted from \citet{kn:kharitonov2008}}
\end{figure}

From symmetry considerations one should expect that
without atomically sharp defects, graphene would exhibit weak
antilocalization (where $\delta \sigma >0$) and no Anderson
localization (see Sec.~\ref{Sec:WAL}).  However, with intervalley scattering,
graphene should have weak localization ($\delta \sigma <0$) and be
insulating at zero temperature.  These conclusions were verified
by \citet{kn:suzuura2002} from a microscopic 
Hamiltonian by calculating the Cooperon and 
obtaining the corrections to the conductivity from the bare Hikami box 
(see Sec.~\ref{Sec:WL_UCF} below for a more complete discussion).  

For the case of no intervalley scattering 
${\mathcal U} = {\mathcal U}^{\rm LR}$,
the resulting Cooperon is
\begin{equation}
C^{\rm LR}_{\vec{k}_\alpha \vec{k}_\beta}(Q) = \frac{n u^2}{A} e^{i (\psi_{\vec{k}_\alpha} - 
\psi_{\vec{k}_\beta})} \frac{1}{(v_{\rm F} \tau Q)^2},
\end{equation}
\noindent with area $A=LW$, $Q = \vec{k}_\alpha + \vec{k}_\beta$ and $ e^{i
  (\psi_{\vec{k}_\alpha} - \psi_{\vec{k}_\beta})} \approx -1$, giving
$\delta \sigma_{\rm LR} = (2e^2/\pi^2 \hbar) \ln[L_\phi/\ell]$.  As
expected for the symplectic class, without intervalley scattering, the
quantum correction to the conductivity is positive.

With intervalley scattering ${\mathcal U} = {\mathcal U}^{\rm SR}$ calculating
the same diagrams gives 
\begin{equation}
  C^{\rm SR}_{\vec{k}_\alpha \vec{k}_\beta}(Q) =  \frac{n u^2}{A} j_{\alpha} j_{\beta}
 e^{i (\psi_{\vec{k}_\alpha} - 
\psi_{\vec{k}_\beta})} \frac{1}{(v_{\rm F} \tau Q)^2},
\end{equation}
with current $j_\alpha = - j_\beta$ and $\delta \sigma_{\rm SR} 
= -e^2/(2\pi^2 \hbar) \ln[L_\phi/\ell]$.  This negative $\delta \sigma$
is consistent with the orthogonal symmetry class.  The explicit microscopic 
calculation demonstrates the crossover from weak antilocalization 
to weak localization induced by atomically sharp microscopic defects 
providing intervalley coupling.

\begin{figure}
\hspace{0.1\hsize}
\begin{center}
\includegraphics[width=.8\columnwidth]{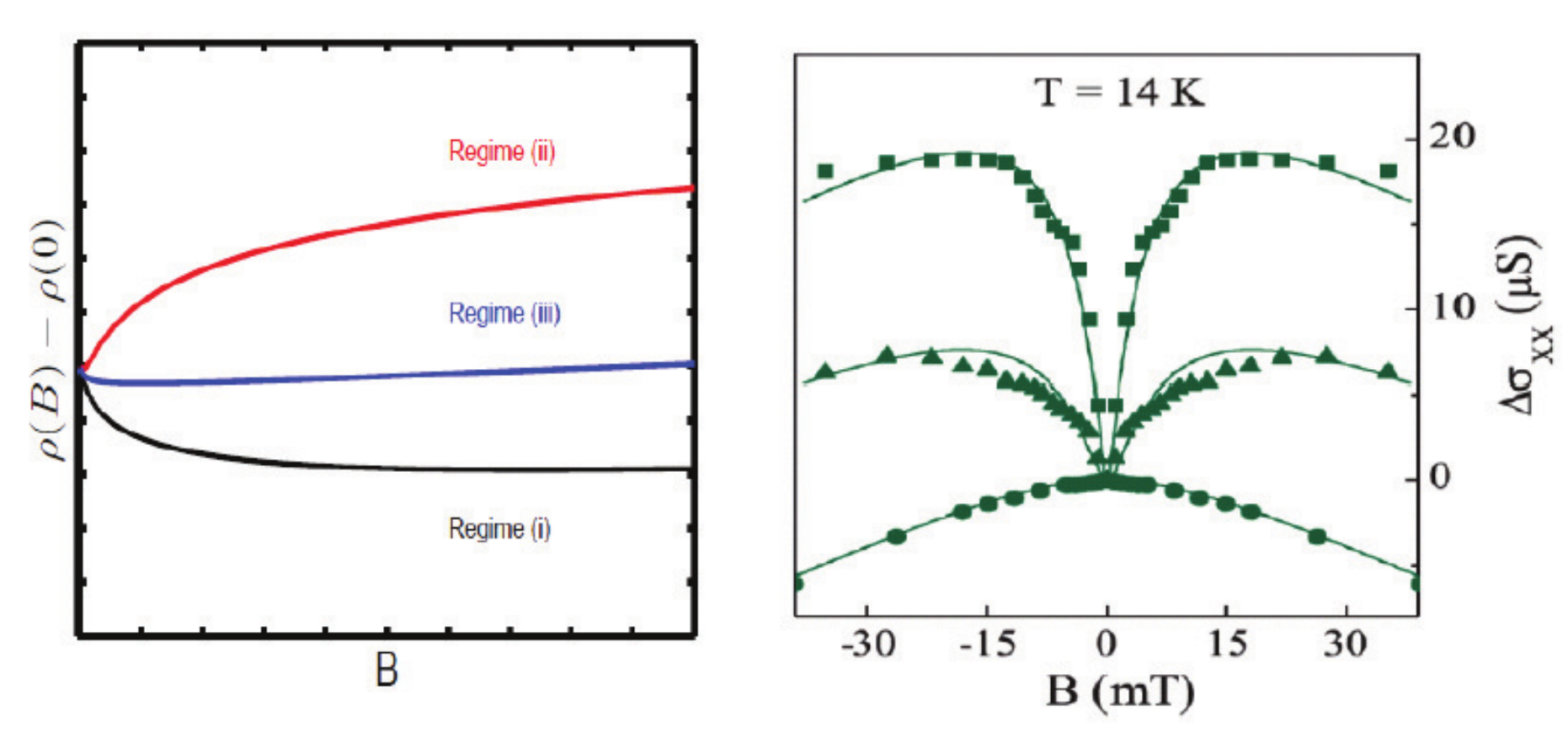}
\end{center}
\caption{\label{SA_Fig_Savchenko1} (Color online)
Left panel: Schematic of 
different magnetoresistance regimes: (i) For strong intervalley
scattering (i.e. $\tau_i \ll \tau_{\phi}$ and $\tau_z \ll
\tau_{\phi}$), Eq.~\ref{SA_Eq:WL} gives weak localization or $\delta
\sigma \sim \rho(B) - \rho(0) <0$.  This is similar to quantum
transport in the usual 2DEG; (ii) For weak intervalley scattering
(i.e. $\tau_i \gg \tau_{\phi}$ and $\tau_z \gg \tau_{\phi}$), one
has weak antilocalization, characteristic of the symplectic symmetry
class; (iii) For $\tau_i \gg \tau_{\phi}$, but $\tau_z \ll
\tau_{\phi}$, Eq.~\ref{SA_Eq:WL} gives a regime of suppressed weak
localization. Right Panel: Experimental realization of these 
three regimes.  Graphene magnetoconductance is shown 
for carrier density (from bottom to top)
$n = 2.2 \times 10^{10}~{\rm cm}^{-2}, 1.1 \times
10^{12}~{\rm cm}^{-2}$ and $2.3 \times 10^{12}~{\rm
  cm}^{-2}$, at T = $14$~{\rm K}.  The lowest carrier density
(bottom curve) has a small contribution from short-range disorder and
shows weak anti-localization (i.e. the zero-field conductivity is
larger than at finite field.  $\sigma(B=0) = \sigma_0 + \delta \sigma$
and $\sigma(B>B^*) = \sigma_0$, with $\delta \sigma >0$.  $B^*$ is the
phase-breaking field).  In contrast, the highest density data (top
curve) has a larger contribution of intervalley scattering and shows
weak localization i.e.  $\delta
\sigma <0$.  Figure taken from \citet{kn:tikhonenko2009}.}
\end{figure}

This crossover was recently observed
experimentally~\cite{kn:tikhonenko2009}.  They noted empirically that
for their samples, the scattering associated with short-range defects is
stronger at high carrier density.  In fact, this is what one expects
from the microscopic theory discussed in 
Sec.~\ref{sec_high_density} below.  Due to the
unique screening properties of graphene, long-range scatterers
dominate transport at low carrier density while short-range
scatterers dominate at high-density.  Assuming that these short-range
defects are also the dominant source of intervalley scattering, then
one would expect to have weak localization at high carrier density
(due to the large intervalley scattering), and weak antilocalization
at low carrier density where transport is dominated by
``atomically smooth'' defects like charged impurities in the
substrate.  This is precisely what was seen experimentally.  Figure
\ref{SA_Fig_Savchenko1} shows a comparison of the magnetoconductance
at three different carrier densities.  At the lowest carrier density,
the data show the weak antilocalization characteristic of the
symplectic symmetry class, while at high density, one finds weak
localization signaling a crossover to the orthogonal universality
class.  

 A second crossover away from the symplectic universality class was
 examined by ~\citet{kn:morpurgo2006}.  As discussed earlier, a
 magnetic field breaks time reversal symmetry and destroys the leading
 quantum corrections to the conductivity $\delta \sigma(B>B^*) = 0$.
 This can also be understood as a crossover from the symplectic (or
 orthogonal) universality class to the unitary class.  The unitary
 class is defined by the absence of time reversal symmetry and hence
 vanishing contribution from the Cooperon.\footnote{The sign of the quantum
 correction in relation to the global symmetry of the Hamiltonian can
 also be obtained from Random Matrix Theory~\cite{kn:beenakker1997}
$\delta \sigma/\sigma_0 =  (1 - 2/\beta)/4$,
where $\beta = 1,2,4$ for the orthogonal, unitary and symplectic
Wigner-Dyson symmetry classes.}

Similar to short-range impurities inducing a crossover from symplectic
to orthogonal classes, \citet{kn:morpurgo2006} asked if there
were other kinds of disorder that could act as pseudo-magnetic fields
and induce a crossover to the unitary symmetry class leading to the
experimental signature of a suppression of weak antilocalization.
This was in part motivated by the first experiments on graphene
quantum transport showing that the weak localization correction was an
order of magnitude smaller than expected~\cite{kn:morozov2006}.  The
authors argued that topological lattice
defects~\cite{ISI:A1995RK99000014} (e.g. pentagons and heptagons) and
non-planarity of graphene (commonly referred to as ``ripples'') would
generate terms in the Hamiltonian that looked like a vector potential
and correspond to a pseudo-magnetic field.

In addition, experiments on both suspended
graphene~\cite{kn:meyer2007} and on a substrate showed that graphene
is not perfectly co-planar.  It is noteworthy, however, that
experiments on SiO$_2$ substrate showed that these ripples were
correlated with the height fluctuations of the substrate and varied by
less than 1~{\rm nm}~\cite{kn:ishigami2007}; while graphene on mica
was even smoother with variations of less than 0.03~{\rm
  nm}~\cite{ISI:000271899300041}.  On the other hand, one could
deliberately induce 
lattice defects~\cite{kn:chen2009} or create controlled ripples by
straining graphene before cooling and exploiting graphene's negative
thermal expansion coefficient~\cite{kn:baladin2008,bao-nn-4-562-2009}. 

Just like a real magnetic field, these terms would break the TRS in a
single valley (while preserving the TRS of the combined system).  If
$\tau_{\rm i} \gg \tau_{\phi}$, the two valleys are decoupled and
these defects would cause a crossover to the unitary symmetry class
and the resulting Cooperon (Fig.~\ref{SA_Fig_DressedHikami}) would vanish.
For example, considering the case of lattice defects, the disorder Hamiltonian
would be given by ${\mathcal U}^{G} = (1/4)[\sigma_x \otimes \sigma_z]
\boldsymbol \nabla (\partial_y u_x(\vec{r}) - \partial_x
u_y(\vec{r}))$, where $\vec{u}(\vec{r})$ is the lattice strain vector
induced by the defect.  One notices that this term in 
the Hamiltonian has the form of an effective magnetic field $+B$ in the $K$
valley and $-B$ in the $K'$ valley~\cite{kn:morpurgo2006}.  In 
the absence of intervalley coupling, this would suppress weak 
antilocalization when the effective magnetic field $|B|$ is larger than the
field $B^*$ discussed in Sec.~\ref{SA_Sec_Intro}.

\subsubsection{Magnetoresistance and mesoscopic conductance fluctuations}
\label{Sec:WL_UCF}
As already discussed, at low energies and in the absence of disorder,
graphene is described by two decoupled Dirac cones located at points
$K$ and $K'$ in the Brillouin zone.  Within each cone, one has a
pseudospin space corresponding to wavefunction amplitudes on the $A$
and $B$ sublattice of the honeycomb lattice.  The two valley
Hamiltonian is then the outer product of two $SU(2)$ spin 
spaces $KK' \otimes AB  $.  The most generic Hamiltonian in this 
space of $4 \times
4$ Hermitian matrices can be parameterized by the generators 
of the group $U(4)$~\cite{kn:mccann2006,kn:aleiner2006,kn:altland2006}

\begin{equation}
\label{SA_Eq_GeneralH}
{\mathcal H} = \hbar v_{\rm F} \boldsymbol \Sigma \vec{p}
+ \openone_4 u_0(\vec{r}) + \sum_{s,l = x,y,z} 
\left. \Sigma_s \Lambda_l u_{sl} (\vec{r}) \right.,
\end{equation}
where $\boldsymbol \Sigma = (\Sigma_x, \Sigma_y, \Sigma_z) = (\sigma_z
\otimes \sigma_x, \sigma_z \otimes \sigma_y, \openone_2 \otimes
\sigma_z)$ is the algebra of the sublattice $SU(2)$ space (recall that
the outer product is in the space $KK' \otimes AB$, and
the $\boldsymbol \Sigma$ operator is diagonal in the $KK'$ space).
Similarly $\boldsymbol \Lambda = (\sigma_x \otimes \sigma_z, \sigma_y
\otimes \sigma_z, \sigma_z \otimes \openone_2)$ forms the algebra of
the valley-spin space (being diagonal in the $AB$ space).   

The Hamiltonian of Eq.~\ref{SA_Eq_GeneralH} can be understood in
simple terms. The first term is just two decoupled Dirac cones and is
equivalent to the disorder free case discussed earlier,
but written here in a slightly modified
basis.  The second term is identical to ${\mathcal U}^{LR}$ and as
discussed previously represents any long-range diagonal disorder.  The
last term parameterized by the nine scattering potentials $u_{sl}
(\vec{r})$ represent all possible types of disorder allowed by the
symmetry of the honeycomb lattice.  For example, a vacancy would
contribute to all terms (including $u_0$) except $u_{xz}$ and
$u_{yz}$; while bond disorder contributes to all terms except
$u_{zz}$~\cite{kn:aleiner2006}.

The ``diagonal'' term $u_0(\vec{r})$ is the dominant scattering
mechanism for current graphene experiments and originates from long
ranged Coulomb impurities and is discussed in more detail in 
Sec.~\ref{sec_high_density}.  Due to the peculiar screening properties of
graphene, such long-range disorder cannot be treated using the
Gaussian white noise approximation.  To circumvent this problem (both
for the long-range $u_0$ and short-range $u_{sl}$ terms) we simply
note that for each kind of disorder, there would be a corresponding
scattering time $\{ \tau_0, \tau_{sl} \}$ that could, in principle,
have very different dependence on carrier density.

For the special case
of Gaussian white noise, i.e. where $\langle u_{sl}(\vec{r})
u_{s'l'}(\vec{r'}) \rangle = u_{sl}^2 \delta_{s,s'} \delta_{l, l'}
\delta(\vec{r}-\vec{r'})$, we have $\hbar \tau_{sl}^{-1} = \pi
D(E_{\rm F}) u_{sl}^2$.)  Moreover, one could assume that after
disorder averaging, the system is isotropic in the $xy$ plane.
Denoting $\{x,y\} \equiv \perp$, the total scattering time is 
given by  
\begin{equation}
\tau^{-1} = \tau_0^{-1} + \tau_{zz}^{-1} + 2\tau_{\perp z}^{-1} + 2\tau_{z
  \perp}^{-1} + 4\tau_{\perp \perp}^{-1}.
\end{equation}
\noindent These five 
scattering times could be viewed as independent microscopic 
parameters entering the theory~\cite{kn:aleiner2006}, or one could
further classify scattering times as being either ``intervalley'' 
$\tau_i^{-1} = 4 \tau_{\perp \perp}^{-1} + 2 \tau_{z \perp}^{-1}$
or ``intravalley'' 
$\tau_z^{-1} = 4 \tau_{\perp z}^{-1} + 2 \tau_{z z}^{-1}$.  A small
contribution from trigonal warping (a distortion to the Dirac cone
at the energy scale of the inverse lattice spacing) could be modeled
by the perturbative term $H_w \sim \Sigma_x (\boldsymbol \Sigma \vec{p})
\Lambda_z \Sigma_x (\boldsymbol \Sigma \vec{p}) \Sigma_x$ which 
acts as an additional source of intravalley scattering~\cite{kn:mccann2006}.

The transport properties of the Hamiltonian (Eq.~\ref{SA_Eq_GeneralH})
are obtained by calculating the two particle propagator.  In general, both 
the classical
contribution (diffusons) and quantum corrections (Cooperons) 
will be $4 \times 4$ matrices defined in terms of the 
retarded (R) and advanced (A) Green's functions 
${\mathcal G}^{R,A}$ as (see also Fig.~\ref{SA_Fig_DressedHikami})
\begin{eqnarray}
{\mathcal D}(\omega, \vec{r}, \vec{r'}) &=& 
\langle {\mathcal G}^R (\epsilon + \omega, \vec{r}, \vec{r'})
\otimes {\mathcal G}^A (\epsilon, \vec{r'}, \vec{r}) \rangle \\
{\mathcal C}(\omega, \vec{r}, \vec{r'}) &=& 
\langle {\mathcal G}^R (\epsilon + \omega, \vec{r}, \vec{r'})
\otimes {\mathcal G}^A (\epsilon, \vec{r}, \vec{r'}) \rangle. \nonumber 
\end{eqnarray}
\noindent As will be discussed in Sec.~\ref{subsec_boltzmann}, the scattering rate is
dominated by the diagonal disorder $\tau \approx \tau_0$.  Since both
this term and the Dirac part of Eq.~\ref{SA_Eq_GeneralH} is
invariant under the valley $SU(2)$, one can classify the diffussons
and Cooperons as ``singlets'' and ``triplets'' in the AB-sublattice
$SU(2)$ space.  Moreover, one finds that for both the diffussons and
Cooperons, only the valley ``singlets'' are gapless, and one can
completely ignore the valley triplets whose energy gap scales as
$\tau_0^{-1}$.  Considering only the sublattice singlet ($j=0$)
and triplet ($j=x,y,z$), one 
finds~\cite{kn:mccann2006,kn:kechedzhi2008,kn:kharitonov2008,fal'ko-ssc-143-33-2007}
\begin{widetext}
\begin{eqnarray}
\left[ -i \omega -\frac{1}{2}v_{\rm F}^2 \tau_0 \left(\nabla- \frac{2e\vec{A}}{c} \right)^2 + \Gamma^j\right] {\mathcal D}^{j} (\vec{r}, \vec{r'}) &=& \delta(\vec{r} - \vec{r'}), \\
\left[ -i \omega -\frac{1}{2}v_{\rm F}^2 \tau_0 \left(\nabla+ \frac{2e\vec{A}}{c} \right)^2  + \Gamma^j  + \tau_{\phi}^{-1} \right] {\mathcal C}^j  (\vec{r}, \vec{r'}) &=& \delta(\vec{r} - \vec{r'}),
\nonumber 
\end{eqnarray}
with $\Gamma^0 = 0$~(singlet); $\Gamma^x = \Gamma^y =  \tau^{-1}_i  + \tau^{-1}_z$, and $\Gamma^z =2 \tau_{i}^{-1}$~(triplet).  This equation captures
all the differences in the quantum corrections to the conductivity 
between graphene and usual 2DEGs. \\
\end{widetext}
The magnetoresistance and conductance fluctuations properties in
graphene follow from this result.  The quantum correction
to the conductivity $\delta \sigma \sim N^C_{t} - N^C_{s}$, where
$N^C_{t}$ is the number of gapless triplet Cooperon modes and
$N^C_{s}$ is the number of gapless singlet Cooperon modes.  In this 
context, gapped modes do not have a divergent quantum correction and 
can be neglected.  Similarly,
the conductance fluctuations are given by $\langle \left[ \delta G
  \right]^2 \rangle = N_{CD} \langle \left[ \delta G \right]^2
\rangle_{\rm 2DEG}$, where $N_{CD}$ counts the total number of gapless
Cooperons and diffusons modes, and $\langle \left[ \delta G \right]^2
\rangle_{\rm 2DEG}$ is the conductance variance for a conventional 2D
electron gas.  For a quasi-1D geometry, 
  $\langle \left[ \delta G \right]^2
\rangle_{\rm 2DEG} = \frac{1}{15}(2e^2/h)^2$~\cite{kn:lee1985a}.

We can immediately identify several interesting regimes that are
shown schematically in Fig.~\ref{SA_Fig_Savchenko1}: \\

\noindent {\em (i)
  Strong intervalley scattering}. Even with strong short-range
disorder (i.e. $\tau_i \ll \tau_{\phi}$ and $\tau_z \ll \tau_{\phi}$),
both the singlet Cooperon ${\mathcal C}^0$ and the singlet diffuson
${\mathcal D}^0$ remain gapless since $\Gamma^0 = 0$.  Contributions
from all triplet Cooperons and difussons vanish.  In this situation,
the quantum corrections to the conductivity in graphene is very
similar to the regular 2DEG.  The Hamiltonian is in the orthogonal
symmetry class discussed earlier and one has weak localization
($\delta \sigma <0$).  Similarly, for conductance fluctuations (typically
measured at large magnetic fields), one would have the same result as
the non-relativistic electron gas. \\

\noindent {\em (ii) Weak short-range disorder}.
For $\tau_i \gg \tau_{\phi}$ and $\tau_z \gg \tau_{\phi}$, all
sublattice Cooperons and diffusons remain gapless at zero magnetic
field.  One then has $\delta \sigma >0$ or weak anti-localization
(symplectic symmetry).  This regime was observed in experiments on
epitaxial graphene~\cite{kn:wu2007}.  The diffuson contribution to the
conductance fluctuations is enhanced by a factor of 4 compared
with conventional metals. \\ 

\noindent {\em (iii) Suppressed localization regime}.  In
the case that there is strong short-range scattering $\tau_z \ll
\tau_{\phi}$, but weak intervalley scattering $\tau_i \gg
\tau_{\phi}$.  The Cooperons ${\mathcal C}^x$ and ${\mathcal C}^y$ will
be gapped, but ${\mathcal C}^z$ will remain and cancel the effect of
the singlet ${\mathcal C}^0$.  In this case one would have the
suppressed weak localization that was presumably seen in the first
graphene quantum transport experiments~\cite{kn:morozov2006}. \\

Although the discussion above captures the main physics, for
completeness we reproduce the results of
calculating the dressed Hikami boxes 
in Fig.~\ref{SA_Fig_DressedHikami}~\cite{kn:aleiner2006,kn:mccann2006,kn:kechedzhi2008,kn:kharitonov2008} and using
known results~\cite{kn:lee1985a}.  The quantum correction 
to the conductance is
  \begin{equation}
\delta g = \frac{2 e^2 D}{\pi \hbar} \int \frac{d^2 q}{(2 \pi)^2} 
\left( C^x + C^y + C^z - C^0 \right),
\end{equation}
\noindent and for the magnetoresistance
\begin{eqnarray}
\label{SA_Eq:WL}
\rho(B) - \rho(0) &=& - \frac{e^2 \rho^2}{\pi \hbar}
\left[ F\left(\frac{B}{B_\phi}\right)
- F\left(\frac{B}{B_\phi + 2 B_i}\right)  \right. \nonumber \\
&& \mbox{} \left.
-2 F\left(\frac{B}{B_\phi + 2 B_z}\right)\right],
\end{eqnarray}
where $B_{\phi, i, z} = (\hbar c/4 D e) \tau^{-1}_{\phi, i, z}$ and
$F(x) = \ln x + \psi(1/2 + 1/x)$, with $\psi$ as the digamma function.
The function $F(x)$ is the same as for 2DEGs~\cite{kn:lee1985a}, however
the presence of three terms in Eq.~\ref{SA_Eq:WL} is unique to graphene.  
The universal conductance fluctuations are
\begin{eqnarray}
\langle \left[ \delta G
  \right]^2 \rangle &=& 
\sum_{C,D} 3 \left[ \frac{g_s g_v e^2}{2 \pi \hbar} \right]^2 
\sum_{i=0}^{3} \sum_{n_x = 1}^{\infty} 
\sum_{n_y = 0}^\infty \nonumber \\
&& \mbox{} 
\frac{1}{\pi^4 L_x^4}
\left[ \frac{\Gamma_i}{\pi^2 D}
+ \frac{n_x^2}{L_x^2} + \frac{n_y^2}{L_y^2}\right]^{-2} \nonumber \\
&=&  N_{CD} \langle \left[ \delta G \right]^2
\rangle_{\rm 2DEG},
\end{eqnarray}
with only diffusions contributing for $B>B_\phi \approx B^*$.

In this section we have assumed Gaussian white noise correlations
to calculate the Green's functions.  Since we know that this
approximation fails for the semi-classical contribution arising
from Coulomb disorder, why can we
use it successfully for the quantum transport?  It turns out that
the quasi-universal nature of weak localization and conductance 
fluctuations means that the exact nature of the disorder potential 
will not change the result.  Several numerical calculations using
long-range Coulomb potential have checked this 
assumption~\cite{kn:yan2008}.  Many of the symmetry arguments discussed
here apply to confined geometries like quantum dots~\cite{kn:wurm2009}.
Finally, the diagrammatic perturbation theory discussed here applies only 
away from the Dirac point.  As discussed in Sec.~\ref{Sec:WAL}, numerically
calculated  weak (anti) localization corrections remain as expected even at the
Dirac point.  However, \citet{kn:rycerz2007} found enhanced
conductance fluctuations at the Dirac point, a possible consequence of 
being in the ballistic to diffusive crossover regime.

As for the experimental situation, in addition to the observation 
of suppressed localization~\cite{kn:morozov2006} and 
anti-localization~\cite{kn:wu2007}, \citet{kn:horsell2009}
made a systematic study of several samples fitting the data
to Eq.~\ref{SA_Eq:WL} to extract $\tau_\phi$, $\tau_i$ and $\tau_z$.
Their data show
a mixture of localization, anti-localization and saturation behavior.
An interesting feature is that the intervalley scattering length $L_i = 
(D \tau_i)^{1/2}$ is strongly correlated to the sample width (i.e. 
$L_i \approx W/2$) implying that the edges are the dominant source
of intervalley scattering.  This feature has been corroborated
by Raman studies that show a strong $D$-peak at the edges, but not
in the bulk~\cite{ISI:000244206500004,kn:chen2009}.  The most 
important finding of \citet{kn:horsell2009} is that the contribution 
from short-range scattering ($\tau_z$) is much larger than one would 
expect (indeed, comparable to $\tau_0$).  The authors showed that
any predicted microscopic mechanisms such as ripples or 
trigonal warping that might contribute to $\tau_z$ (but not $\tau_i$)
were all negligible and could not explain such a large $\tau_z^{-1}$.  It remains 
an open question as to why $\tau_z^{-1} \gg \tau_i^{-1}$ in the experiments.

\subsubsection{Ultraviolet logarithmic corrections}
\label{SA_Sec_uvlc}
The semiclassical Boltzmann transport theory treats the impurities 
within the first Born approximation.  In 
a diagrammatic perturbation theory, this is the leading order term in 
an expansion of $n_{\rm imp} \rightarrow 0$.  Typically
for other conventional metals and semiconductors, one makes a
better approximation by trying to include more diagrams that capture
multiple scattering off the same impurity.  For example, in the 
Self-Consistent Born Approximation (SCBA) one replaces bare 
Green's functions with dressed ones to obtain a self-consistent 
equation for the self energy~\cite{kn:bruus2004}.

In practice, for graphene, one often finds that attempts to go beyond
the semiclassical Boltzmann transport theory described in
Sec.~\ref{subsec_boltzmann} fare far worse than the simple theory.  The
theoretical underpinnings for the failure of SCBA was pointed out by
\citet{kn:aleiner2006} where they argued that the SCBA (a
standard technique for weakly disordered metals and superconductors),
is not justified for the Dirac Hamiltonian.  They demonstrated this
by calculating terms to fourth order in perturbation theory, showing 
that SCBA neglects most terms of equal order.  This could have
severe consequences.  For example, considering only diagonal 
disorder, 
the SCBA breaks time reversal symmetry.  To further illustrate their point,
\citet{kn:aleiner2006} argued that for the full disorder
Hamiltonian (Eq.~\ref{SA_Eq_GeneralH}), considering three impurity
scattering, there are 54 terms to that order, and only 6 are captured
by the SCBA.

These terms provide a new divergence in the diagrammatic perturbation
series that is distinct from the weak localization discussed 
in Sec.~\ref{Sec:WL_UCF}.  Unlike weak localization that for 2D 
systems diverges as the size ($\delta \sigma \sim \ln[L/\ell]$),
this additional divergence occurs at all length scales, and was called
``ultraviolet logarithmic corrections''.  The consequences of
this divergence include the 
logarithmic renormalization of the bare disorder parameters which 
was studied using the Renormalization Group (RG) 
in \citet{kn:foster2008}.  For the experimentally relevant case of 
strong diagonal disorder, the renormalization does not change the
physics.  However, when all disorder couplings (i.e. intervalley and
intravalley) are comparable e.g.  relevant for graphene after ion
irradiation, the system could flow to various strong coupling fixed
points depending on the symmetry of the disorder potential.

In addition to these considerations, interaction effects could also
affect quantum transport (e.g. the Altshuler-Aronov phenomena),
particularly in the presence of disorder.  Although such interaction
effects are probably relatively small in monolayer graphene, they may
not be negligible.  Interaction effects may certainly be important in
determining graphene transport properties near the charge neutral
Dirac
point~\cite{kashuba2008,kn:fritz2008,kn:bistritzer2009,kn:muller2008}.

\setcounter{sub3section}{0}


\section{Transport at high carrier density} \label{sec_high_density}

\subsection{Boltzmann transport theory} \label{subsec_boltzmann}

In this section we review graphene transport
for large carrier densities ($n \gg n_i$, $n_i$ being the impurity
density), where the system is homogeneous.
We discuss in detail the microscopic transport
properties at high carrier density using the semiclassical Boltzmann transport
theory.

\begin{figure}
\includegraphics[width=\columnwidth]{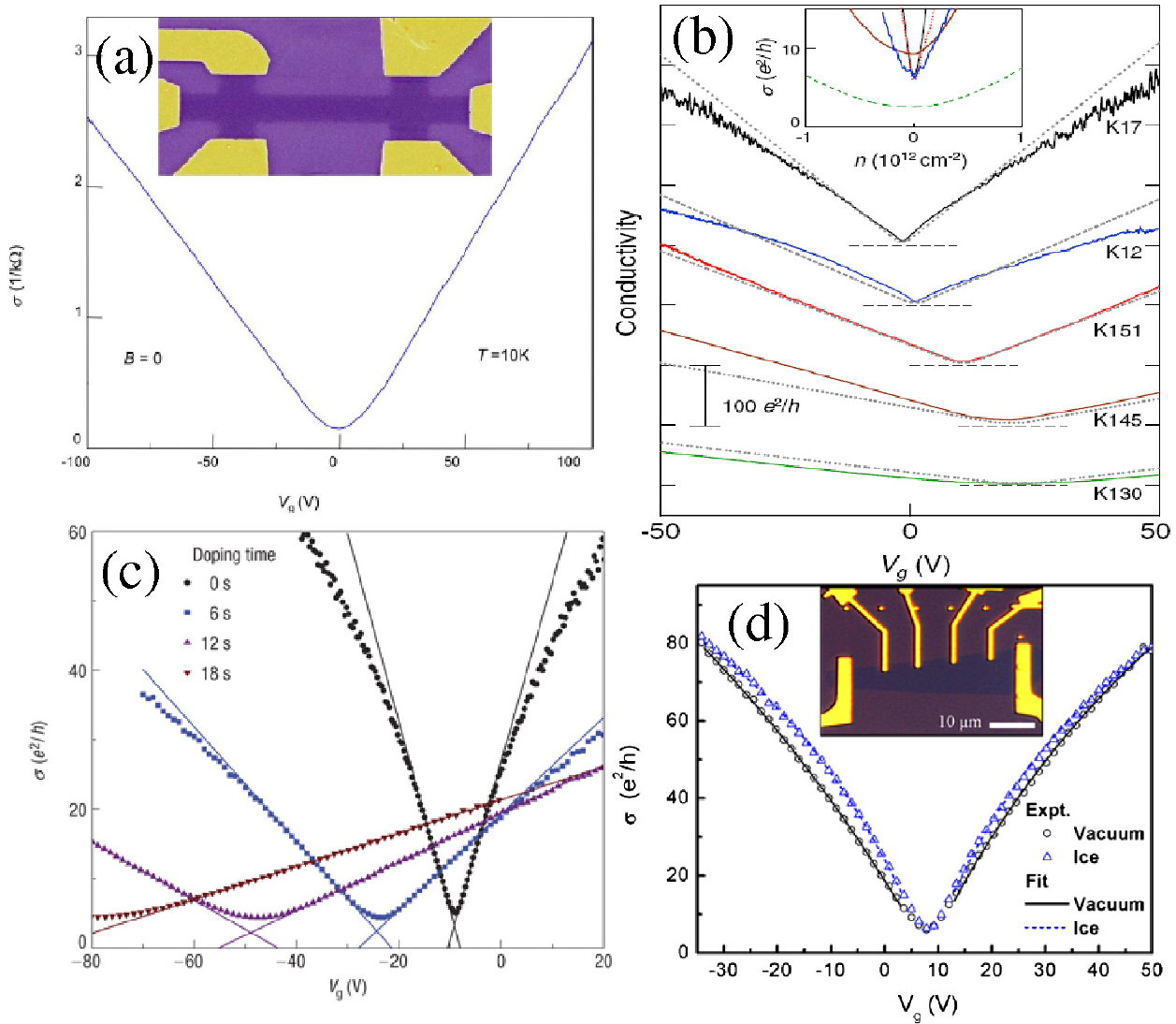}
\caption{\label{fig_high_den_1} 
(color online) (a) The measured conductivity $\sigma$ of graphene as a
  function of 
  gate voltage  $V_g$ (or carrier density). The conductivity increases
  linearly with 
the density. Adapted from \citet{kn:novoselov2005}.
(b) $\sigma$ as a function of $V_g$ for
five different samples  
For clarity, curves are vertically displaced. 
The inset shows the detailed view of the
density-dependent conductivity near the Dirac point for the data
in the main panel. Adapted from \citet{kn:tan2007}.
(c) $\sigma$ versus $V_g$ for the
pristine sample and three different doping concentrations.
Adapted from \citet{kn:chen2008}.
(d) $\sigma$ as a
function $V_g$ for pristine graphene (circles) and
after deposition of 6 monolayers of ice (triangles). 
Inset: Optical microscope image of the
device. Adapted from \citet{kn:jang2008}.
}
\end{figure}

\begin{figure}
\includegraphics[width=\columnwidth]{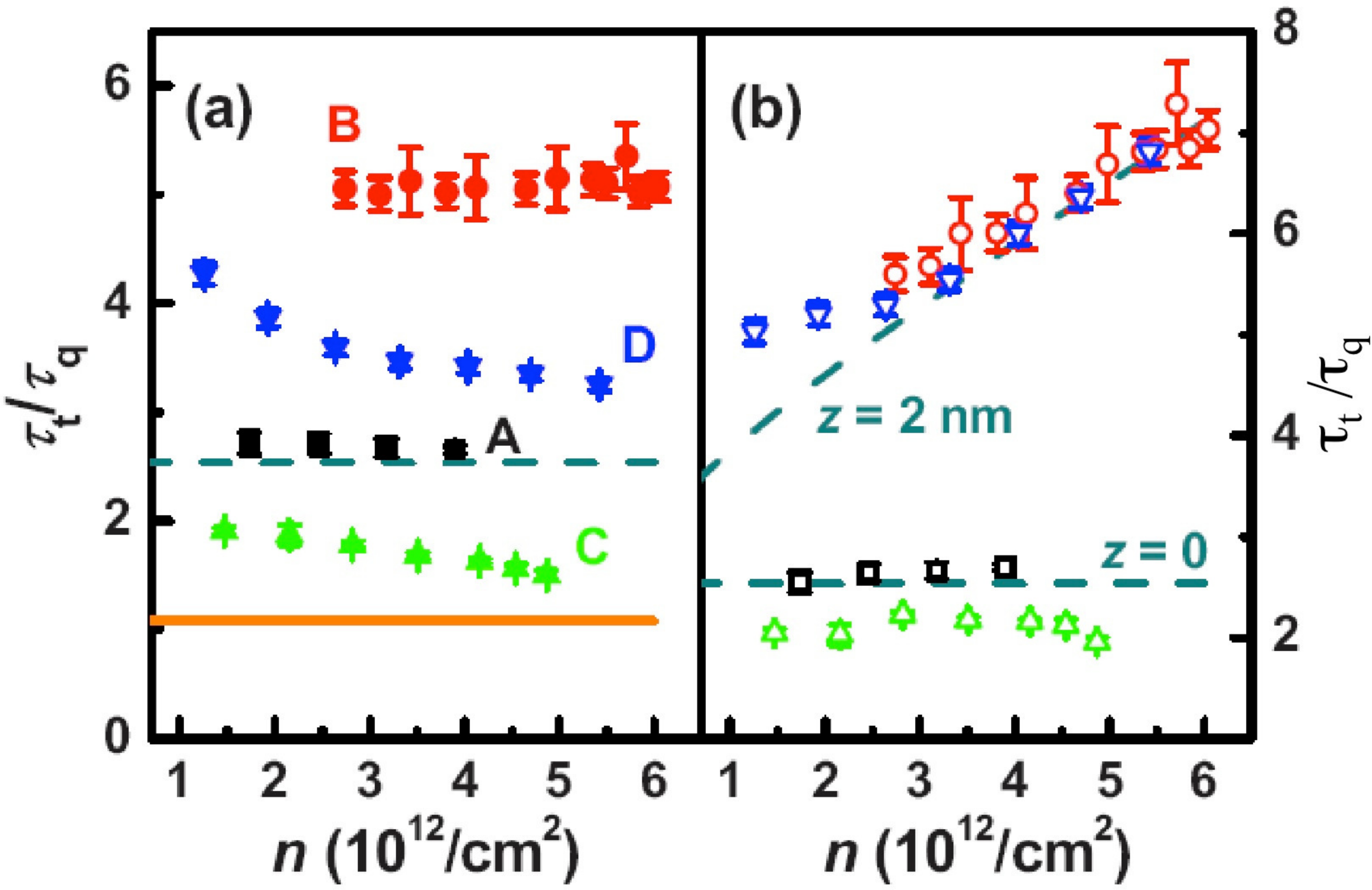}
\caption{(color online) 
The ratio of transport scattering time ($\tau_t$) to quantum
scattering time ($\tau_q$) as a function of density for different
samples. 
Dashed (solid) lines indicate the  theoretical calculations with
Coulomb disorder ($\delta$-range disorder) \citep{kn:hwang2008s}. 
Adapted from \citet{kn:hong2009b}. 
}
\label{fig_hong_2009}
\end{figure}

It was predicted that the graphene conductivity
limited by the short 
ranged scatterers (i.e., $\delta$-range disorder) is independent
of the carrier density,  because of the linear-in-energy density of
states  \citep{kn:shon1998}. However, the  
experiments 
(Fig.~\ref{fig_high_den_1}) show   
that the conductivity increases linearly  in the
carrier density concentration. To explain this linear-in density
dependence of experimental conductivity, the long range Coulomb disorder was
introduced
\citep{kn:ando2006,kn:cheianov2006,kn:nomura2006a,kn:nomura2007,kn:hwang2006c,kn:katsnelson2009,kn:trushin2008}.
The long range Coulomb disorder also successfully explains several
recent transport experiments. 
\citet{kn:tan2007} have found the correlation of the sample mobility
with the shift of the Dirac point and minimum conductivity plateau
width,  showing qualitative
and semi-quantitative agreement with the calculations with long range
Coulomb disorder (Fig.~\ref{fig_high_den_1}(b)).
\citet{kn:chen2008} investigated the effect of
Coulomb scatterers on graphene conductivity
by intentionally adding potassium ions to graphene in ultrahigh
vacuum, observing qualitatively the prediction of the transport theory
limited by Coulomb disorder (Fig.~\ref{fig_high_den_1}(c)). 
\citet{kn:jang2008}
tuned graphene's fine structure constant by depositing ice on the top of
graphene and observed an enhancement in mobility which is predicted in
the Boltzmann theory with Coulomb disorder (Fig. \ref{fig_high_den_1}(d)
and \onlinecite{chen-nl-9-1621-2009,chen-nl-9-2571-2009,kim:062107}).
The role of remote impurity scattering was further confirmed in the 
observation of drastic improvement of mobility by
reducing carrier scattering in suspended graphene through current annealing
\citep{kn:bolotin2008b,kn:du2008}.
Recent measurement of the ratio of transport
scattering time to the quantum scattering time by \citet{kn:hong2009b}
also strongly supports the long range Coulomb disorder as the main
scattering mechanism in graphene (Fig. \ref{fig_hong_2009}) .

The
conductivity $\sigma$ (or mobility $\mu = \sigma/ne$) is calculated in the
presence of randomly distributed Coulomb impurity 
charges  with the electron-impurity interaction being
screened by the 2D electron gas in the random phase approximation
(RPA). Even though the screened Coulomb scattering is the most important
scattering mechanism, there are additional
scattering mechanisms (e.g., neutral point defects) unrelated to the
charged impurity  scattering.
The Boltzmann formalism can treat both effects, where zero
range scatterers are treated with an effective point defect density of 
$n_d$. Phonon scattering effects, important at higher temperatures,
are treated in the next section.
We also discuss other scattering mechanisms which could contribute to
graphene transport.

We start by assuming the system to be a homogeneous 2D carrier system
of electrons (or holes) with a carrier density $n$ induced by the
external gate voltage $V_g$.
When the external electric field is  weak and the displacement of the
distribution function from thermal equilibrium is small, we
may write the distribution function to the lowest order in the applied
electric field ({\bf E}) $f_{\vk} = f(\epsilon_{\vk}) + \delta f_{\vk}$,
where $\epsilon_{\vk}$ is the carrier energy and
$f(\epsilon_{\vk})$ is the equilibrium Fermi distribution function 
and $\delta f_{\vk}$ is proportional to the field.
When the relaxation time approximation is valid, we have
$\delta f_{\vk} = -\frac{\tau(\epsilon_{\vk})}{\hbar}e{\bf
  E\cdot}{\bf v}_{\vk}\frac{\partial f(\epsilon_{\vk})}{\partial
  \epsilon_{\vk}},$
where ${\bf v}_{\vk}=d\epsilon_{\vk}/d\vk$ is the velocity of carrier and
$\tau(\epsilon_{\vk})$ is the relaxation time or the transport 
scattering time, and is given by
\begin{eqnarray}
\frac{1}{\tau(\epsilon_{\vk})} & = & \frac{2\pi}{\hbar}  \sum_{a} \int
dz n_i^{(a)}(z) \int \frac{d^2
  k'}{(2\pi)^2} \nonumber \\ 
&\times& |\langle V_{\vk,\vk'}(z)\rangle |^2  
[1-\cos\theta_{\vk\vk'}]
\delta\left (\epsilon_{\vk} - \epsilon_{\vk'} \right ),
\label{tau_ek}
\end{eqnarray}
where $\theta_{{\bf kk}'}$ is the scattering angle between the scattering in-
and out- wave vectors ${\bf k}$ and ${\bf k}'$, $n_i^{(a)}(z)$ is the
concentration of the $a$-th kind of impurity, and $z$ represents the
coordinate of normal direction to the 2D plane.
In Eq.~(\ref{tau_ek}) $\langle V_{\vk,\vk'}(z) \rangle$ is the matrix
element of the 
scattering potential associated with impurity disorder in the
system environment.
Within Boltzmann transport theory by averaging over energy
we obtain the conductivity
\begin{equation}
\sigma = \frac{e^2}{2} \int d\epsilon D(\epsilon) {\bf v}_{\bf
  k}^2\tau(\epsilon) \left ( 
  -\frac{\partial f}{\partial \epsilon} \right ),
\label{sigma}
\end{equation}
and the corresponding temperature dependent resistivity is given by
$\rho(T) = 1/\sigma(T)$. 
Note that 
$f(\epsilon_k) =\{ 1+\exp[(\epsilon_k-\mu)]/k_B T \}^{-1}$ 
where the finite temperature chemical potential, $\mu(T)$, is
determined  
self-consistently to conserve the total number of electrons. At $T=0$,
$f(\epsilon)$ is a step function at the Fermi energy $E_F \equiv
\mu(T=0)$, and we then recover the usual conductivity formula: 
\begin{equation}
\sigma
= \frac{e^2 v_F^2}{2} D(E_F)\tau(E_F),
\label{eq:sigma}
\end{equation}
where $v_F$ is the carrier velocity at the Fermi energy.

\subsection{Impurity scattering} \label{subsec_impurity}

The matrix element of the scattering potential is determined by the
configuration of the 2D systems and the spatial distribution of the
impurities. In general, impurities are located in the environment of the
2D systems. But for simplicity we consider the impurities are
distributed completely at random in the plane parallel to the 2D
systems located at $z=d$. The location `$d$' is a single parameter
modeling the impurity configuration.
Then the matrix element of the scattering potential
of randomly distributed screened impurity charge centers
is given by 
\begin{equation}
\int dz n_i^{(a)}(z)|\langle V_{\vk,\vk'}(z) \rangle |^2 = n_i\left |
  \frac{v_i(q)}{\varepsilon(q)} \right |^2 F(q)
\label{vkk}
\end{equation} 
where $q = |{\bf k} - {\bf k}'|$, $\theta \equiv \theta_{\vk \vk'}$,
$n_i$ the number of impurities per unit area,
$F(q)$ is the form factor associated with the carrier wave function of
the 2D system, and 
$v_i(q)=2\pi e^2/(\kappa q) e^{-qd}$ is the Fourier transform 
of the 2D Coulomb potential in an effective background lattice
dielectric constant $\kappa$. 
The form factor $F(q)$ in Eq.~(\ref{vkk}) comes from the overlap of
the wave function. In 2D semiconductor systems it is related to the
quasi-2D nature of systems, i.e. finite width of the 2D systems. The
real functional form depends on the details of the quantum
structures (i.e. heterostructures, square well, etc.). $F(q)$ becomes
unity in the two dimensional limit (i.e. $\delta$-layer). However, in
graphene the form factor is related to the chirality, not to the
quantum structure  since graphene is strictly a 2D layer.
In Eq.~(\ref{vkk}),
$\varepsilon(q)\equiv \varepsilon(q,T)$ is the 2D finite temperature
static RPA dielectric (screening) function appropriate 
and is given by 
\begin{equation}
\varepsilon(q,T) = 1 + v_c(q) \Pi(q,T),
\label{eq:screening}
\end{equation}
where $\Pi(q,T)$ is the irreducible
finite-temperature polarizability function and $v_c(q)$ is the Coulomb
interaction. 
For short-ranged disorder we have
\begin{equation}
\int dz n_i^{(a)}(z)|\langle V_{\vk,\vk'}(z) \rangle |^2 = 
n_d V_0^2 F(q),
\label{vkk_s}
\end{equation} 
where $n_d$ the 2D impurity density and
$V_0$ a constant short-range (i.e. a
$\delta$-function in real space) potential strength.

One can also consider the effect on carrier transport by scattering
from cluster of correlated charged impurities
\citep{kn:katsnelson2009}, as was originally done for 2D
semiconductors by Kawamura and Das Sarma
\citep{kn:kawamura1996,kn:dassarma1998}. Without any detailed knowledge
about the 
clustering correlations, however, this is little more than arbitrary
data fitting.

Because the screening effect is known to be of vital importance
for charged impurities \citep{kn:ando2006,kn:hwang2006c}, we first provide
the static polarizability function. It is known that the screening
has to be considered to explain the density and temperature dependence
of the conductivity of 2D semiconductor systems
\citep{kn:dassarma1999,kn:dassarma2005} and
the screening property in graphene exhibits 
significantly different behavior \citep{kn:hwang2006b} from that in
conventional 2D metals.
Also significant temperature dependence of the scattering time $\tau$
may arise from the screening function in Eq. (\ref{eq:screening}). Thus, 
before we discuss the details of  conductivity we first
review  screening in graphen and in 2D semiconductor systems.

\subsubsection{Screening and polarizability}

\sub3section{\label{sec:slg_pol}graphene}

The polarizability
is given by the bare bubble diagram
\citep{kn:ando2006,kn:hwang2006b,kn:wunsch2006} 
\begin{equation}
\Pi(q,T) =-\frac{g}{A}
\sum_{{\bf k}ss'}\frac{f_{s{\bf k}}-f_{s'{\bf k}'}}
{\varepsilon_{s{\bf k}}-\varepsilon_{s'{\bf k}'} }F_{ss'}({\bf
  k},{\bf k}'),
\label{eq:pol}
\end{equation}
where $s=\pm 1$ indicate the conduction ($+1$) and valence ($-1$) bands,
respectively,  ${\bf k}'={\bf k} + {\bf q}$, $\varepsilon_{s{\bf k}}=s\hbar v_F
|{\bf k}|$, $F_{ss'}({\bf k},{\bf k}')=(1+\cos\theta)/2$, and
$f_{s{\bf k}} = [\exp \{\beta(\varepsilon_{s{\bf k}}-\mu)\} +
1]^{-1}$ with $\beta = 1/k_B T$.

\begin{figure*}
\epsfysize=3.7in
\epsffile{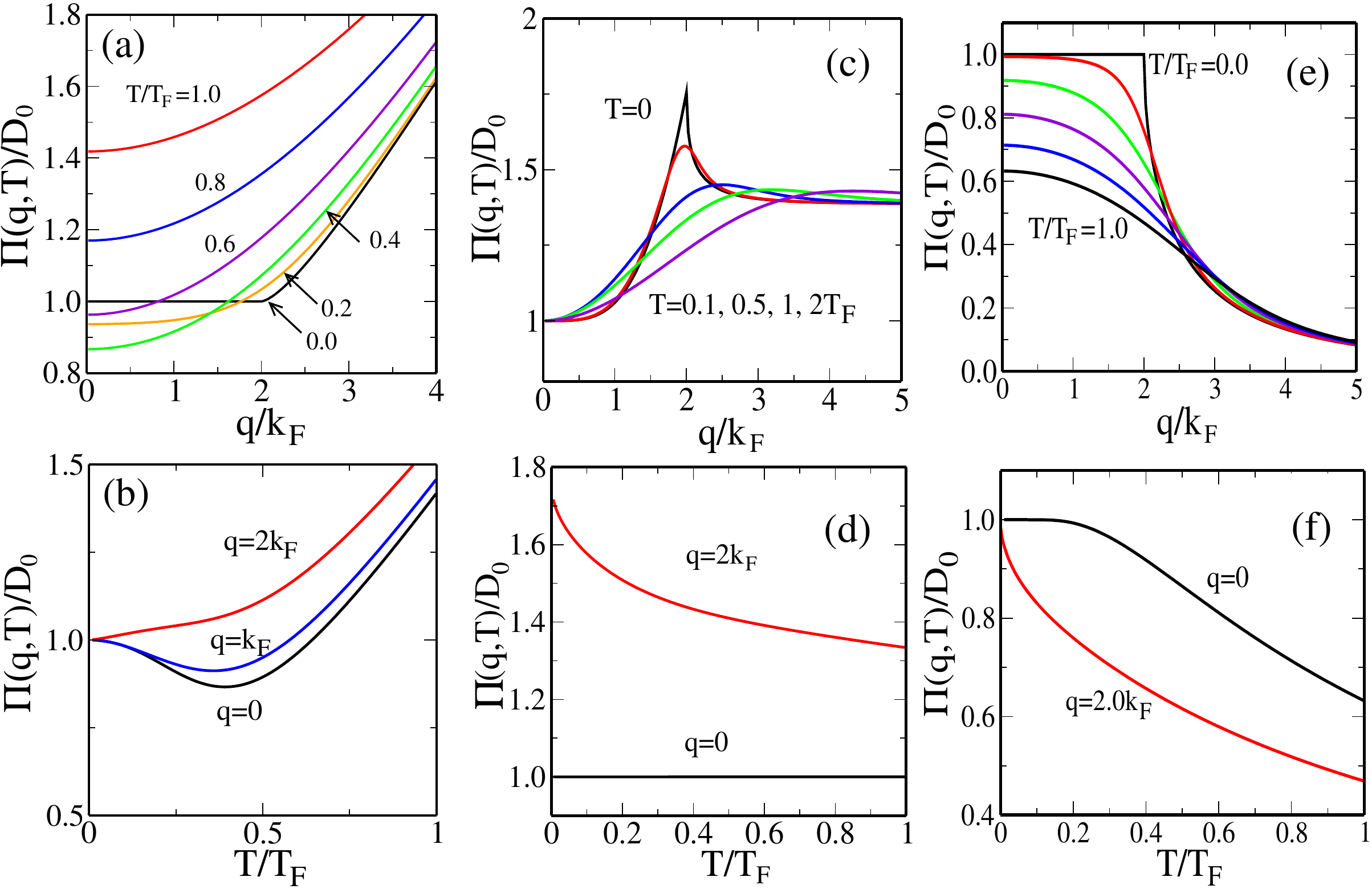}
\caption{(color online) 
Polarizability $\Pi(q,T)$ in units of the
density of states at the Fermi level $D_0$.
(a) and (b) show $\Pi(q,T)$ of monolayer graphene (a) as a
  function of wave  
vector for different temperatures 
and (b) as a function of 
temperature for different wave vectors.
(c) and (d) show BLG polarizability.
(e) and (f) show the 2DEG polarizability.
}
\label{fig_pol}
\end{figure*}

After performing the summation over $ss'$ 
it is useful to rewrite the polarizability as the sum of intraband and
interband polarizaibility
$\Pi(q,T) = \Pi^+(q,T) + \Pi^-(q,T)$.
At $T=0$, the intraband ($\Pi^+$) and interband
($\Pi^-$) polarizability becomes \citep{kn:ando2006,kn:hwang2006b} 
\begin{subequations}
\begin{equation}
\tilde{\Pi}^+(q)  = \left \{
\begin{matrix}
1 - \frac{\pi q}{8k_F}, & q \le 2k_F \cr
1-\frac{1}{2}\sqrt{1-\frac{4k_F^2}{q^2}} -
\frac{q}{4k_F}\sin^{-1}\frac{2k_F}{q} , &
q > 2k_F  \cr
\end{matrix} 
\right .
\end{equation}
\begin{equation}
\tilde{\Pi}^-(q) = \frac{\pi q}{8k_F},
\end{equation}
\end{subequations}
where
$\tilde{\Pi}^{\pm}=\Pi^{\pm}/D_0$, where $D_0 \equiv g E_F/2\pi
\hbar^2 v_F^2$ is 
the DOS at Fermi level.
Intraband $\Pi^+$ (interband $\Pi^-$) polarizability decreases (increases) 
linearly as q increases and these two effects exactly cancel out up to
$q=2k_F$, which gives rise to 
the total static polarizability being constant for $q<2k_F$ as in the
2DEG \citep{kn:stern1967}, i.e. $\Pi(q) =\Pi^+(q) + \Pi^-(q) = D(E_F)$ for
$q \le 2k_F$.  
In Fig. \ref{fig_pol} we show the calculated graphene static polarizability
as a function of wave vector. 
In the large momentum transfer regime, $q > 2k_F$, the static screening
increases linearly with $q$
due to the interband transition.
In a normal 2D system  the static polarizability falls
off rapidly for $q >2k_F$ with a cusp at $q=2k_F$ \citep{kn:stern1967}. 
The linear increase of the
static polarizability with $q$ gives rise to 
an enhancement of the effective dielectric constant
$\kappa^* (q \rightarrow \infty)= \kappa (1 + g_sg_v\pi r_s/8)$ in
graphene.  Note that in a normal 2D system
$\kappa^* \rightarrow \kappa$ as $q \rightarrow \infty$. Thus, the
effective interaction in 2D
graphene decreases at short wave lengths due to interband polarization
effects. This large wave vector screening behavior is typical of an
insulator. Thus, 2D graphene screening is a combination of
``metallic'' screening (due to $\Pi^{+}$) and ``insulating''
screening (due to $\Pi^-$), leading, overall, to rather strange
screening property, all of which can be traced back to the zero-gap
chiral relativistic nature of graphene.


It is interesting to note that
the non-analytic behavior of graphene polarizability at $q=2k_F$ occurs
in the second derivative, $d^2\Pi(q)/dq^2 \propto
1/\sqrt{q^2-4k_F^2}$, i.e., the total polarizability, as well as its first
derivative are continuous at $q=2k_F$. 
This leads to an oscillatory decay of the screened potential in the
real space (Friedel oscillation) which goes as 
$\phi(r) \sim \cos(2k_Fr)/r^3$ \citep{kn:cheianov2006,kn:wunsch2006}.
This is in contrast to the behavior of a 2DEG, where Friedel
oscillations scale like $\phi(r) \sim  \cos (2k_Fr)/r^2$. 
The polarizability also determines the  RKKY interaction 
between two magnetic impurities as well as the induced spin density
due to a magnetic 
impurity, both quantities being proportional to the Fourier transform
of $\Pi(q)$. 
Like for the screened potential, 
the induced spin density decreases  as $r^{-3}$ for large
distances. Again, this contrasts 
with the $r^{-2}$ behavior found in a 2DEG. For the particular case
of intrinsic graphene 
the Fourier transform of interband polarizability ($\Pi^-(q)$) diverges
[even though $\Pi(r)$ formally scales as $r^{-3}$, its
magnitude does not converge], which means that intrinsic
graphene is susceptible to ferromagnetic ordering in the presence
of magnetic impurities due to the divergent
RKKY coupling \citep{kn:brey2007}.


Since the explicit temperature dependence of screening gives rise
to significant temperature dependence of the conductivity,
we consider the properties of the polarizability at finite temperatures.
The asymptotic form of polarizability is given by 
\begin{subequations}
\begin{eqnarray}
\tilde{\Pi}(q,T \gg T_F)  & \approx & \frac{T}{T_F} \ln4  +
\frac{q^2}{24k_F^2}\frac{T_F}{T}, \\
\tilde{\Pi}(q,T \ll T_F) & \approx & \frac{\mu(T)}{E_F} = 1 -
\frac{\pi^2}{6} \left ( \frac{T}{T_F}  \right )^2,
\end{eqnarray}
\end{subequations}
where $T_F = E_F/k_B$ is the Fermi temperature.
In addition, the finite temperature Thomas-Fermi wave vector in the
$q\rightarrow 0$ long wavelength limit is given by
\citep{kn:ando2006,kn:hwang2008b} 
\begin{subequations}
\begin{eqnarray}
q_s(T \gg T_F) & \approx & 8\ln(2)r_sk_F \left ( \frac{T}{T_F} \right ) \\
q_s(T \ll T_F) & \approx & 4r_s k_F\left [ 1-\frac{\pi^2}{6}\left (
  \frac{T}{T_F}  \right )^2 
 \right ].
\end{eqnarray}
\label{eq:slg_tf}
\end{subequations}
The screening wave vector increases linearly with temperature at high
temperatures ($T \gg T_F$), but becomes a constant with a small
quadratic correction at low temperatures ($T \ll T_F$). 
In Fig. \ref{fig_pol}  we show the finite temperature polarizability
$\Pi(q,T)$.

\sub3section{bilayer graphene}

For bilayer graphene we have the polarizability of Eq. (\ref{eq:pol})
with  $\varepsilon_{s{\bf k}}=s {\bf k}^2/2m$ and $F_{ss'}({\bf
  k},{\bf k}') = (1+ss'\cos2\theta)/2$ due to the chirality of bilayer
graphene. 
At $T=0$ the
polarizability of bilayer graphene \citep{kn:hwang2008c} is given by
\begin{equation}
\Pi(q) = D_0 \left [ f(q) - g(q) \theta(q-2k_F) \right ],
\label{blg_piq}
\end{equation}
where $D_0=g_s g_v m/2\pi \hbar^2$ is the BLG density of states at the
Fermi level 
and
\begin{subequations}
\begin{eqnarray}
f(q) & = &
(1+\frac{\tilde{q}^2}{2})\sqrt{1-\frac{4}{\tilde{q}^2}} + \log
\frac{\tilde{q}-\sqrt{\tilde{q}^2-4}}{\tilde{q}+\sqrt{\tilde{q}^2-4^2}}  \\
g(q) & = & \frac{1}{2}\sqrt{4+\tilde{q}^4} -\log \left [
  \frac{1+\sqrt{1+\tilde{q}^4/4}}{2} \right ],
\end{eqnarray}
\end{subequations}
where $\tilde{q}=q/k_F$.

In Fig. \ref{fig_pol} the wave vector
dependent BLG polarizability is shown.
For MLG intraband and interband effects in polarizability 
exactly cancel out up to $q=2k_F$, 
which gives rise to 
the total static polarizability being constant for $q<2k_F$.
However,
for BLG the cancellation of two
polarizability functions is not exact
because of the enhanced backscattering, so the
total polarizability increases as $q$ approaches  $2k_F$, which
means screening increases as $q$ increases.
Thus BLG, in spite of having the same parabolic carrier energy
dispersion of 2DEG 
systems, does not have a constant 
Thomas-Fermi screening up to $q=2k_F$ \citep{kn:borghi2009,kn:hwang2008c}
as exists in MLG and 2DEG. 
In the large momentum transfer regime, $q > 2k_F$, the BLG
polarizability 
approaches a constant value, i.e.,
$\Pi(q) \rightarrow  N_0 \log 4$, 
because the interband transition dominates over the intraband contribution
in the large wave vector limit. 
For $q >2k_F$ the static polarizability falls
off rapidly ($\sim 1/q^2$) for 2DEG \citep{kn:stern1967} 
and for MLG it increases linearly with $q$ (see Sec. \ref{sec:slg_pol}).

The long-wavelength ($q\rightarrow 0$) Thomas-Fermi screening can be
expressed as $q_{TF}=g_s g_v m e^2 /\kappa \hbar^2$,
which is the same form as a regular 2D system and 
independent of electron concentration.
The screening at $q=2k_F$ is given by
$q_{s}(2k_F)=q_{TF}{\kappa \hbar^2} \left [ \sqrt{5}-\log
  \left \{ (1+\sqrt{5})/2 \right \} \right ]$.
Screening at $q=2k_F$ is about 75~\% larger than normal 2D TF
screening, which indicates that 
in bilayer graphene the scattering by screened Coulomb
potential is  much reduced due to the enhanced screening.

A qualitative difference between MLG and BLG polarizability functions
is at $q=2k_F$. 
Due to the suppression of $2k_F$ backward
scattering in MLG, the total polarizability as well as its first
derivative are continuous. 
In BLG, however, the large angle scattering is 
enhanced due to chirality [i.e. the overlap factor $F_{ss'}$ in
Eq.~(\ref{eq:pol})], which gives rise to the 
singular behavior of polarizability at $q=2k_F$. Even though the
BLG polarizability is continuous at $q=2k_F$,  it has a sharp cusp
and its derivative is discontinuous at $2k_F$.
As $q \rightarrow 2k_F$, $d\Pi(q)/dq \propto
1/\sqrt{q^2-4k_F^2}$. This behavior is exactly the same as that of
the regular 2DEG, which also has a cusp at $q=2k_F$.
The strong cusp in BLG $\Pi(q)$ at $q=2k_F$ leads to Friedel oscillations
in contrast to the MLG behavior.
The leading oscillation term in the screened potential at large
distances can be calculated as 
\begin{equation}
\phi(r) \sim -\frac{e}{\kappa}\frac{4q_{TF}k_F^2}{(2k_F+Cq_{TF})^2}
\frac{\sin(2k_Fr)}{(2k_Fr)^2},
\end{equation}
where $C=\sqrt{5}-\log[(1+\sqrt{5})/2]$, which is similar to the 2DEG 
except for the additional constant $C$ ($C=1$ for  2DEG), but 
different from 
MLG where Friedel oscillations scale as
$\phi(r) \sim \cos(2k_Fr)/r^3$ \citep{kn:cheianov2006,kn:wunsch2006}.


The enhanced singular behavior of the BLG screening function at $q=2k_F$
has other interesting 
consequences related to Kohn anomaly \citep{kn:kohn1959} and
RKKY interaction.
For intrinsic BLG the Fourier transform of
$\Pi(q)$ simply becomes a $\delta$-function, which
indicates that the
localized magnetic moments are not correlated by the long
range interaction and there is no net magnetic moment. 
For extrinsic BLG, the oscillatory term in RKKY
interaction is restored due to the singularity of polarizability
at $q =2k_F$, and the oscillating behavior dominates at
large $k_Fr$. At large distances $2k_Fr \gg 1$, the dominant
oscillating term in $\Pi(r)$ is given by
$\Pi(r) \propto \sin(2k_Fr)/(k_Fr)^2$.
This is the same RKKY interaction as in a regular 2DEG.

In Fig. \ref{fig_pol} the wave vector
dependent BLG polarizability is shown for different temperatures.
Note that at $q=0$, $\Pi(0,T) = N_F$ for all temperatures.
For small $q$, $\Pi(q,T)$ increases as $q^4$.
The asymptotic form of
polarizability  becomes
\begin{subequations}
\begin{eqnarray}
\tilde{\Pi}(q,T \gg T_F) & \approx &  1+
\frac{q^2}{6k_F^2}\frac{T_F}{T}, \\
\tilde{\Pi}(q,T \ll T_F) & \approx & 1+\frac{1}{16}\frac{q^4}{k_F^4}+\frac{\pi^2}{16}
\left ( \frac{T}{T_F} \right )^2 \frac{q^4}{k_F^4}.
\end{eqnarray}
\end{subequations}
More interestingly, the polarizability at $q=0$ is temperature
independent, i.e., the finite temperature Thomas-Fermi wave vector is
constant for all temperatures,
\begin{equation}
q_s(T) = q_{TF}.
\label{eq:blg_tf}
\end{equation}
In BLG polarizability at $q=0$ two temperature effects from
the intraband  and the interband transition exactly cancel out, which
gives rise to  
the total static polarizability at $q=0$ being constant for all temperatures.

\sub3section{2D semiconductor systems}

The polarizability of ordinary 2D system was first calculated by Stern
and all 
details can be found in the literature
\citep{kn:ando1982,kn:stern1967}. Here 
we provide the 2D polarizability for comparison with graphene.
The 2D polarizability can be calculated with
$\epsilon_{s\vk}=\hbar^2 \vk^2/2m$ and
$F_{ss'}=\delta_{ss'}/2$ because of the non-chiral property of the
ordinary 2D systems. $\Pi(q)$ at $T=0$ becomes \citep{kn:stern1967}
\begin{equation}
\Pi(q)  = D_0\left [ 1 - \sqrt{1-\left ( {2k_F}/{q} \right )^2}
  \theta(q-2k_F)        \right ],
\label{2D_piq}
\end{equation}
where $D_0=g_sg_vm/2\pi\hbar^2$ is the 2D density of states at the Fermi level.
Since the polarizability is a constant for $q<2k_F$ both the long wave
length TF screening and $2k_F$ screening are same, which is given by 
$q_s = q_{TF}=g_s g_v m e^2 /\kappa \hbar^2$.

The asymptotic form for the regular 2D polarizability are given by
\begin{subequations}
\begin{eqnarray}
\tilde{\Pi} (q=0,T \ll T_F) & \approx & 1- e^{-T_F/T}, \\
\tilde{\Pi}(q,T \gg T_F)  & \approx &  \frac{T_F}{T} \left [1 -
  \frac{q^2}{6k_F^2} 
  \frac{T_F}{T} \right ].
\end{eqnarray}
\end{subequations}
For $q=0$, in the $T \gg T_F$ limit, we get the usual Debye
screening for the regular 2D electron gas system
\begin{equation}
q_s(T \gg T_F) \approx q_{TF}\frac{T_F}{T}.
\label{eq:2D_tf}
\end{equation}
A comparison of Eq.~(\ref{eq:2D_tf}) with  Eqs.~(\ref{eq:slg_tf}) and
(\ref{eq:blg_tf}) shows that the high-temperature Debye
screening behaviors are different in all three systems
just as the low-temperature screening behaviors, i.e.,  the high
temperature screening wave vector $q_s$ in
semiconductor 2D systems decreases linearly with temperature while $q_s$
in MLG increases linearly with temperature and $q_s$ in BLG is
independent of temperature.

In Fig.~\ref{fig_pol} we show the
corresponding parabolic 2D polarizability normalized by the density of
states at Fermi level, $D_0 = g m/\hbar^2 2\pi$. 
Note that the temperature
dependence of 2D polarizability  at $q=2k_F$ is much stronger
than that of graphene polarizability. 
Since in normal 2D systems the $2k_F$ scattering event 
is most important for the electrical resistivity, 
the temperature dependence of polarizability at $q=2k_F$ completely
dominates at low temperatures ($T \ll T_F$).
It is known that the strong temperature
dependence of the polarizability function at $q=2k_F$
leads to the anomalously strong temperature
dependent resistivity in ordinary 2D systems
\citep{kn:stern1980,kn:dassarma1999}.  

In the next section the temperature-dependent 
conductivities  are provided due to the scattering by
screened Coulomb impurities 
using the temperature dependent screening properties of
this section.

\subsubsection{Conductivity}
\setcounter{sub3section}{0}

\sub3section{Single layer graphene}

The eigenstates of single layer graphene are given
by the plane wave 
$\psi_{s\vk}({\bf r})=\frac{1}{\sqrt{A}}\exp(i\vk \cdot {\bf r})
F_{s\vk}$,  
where $A$ is the area of the system, $s = \pm 1$ indicate the
conduction ($+1$) and valence ($-1$) bands, respectively, and
$F_{s\vk}^{\dagger} = \frac{1}{\sqrt{2}}(e^{i\theta_{\vk}},s)$
with $\theta_{\vk} = \tan(k_y/k_x)$ being the polar angle of the
momentum $\vk$.
The corresponding energy of graphene for
2D wave vector $\vk$ is given by 
$\epsilon_{s\vk}= s \hbar v_F |\vk|$, and
the density of states (DOS) is given
by $D(\epsilon) = g|\epsilon|/(2\pi \hbar^2 v_F^2)$, where $g=g_s g_v$
is the total degeneracy ($g_s = 2, g_v = 2$ being the spin and valley
degeneracies, respectively). 
The corresponding form factor $F(q)$ in the matrix elements of
Eqs.~(\ref{vkk}) and (\ref{vkk_s})
arising from the sublattice symmetry (overlap of wave
function) \citep{kn:ando2006,auslender-prb-76-235425-2007} becomes
$F(q) = (1+\cos\theta)/2$,
where $q = |{\bf k} - {\bf k}'|$, $\theta \equiv \theta_{\vk \vk'}$.
The matrix element of the scattering potential
of randomly distributed screened impurity charge centers in graphene
is given by 
\begin{equation}
|\langle V_{s\vk,s\vk'} \rangle |^2 = \left |
  \frac{v_i(q)}{\varepsilon(q)} \right |^2 \frac{1+\cos \theta}{2},
\label{vkk_slg}
\end{equation} 
and the matrix element of the 
short-ranged disorder is
\begin{equation}
|\langle V_{s\vk,s\vk'} \rangle |^2 = V_0^2 (1+\cos \theta)/{2},
\label{vkk_slg0}
\end{equation} 
where $V_0$ is the strength of the short-ranged disorder potential measured in eVm$^2$.
The factor $(1-\cos\theta)$ in Eq.~(\ref{tau_ek}) weights the
amount of backward scattering of the electron by the impurity. 
In normal parabolic 2D systems \citep{kn:ando1982} the factor $(1-\cos\theta)$
favors large angle scattering events.
However, in graphene the large
angle scattering is  suppressed due to the wave function overlap
factor $(1+\cos\theta)$, which arises from the sublattice symmetry
peculiar to graphene. The energy dependent scattering time in graphene
thus gets weighted by an angular
contribution factor of $(1 - \cos \theta)(1+\cos\theta)$, which
suppresses both small-angle 
scattering and large-angle
scattering contributions in the scattering rate.

Assuming random distribution of charged centers
with density $n_{i}$,
the scattering time $\tau$ at $T=0$ is given by
\citep{kn:hwang2008s,kn:adam2007a} 
\begin{equation} 
\frac{1}{\tau} =  \frac{r_s^2}{\tau_{0}} \left \{
\frac{\pi}{2}-4\frac{d}{dr_s}\left [ r_s^2g(2r_s) \right ]
\right \}
\label{eq:frs}
\end{equation}
where 
${\tau_{0}^{-1}} = {2\sqrt{\pi} n_{i} v_F}/{\sqrt{n}}$, 
and $g(x) = -1 + \frac{\pi}{2}x+(1-x^2)f(x)$ with
\begin{eqnarray}
f(x) = \left \{ 
 \begin{array}{cl} \frac{1}{\sqrt{1-x^2}} \cosh^{-1}\frac{1}{x}
                  & \mbox{for  $x < 1$} \\
                   \frac{1}{\sqrt{x^2-1}}\cos^{-1}\frac{1}{x}
                  & \mbox{for $x > 1$} \end{array} 
\right. .
\label{fx}
\end{eqnarray}
Since $r_s$ is independent of the carrier density the scattering time
is simply given by $\tau \propto \sqrt{n}$. With
Eq.~\ref{eq:sigma} we find the density dependence of graphene  conductivity
$\sigma(n) \propto n$ because $D(E_F) 
\propto \sqrt{n}$. For graphene on SiO$_2$ substrate the interaction
parameter $r_s \approx 0.8$, then the conductivity is given by
$\sigma(n) \approx \frac{20 e^2}{h}\frac{n}{n_i}$. \citep{kn:adam2007a}
On the other hand the corresponding energy dependent scattering time
of short-ranged disorder is
\begin{equation}
\label{eq:scattime0}
\frac{1}{\tau}  =  \frac{ n_dV_0^2}{\hbar}
\frac{E_F}{4 (\hbar v_F)^2} .
\end{equation}
Thus, the density dependence of scattering time due to the short-range
disorder scattering is given by $\tau(n) \propto n^{-1/2}$. With
Eq.~\ref{eq:sigma} we find the conductivity to be independent of
density for short-range scattering, i.e., $\sigma(n) \propto n^0$,
in contrast to charged
impurity scattering which produces a conductivity linear
in $n$.

\begin{figure}
\includegraphics[width=1.0\columnwidth]{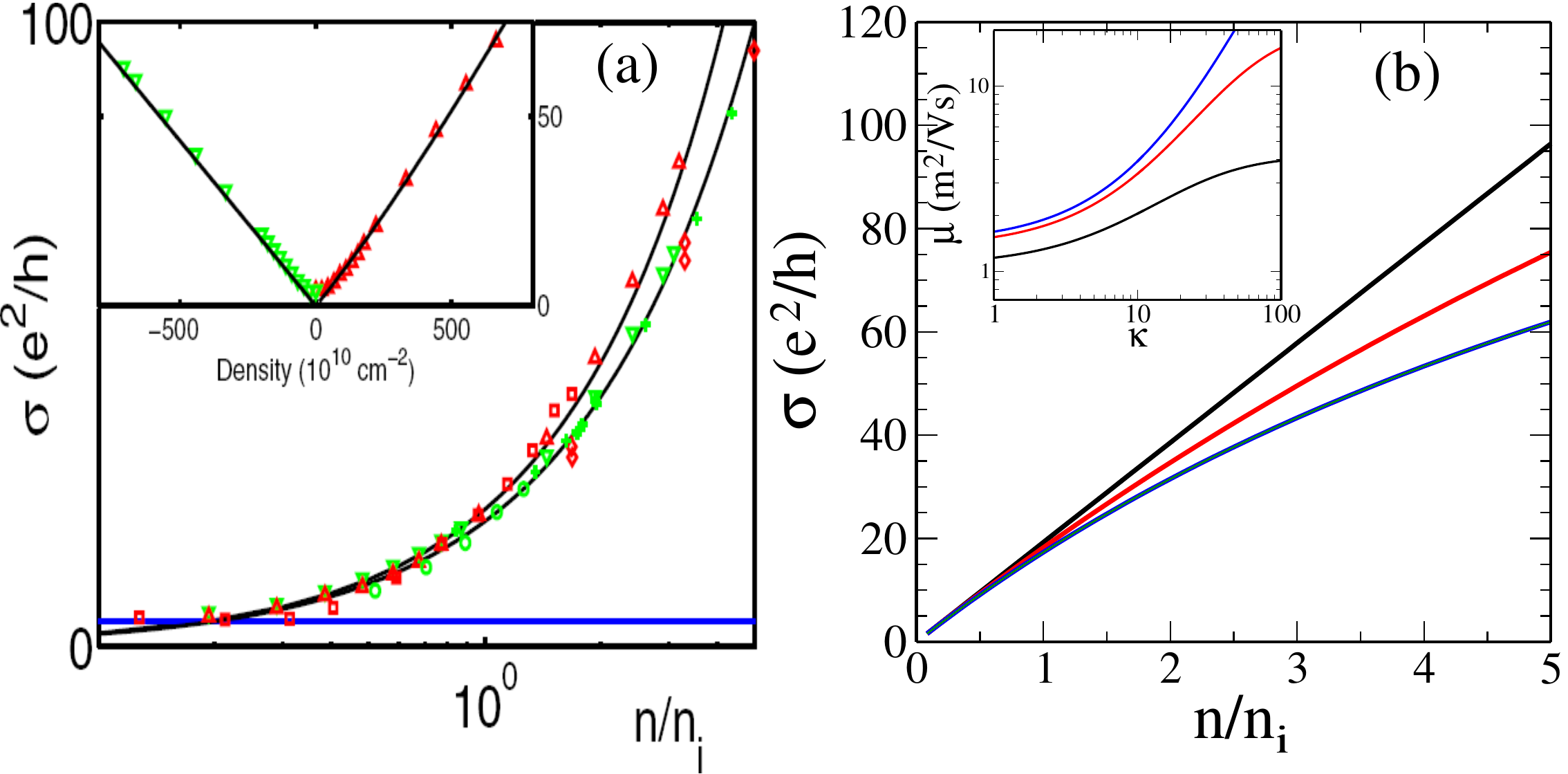}
\caption{\label{fig_hwang_2007}
(color online) (a) Calculated graphene conductivity as a  function of 
carrier density ($n_i$ is an impurity density)
limited by Coulomb scattering with experimental data.
Solid lines (from bottom to top) show the minimum conductivity of
$4e^2/h$, theory for $d=0$, and $d=0.2$ nm. The inset shows the
results in
a linear scale assuming that the impurity shifts by $d=0.2$ nm for
positive voltage bias. 
Adapted from \citet{kn:hwang2006c}.
(b) Graphene conductivity calculated using
a combination of short and long range disorder. 
In the calculation, $n_d/n_i=0$, 0.01, 0.02 (top to bottom) are used.
In inset the graphene mobility as a function of dielectric
constant ($\kappa$) of substrate is shown for different carrier densities
$n = 0.1$, 1, $5\times 10^{12}$ cm$^{-2}$ (from top to bottom) 
in the presence of both long ranged
charged impurity ($n_i = 2\times 10^{11}$ cm$^{-2}$) and
short-ranged neutral impurity ($n_d = 0.4\times 10^{10}$ cm$^{-2}$). 
$V_0=10$ eVnm$^2$ is used in the calculation.
}
\end{figure}

In Fig.~\ref{fig_hwang_2007}(a) the
calculated graphene conductivity limited by screened charged
impurities is shown along with the experimental data
\citep{kn:tan2007,kn:chen2008}.  
In order to get quantitative agreement with
experiment, the screening effect must be included. 
The effect of remote scatterers which are
located at a distance $d$ from the interface is also shown.  
The main effect of
remote impurity scattering is that the conductivity deviates from
the linear behavior with density and
increases with both the distance $d$  and  $n/n_i$ \citep{kn:hwang2006c}.

For very high mobility samples, 
a sub-linear conductivity, instead of
the linear behavior with density, is found in experiments
\citep{kn:tan2007,kn:chen2008}.  Such  
high quality samples presumably
have a small charge impurity concentration $n_i$ and 
it is therefore likely that  short-range disorder plays a more dominant role.  
Fig.~\ref{fig_hwang_2007}(b) shows the graphene conductivity calculated
including both charge impurity  and short range disorder
for different values of $n_d/n_i$. For
small $n_d/n_i$  the conductivity is linear in density, which
is seen in most experiments, and for large $n_d/n_i$ 
the total conductivity  shows the sub-linear behavior. 
This high-density flattening of the graphene conductivity
is a non-universal crossover behavior arising from the competition
between two kinds of scatterers. In general this crossover occurs when
two scattering potentials are equivalent, that is, $n_iV_i^2  \approx
n_d V_0^2$.  
In the inset of Fig.~\ref{fig_hwang_2007}(b) 
the mobility in the presence of both charged 
impurities and short-ranged impurities is shown as a function of $\kappa$.
As the scattering limited by the short-ranged impurity dominates over that
by the long-ranged impurity (e.g. $n_d V_0^2 \gg n_iV_i^2$)   
the mobility is no longer linearly
dependent on the charged impurity and approaches its limiting value
\begin{equation}
\mu = \frac{e}{4\hbar}\frac{(\hbar v_F)^2}{n}\frac{1}{n_{d}V_0^2}.
\end{equation}
The limiting mobility depends only on neutral impurity concentration 
$n_{d}$ and carrier density, i.e. long-range Coulomb scattering is
irrelevant in this high density limit.

\begin{figure}
\bigskip
\epsfxsize=1.\hsize
\hspace{0.0\hsize}
\epsffile{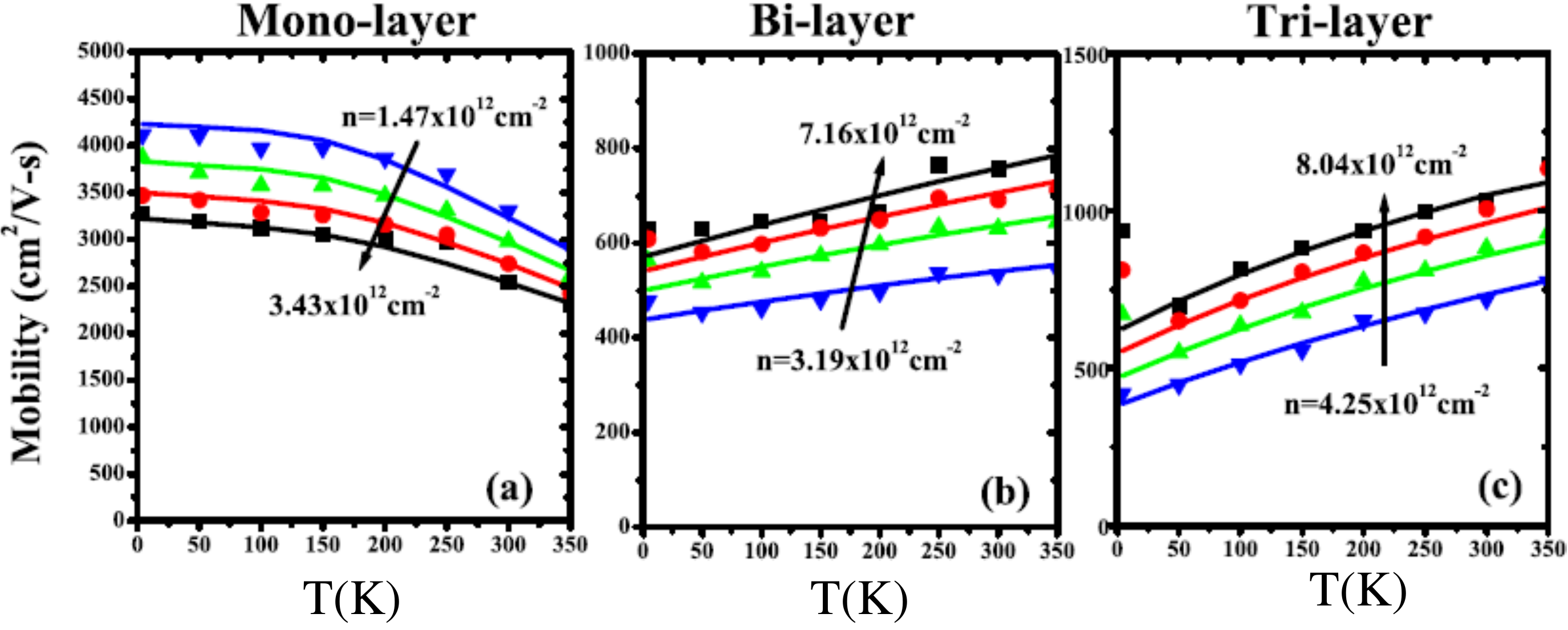}
\caption{\label{fig_zhu_2009} 
(color online) Hall mobility as a function of temperature for different
hole densities in (a) monolayer graphene, (b) 
bilayer graphene, and (c) trilayer graphene. The symbols are the
measured data and the lines are fits. 
Adapted from \citet{kn:zhu2009}.
}
\end{figure}

The temperature dependent conductivity of graphene
arising from screening and  the energy
averaging defined in Eq. (\ref{sigma}) is given
at low temperatures ($T \ll T_F$)
${\sigma(T)}/{\sigma_0}  \approx  1 - C_1 
( {T}/{T_F} )^2$, 
where $\sigma_0 = e^2v_F^2 D(E_F) \tau_0/2$ and
$C_1$ is a positive constant depending only on $r_s$ \citep{kn:hwang2008b}. 
The conductivity decreases quadratically as the temperature
increases and shows typical metallic temperature dependence.
On the other hand, at high temperatures ($T/T_F \gg 1$) it becomes
${\sigma(T)}/{\sigma_0} \approx C_2  ({T}/{T_F}  )^2$,
where $C_2$ is a positive constant. 
The temperature dependent conductivity increases
as the temperature increases in the high temperature regime,
characteristic of an insulating system. 
$\sigma_s(T)$ of graphene due to the
short-range disorder (with scattering strength $V_0$) is given by
$\sigma_{s}(T) = \frac{\sigma_{s0}}{1+e^{-\beta \mu}},$
where $\sigma_{s0} = e^2 v_F^2 D(E_F) \tau_s/2$ with $\tau_s =
\frac{n_d}{4\hbar} E_F V_0^2/(\hbar v_F)^2$.
In the low temperature limit the temperature dependence of
conductivity is exponentially suppressed, but the high temperature
limit of the conductivity
approaches $\sigma_{s0}/2$ as $T \rightarrow \infty$, i.e., the 
resistivity at high temperatures increases up to a factor of two
compared with the low temperature 
limit resistivity.

Recently, the temperature dependence of resistivity of graphene has been
investigated experimentally
\citep{kn:tan2007b,kn:chen2008b,kn:bolotin2008b,kn:zhu2009}. In
Fig.~\ref{fig_zhu_2009}(a) the  graphene mobility is
shown as a function of temperature. An effective
metallic behavior at high density is observed as explained with
screened Coulomb impurities. However, it is not obvious
whether the temperature dependent correction is quadratic because 
phonon scattering also gives rise to a temperature dependence (see
Sec.~\ref{subsec_phonons}).

\sub3section{bilayer graphene}

The eigenstates of bilayer graphene can be written as
$\psi_{s{\bf k}} =e^{i{\bf k r}}
(e^{-2i\theta_{\bf k}},s)/\sqrt{2}$ and the corresponding energy is
given by 
$\epsilon_{s{\bf k}}=s\hbar^2k^2/2m$, where $\theta_{\bf k} = \tan^{-1}(k_y/k_x)$
and $s=\pm 1$ denote the band index. 
The corresponding form factor $F(q)$ of Eqs.~(\ref{vkk}) and
(\ref{vkk_s}) in the matrix elements
arising from the sublattice symmetry of bilayer graphene becomes
$F(q) = (1+\cos2\theta)/2$,
where $q = |{\bf k} - {\bf k}'|$, $\theta \equiv \theta_{\vk \vk'}$.
Then the matrix element of the scattering potential
of randomly distributed screened impurity charge centers in graphene
is given by~\citep{kn:adam2008a,koshino-prb-73-245403-2006,kn:katsnelson2007b,kn:nilsson2006b,kn:nilsson2008}
\begin{equation}
|\langle V_{s\vk,s\vk'} \rangle |^2 = \left |
  {v_i(q)}/{\varepsilon(q)} \right |^2 ({1+\cos 2\theta})/{2}.
\end{equation} 
The matrix element of the short-ranged disorder
is given by 
$|\langle V_{s\vk,s\vk'} \rangle |^2 = V_0^2 (1+\cos 2\theta)/{2}$,
and the corresponding energy dependent scattering time 
becomes
${\tau^{-1}(\epsilon_k)}  =  { n_dV_0^2 m}/{\hbar^3}$.
The density dependent conductivity is given by $ \sigma(n) \sim n^2$
in the weak screening limit ($q_0 =q_{TF}/2k_F =\ll 1$)
or for the unscreened Coulomb disorder, and in the strong screening limit
($q_0 \gg 1$)  $ \sigma(n) \sim n$. 
In general for screened Coulomb disorder $ \sigma(n) \sim
n^\alpha$, \citep{kn:dassarma2009c} 
where $\alpha$ is density dependent and 
varies slowly changing from 1 at low density to 2  
at high density.
Increasing temperature, in general, suppresses screening, leading to 
a slight enhancement of the exponent $\alpha$. 
For short range disorder $ \sigma(n) \sim n$. 

\begin{figure}
\includegraphics[width=5.cm]{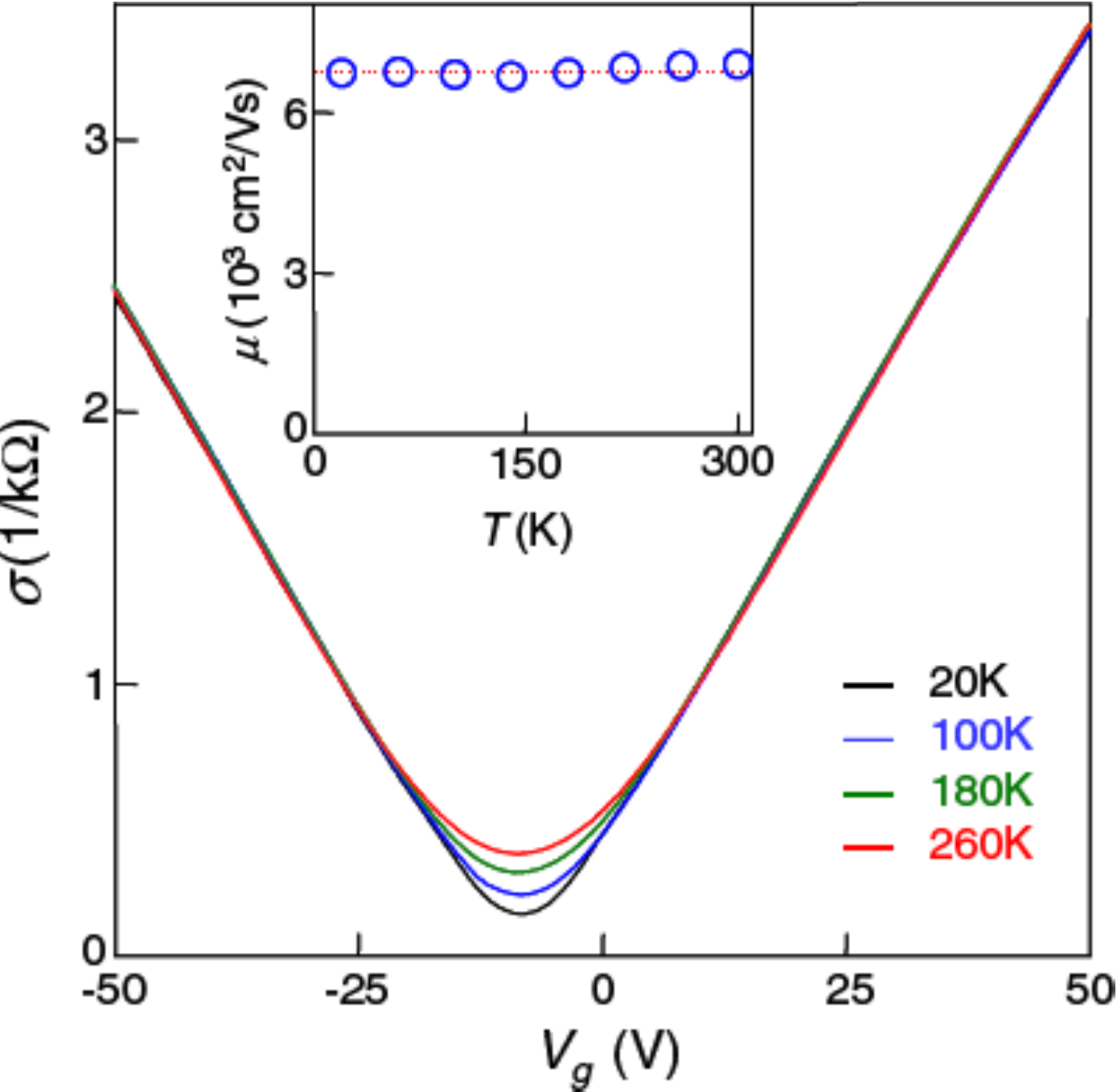}
\caption{(color online)
The measured conductivity of bilayer graphene as a function
  of gate voltage 
$V_g$ (or carrier density). The measured conductivity increases linearly with
the density. Adapted from \citet{kn:morozov2008}.
}
\label{fig:morozov2008}
\end{figure}

\begin{figure}
\includegraphics[width=\columnwidth]{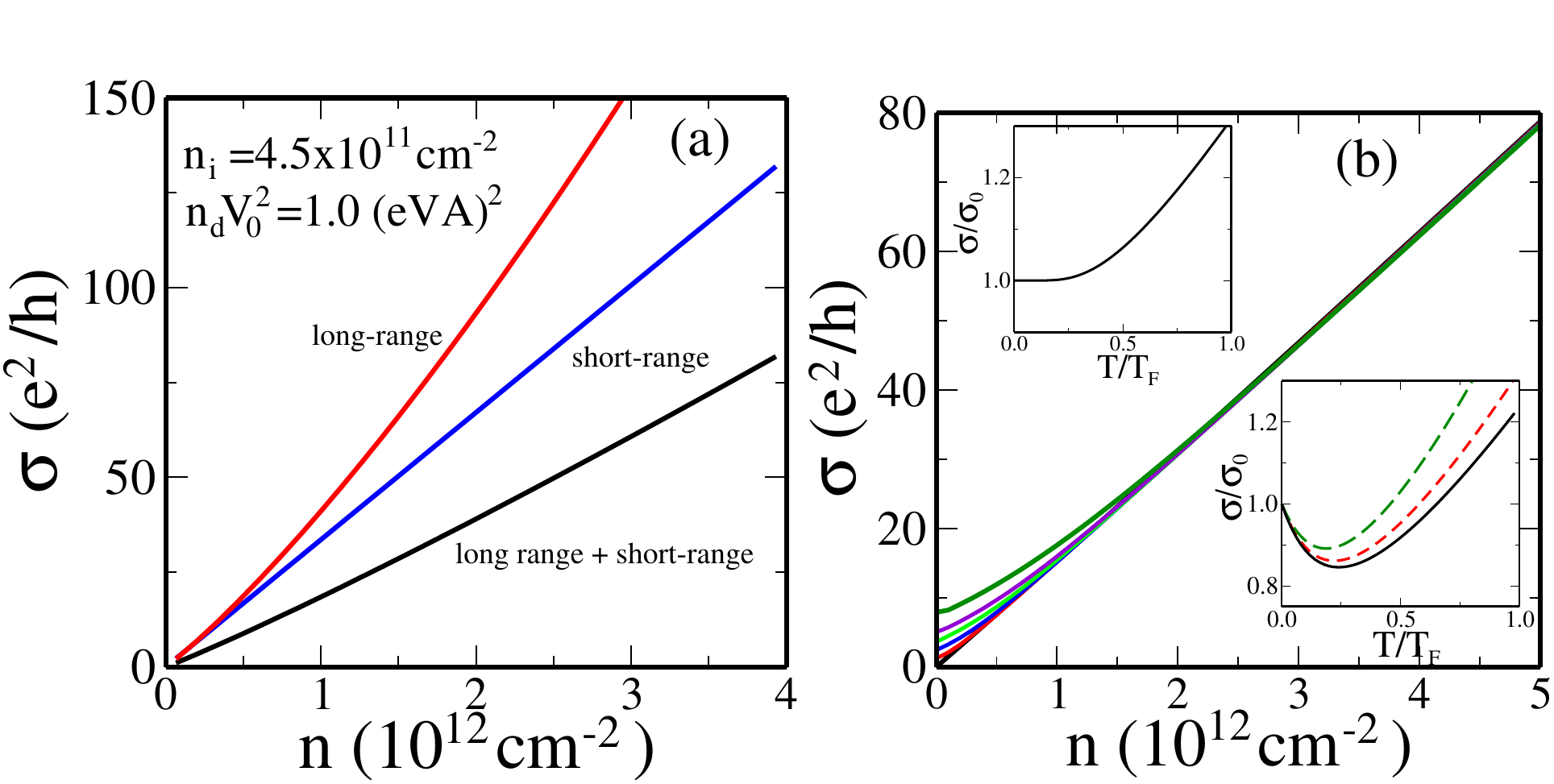}
\caption{(color online) (a) Density dependence of bilayer graphene
  conductivity with two 
  scattering sources: screened long-range Coulomb disorder and
  short-ranged neutral disorder. 
(b) Density dependence of BLG
  conductivity for different 
  temperatures, T=0~K, 50~K, 100~K, 150~K, 200~K, 300~K (from bottom to top).
Top inset shows $\sigma$ as a function of $T$ in presence of short
range disorder. 
Bottom inset shows $\sigma$ as a function of $T$ in presence of
screened Coulomb disorder for different densities
$n= [5 , 10, 30]~\times 10^{11}{\rm cm}^{-2}$ (from bottom to top).
Adapted from \citet{kn:dassarma2009c}.
\label{fig:blg_con1}
}
\end{figure}

Fig.~\ref{fig:morozov2008} shows the experiment of BLG conductivity. 
In Fig.~\ref{fig:blg_con1}(a) the density dependent conductivities
both for screened 
Coulomb disorder and for short range disorder are shown. For screened
Coulomb disorder the conductivity shows super-linear behavior, 
which indicates that pure
Coulomb disorder which dominates mostly in MLG transport can not
explain the density dependent conductivity as seen experimentally (see
Fig.~\ref{fig:morozov2008}) 
\citep{kn:morozov2008,kn:xiao2009}. 
The density dependence of conductivity with both disorders
is approximately linear over  
a wide density range,
which indicates that BLG carrier transport is controlled 
by two distinct and independent physical scattering
mechanisms, i.e.  screened Coulomb disorder due to random
charged impurities in the environment and a short-range disorder.
The weaker scattering rate of screened Coulomb disorder
for BLG than for MLG
is induced by the stronger BLG screening 
than MLG screening, rendering the effect
of Coulomb scattering relatively less important in BLG (compared
with MLG).

The temperature dependent
conductivity due to screened Coulomb
disorder~\citep{kn:adam2010,kn:dassarma2009c,kn:lv2009,kn:hwang2010} is given by
${\sigma(T)}/{\sigma_0} \approx 1-C_0 ({T}/{T_F})$ at low temperatures, 
where $C_0=4\log2/(C+1/q_0)$ with $q_0 = q_{TF}/2k_F$, and
$\sigma(T) \approx \sigma_1  ( {T}/{T_F}  )^2$ at high temperatures.
When the dimensionless temperature is very small $(T/T_F\ll 1)$
a linear-in $T$ metallic $T$ dependence arise from the 
temperature dependence of the screened charge impurity scattering, i.e. 
the thermal suppression of the $2k_F$-peak associated with
back-scattering (see Fig.~\ref{fig_pol}).
For the short-ranged scattering the temperature
dependence only comes from the energy averaging and the conductivity
becomes 
$\sigma(T) = \sigma(0)  [ 1 + t \ln  ( 1 + e^{-1/t} 
  )  ]$,
where $t = T/T_F$. 
At low temperatures the conductivity is exponentially suppressed, but
at high temperatures it increases linearly.

Fig.~\ref{fig:blg_con1}(b) shows the
finite temperature BLG conductivity as a function of $n$. 
The temperature dependence is very weak at higher densities as
observed in recent experiments \citep{kn:morozov2008}.
At low densities, where $T/T_F$ is not too small, there is a strong
insulating-type $T$ dependence arising from the thermal excitation
of carriers (which is exponentially suppressed at higher densities) and 
energy averaging, as observed experimentally \citep{kn:morozov2008}.
Note that for BLG $T_{F} = 4.23 \tilde{n}$ K, where $\tilde{n} =
n/(10^{10} cm^{-2})$. 
In bottom inset the conductivity due to screened Coulomb disorder is
shown as a function of temperature for different densities.
At low temperatures $(T/T_F\ll 1)$
the conductivity decreases linearly with temperature, but $\sigma(T)$
increases quadratically in high temperature limit.
By contrast, for the short-range
disorder $\sigma$ always increases with $T$, as shown in the upper
inset of Fig.~\ref{fig:blg_con1}(b). 
Thus for bilayer graphene the metallic behavior due to screening
effects is expected at very low temperatures for
low mobility samples, in which the screened Coulomb disorder dominates.  
In Fig.~\ref{fig_zhu_2009}(b) the temperature dependence of mobility
for bilayer graphene is shown. As we expect the metallic behavior
shows up at very low temperatures ($T < 100~{\rm K}$).

We conclude  this section by emphasizing the similarity and the difference
between  BLG and MLG transport at high densities from the perspective
of Boltzmann transport theory considerations. 
In the MLG the linear density dependent conductivity arises entirely
from Coulomb disorder. However, 
in the BLG the existence of  short-range
disorder scattering must be included to explain the linearity
because the Coulomb disorder gives rise to a higher power density
dependence in conductivity. 
The importance of short-range scattering in BLG
compared with MLG is understandable based on BLG 
screening being much stronger than MLG screening leading to the
relative importance of short-range scattering 
in BLG.

\sub3section{2D semiconductor systems}

Transport properties of 2D semiconductor based parabolic 2D 
systems (e.g. Si MOSFETs, GaAs heterostructures and quantum wells,
SiGe-based 2D structures) have been studied extensively over the
last forty years \citep{kn:ando1982,kn:abrahams2001,kn:kravchenko2004}. 
More recently,  2D transport properties have 
attracted much attention because of
the experimental observation of an apparent metallic behavior 
in high-mobility low-density electron inversion layer in
Si metal-oxide-semiconductor-field-effect transistor (MOSFET)
structures \citep{kn:kravchenko1994}. 
However, in this review 
we do not make any attempt at reviewing the whole 2D MIT
literature. Early comprehensive reviews of 2D MIT can be found in
the literature \citep{kn:abrahams2001,kn:kravchenko2004}. More recent
perspectives can be found in \citet{kn:dassarma2005,kn:spivak2009}.  
Our goal in this review is to provide a direct comparison of the
transport properties of 2D semiconductor systems with those of MLG and
BLG, emphasizing similarities and differences.

\begin{figure}
\includegraphics[width=\columnwidth]{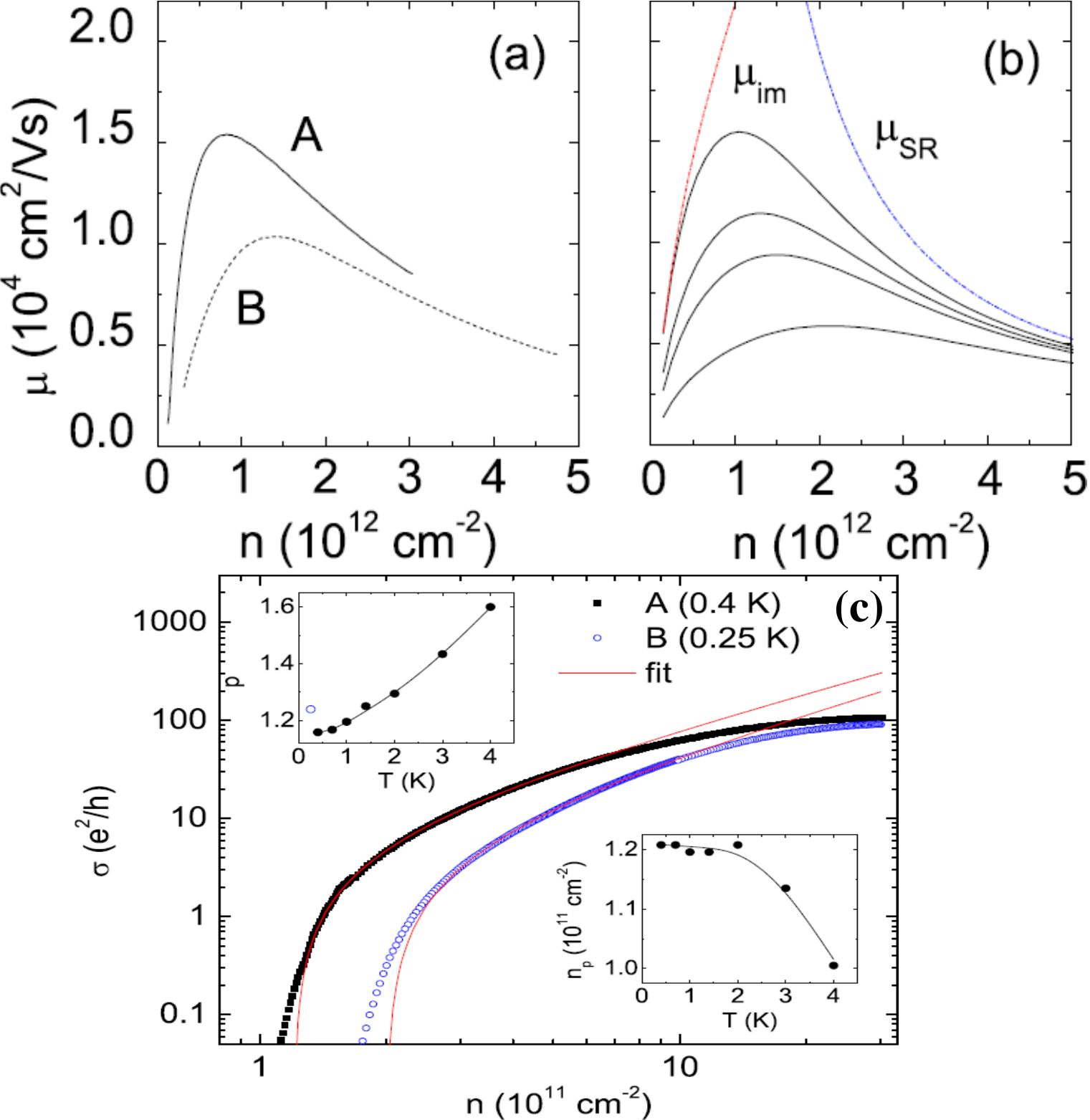}
\caption{ (Color online)
(a) Experimental mobility $\mu$ as a function of
  density for two Si-MOSFET samples at a temperature T=0.25 K. (b)
  Calculated mobility with two different scatterings, i.e.  charge
  impurities and surface-roughness scatterings. 
(c) Measured conductivity
$\sigma(n)$ for Si-MOSFET as a function of electron density $n$
for two different 
samples. The solid lines are fits to the data of the form
$\sigma(n) \propto A(n - n_p)^p$. The
upper and lower insets show the exponent $p$ and critical density $n_p$,
respectively, as a function of temperature. Solid lines
are a guides for the eyes. 
Adapted from  \citet{kn:tracy2009}.     
}
\label{fig_tracy_2009}
\end{figure}

It is well known that the long-range charged impurity scattering
and the short-range surface-roughness scattering dominate,
respectively, in the low and the high carrier density regimes of
transport in 2D semiconductor systems.  
In Fig.~\ref{fig_tracy_2009} the experimental mobility of Si-MOSFETs is
shown as a function of density. As density increases,
the measured mobility first increases 
at low densities and after reaching the maximum mobility it decreases
at high densities. This behavior is typical for all 2D semiconductor
systems, even though the mobility of GaAs systems 
decreases very slowly at high densities. This mobility behavior in
density can be explained with mainly two scattering mechanisms as shown
in Fig.~\ref{fig_tracy_2009}(b). 
In the low temperature region phonons do not play much of a
role in resistive scattering.   
At low carrier densities  long-range
Coulomb scattering by unintentional random charged impurities
invariably present in the environment of 2D semiconductor systems
dominates the 2D mobility \citep{kn:ando1982}.
However, at high densities as more carriers are
pushed to the interface the surface roughness scattering becomes more
significant. 
Thus transport in 2D semiconductor systems is limited by the same
mechanisms as in graphene even though at high densities the unknown
short-range disorder in graphene is replaced by the surface
roughness scattering in 2D semiconductor systems.
The crucial difference between 2D transport and graphene transport is
the existence of the insulating behavior of 2D semiconductor systems
at very low densities which arises from the gapped nature of 2D
semiconductors. 
However, the high density 2D semiconductor transport is not
qualitatively different 
from graphene transport since charged impurity scattering dominates
carrier transport in both cases.

\begin{figure}
\includegraphics[width=\columnwidth]{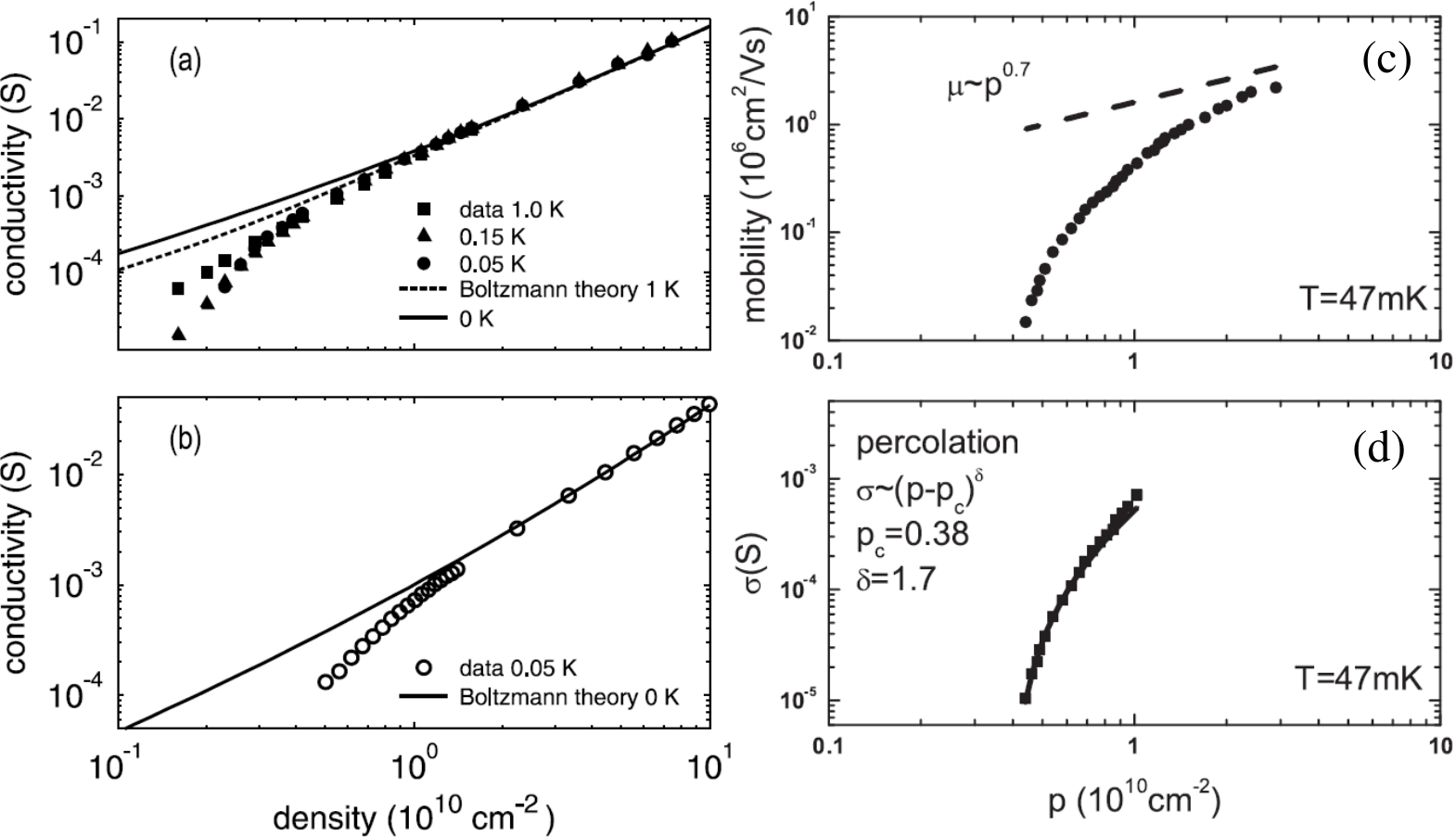}
\caption{ (a),(b)
Experimentally measured (symbols) and calculated (lines)
conductivity of two different n-GaAs samples. The high density conductivity
limited by the charged impurities fit well to the experimental
data. Adapted from 
  \citet{kn:dassarma2005b}.     
(c) Mobility of p-GaAs 2D system vs. density at fixed temperature T
  =47 mK. (d) 
The corresponding conductivity vs. density (solid squares) along 
with the fit generated assuming a percolation transition. The
dashed line in (c) indicates the $\mu \sim p^{0.7}$ behavior.
Adapted from
  \citet{kn:manfra2007}.     
}
\label{fig_dassarma_2005a}
\end{figure}

The experimentally measured conductivity and mobility for three
different systems as a function of density are shown in
Figs.~\ref{fig_tracy_2009}-\ref{fig_dassarma_2005a}. 
At high densities, the conductivity depends
on the density approximately as $\sigma\propto n^{\alpha}$ with $1<\alpha <2$,
where $\alpha(n)$  depends weakly on the density for a given system
but varies strongly from one system (e.g., Si-MOSFET) to another (e.g. GaAs). 
At high densities, before surface
roughness scattering sets in the conductivity 
is consistent with 
screened charged impurity scattering for all three systems.
As $n$ decreases, $\sigma(n)$ starts decreasing faster with decreasing
density and the experimental conductivity exponent $\alpha$ becomes
strongly density dependent with its value increasing 
substantially, and the conductivity vanishes as the density further
decreases.  To explain this behavior 
a density-inhomogeneity-driven percolation transition was proposed
\citep{kn:dassarma2005b}, 
i.e.  the
density-dependent conductivity vanishes as $\sigma(n) \propto  (n-n_p)^p$
with the exponent $p=1.2$ being consistent with a 
percolation transition.
At the lowest density, linear
screening in a homogeneous electron gas
fails qualitatively in explaining the $\sigma(n)$ behavior 
whereas it gives quantitatively accurate results at
high densities. As has been found from direct numerical
simulations \citep{kn:shi2002,kn:efros1988,kn:nixon1990} homogeneous
linear screening of charged impurities 
breaks down at low carrier densities with the 2D
system developing strong inhomogeneities leading to a percolation
transition at $n <n_p$. Nonlinear screening dominates transport in
this inhomogeneous low carrier density regime. For $n<n_p$, the system
is an 
insulator containing isolated puddles of electrons with no
metallic conducting path spanning through the whole
system. By contrast, graphene, being gapless,
goes from being an electron metal to a hole metal, i.e.,
the conductivity is always finite for all densities, as the chemical
potential passes through the puddle region.

Except for being an insulator at very low densities the
transport behavior of 2D semiconductor systems 
is not qualitatively different
from graphene transport because both systems are  governed by the charged
impurities.
To understand the $\rho$ (or $\sigma$) behavior at high density 
we start with
the Drude-Boltzmann semiclassical formula, Eq. (\ref{sigma}), for 2D
transport limited by 
screened charged impurity scattering \citep{kn:dassarma1999}. 
However, due to the finite extent in the $z$-direction of the real 2D
semiconductor system the 
Coulomb potential has a form factor depending on the details of the 2D
structure. For comparison with graphene we consider the
simplest case of 2D limit, i.e. $\delta$-layer. 
For $\delta$-layer 2D systems with parabolic band
the scattering times at $T=0$  for charged impurity centers
with impurity density $n_{i}$ located at the 2D systems is calculated by
\begin{equation}
\frac{1}{\tau} =  \frac{1}{\tau_0} \left \{
\pi-2\frac{d}{dq_0}\left[q_0^2f(q_0)\right ] \right \} 
\end{equation}
where 
$\tau_0^{-1} = {2\pi\hbar} \frac{n_{i}}{m} (\frac{2}{g_sg_v})^2q_0^2$,
$q_0 = q_{TF}/2k_F$ ($q_{TF}$ is a 2D Thomas-Fermi wave
vector), and $f(x)$ is given in Eq. (\ref{fx}).  Then, the density
dependence of conductivity can be expressed as $\sigma(n) \propto
n^{\alpha}$ with $1<\alpha <2$.
In the strong screening limit ($q_0 \gg 1$) the scattering time
becomes $\tau^{-1} \propto q_0^2 \propto n^{-1}$, then the
conductivity  behaves as $\sigma(n) \propto n^2$. In the weak screening
limit $\tau^{-1} \propto q_0^0 \propto n^0$ and $\sigma(n) \propto n$.
These conductivity behaviors are common for  2D systems with parabolic
bands and are qualitatively similar to graphene where $\sigma \propto
n$ behavior is observed.
However, due to the complicated impurity configuration (spatial
distribution of impurity centers) and finite width effects
of real 2D semiconductor systems the exponent $\alpha$ varies with
systems. In general, modulation doped GaAs systems have larger $\alpha$
than Si-MOSFETs due to the configuration of impurity centers.

\begin{figure}
\includegraphics[width=\columnwidth]{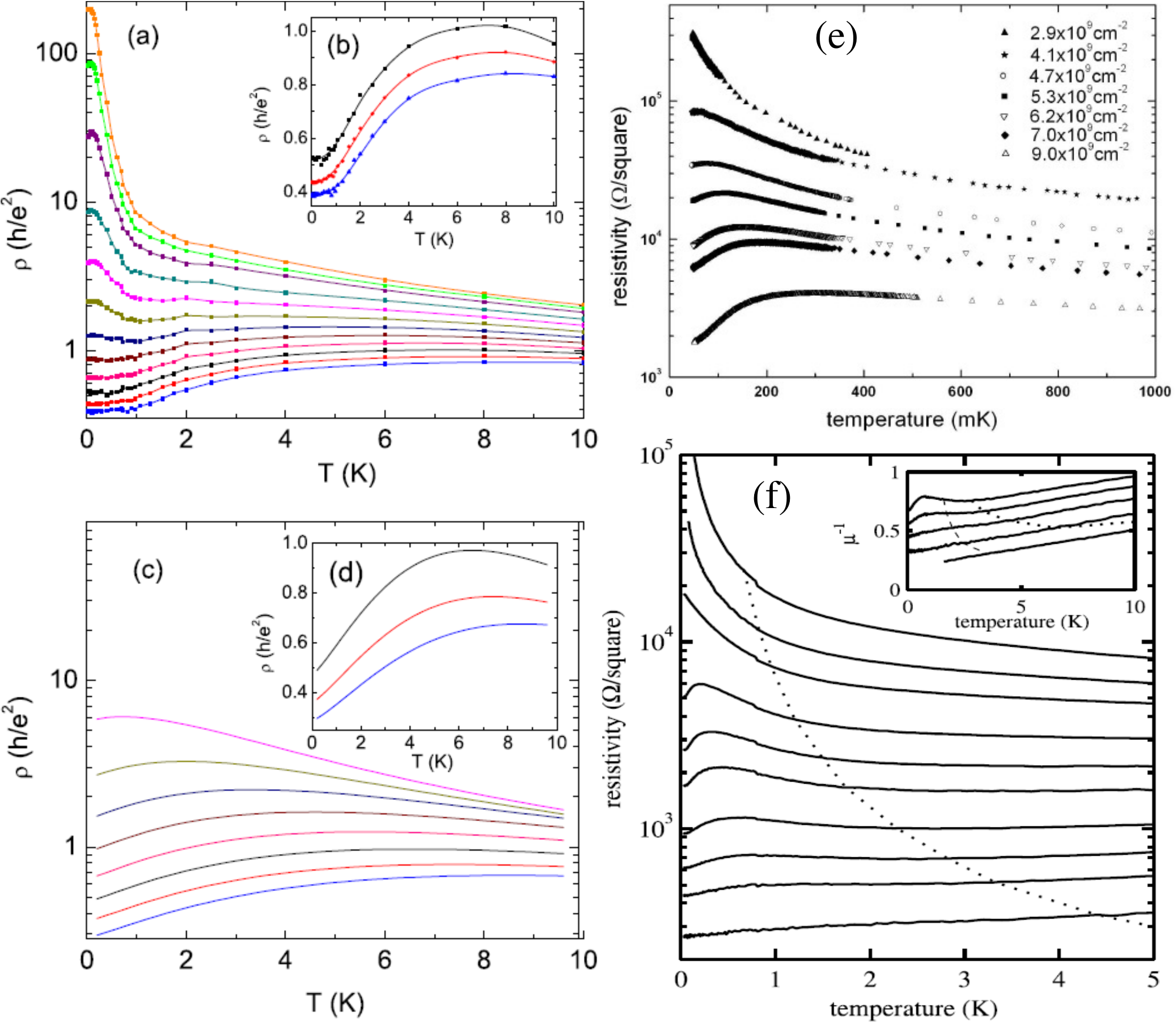}
\caption{ (Color online)
(a) Experimental resistivity $\rho$ of Si-MOSFET 
as a function of temperature at 2D electron densities
(from top to bottom) $n=[1.07$, 1.10, 1.13, 1.20, 1.26, 1.32,
1.38, 1.44, 1.50, 1.56, 1.62, and $1.68]~\times 10^{11}~{\rm cm}^{−2}$. Inset
(b) shows $\rho$ for   
$ n=[1.56$,
1.62, and $1.68]~\times 10^{11}{\rm cm}^{−2}$. 
(c) Theoretically calculated temperature and density-dependent resistivity
for sample A for densities $n=[1.26$, 1.32, 1.38, 1.44, 1.50,
1.56, 1.62, and $1.68]~\times10^{11}~{\rm cm}^{−2}$ (from top to bottom). 
Adapted from \citet{kn:tracy2009}.
(e) Experimental $\rho(T)$ for n-GaAs (where $n_c
= 2.3 \times 10^9~{\rm cm}^{-2}$). 
The density ranges 
from 0.16$\times 10^{10}cm^{-2}$ to
1.06$\times 10^{10}cm^{-2}$. 
Adapted from \citet{kn:lilly2003}.
(f) Temperature dependence of the resistivity for p-GaAs systems for
densities ranging from 9.0$\times 10^9~{\rm  cm}^{-2}$ to 2.9$\times 10^9
~{\rm cm}^{-2}$. Adapted from \citet{kn:manfra2007}.
}
\label{fig:tracy_2009c}
\end{figure}

An interesting transport property of 2D Semiconductor systems is 
the remarkable  observation of the
extremely strong anomalous  metallic
(i.e., $d\rho/dT>0$) temperature dependence of the resistivity
$\rho(T)$ in the density range just above a critical carrier density
$n_c$ where $d\rho/dT$ changes its sign at low temperatures (see
Fig.~\ref{fig:tracy_2009c}), which is not seen
in graphene.
Note that the experimentally measured $\rho(T)$ of graphene shows very weak
metallic behavior at high density due to the weak temperature
dependence of screening function. 
It is suggested \citep{kn:dassarma1999} that the anomalously strong
metallic temperature dependence discovered in 2D semiconductor systems
arises from the physical mechanism of temperature, density, and wave
vector dependent screening of charged impurity scattering in 2D
semiconductor structures, leading to a strongly
temperature dependent 
effective quenched disorder controlling $\rho(T,n)$ at low
temperatures and densities. Interaction effects also lead to a
linear-T conductivity in 2D semiconductors \citep{kn:zala2001} .

With temperature dependent screening function $\varepsilon(q,T)$ in
Eq. (\ref{eq:screening}), 
the asymptotic low \citep{kn:dassarma2003} and high \citep{kn:dassarma2004}
temperature behaviors of 2D conductivity 
 are given by:
\begin{subequations}
\begin{eqnarray}
\sigma(t \ll 1) & \approx & \sigma_0^{2D} \left [ 1- C_1 \left (
    {T}/{T_F} \right ) 
\right ], \\
\sigma(t \gg 1) & \approx & \sigma_1^{2D} \left [{T}/{T_F}
 + ({3
    \sqrt{\pi} q_0}/{4}) \sqrt{{T_F}/{T}} \right ],
\end{eqnarray}
\end{subequations}
where $t=T/T_F$, $\sigma_0^{2D} \equiv \sigma(T=0)$, $C_1 = 2
q_0/(1+q_0)$, and $\sigma_1^{2D} =
(e^2/h) (n/n_i) \pi q_0^2$. 
Here an ideal 2D electron gas with zero thickness is considered in
order to compare with the 2D graphene sheet which also has a zero
thickness.
It is important to include the temperature dependent
polarizability of Fig.~\ref{fig_pol} in the calculation in order to get
strong temperature dependent resistivity. 
Since the most dominant scattering occurs at
$q=2k_F$ and the temperature dependence of screening function
at $2k_F$ is strong, the calculated 2D resistivity shows the strong
anomalous linear $T$ metallic behavior, which is observed in many
different 
semiconductor systems (e.g., Si-MOSFET \citep{kn:kravchenko1994}, $p$-GaAs
\citep{kn:noh2003,kn:manfra2007}, $n$-GaAs \citep{kn:lilly2003}, SiGe
\citep{kn:senz2002}, AlAs \citep{kn:papadakis1998}). 
In addition, for the  observation of a large
temperature-induced change in resistivity 
it is required to have a  
comparatively large change in the value of the dimensionless
temperature $t=T/T_F$ and
the strong screening condition, $q_{TF} \gg 2k_F$, which explains why
the Si MOS 2D electron system exhibits
substantially stronger metallic behavior than the GaAs 2D electron system, as
is experimentally observed, since
$(q_{TF}/2k_F)_{Si} \approx 10 (q_{TF}/2k_F)_{GaAs}$ at similar
density.


Before concluding this basic transport theory section of this review we
point out 
the key qualitative similarities and differences in the transport
theory of all systems (i.e. graphene, bilayer  graphene, and 2D
semiconductor based parabolic 2D systems).
First, the graphene conductivity is qualitatively similar to that of
2D semiconductor systems in the sense that the conductivity at high
density of both systems follows  the power law in terms of density,
$\sigma(n) \sim n^{\alpha}$. The formal Boltzmann
theory for the scattering times is the same in all systems except for
the different angular factor arising from chiral properties of graphene.
This angular part does not play a role in the density dependence of
conductivity, but significantly affects the temperature dependence
of conductivity.
The explicit differences
in the density of states $D(\varepsilon)$ 
and the dielectric function $\epsilon(q,T)$  also 
lead to different temperature dependent conductivities in these
systems.
The most important qualitative difference between graphene and
semiconductor 2D systems occurs at low carrier densities, in which
semiconductor 2D systems become insulators, but graphene conductivity
is finite for all densities.  

\subsection{Phonon Scattering in Graphene} \label{subsec_phonons}

In this section we review the phonon scattering limited carrier
transport in graphene.
Lattice vibrations are inevitable sources of scattering and can
dominate transport near room temperature.
It is an intrinsic scattering source of the system, i.e.,
it limits mobility at finite temperatures when all extrinsic
scattering sources are removed.
In general, three different types of phonon scattering are considered,
i.e., intravalley acoustic (optical) phonon 
scattering which induce the electronic transition
within a single valley via a acoustic (optical) phonons, and
intervalley scattering, i.e. electronic transition between
different valleys. 

\begin{figure}
\includegraphics[width=\columnwidth]{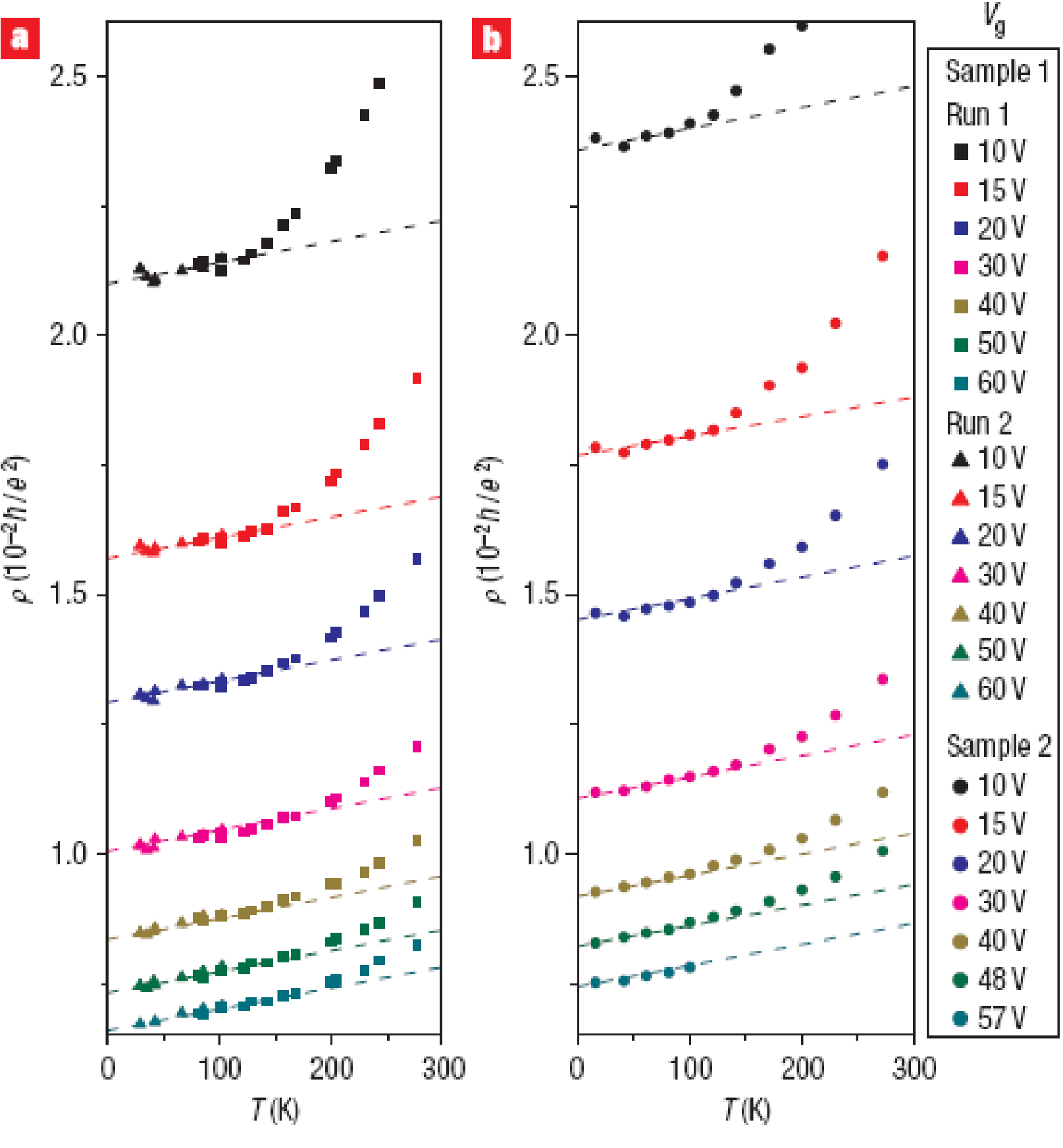}
\caption{ (color online)
Temperature-dependent resistivity of graphene on SiO$_2$.
Resistivity of two graphene samples as a
function of temperature for different gate voltages.
Dashed lines are fits
to the linear T-dependence with Eq.~(\ref{eq:rho_ph}). 
Adapted from \citet{kn:chen2008b}.
\label{fig_chen_2008b}
}
\end{figure}

\begin{figure}
\epsfysize=2.5in
\epsffile{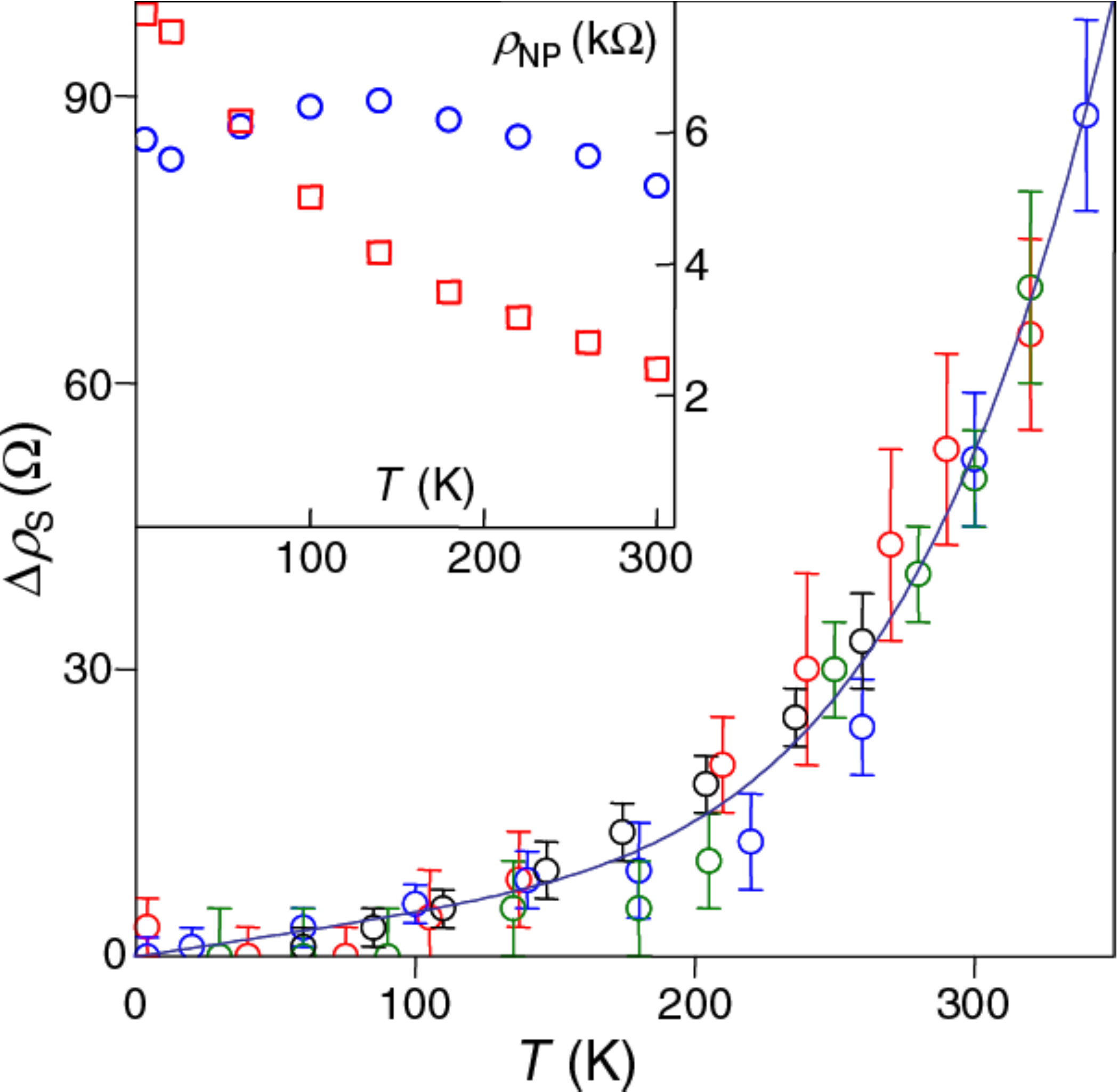}
\caption{ (color online)
Temperature dependent resistivity for four different  MLG
samples (symbols). 
The solid curve
is the best fit by using a combination of T and $T^5$ functions.
The inset
shows T dependence of maximum resistivity at the neutrality
point for MLG and BLG (circles and squares, respectively).
Adapted from \citet{kn:morozov2008}.
\label{fig_morozov_2008}}
\end{figure}

The intravalley acoustic phonon scattering is induced by 
low energy phonons and is considered an elastic process. 
The temperature dependent phonon-limited
resistivity \citep{kn:hwang2008,kn:stauber2007}  
was found to be linear (i.e. $\rho_{ph} \propto T$) for
$T>T_{BG}$ where $T_{BG}$ is 
the Bloch-Gr\"{u}neisen (BG) temperature \citep{kn:kawamura1992},
and $\rho_{ph}(T) \sim T^4$ 
for $T<T_{BG}$. The acoustic phonon
scattering gives a quantitatively small contribution in graphene even
at room temperature due to the high Fermi temperature of graphene in contrast to
2D semiconductors where 
room-temperature transport is dominated by phonon scattering 
\citep{kn:kawamura1990,kn:kawamura1992}. 
The intravalley optical phonon scattering is
induced by optical phonons of low momentum ($q\approx 0$) and very high energy
($\omega_{OP} \approx 200$ meV in graphene ) and is 
negligible.
The intervalley scattering can be induced by the emission and
absorption of high momentum, high energy acoustic or optical phonons.
In graphene intervalley
scattering may be important at high temperatures because of relatively
low phonon energy ($\approx 70$ meV, the out-of-plane acoustic (ZA)
phonon mode at the K 
point) \citep{kn:mounet2005,kn:maultzsch2004}. Even though the effects of
inter-valley phonon scattering can explain
a crossover (Figs. \ref{fig_chen_2008b} and
\ref{fig_morozov_2008}) seen in experiments in the 
$150~{\rm K}$ to $250~{\rm K}$ range,  
more work is needed to validate the model of combined acoustic phonon
and ZA phonon 
scattering contributing to the temperature dependent graphene
resistivity.

The remote interface polar optical phonons in
the substrate (i.e. SiO$_2$) have recently been considered
\citep{kn:fratini2008,kn:chen2008b}.  
Even though these modes are  known to be not very important in Si-MOSFETs
\citep{kn:hess1979,kn:moore1980} 
their role in graphene transport seems to be important \citep{kn:chen2008b,kn:dasilva2010}.
Another possibility considered in  \citet{kn:morozov2006} is that
the thermal fluctuations 
(ripplons) of the mechanical ripples invariably present in graphene
samples contribute to the graphene resistivity.  In addition
\citet{kn:mariani2008} investigated the role of 
the flexural 
(out-of-plane) phonons of 
free standing graphene membranes which arise from the
rotation and reflection symmetries.
Flexural phonons make a 
contribution to the resistivity at low temperatures with 
an anomalous temperature dependence $ ρ \propto
T^{5/2} \ln T$. 

Before we discuss the phonon transport theory we mention the
phonon contribution obtained from the experimental resistivity data.
Since the experimentally measured resistivity in the current graphene
samples is completely dominated by extrinsic scattering (impurity
scattering described in Sec. \ref{subsec_impurity}) even at room temperatures 
the experimental extraction of
the pure phonon contribution to graphene resistivity is not unique.
In particular, the impurity contribution to resistivity also
has a temperature dependence arising from Fermi statistics and
screening which, although weak, cannot be neglected in extracting the
phonon contribution (particularly since the total phonon contribution
itself is much smaller than the total extrinsic contribution). 
In addition, the experimental phonon contribution is obtained assuming
Matthiessen's rule, i.e. $\rho_{tot} = \rho_{ph}+\rho_{i}$ where
$\rho_{tot}$ is the total resistivity contributed by impurities and
defects ($\rho_{i}$) and phonons ($\rho_{ph}$), which is not
valid at room temperature \citep{kn:hwang2008}. 
Thus, two different groups \citep{kn:chen2008b,kn:morozov2008} 
(Figs. \ref{fig_chen_2008b} and \ref{fig_morozov_2008}) have
obtained totally different  behavior 
of phonon contribution of resistivity. In \citet{kn:morozov2008}
it is found that the 
temperature dependence is a rather high power ($T^5$) at room
temperatures,  and the phonon contribution is
independent of carrier density. In \citet{kn:chen2008b} the
extracted phonon 
contribution is  strongly density dependent  and is fitted  with both
linear T from acoustic phonons and  Bose-Einstein distribution.   
Therefore the phonon contribution, as determined by a
simple subtraction, could have large errors due to the dominance of
extrinsic scattering.

In this section 
we describe transport only due to the longitudinal acoustic (LA)
phonons since either the coupling to other graphene lattice 
modes is too weak or the energy scales of these (optical) phonon
modes are far too high for them to provide an effective scattering
channel in the temperature range ($5~{\rm K}$ to $500~{\rm K}$) of our interest.
Since graphene is a non-polar material the most important scattering 
arises from the deformation potential due to quasi-static deformation
of the lattice. Within the Boltzmann transport theory
\citep{kn:kawamura1990,kn:kawamura1992} 
the relaxation time due to deformation potential
coupled acoustic phonon mode is given by
\begin{equation}
\frac{1}{\tau(\varepsilon)} = \sum_{\vk'}(1-\cos\theta_{\vk \vk'}) W_{\vk
  \vk'}\frac{1 - f(\varepsilon')}{1-f(\varepsilon)}
\end{equation}
where $\theta_{\vk \vk'}$ is the scattering angle between $\vk$ and
$\vk'$, $\varepsilon = \hbar v_F |{\vk}|$, and
$W_{\vk \vk'}$ is the transition probability from the state with 
momentum $\vk$ to the state with momentum $\vk'$ and is given by
\begin{equation}
W_{\vk \vk'}=\frac{2\pi}{\hbar}\sum_{\vq}|C(\vq)|^2
\Delta(\varepsilon,\varepsilon')   
\end{equation}
where $C(\vq)$ is the matrix element for scattering by acoustic phonons,
and 
$\Delta(\varepsilon,\varepsilon')$ is given by
\begin{equation}
\Delta(\varepsilon,\varepsilon') = N_q
\delta(\varepsilon-\varepsilon'+\omega_{\vq}) + (N_q + 1) 
\delta(\varepsilon-\varepsilon'-\omega_{\vq}),
\label{delta}
\end{equation}
where $\omega_{\vq}= \hbar v_{ph} \vq$ is the acoustic phonon energy with
$v_{ph}$ being the phonon velocity  and
$N_q$  the phonon 
occupation number
$N_q = {1}/({\exp(\beta \omega_{\vq}) -1})$.
The first (second) term is Eq. (\ref{delta}) corresponds to the
absorption (emission) of an acoustic phonon of wave vector $\vq = \vk-\vk'$.
The matrix element $C(\vq)$ is independent of the phonon
occupation numbers.
The matrix element $|C(\vq)|^2$ for the deformation potential is given by
\begin{equation}
|C(\vq)|^2 = \frac{D^2\hbar q}{2A\rho_m v_{ph}}\left [ 1- \left (
    \frac{q}{2k} \right )^2 \right ],
\end{equation}
where $D$ is the deformation potential coupling constant, $\rho_m$
is the graphene mass density, and $A$ is the area of the sample.

The scattering of electrons by acoustic phonons may be considered
quasi-elastic since $\hbar \omega_{\vq} \ll E_F$, where $E_F$ is the
Fermi energy. 
There are two transport regimes, which apply to the
temperature regimes $T \ll T_{BG}$ and $T \gg T_{BG}$, depending on
whether the phonon system is degenerate (Bloch-Gr\"{u}neisen, BG) or
non-degenerate (equipartition, EP). The
characteristic temperature $T_{BG}$ is defined as $k_B T_{BG} = 2
\hbar k_F
v_{ph}$, which is given, in graphene, by $T_{BG} = 2 v_{ph}k_F/k_B \approx 54
\sqrt{n}$ K with density 
measured in unit of $n=10^{12}cm^{-2}$.  
The relaxation
time in the EP regime is calculated to be
\citep{kn:hwang2008,kn:stauber2007,kn:vasko2007} 
\begin{equation}
\frac{1}{\tau(\varepsilon)} =
\frac{1}{\hbar^3}\frac{\varepsilon}{4v_F^2}\frac{D^2}{\rho_m v_{ph}^2}
k_BT.
\label{eq:rho_ph}
\end{equation}
Thus, in the non-degenerate EP regime ($\hbar \omega_{\vq} \ll k_B T$) the
scattering rate [$1/\tau(\varepsilon)$] depends linearly on the
temperature. At low 
temperatures ($T_{BG} \ll T \ll E_F/k_B$) 
the calculated conductivity is independent of 
electron density. Therefore the electronic
mobility in graphene is inversely proportional to the carrier density,
i.e. $\mu \propto 1/n$. 
The linear temperature dependence of the scattering time has been
reported for nanotubes \citep{kn:kane1998b} and graphites
\citep{kn:pietronero1980,kn:woods2000,kn:suzuura2002b}. 

In BG regime the scattering rate is strongly reduced by the
thermal occupation factors because the phonon population 
decreases exponentially, and the phonon emission is prohibited by the sharp
Fermi distribution. 
Then, in the low temperature limit $T \ll T_{BG}$ the scattering time becomes
\citep{kn:hwang2008}
\begin{equation}
\frac{1}{\langle \tau \rangle} \approx  \frac{1}{\pi} \frac{1}{
  E_F} \frac{1}{k_F}\frac{D^2}{2\rho_mv_{ph}}\frac{4! \zeta(4)}{(\hbar
  v_{ph})^4}(k_BT)^4.
\label{tau_t}
\end{equation}
Thus, the temperature dependent resistivity in BG regime
becomes $\rho \propto T^{4}$. 
Even though the resistivity in EP regime is density independent,
Eq. (\ref{tau_t}) indicates that the calculated resistivity in BG regime
is inversely proportional to the density, i.e. $\rho_{BG} \propto
n^{-3/2}$ since $\rho \propto [D(E_F) \langle \tau
  \rangle]^{-1}$. 
More experimental and theoretical work would be needed for a precise
quantitative understanding of phonon scattering effect on graphene
resistivity.

\subsection{Intrinsic mobility}

\begin{figure}
\epsfysize=1.6in
\epsffile{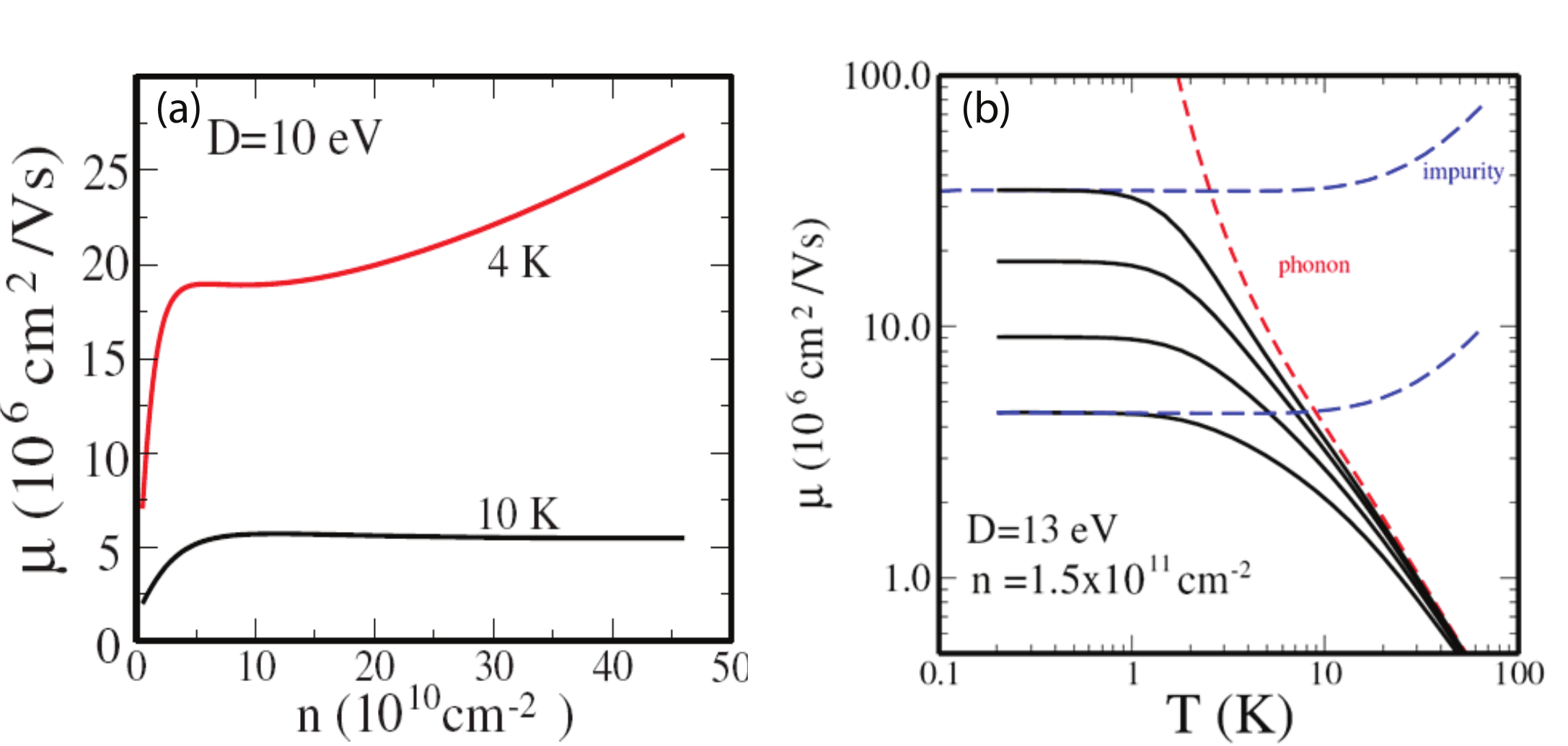}
\caption{ (color online)
(a) Acoustic phonon-limited mobility of n-GaAs 2D system as a function
  of density for two different temperatures.
(b) Calculated n-GaAs mobility as a function of temperature for different impurity
  densities. At low
temperatures ($T < 1 K$) the mobility is completely limited by
impurity scattering.   
Adapted from \citet{kn:hwang2008e}.
\label{fig_hwang_2008a}}
\end{figure}

Based on the results of previous sections one can extract the
possible (hypothetical) intrinsic mobility of 2D systems when all
extrinsic impurities 
are removed. In Fig.~\ref{fig_hwang_2008a} the acoustic phonon-limited
mobility is shown for 2D n-GaAs system.
For lower temperatures, $\mu(T)$  increases by a
large factor ($\mu \propto T^{-7}$ for deformation-potential scattering and
$\mu \propto T^{-5}$ for piezoelectric scattering) since one is in the
Bloch-Gr\"{u}neisen regime where phonon occupancy is suppressed
exponentially \citep{kn:kawamura1990,kn:kawamura1992}. Thus the
intrinsic mobility of semiconductor systems is extremely high at low
temperatures ($T <T_{BG}$). For currently available
semiconductor samples the mobility below $T_{BG}$ 
is completely limited by extrinsic impurity scattering in 2D
systems. 
Above the BG regime (or $T>4~{\rm K}$)
the mobility 
is dominated by phonons.
In this limit the mobility limited by phonon scattering is much lower
than that for charged impurity scattering. Therefore
it will be impossible to raise 2D mobility (for $T>4~{\rm K}$) by removing
the extrinsic 
impurities since
acoustic phonon scattering sets the intrinsic limit at these
higher temperatures (for $T>100~{\rm K}$, optical phonons become dominant)
\citep{kn:pfeiffer1989}.

\begin{figure}
\epsfysize=1.35in
\epsffile{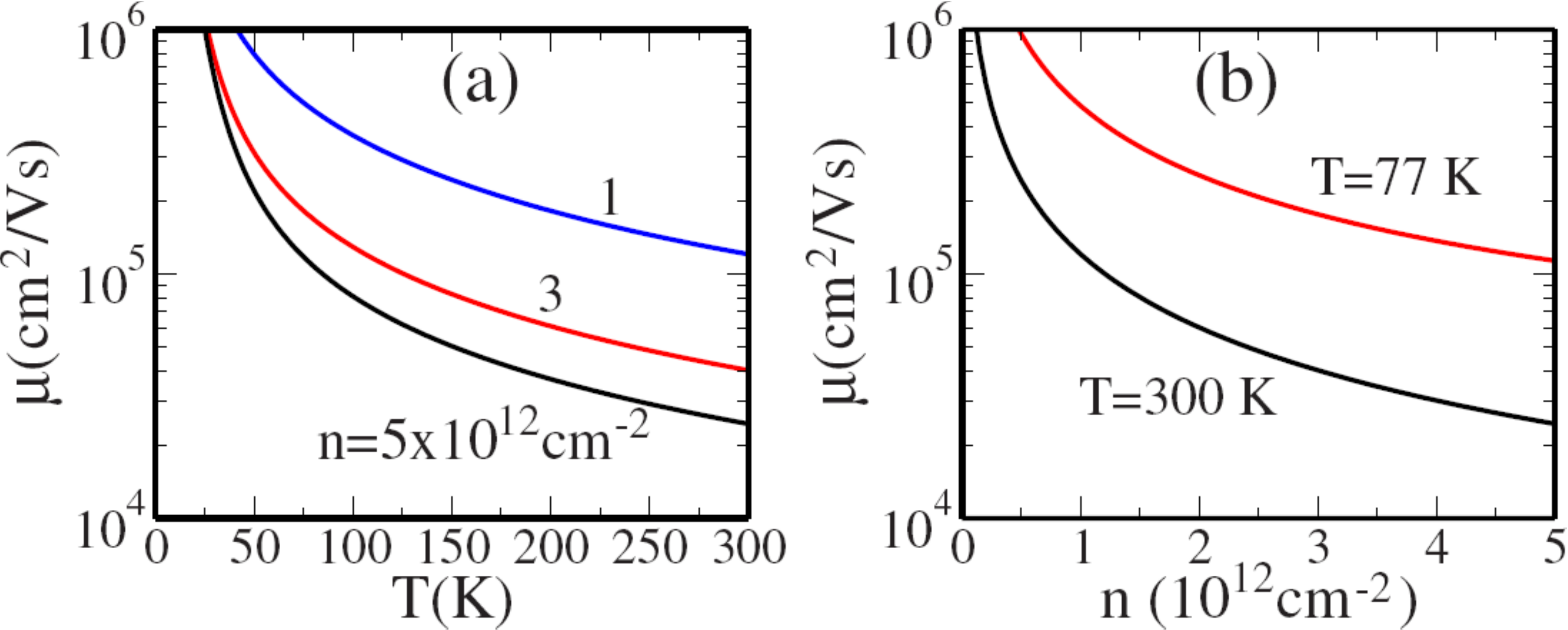}
\caption{ (color online)
Calculated graphene mobility limited by the acoustic phonon with the
deformation potential coupling constant $D=19$ eV (a) as a function of
temperature 
and (b) as a function of density.
Adapted from \citet{kn:hwang2008}.
\label{fig_hwang_2008}}
\end{figure}

In Fig.~\ref{fig_hwang_2008}, the acoustic phonon-limited graphene
mobility, $\mu 
\equiv (en\rho)^{-1}$, is shown as functions of temperature and carrier
density, which is given by $\mu \agt
10^{10}/D^2 \tilde{n} T$ cm$^2/Vs$ where $D$ 
is measured in eV, the temperature $T$ in K, and $\tilde{n}$ 
carrier density measured in 
units of $10^{12}$ cm$^{-2}$. 
Thus, the acoustic phonon scattering
limited graphene mobility is inversely proportion to $T$ and $n$
for $T > T_{BG}$. 
Also with the generally accepted values in the literature for the
graphene sound velocity and deformation coupling \cite{kn:chen2008b} (i.e.,
$v_{ph} =  2\times 10^6~{\rm cm}/{\rm s}$, and deformation potential
$D = 19~{\rm eV}$)  
$\mu$ could reach values as high as $10^5$ cm$^2/{\rm Vs}$ for lower carrier
densities ($n \alt 10^{12}$ cm$^{-2}$) at $T=300 K$
\citep{kn:shishir2009,kn:hwang2008}.  For larger
(smaller) values of 
$D$, $\mu$ would be smaller (larger) by a factor of $D^2$. It may be
important to emphasize here that we know of no other system where the
intrinsic room-temperature carrier mobility could reach a value as
high as $10^5$ cm$^2/{\rm Vs}$, which is also consistent with the
experimental conclusion by
\citep{kn:morozov2008,kn:chen2008b,kn:hong2009}. This would, however,
require the elimination of all extrinsic scattering, and first steps
in this direction have been taken in fabricating suspended graphene
samples \citep{kn:bolotin2008,kn:du2008}.
Finally we point out the crucial difference
between graphene and 2D GaAs in phonon limited mobility. 
In the 2D GaAS system the acoustic phonon scattering is important
below $T=100~{\rm K}$ and polar optical phonon scattering becomes
exponentially more important for $T \agt 100~{\rm K}$ 
whereas in graphene a resistivity linear in $T$ is observed 
upto very high temperatures ($\approx 1000~{\rm K}$) since the relevant optical
phonons have very high energy ($\approx 2000~{\rm K}$) and are simply irrelevant
for carrier transport.

\subsection{Other scattering mechanisms} \label{subsec_other_scattering}
 
\subsubsection{Midgap states}
The Boltzmann transport theory developed in Sec.~\ref{subsec_boltzmann}
 considered
the limit of weak scattering.  One can ask about the opposite limit
of very strong scattering.  The unitarity of the wavefunctions implies
that a potential scatterer can only cause a phase-shift in the outgoing
wave.  Standard treatment of s-wave elastic scattering gives the
scattering time 
\begin{equation}
\frac{\hbar}{\tau_k} = \frac{8 n_{\rm d}}{\pi D(E_k)} \sin^2(\delta_k),
\end{equation}
where the conductivity is then given by the Einstein relation
$\sigma = (2e^2/h) v_{\rm F} k_{\rm F} \tau_{k_{\rm F}}$.

To model the disorder potential induced by a vacancy, 
\citet{kn:hentschel2007} assumed a circularly symmetric 
potential with $V(0<r<R') = \infty$, $V(R'<r<R) = \mbox{const}$,
and  $V(R>r) = 0$.  This corresponds to a circular void of 
radius $R'$, and appropriate boundary conditions are chosen to 
allow for zero energy states (also called mid-gap states).  By matching
the wavefunctions of incoming and outgoing waves, 
the scattering phase shift can be calculated 
as~\cite{kn:hentschel2007,kn:guinea2008b}
\begin{equation}
\delta_k = -\arctan \left(\frac{J_0(k R')}{Y_o(k R')} \right)
\stackrel{k \rightarrow 0}{\longrightarrow} -\frac{\pi}{2} \frac{1}{\ln (kR')},
\end{equation}
where $J_0(x)$ [$Y_0(x)$] is the zeroth order Bessel function of the first (second)
kind. 
\noindent Expanding for small carrier density, 
one then finds for the conductivity~\cite{kn:stauber2007}
\begin{equation}
\label{SA_Eq:Midgap}
\sigma = \frac{2 e^2}{\pi h} \frac{n}{n_{\rm d}} \ln^2(k_{\rm F} R'),
\end{equation}
which other than the logarithmic factor, mimics the 
behavior of charged impurities, and is linear in carrier density.

\begin{figure}
\bigskip
\includegraphics[width=\columnwidth]{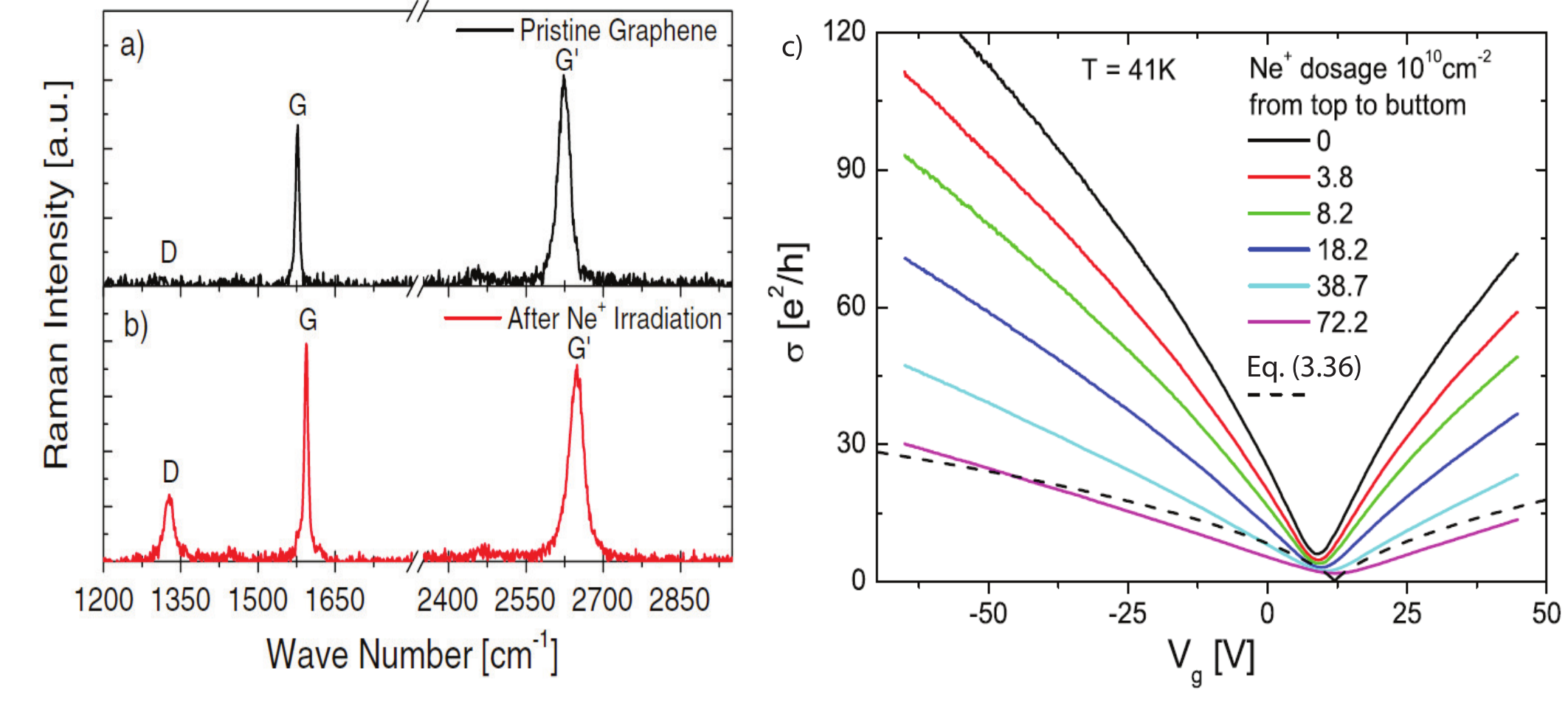}
\caption{\label{SA_Fig_Raman}
(Color online) 
Left panel: Raman spectra (wavelength 633~{\rm nm}) 
for 
(a) pristine graphene and (b) graphene irradiated 
by 500~{\rm eV} Ne$^+$ ions that are known to cause 
vacancies in the graphene lattice.  Right panel:
Increasing the number of 
vacancies by ion irradiation caused a transition from 
the pristine graphene (where Coulomb scattering dominates)
to the lower curves where scattering from vacancies 
dominate.  Also shown is a fit to Eq.~\protect{\ref{SA_Eq:Midgap}}
from \citet{kn:stauber2007} that describes 
scattering off vacancies that have midgap states.
Figures taken from \citet{kn:chen2009}.}
\end{figure}

In recent experimental work, \citet{kn:chen2009} irradiated
graphene with He and Ne ions to deliberately create large vacancies in
the graphene sheet.  They further demonstrated that these vacancies
induced by ion irradiation gave rise to a strong D-peak in the Raman
spectra, inferring that the absence of such a D-peak in the pristine
graphene signalled the lack of such defects (Fig.~\ref{SA_Fig_Raman}).
Moreover, they demonstrated that while transport in
pristine graphene is dominated by charged impurities, after
ion irradiation the electron scattering off these vacancies
appears consistent with the theory including midgap states 
(Eq.~\ref{SA_Eq:Midgap}).  We mention that in this review, we
consider only the case where the disorder changes graphene's 
transport properties without modifying 
its fundamental 
chemical structure~\cite{kn:hwang2006e,kn:schedin2006}.  
The subject of transport in 
graphane~\cite{PhysRevB.75.153401,Elias01302009} and other chemical 
derivatives of graphene is beyond the scope of this 
work, see e.g.~\cite{geim-s-324-1530-2009,wehling-prb-80-085428-2009,wehling-cpl-476-125-2009,bostwick-prl-103-056404-2009,cheianov-ssc-149-1499-2009,kn:robinson2008}.    

\subsubsection{Effect of strain and corrugations}

While graphene is often assumed to be an atomically perfect 2D sheet,
in reality, graphene behaves more like a membrane.  When placed on a
substrate, graphene will conform to the surface roughness developing ripples.
Even without a substrate, experiments reveal significant
deformations~\cite{kn:meyer2007}, although the theoretical picture is
still contentious~\cite{kn:fasolino2007,kn:thompson2009,pereira-prb-80-045401-2009}. It is
nonetheless an important theoretical question to address the nature
of electronic scattering off such ripples.  Ripples, by their very
nature, are correlated long-range fluctuations across the entire
sample (i.e. most experiments measuring ripples calculate a
height-height correlation function).  Yet, for electronic transport,
one would like to isolate a ``single ripple'' and calculate its
scattering cross-section (assuming that the rest of the sample is
flat), and then treat the problem of electrons scattering off ripples
as that of random uncorrelated impurities with the cross-section of a
single ripple.  This was the approach followed by
\citet{kn:katsnelson2008} and
\citet{kn:guinea2008b,kn:leon2009}.

With this qualitative picture in mind, one could estimate the 
transport time due to ripples as
\begin{equation}
\frac{\hbar}{\tau} \approx 2 \pi D(E_{\rm F}) 
\langle V_q V_{-q} \rangle, 
\end{equation}
where $V_q$ is the scattering potential caused by the strain-fields
of a single ripple.

Introducing a height field $h(\vec{r})$ (that measures displacements
normal to the graphene sheet), one can 
approximate~\cite{kn:katsnelson2008}
\begin{eqnarray}
\langle V_q V_{-q} \rangle &\approx& \left(\frac{\hbar v_{\rm F}}{a} \right)^2
\sum_{\vec{q_1},\vec{q_2}} \langle h_{\vec{q} - \vec{q_1}} h_{\vec{q}} h_{-\vec{q}
+\vec{q_2}} h_{-\vec{q_2}}\rangle \nonumber \\
&& \mbox{} \left[ (\vec{q} - \vec{q_1})\cdot \vec{q_1} \right]
 \left[ (\vec{q} - \vec{q_2})\cdot \vec{q_2} \right] ,
\end{eqnarray}
where $a$ is the lattice spacing.  Following
\citet{kn:ishigami2007} ripple correlations can be parametrized
as $\langle [h(r) -
  h(0)]^2 \rangle = r^{2H}$, where the exponent $H$ provides information
about the origin of the ripples.  An exponent $2H=1$ indicates that
height fluctuation domains have short-range correlations implying that
graphene conforms to the morphology of the underlying substrate, while
$2H=2$ suggests a thermally excitable membrane only loosely bound by
Van der Waals forces to the substrate.  \citet{kn:ishigami2007}
found experimentally that $2H \approx 1.11 \pm 0.013$ implying that
graphene mostly conforms to the substrate, but with some intrinsic
stiffness.  \citet{kn:katsnelson2008} showed that this 
has consequences for  transport properties, where 
for $2H=1$, $\sigma \sim 1/\ln^2[k_{\rm F} a]$; and for $2H > 1$, 
$\sigma(n) \sim n^{2H-1}$.  For the special case of $2H =2$ (flexural
ripples), this scattering mimics the long-range Coulomb scattering discussed 
in Sec.~\ref{subsec_boltzmann}.  For the experimentally relevant case 
of $2H \gtrsim 1$, electron scattering off ripples would mimic short-range
disorder also discussed in Sec.~\ref{subsec_boltzmann}.  Thus, ripple
scattering in graphene for $2H \approx 1$ mimic surface roughness
scattering in Si MOSFET \citep{kn:ando1982}. We should
emphasize that these conclusions are at best qualitative, since the 
approximation of treating the ripples as uncorrelated single impurities
is quite drastic.  A complete theory for scattering off ripples in graphene 
is an interesting, and at present,  open problem.  
Ripple scattering effects on graphene transport have a formal similarity
to the well-studied problem of interface roughness scattering effects on
carrier transport in Si-SiO$_2$ 2D 
electron systems~\cite{kn:ando1982,kn:adam2007b,kn:tracy2009}.

\setcounter{sub3section}{0}


\section{Transport at low carrier density}
\label{sec_low}

\subsection{Graphene minimum conductivity problem}

\subsubsection{Intrinsic conductivity at the Dirac point}

One of the most discussed issues in the context of fundamental
graphene physics has been the so-called minimum (or minimal)
conductivity problem (or puzzle) for intrinsic graphene.
In the end, the graphene minimum conductivity problem turns out
to be an ill-posed problem, which can only be solved if the real
physical system underlying intrinsic (i.e. undoped) graphene
is taken into account. An acceptable and reasonably quantitatively
successful theoretical solution of the minimum conductivity problem
has only emerged in the last couple of years, where the theory
has to explicitly incorporate carrier transport in the highly
inhomogeneous electron-hole landscape of extrinsic graphene,
where density fluctuations completely dominate transport properties
for actual graphene samples.

The graphene minimum conductivity problem is the dichotomy between
the theoretical prediction of a universal Dirac point conductivity
$\sigma_D$ of undoped intrinsic graphene and the actual experimental
sample-dependent non-universal minimum of conductivity observed in
gated graphene devices at the charge neutrality point with the
typical observed minimum conductivity being much larger than
the universal prediction.

Unfortunately $\sigma_D$
is ill-defined, and depending on the theoretical methods
and approximation schemes, many different universal results have been predicted 
\cite{kn:fradkin1986,kn:ludwig1994,kn:tworzydlo2006,kn:bardarson2007,kn:aleiner2006,kn:altland2006,kn:peres2006,kn:fritz2008,kashuba2008}
$$
 \sigma_D = \frac{4e^2}{\pi h};\quad 
            \frac{\pi e^2}{2h};\quad
            0;\quad
            \infty
$$ 
and other values.
The conductivity, $\sigma(T,\omega, \epsilon_F,\Gamma,\Delta,L^{-1})$,
is in general a function of many variables: temperature $(T)$,
frequency $(\omega)$, Fermi energy or chemical potential $(\epsilon_F)$,
impurity scattering strength or broadening $(\Gamma)$, intervalley
scattering strength $(\Delta)$, system size $(L)$.
The Dirac point conductivity of clean graphene,
$\sigma_D(0,0,0,0,0,0)$, is obtained in the limit of all the
independent variables being zero, and the result depends 
explicitly on how and in which order these limits are taken.
For example, $\omega\to 0$ and $T\to 0$ limit is not
necessarily interchangeable with the $T\to 0$ and $\omega\to 0$
limit! In addition, the limit of vanishing impurity scattering
$(\Gamma\to 0)$ and whether $\Gamma=0$ or $\Gamma\neq 0$ also may matter.
In the ballistic limit $(\Gamma=0)$, the mesoscopic physics
of the system size being finite $(1/L\neq 0)$ or infinite
$(1/L= 0)$ seems to matter. The intervalley scattering
being finite $(\Delta\neq 0)$ or precisely zero $(\Delta\equiv 0)$
seems to matter a great deal because the scaling theory of
localization predicts radically different results for
$\sigma_D$, $\sigma_D=0$ for $\Delta\neq 0$,  $\sigma_D=\infty$ for $\Delta= 0$,  
in the presence of any finite disorder $(\Gamma\neq 0)$.

A great deal of the early discussion on the graphene minimum of 
conductivity problem has been misguided by the existing theoretical
work which considered the strict $T=0$ limit and then taking
the $\omega\to 0$ limit. 
Many theories claim $\sigma_D=4e^2/(\pi h)$ in this limit,
but the typical experimentally measured value is much larger
(and sample-dependent), leading to the so-called 
``problem of the missing pi''. The limit
$\lim_{\omega\to 0} \sigma(\omega, T=0)$ is, in fact, 
experimentally irrelevant since for experimental
temperatures (even 10~mK), $k_BT \gg \hbar\omega$,
and thus the appropriate limiting procedure
for dc conductivity is $\lim_{T\to 0} \sigma(\omega=0, T)$.
There is an intuitive way of studying this limit theoretically,
which, however, can only treat the ballistic (and therefore,
the completely unrealistic disorder-free) limit.
Let us first put $\omega=0$ and assume $\mu=0$, i.e. intrinsic graphene.
It is then easy to show that at $T\neq 0$, there will be a finite
carrier density $n_e=n_h\propto T^2$ thermally excited from
the graphene valence band to the conduction band. The algebraic
$T^2$ dependence of thermal carrier density, rather than the
exponentially suppressed thermal occupancy in semiconductors,
of course follows from the non-existence of a band gap in graphene.
Using the Drude formula for dc conductivity, we write $
\sigma_D=ne^2\tau/m\propto T^2\tau(T)/m(T)$, where $\tau$, m
are respectively the relaxation time and the effective mass.
Although graphene effective mass is zero due to its linear
dispersion, an effective definition of effective mass follows
from writing $\epsilon=\hbar v_F k =(\hbar^2 k_F^2)/(2m)$,
which leads to $m\propto\sqrt{n}\propto T$ (which vanishes as $T\to 0$)
by using $k\propto\sqrt{n}$. This then leads to $\sigma_D\sim T\tau(T)$.
In the ballistic limit, the only scattering mechanism is the electron-hole
scattering, where the thermally excited electrons and holes 
scatter from each other due to mutual Coulomb interaction.
This inelastic electron-hole scattering rate $1/\tau$
is given by the imaginary part of the self-energy 
which, to the leading order, is given by $1/\tau\sim T$,
leading to $\sigma_D\sim T (1/T)\sim$~a~constant in the ballistic
limit. There are logarithmic sub-leading terms which indicate
that $\sigma_D(T\to 0)$ grows logarithmically at low temperature
in the ballistic limit. The conductivity in this picture, 
where interaction effects are crucial, is non-universal
even in the ballistic limit, depending logarithmically on temperature
and becoming infinite at $T=0$. The presence of any finite
impurity disorder modifies  the whole picture  completely.
More details along this idea can be found in the literature
\cite{kashuba2008,kn:fritz2008,kn:muller2008,kn:foster2008,foster2009}.

\subsubsection{Localization}

A fundamental mystery in graphene transport is the absence of any
strong localization-induced insulating phase at low carrier density
around the Dirac point, where $k_F l \ll 1$ since $k_F \approx 0$ at
the charge neutrality point and the transport mean free path $l$ is
finite (and small). This is a manifest violation of the Ioffe-Reggel
criterion which predicts strong localization for $k_F l \alt 1$. By
contrast, 2D semiconductor systems {\it always} go insulating in the
low density regime. It is conceivable, but does not seem likely, that
graphene may go insulating due to strong localization at lower
temperatures. Until that happens, the absence of any signature of
strong localization in graphene is a fundamental mystery deserving
serious experimental attention. Two noteworthy aspects stand out in
this context.  First, no evidence of strong localization is observed in 
experiments that deliberately break the A-B sublattice 
symmetry~\cite{kn:chen2009}.
Thus, the absence of localization in graphene cannot be attributed to the
chiral valley symmetry of the Dirac fermions. Second, the opening of
an intrinsic spectral gap in the graphene band structure by using
graphene nanoribbons
\cite{kn:adam2008c,kn:han2007}
or biased BLG \cite{kn:oostinga2007,zhang-n-2009} immediately 
introduces an insulating phase 
around the charge
neutrality point. These two features indicate that the insulating
behavior in graphene and 2D semiconductors is connected more with the
existence of a spectral gap than with the quantum localization
phenomena.

\subsubsection{Zero density limit}

It is instructive to think about the intrinsic conductivity as the
zero-density limit of the extrinsic conductivity for gated
graphene. Starting with the Boltzmann theory high density result of
Sec.~\ref{sec_high_density} we see that
\begin{eqnarray}
\sigma_D \equiv \sigma(n\rightarrow 0) =
\left \{ 
 \begin{array}{cl} 0
                  & \mbox{Coulomb scattering} \\
                   C_i
                  & \mbox{zero-range scattering}
 \end{array} 
\right. ,
\end{eqnarray}
where the non-universal constant $C_i$ is proportional to the strength of the
short-range scattering in the system. We note that the vanishing of
the Boltzmann conductivity in the intrinsic zero-density limit for
Coulomb scattering is true for both unscreened and
screened Coulomb impurities. The non-vanishing of graphene Boltzmann
conductivity for zero-range $\delta$-function scattering potential in the
zero carrier density intrinsic limit follows directly from the gapless
linear dispersion of graphene carriers. We emphasize, however, that
$\sigma_D$ is non-universal for zero-range scattering. 

For further insight into the zero-density Boltzmann limit the $T=0$
for $\sigma$ let's consider Eq.~(\ref{eq:sigma}).
In general, $\tau^{-1}(E) \sim D(E)$ since the availability of
unoccupied states for scattering should be proportional to the density
of states. This immediately shows that the intrinsic limit, $E_F(n
\rightarrow 0) \rightarrow 0$, is extremely delicate for graphene because
$D(E\to 0)\to 0$, and the product $D\tau$ becomes ill-defined at the Dirac
point. 

We emphasize in this context, as discussed in Sec.~\ref{sec_intro} 
that as a function of carrier density (or
gate voltage), graphene conductivity (at high carrier density) is
qualitatively identical to that of semiconductor-based 2DEG. This
point needs emphasis because it seems not to be appreciated much in
the general graphene literature. In particular, $\sigma(n) \sim
n^{\alpha}$ for both graphene and 2DEG with $\alpha=1$ for graphene at
intermediate density and $\alpha \approx 0.3$~to~$1.5$ in 2DEG depending on
the semiconductor system. At very high density, $\alpha \approx 0$ (or
even negative) for both graphene and 2DEG. The precise nature of
density dependence (i.e. value of the exponent $\alpha$) depends
strongly on the nature of scattering potential and screening, and
varies in different materials with graphene ($\alpha \approx 1$) falling
somewhere in the middle between Si-MOSFETs ($\alpha \approx 0.3$) and
modulation doped 2D n-GaAs ($\alpha \approx 1.5$). Thus, from the
perspective of high-density low-temperature transport properties,
graphene is simply a rather low-mobility (comparable to Si MOSFET, but
much lower mobility than 2D GaAs) 2D semiconductor system.



\subsubsection{Electron and hole puddles}

The low-density physics in both graphene and 2D semiconductors is
dominated by strong density inhomogeneity (``puddle'') arising from
the failure of screening. This inhomogeneity is mostly due to the
random distribution of unintentional quenched charged impurity centers
in the environment. (In graphene, ripples associated with either intrinsic
structural wrinkles or the substrate interface roughness may also make
contribution to the inhomogeneity.) At low density, the
inhomogeneous puddles control transport phenomena in graphene as well
as in 2D semiconductors. 
Inhomogeneous puddles would form also in doped 3D semiconductors at low carrier densities
\cite{shklovskii1984}.

In section IV B we discuss the details of electron-hole puddle
formation in graphene around the charge neutrality point and describe
its implications for graphene transport properties. Here we emphasize
the qualitative difference between graphene and 2D semiconductors with
respect to the formation of inhomogeneous puddles. In 2D
semiconductors, depending on whether the system is electron-doped or
hole-doped, there are only just electron- or just hole-puddle. At low
density, $n \approx 0$, therefore most of the macroscopic sample has
little finite carrier density except for the puddle regime. From the
transport perspective, the system becomes the landscape of mountains
and lakes for a boat negotiating a hilly lake. When percolation
becomes impossible, the system becomes an insulator. In graphene,
however, there is no gap at the Dirac point, and therefore, the
electron (hole) lakes are hole (electron) mountains, and one can
always have transport even at zero carrier density.
This picture breaks down when a spectral gap is introduced, and gapped
graphene should manifest an insulating behavior around the charge
neutrality point as it indeed does experimentally.

\subsubsection{Self-consistent theory}

The physical puddle picture discussed above enables one to develop a
simple theory for graphene transport at low densities using a
self-consistent approximation where the graphene puddle density is calculated
by considering the potential and density fluctuations induced by the
charged impurities themselves. Such a theory was developed by 
\citet{kn:adam2007a}.
The basic idea is to realize that at low carrier density
$|n| < |n_i|$, the self-consistent screening adjustment between the
impurities and the carriers could physically lead to an
approximate pinning of the carrier density at $n=n^*\approx n_i$.  A calculation
within the RPA approximation yields~\cite{kn:adam2007a}
\begin{eqnarray}
\label{Eq:CORPA}
\frac{n^*}{n_{\rm imp}} &=& 
2 r_s^2 C_0^{\rm RPA}(r_s, a = 4 d \sqrt{\pi n^*}), \\ 
 C_0^{\rm RPA}(r_s, a) &=& -1 + \frac{4 E_1(a)}{(2 + \pi r_s)^2}
+ \frac{2 e^{-a} r_s }{1 + 2 r_s}  \nonumber \\
&& \mbox{\hspace{-1in}} + (1 + 2 r_s a)~e^{2 r_s a} (E_1[2 r_s a] - E_1[a (1+ 2 r_s)]),
\nonumber 
\end{eqnarray}
where $E_1(z)= \int_z^\infty t^{-1} e^{-t} dt$ is the exponential 
integral function.   This density pinning then leads to an
approximately constant minimum graphene conductivity which can be
obtained from the high-density Boltzmann theory by putting in a
carrier density of $n^*$. This simple intuitive self-consistent theory
is found to be in surprisingly good agreement with all experimental
observations \cite{kn:adam2007a,kn:chen2008}.
In the next section we describe a more elaborate density
functional theory and an effective medium approximation to calculate
the puddle electronic structure and the resultant transport
properties \cite{kn:rossi2008,kn:rossi2008b}.

\subsection{Quantum to classical crossover}
\label{SA_SEC_QC_crossover}

The starting point for the quantum transport properties 
at the Dirac point discussed in Sec.~\ref{Sec:WAL} is the ballistic 
universal minimum conductivity $\sigma_{\rm min} = 4e^2/(\pi h)$ for clean
graphene.  The addition of disorder i.e. including 
potential fluctuations (given by Eq.~\ref{SA_Eq:GaussDis}) that are smooth 
on the scale of the lattice spacing {\em increases} the
conductivity through weak anti-localization.  This picture is
in contrast to the semi-classical picture discussed above 
where the transport properties are calculated at high density 
using the Boltzmann transport theory and the self-consistent theory
is used to handle the inhomogeneities of the carrier 
density around the Dirac point.  This theory predicts that
the conductivity {\em decreases} with increasing disorder strength.
Given their vastly different starting points, it is perhaps not 
surprising that the two approaches disagree.

A direct comparison between the two approaches has not been possible mainly 
because the published predictions of the Boltzmann approach include screening 
of the Coulomb disorder potential, whereas the fully quantum-mechanical 
calculations are for a non-interacting model using Gaussian disorder.  
Notwithstanding the fact that screening and Coulomb scattering play crucial 
roles in transport of real electrons through real graphene, the important 
question of the comparison between quantum and Boltzmann theories, even for 
Gaussian disorder, was addressed only recently \citet{kn:adam2008d} where 
they considered non-interacting Dirac electrons at zero temperature
with potential fluctuations of the form shown in Eq.~\ref{SA_Eq:GaussDis}.
They numerically solved the full quantum problem for a sample 
of finite size $L\gg \xi$ (where $\xi$ is the correlation length of
the disorder potential in Eq.~\ref{SA_Eq:GaussDis}), for a range of
disorder strengths parameterized by $K_0$.\footnote{Quantum effects
  are a small correction to the conductivity only if the carrier
  density $n$ is increased at fixed sample size $L$.  This is the
  experimentally relevant limit.  If the limit $L \rightarrow \infty$
  is taken at fixed $n$, quantum effects dominate (see Eq.~\ref{SA_Eq:WAL}), 
where the semiclassical theory does not capture the logarithmic scaling of
  conductivity with system size.  Here, we are not considering the
  conceptually simple question of how quantum transport becomes
  classical as the phase coherence length decreases, but the more
  interesting question of how this quantum-Boltzmann crossover depends
  on the carrier density and disorder strength.}

\begin{figure}
\bigskip
\includegraphics[scale=0.3]{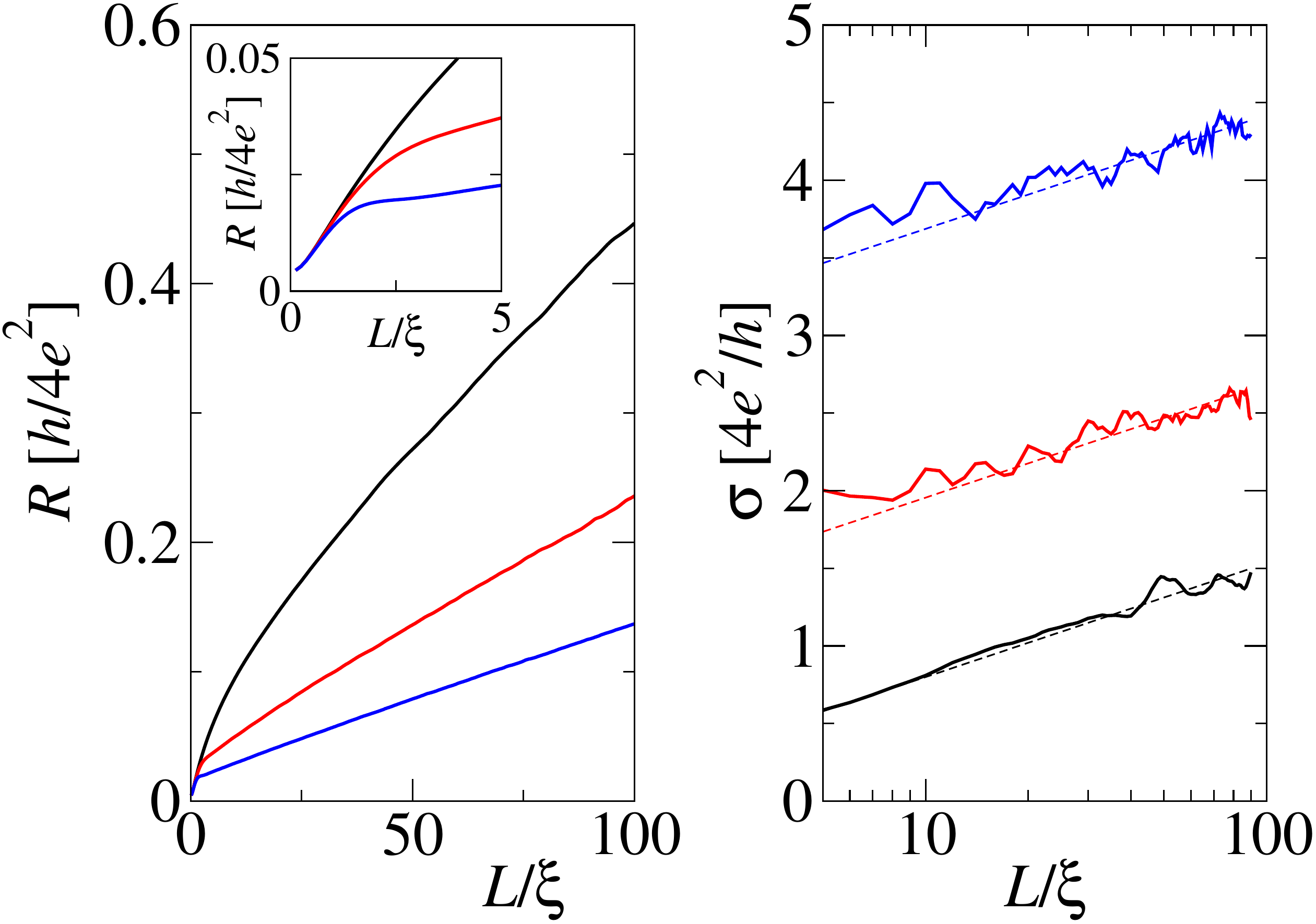}
\caption{\label{SA_Fig_Adam1} (Color online) Resistance,
$R = 1/G$ (left) and conductivity (right) obtained using
$\sigma = [W dR/dL]^{-1}$, as a function of sample length $L$.
The three curves shown are for $W/\xi=200$,
$K_0=2$ and $\pi n \xi^2=0$, $0.25$, and
$1$ [from top to bottom (bottom to top) in left (right) panel].
Dashed lines in the right panel show d$\sigma/d \ln L = 4 e^2/\pi h$. 
The inset in the left panel shows the crossover 
to diffusive transport ($L \gg \xi$).}
\end{figure}

\begin{figure}
\bigskip
\includegraphics[scale=0.35]{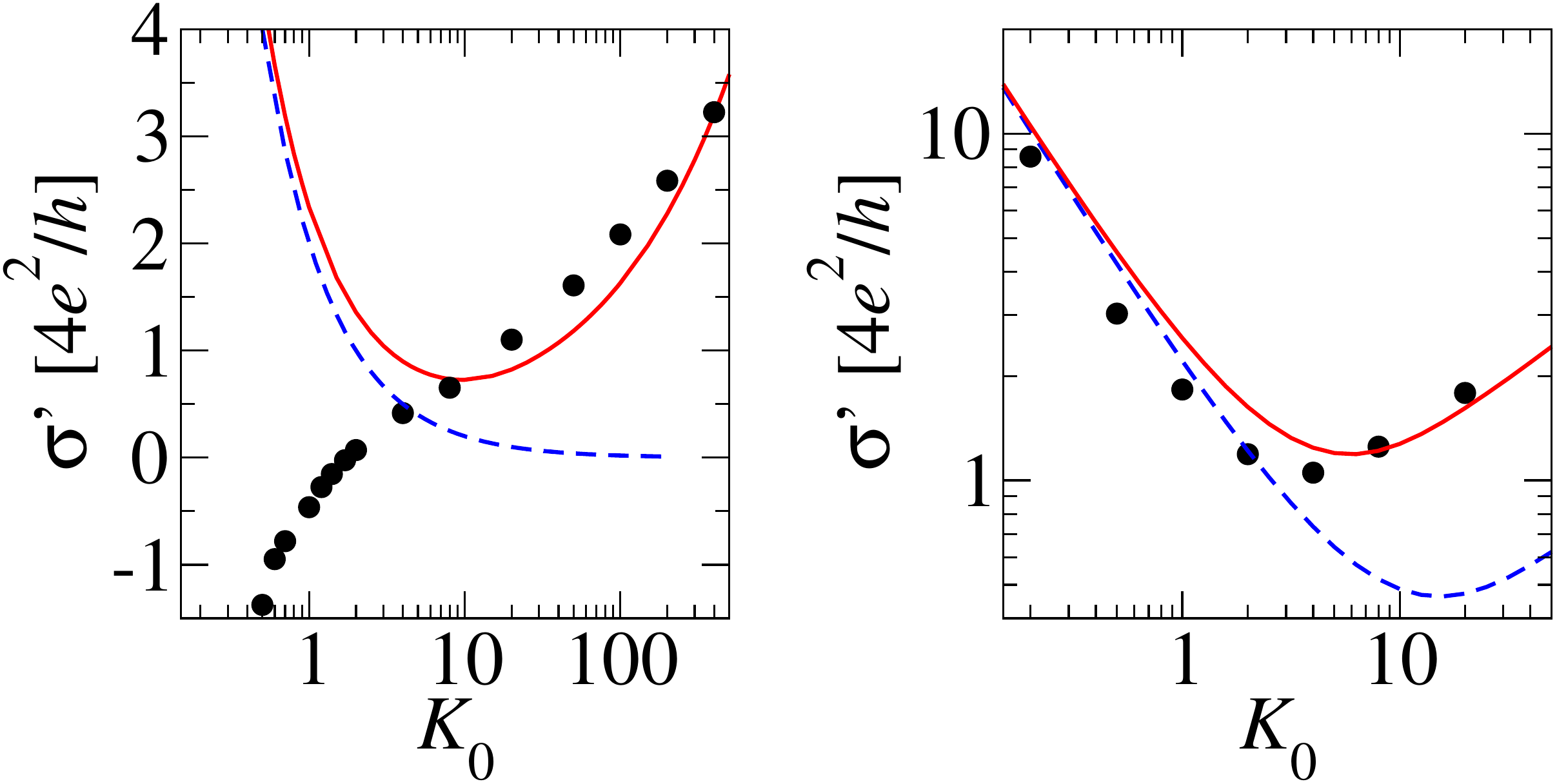}
\caption{\label{SA_Fig_Adam2}
(Color online) Semiclassical conductivity  
$\sigma' = \lim_{L \to \infty} [\sigma(L) -
\pi^{-1} \ln(L/\xi)]$ versus disorder strength at the
  Dirac point (left) and at carrier density $ \pi n=K_0/(\pi \xi)^2$, 
  corresponding to the edge of
  the minimum conductivity plateau of \citet{kn:adam2007a} (right). 
  Data points are
  from the numerical calculation for $L = 50 \xi$ and the (solid)
  dashed curves represent the (self-consistent) Boltzmann theory.}
\end{figure}

Typical results
for the quantum transport are shown in Fig.~\ref{SA_Fig_Adam1}.  For
$L\lesssim \xi$, the transport is ballistic and the conductivity
given by the universal value $\sigma_{\rm min} = 4e^2/(\pi h)$.  For
$L \gg \xi$, one is in the diffusive transport regime.  For the 
diffusive regime, \citet{kn:adam2008d} demonstrated that
away from the Dirac point, both the Boltzmann theory and 
the full quantum theory agree to leading order with 
\begin{equation}
\sigma(n) = \frac{2 \sqrt{\pi} e^2}{K_0 h} \left[ (2 \pi n \xi^2)^{3/2} 
+ {\mathcal O}(n \xi^2)^{1/2} \right].
\end{equation}      
\noindent
While this agreement is perhaps not surprising, it validates the
assumptions of both theories and demonstrates that they are compatible
at high carrier density.  More interesting are the results at 
the Dirac point.  Generalizing the self-consistent Boltzmann theory to
the case of a Gaussian correlated disorder 
potential (Eq.~\ref{SA_Eq:GaussDis}), one finds
\begin{equation}
\label{SA_Eq:SCG}
\sigma_{\rm min}^{\rm SC} = \frac{2e^2}{\pi h} \left( 
\exp\left[\frac{-K_0}{2 \pi}\right] I_1 \left[\frac{K_0}{2 \pi}\right]
\right)^{-1},   
\end{equation}
\noindent  where $I_1$ is the modified Bessel function.  Shown in the
left panel of Fig.~\ref{SA_Fig_Adam2} is a comparison of the numerical
fully quantum Dirac point conductivity where the weak antilocalization
correction has been subtracted $\sigma' = \lim_{L \rightarrow \infty}
[\sigma(L) - \pi^{-1} \ln(L/\xi)]$ with the semiclassical result
(Eq.~\ref{SA_Eq:SCG}).  

The right panel shows the conductivity slightly away from the Dirac point
(i.e. at the edge of the minimum conductivity plateau).  The numerical
calculations at the edge of the plateau are in good quantitative
agreement with the self-consistent Boltzmann theory.  At the Dirac
point, however, the quantum conductivity
$\sigma(K_0)$ is found to increase with $K_0$ for the
entire parameter range considered, which differs from the Boltzmann
theory at small $K_0$.  At large $K_0$ the numerical data follow the
trend of the self-consistent theory which predicts $\sigma \sim 2e^2
K_0^{1/2}/(\pi h)$ for $K_0 \gg 10$.  This implies that even at the
Dirac point, for large enough disorder, the transport is semi-classical
and described by the self-consistent Boltzmann transport theory.

For smaller $K_0$, Fig.~\ref{SA_Fig_Adam2} shows that upon reducing
$K_0$ below unity, the conductivity first decreases sharply consistent
with a renormalization of the mean free path due to the ultraviolet
logarithmic divergences discussed in Sec.~\ref{SA_Sec_uvlc}.  Upon
reducing $K_0$ further, the Dirac point conductivity saturates
at the ballistic value $\sigma_{\rm min} = 4e^2/\pi h$ (discussed 
in Sec.~\ref{SA_Sec_UBMC}).         
    
In a closely related work, \citet{kn:lewenkopf2008} did numerical 
simulations of a tight-binding model to obtain the conductivity
and shot noise of graphene at the Dirac point using a recursive
Green's function method.  This method was then generalized 
to calculate the metal-insulator transition in graphene nanoribbons
where, as discussed in Sec.~\ref{SA_Sec_symplectic_crossover}, edge 
disorder can cause the 
Anderson localization of electrons~\cite{kn:mucciolo2008}.

The important conclusion of this section is that it
provides the criteria for when one needs a full quantum mechanical
solution and when the semi-classical treatment is sufficient.  For 
either sufficiently weak disorder, or when the source and
drain electrodes are closer than the scattering 
mean free path, then the quantum nature of the 
carriers dominates the transport.  On the other hand, for sufficiently
large disorder, or away from the Dirac point, the electronic transport
properties of graphene are semi-classical and the Boltzmann theory 
correctly captures the most of graphene's transport properties.

\subsection{Ground state in the presence of long-range disorder}
\label{subsec:ground_state}
In the presence of long-range disorder that does not mix the degenerate valleys
the physics of the graphene fermionic  excitations is described by the following Hamiltonian:
\begin{align}
 {\cal H} = &\int d^2r\Psi_{\rr\alpha}^\dagger [-i\hbar v_F{\bm \sigma}_{\alpha\beta}\cdot\nabla_{\rr}-\mu{\bf 1}]\Psi_{\rr\beta} + \nonumber \\
            &\frac{e^2}{2\kappa}\int d^2rd^2r' \Psi_\rr^\dagger \Psi_{\rr\alpha} V(|{\bf r-r'}|)\Psi_{\rr'\alpha}^\dagger \Psi_{\rr'\alpha}+\nonumber \\
            &\frac{e^2}{2\kappa}\int d^2r V_D(\rr)\Psi_{\rr\alpha}^\dagger \Psi_{\rr\alpha}
 \label{eq:ln:ham}
\end{align}
where $v_F$ is the bare Fermi velocity, $\Psi_{\rr\alpha}^\dagger$, $\Psi_{\rr\alpha}$
are the creation annihilation spinor operators for a fermionic excitation
at position  $\rr$ and pseudospin $\alpha$, ${\bm \sigma}$ is the 2D vector formed by
the $2\times 2$ Pauli matrices $\sigma_x$ and $\sigma_y$ acting in pseudospin space, 
$\mu$ is the chemical potential, ${\bf 1}$ is the $2\times 2$ identity matrix,
$\kappa$ is the effective static dielectric constant
equal to the average of the dielectric constants of the materials surrounding
the graphene layer, $V(|{\bf r-r'}|)=1/||{\bf r-r'}|$ is the Coulomb interaction  
and  $V_D(\rr)$ is the bare disorder potential.
The Hamiltonian \ceq{eq:ln:ham} is valid as long as the energy of the
fermionic excitations is much lower than the graphene bandwidth $\approx 3$~eV.
Using \ceq{eq:ln:ham} if we know $V_D$ we can 
characterize the ground state carrier density probability close to the Dirac point.
In this section we focus on the case when $V_D$ is a disorder potential
whose spatial autocorrelation decays algebraically, such as the disorder
induced by ripples or charge impurities.  


\subsubsection{Screening of a single charge impurity}

The problem of screening at the Dirac point of a single charge impurity placed in (or close to)
the graphene layer illustrates some of the unique features of the screening properties
of massless Dirac fermions. In addition the problem provides a condensed
matter realization of the QED phenomenon of ``vacuum polarization'' induced by
an external charge
\cite{case1950,zeldovich1972,pomeranchuk1945,darwin1928,gordon1928}.
In the context of graphene the problem was first studied by \citet{divincenzo1984}
and recently more in detail by several authors 
\cite{kn:pereira2007, kn:shytov2007, kn:novikov2007,kn:fogler2007,kn:biswas2007,shytov2007b,fistul-prl-98-256803-2007,kn:terekhov2008}.
The parameter $\beta \df Ze^2/(\kappa \hbar v_f)=Zr_s$
quantifies the strength of the coupling between the Coulomb impurity and
the massless Dirac fermions in the graphene layer.
Neglecting $e-e$ interactions for $|\beta|<1/2$ the Coulomb impurity induces a screening
charge that is localized on length scales of the order of the
size of the impurity itself (or its distance $d$ from the graphene layer).
Even in the limit $|\beta|<1/2$ the inclusion of the $e-e$ interactions induces
a long-range tail in the screening charge with sign equal
to the sign of the charge impurity \cite{kn:biswas2007}.
For $|\beta|>1/2$ the Coulomb charge is supercritical, the induced potential is singular \cite{kn:landau1977},  
and the solution of the problem depends on the regularization of the wavefunction at the site of the impurity, $r\to 0$.
By setting the wavefunction to be zero at $r=a$ the induced electron density in 
addition to a localized $(\delta(\rr))$ term, acquires a long-range tail
$\sim 1/r^2$ (with sign opposite to the sign of the charge impurity) 
\cite{kn:shytov2007, kn:pereira2007,novikov2007} and
marked resonances appear in the spectral density
\cite{shytov2007b,fistul-prl-98-256803-2007} that should also
induce clear signatures in the transport coefficients.
Up to now neither the oscillations in the LDOS nor the prdicted signatures in the conductivity \cite{shytov2007b}
have been observed experimentally. It is likely that in the experiments so
far the supercritical regime $|\beta|>1/2$ has not been reached because
of the low $Z$ of the bare charge impurities and renormalization effects.
\citet{kn:fogler2007} pointed out however that the predicted effects
for $|\beta|>1/2$ are intrinsic to the massless Dirac fermion model that however
is inadequate  when the
the small scale cut-off ${\rm min}[d, a]$ is smaller
than $a r_s\sqrt{Z}$.

\subsubsection{Density functional theory}
\label{sec:ln:lda-dft}

\begin{figure}[!h]
 \begin{center}
  \includegraphics[width=8.5cm]{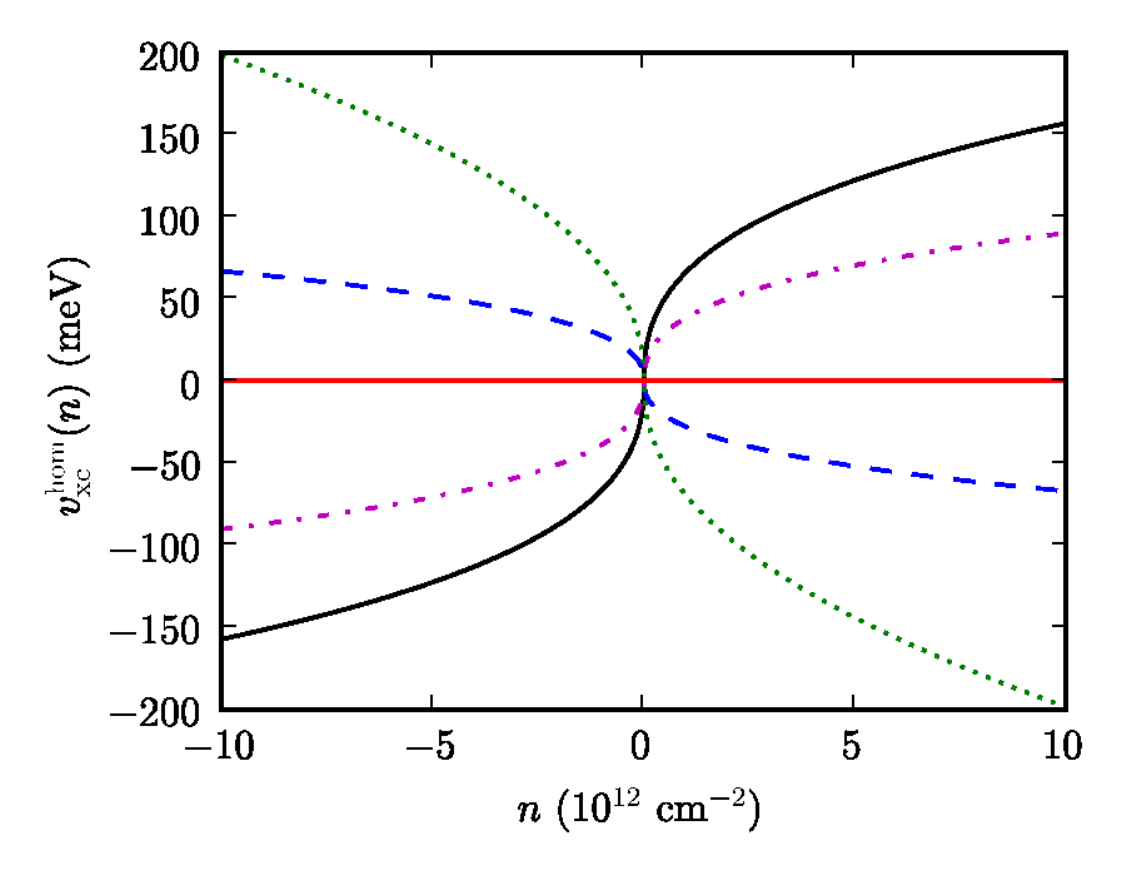}
  \caption{(Color online). 
    Exchange, black solid line,  and RPA correlation potentials
    blue dashed line, as functions of the density $n$
    for $r_s=0.5$. The
    magenta dash-dotted line shows the full
    exchange-correlation potential, $V_{xc}$.
    The green dotted line
    is the quantum Monte Carlo exchange-correlation potential of a
    standard parabolic-band 2D electron
    gas~\cite{attaccalite2002} with effective mass $0.067~m_e$
    placed in  background with dielectric constant $4$.
    Adapted from \citet{kn:polini2008}.
  } 
  \label{fig:ln:xc_n}
 \end{center}
\end{figure}

Assuming that the ground state does not have long range order
\cite{kn:peres2005,kn:min2008b,dahal2006}
a practical approach to many-body problems is the 
Density Functional Theory, DFT, \cite{hohenberg1964,kohn1965,kohn1999,giuliani2005}.
In this approach the interaction term in the Hamiltonian is replaced
by an effective Kohn-Sham potential $V_{KS}$ that is a functional of the 
ground-state density $n(\rr)=\sum_{\sigma} \Psi_{\rr \sigma}^\dagger\Psi_{\rr \sigma}$.
\begin{align}
 {\cal H} = &\int d^2r\Psi_{\rr\alpha}^\dagger [-i\hbar v_F{\bf \sigma}_{\alpha\beta}\cdot\nabla_\rr-\mu{\bf 1}]\Psi_{\rr\beta} +\nonumber \\
            &\int d^2r\Psi_{\rr\alpha}^\dagger V_{KS}[n(\rr)]\Psi_{\rr\alpha}
 \label{eq:ln:hamks}
\end{align}
The Kohn-Sham potential is given by the sum
of the external potential, the Hartree part of the interaction, $V_H$, and 
an exchange-correlation potential, $V_{xc}$, that can only be known
approximately. 
In its original form
$V_{xc}$ is calculated within the local density approximation, LDA \cite{kohn1965},
i.e., $V_{xc}$ is calculated for a uniform liquid of electrons.
The DFT-LDA approach can be justified and applied to the study of interacting
massless Dirac fermions \cite{kn:polini2008}. For graphene the LDA exchange-correlation
potential, within the RPA approximation is given with very good accuracy by the following expression 
\cite{gonzalez1999, kn:katsnelson2006, kn:barlas2007,kn:hwang2007,kn:polini2008,mishchenko2007,vafek2007}:
\begin{align}
 V_{xc}(n)= &+\frac{r_s}{4}\sqrt{\pi|n|}\sgn(n)\ln\frac{4k_c}{\sqrt{4\pi n}}  \nonumber \\
           &-\frac{g r_s^2\xi(gr_s)}{4}\sqrt{\pi|n|}\sgn(n)\ln\frac{4k_c}{\sqrt{4\pi n}}
 \label{eq:ln:vxc}
\end{align}
where $g$ is the spin and valley degeneracy factor ($g=4$), and 
$k_c$ is an ultraviolet wave-vector cut-off, fixed by the range
of energies over which the pure Dirac model is valid. Without loss of generality
we can use $k_c=1/a$, where $a$ is the graphene lattice constant,
corresponding to an energy cut-off $E_c\approx 3\;{\rm eV}$.
Equation \ceq{eq:ln:vxc} is valid for $k_F=\sqrt{\pi |n|}\ll k_c$.
In \ceq{eq:ln:vxc} $\xi$ is a constant that depends on $r_s$ giben by \cite{kn:polini2008}:
\beq
 \xi(g r_s) = \frac{1}{2}\int_0^{+\infty}
 \frac{dx}{(1+x^2)^{2}(\sqrt{1+x^2}+\pi g r_s/8)}.
 \label{eq:ln:xi}
\enq
The terms on the r.h.s. of \ceq{eq:ln:vxc} are the exchange and the correlation 
potential respectively. Notice that exchange and correlation potentials
have opposite signs.
In Fig.~\ref{fig:ln:xc_n} the exchange and correlation potential and their sum, $V_{xc}$, are plotted 
as a function of $n$ for $r_s=0.5$. We see that in graphene the correlation potential is smaller than the exchange
potential but, contrary to the case of regular parabolic band 2DEGs, it is not negligible.
However from \ceq{eq:ln:vxc} we have that exchange and correlation scale with $n$ in the same way.
As a consequence in graphene the correlation potential
can effectively be taken into account
by simply rescaling the coefficient of the exchange potential. 
In Fig.~\ref{fig:ln:xc_n} the green dotted line shows $V_{xc}$ for a regular
parabolic band 2DEG with effective mass $0.067m_e$ in a background with $\kappa=4$.
The important qualitative difference is that $V_{xc}$ 
in graphene has the opposite sign than in regular 2DEGs: due to interlayer
processes in graphene
the exchange correlation potential penalizes density inhomogeneities
contrary to what happens in parabolic-band electron liquids.

Using the DFT-LDA approach
\citet{kn:polini2008} calculated the graphene 
ground-state carrier density for single disorder realizations of charge impurities and small samples
(up to $10\times 10$~nm). The size of the samples is limited by the high computational
cost of the approach.
For single disorder realizations the results of \citet{kn:polini2008}
show that, as it had been predicted \cite{kn:hwang2006c}, at the Dirac
point the carrier density breaks up in electron-hole puddles and that the exchange-correlation
potential suppresses the amplitude of the disorder induced density fluctuations.
Given its computational cost, the DFT-LDA approach does not allow the 
calculation of disordered averaged quantities.
%
%

\subsubsection{Thomas Fermi Dirac theory}

An approach similar in spirit to the LDA-DFT is the 
Thomas-Fermi, TF, \cite{fermi1927,thomas1927,spruch1991,giuliani2005} theory. 
Like DFT the TF theory is a density functional theory:
in the Thomas-Fermi theory the kinetic term is also approximated 
via a functional of the local density $n(\rr)$. 
By Thomas-Fermi-Dirac, TFD, theory we refer to a modification of the TF theory
in which the kinetic functional has the form appropriate
for Dirac electrons and in which exchange-correlation terms are
included via the exchange-correlation potential proper for Dirac electron liquids 
as described above for the DFT-LDA theory.
The TF theory relies on the fact that if the carrier density varies slowly in space
compared to the Fermi wavelength, then the kinetic energy of a small
volume with density $n(\rr)$ is equal, with good approximation, to the 
kinetic energy of the same volume of a homogeneous electron liquid with density $n=n(\rr)$.
The condition for the validity of the TFD theory is given by the inequality \cite{giuliani2005,kn:brey2009}:
\beq
 \frac{|\nabla_\rr n(\rr)|}{n(\rr)}\ll k_F(\rr).
 \label{eq:ln:tfineq}
\enq
Whenever inequality \ceq{eq:ln:tfineq} is satisfied 
the TFD
theory is a computationally efficient alternative to the DFT-LDA approach to
calculate the ground state properties of graphene in presence of disorder.
The energy functional, $E[n]$ in the TFD theory is given by:
\begin{align}
 E[n] = & \hbar v_F\left[\frac{2\sqrt{\pi}}{3}\int d^2 r\sgn(n)|n|^{3/2} \right. \nonumber \\
        & + \frac{r_s}{2}\int d^2 r\int d^2 r'\frac{n(\rr)n(\rr')}{|\rr - \rr'|} 
          + \int d^2 r V_{xc}[n(\rr)]n(\rr) \nonumber \\ 
        & + \left. r_s\int d^2 r V_D(\rr)n(\rr) - 
          \frac{\mu}{\hbar v_F}\int d^2 r n(\rr)\right]
 \label{eq:ln:tfden}
\end{align}
where the first term is the kinetic energy, the second is the Hartree part of
the Coulomb interaction, the third is the term due to exchange and correlation,
and the fourth is the term due to disorder. The expression for the
exchange-correlation potential is given in Eq.~\ceq{eq:ln:vxc}.
The carrier ground state distribution is then calculated by minimizing $E[n]$ with respect to $n$.
Using \ceq{eq:ln:tfden} the condition $\delta E/\delta n=0$ requires
\begin{align}
 &\sgn(n)\sqrt{|\pi n|} 
  + \frac{r_s}{2}\int d^2 r\frac{n(\rr')}{|\rr - \rr'|} 
  + V_{xc}[n(\rr)] \nonumber \\
 &+ r_sV_D(\rr) - \frac{\mu}{\hbar v_F}=0.
 \label{eq:ln:tfddedn}
\end{align}
Equation \ceq{eq:ln:tfddedn} well exemplifies the nonlinear nature of
screening in graphene close to the Dirac point: because in graphene, due
to the linear dispersion, the kinetic energy per carrier, 
the first term in \ceq{eq:ln:tfddedn}, scales with $\sqrt{n}$ when
$\nav=0$ the relation between the density fluctuations $\delta n$
and the external disorder potential is not linear even when
exchange and correlation terms are neglected. 



We now consider the case when the disorder potential is due to random Coulomb impurities.
In general the charge impurities will be a 3D distribution, $C(\rr)$, however
${\rm SiO_2}$ 
we can assume to a very good approximation $C(\rr)$ to be
effectively 2D. The reason is that for normal substrates such as
${\rm SiO_2}$ the charge traps migrate to the surface of the
oxide, moreover any additional impurity charge introduced during
the graphene fabrication will be located either on the graphene top
surface or trapped between the graphene layer and the substrate.
We then assume $C(\rr)$ to be an effective 2D random distribution
located at the average distance $d$ from the graphene layer. 
An important advantage of this approach is that it limits to two
the number of unknown parameters
that enter the theory:
charge impurity density, $\nimp$, and $d$.
With this assumption we have:
\beq
 V_D(\rr) = \int d\rr'\frac{C(\rr')}{[|\rr - \rr']^2+d^2]^{1/2}}.
 \label{eq:ln:Vd}
\enq
The correlation properties of the distribution $C(\rr)$ of the charge impurities
are a matter of long-standing debate in the semiconductor community. Because the
impurities are charged one would expect the positions of the impurities to have some
correlation, on the other hand the impurities are quenched (not annealed), they are either imbedded
in the substrate or between the substrate and the graphene layer or in the graphene itself.
This fact makes very difficult to know the precise correlation
of the charge impurity positions but it also ensures that to good approximation
the impurity positions can be assumed to be uncorrelated:
\begin{equation}
 \langle C(\rr)\rangle = 0; \hspace{0.5cm}
 \langle C(\rr_1) C(\rr_2)\rangle = n_{imp}\delta(\rr_2 - \rr_1).
 \label{eq:ln:C_stat}
\end{equation}
where the angular brackets denote averaging over disorder realizations.
A non zero value of $\langle C(\rr)\rangle$
can be taken into account simply by a shift of the chemical potential $\mu$.
It is easy to generalize the theory to correlated impurities (e.g. impurity clusters)
if the correlation function is known.

The parameters $\nimp$ and $d$ that enter the theory are reliably fixed by
the transport results, see Sec.~\ref{subsec_boltzmann}, at high doping.
Transport results at high density indicate that $d$ is of the order
of 1~nm whereas $\nimp$ varies depending on the sample quality
but in general is in the range $n_{imp}=10^{10}-10^{12}$~${\rm cm^{-2}}$,
where the lowest limit applies to suspended graphene.
The distance $d$ is the physical cutoff for the length-scale of the
carrier density inhomogeneities. 
Therefore to solve numerically
Eq.~\ceq{eq:ln:tfddedn} one can use a spatial discretization with unit
step of the order of $d$. For the TFD results presented below
it was assumed $d=1$~nm and therefore a spatial step $\Delta x=\Delta y=1$~nm
was used. 

\begin{figure}[htb]
 \begin{center}
   \includegraphics[width=8.5cm]{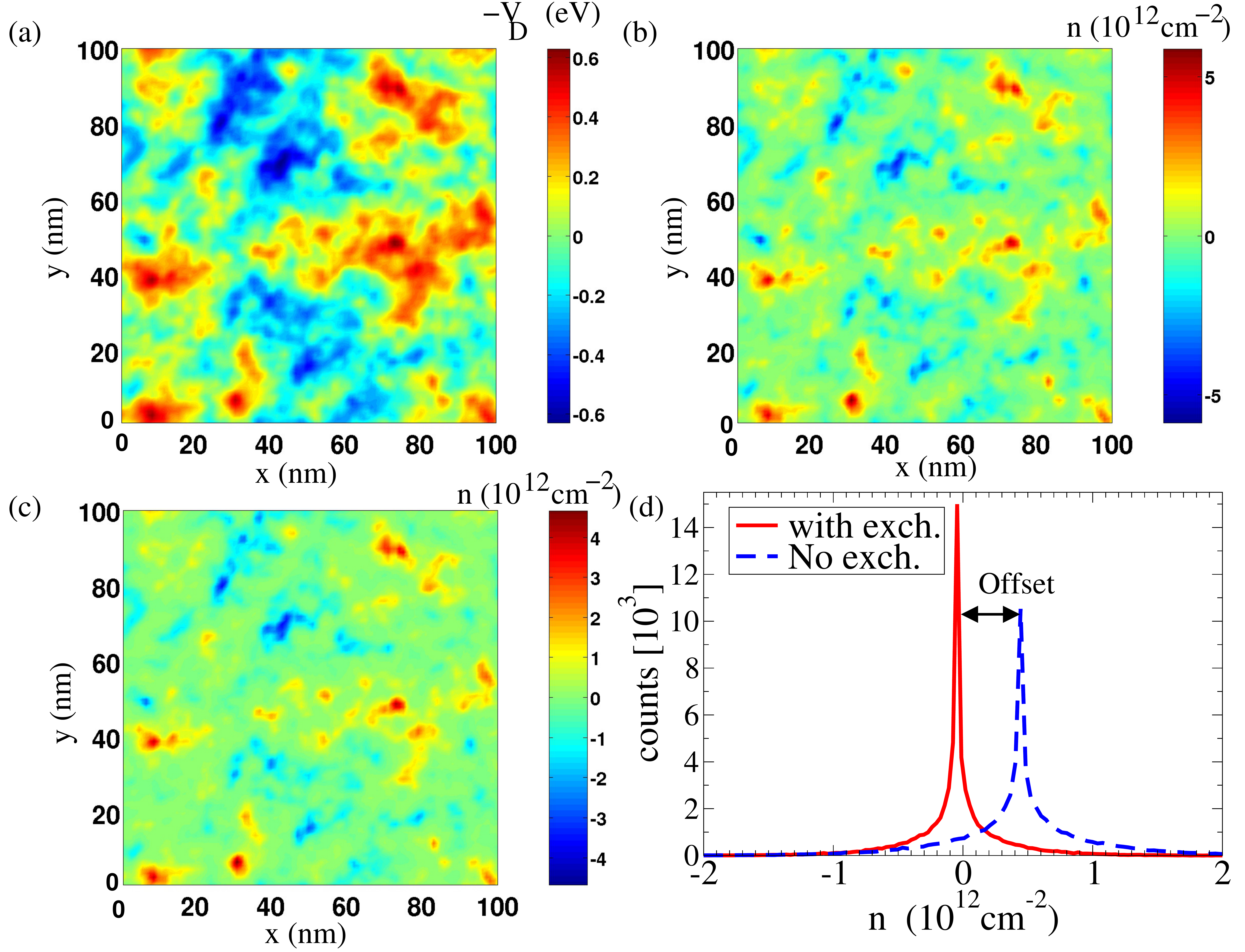}
  \caption{
           (Color online).
           TFD results as a function of position for a single disorder realization.
           (a) Bare disorder potential, $V_D$. 
           (b) Carrier density obtained neglecting exchange-correlation terms.
           (c) Carrier density obtained including exchange-correlation terms.
           (d) Probability density distribution at the CNP with exchange 
           (solid line) and without (dashed line).
           Adapted from \citet{kn:rossi2008}
          } 
  \label{fig:ln:single}
 \end{center}
\end{figure}

Fig.~\ref{fig:ln:single} shows the TFD results for the carrier density distribution
at the Dirac point in the presence of charge impurity disorder
for a single disorder realization
with $\nimp=10^{12}\;{\rm cm^{-2}}$ and $\kappa=2.5$ corresponding to graphene on ${\rm SiO_2}$
with the top surface in vacuum (or air).
It is immediately clear that as was predicted \cite{kn:hwang2006c},
close to the Dirac point the disorder induced by the charge impurities
breaks up the carrier distribution in electron ($n>0$) and hole ($n<0$) puddles.
The electron-hole puddles are separated by disorder-induced $p-n$ junctions, PNJ.
Apart from the PNJ the carrier density
is locally always different from zero even though the average density, $\nav$,
is set equal to zero.
For this reason, in the presence of disorder it is more correct to refer to the
value of the gate voltage for which $\nav=0$ as the charge neutrality point,
CNP, rather than Dirac point: the presence of long-range disorder
prevents the probing of the physical properties of the Dirac point,
i.e. of intrinsic graphene with exactly half-filling, zero-density, everywhere.
%
The important qualitative results that
can be observed even for a single disorder by comparing the results
of panels (b) and (c) is that the exchange-correlation term suppresses
the amplitude of the density fluctuations. This fact is clearly visible
in panel (d) from which we can see that in the presence of exchange-correlation
the density distribution is much narrower and more peaked around zero.
This result, also observed in the DFT-LDA results \cite{kn:polini2008}, is a consequence
of the the fact that as discussed in Sec.~\ref{sec:ln:lda-dft} the 
exchange-correlation potential in graphene, contrary to parabolic-band
Fermi liquids, penalizes density inhomogeneities.

\begin{figure}[htb]
 \begin{center}
  \includegraphics[width=8.5cm]{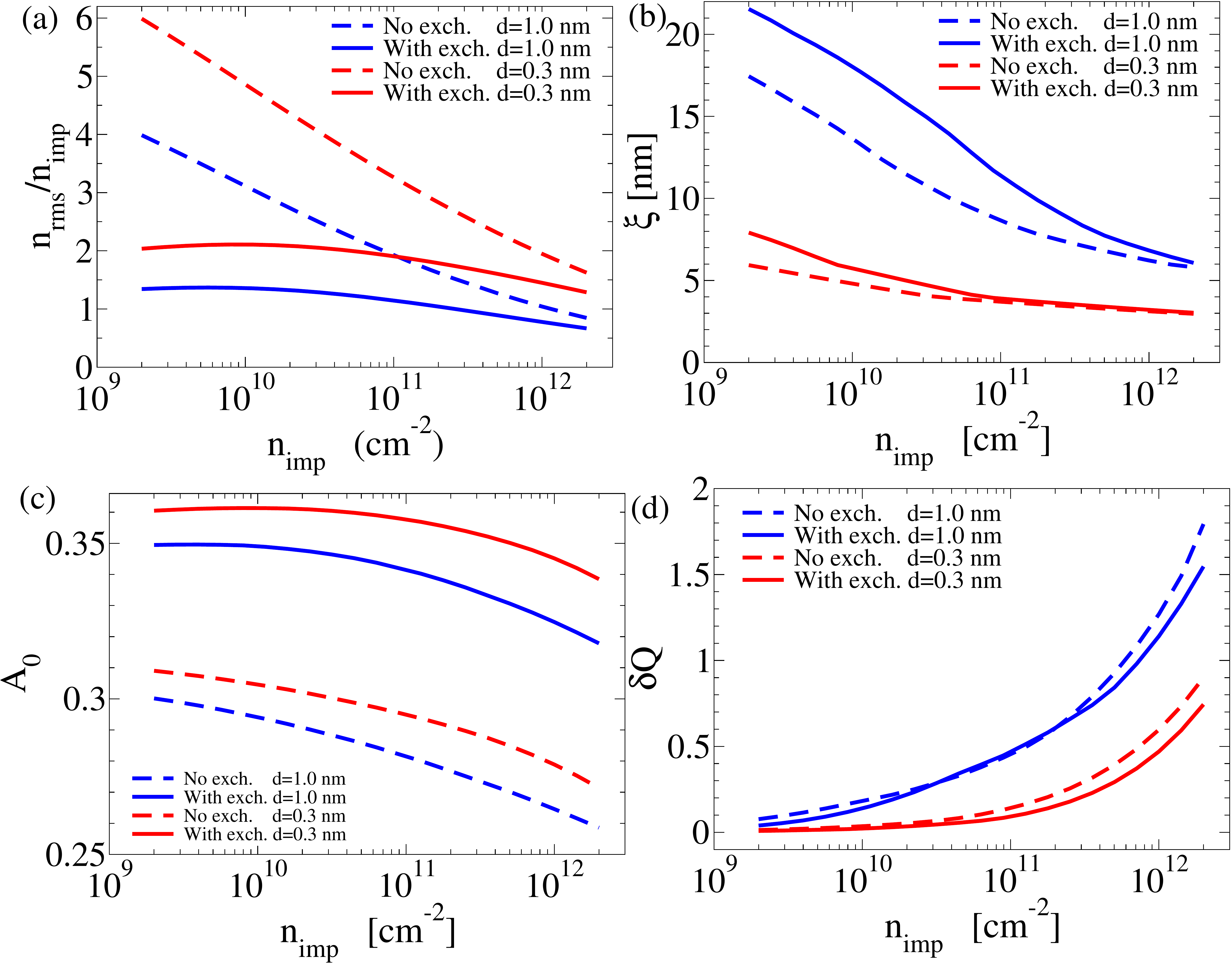}
  \caption{
           (Color online). 
           TFD disorder averaged results.
           The solid (dashes) lines show the results obtained
           including (neglecting) exchange and correlation terms
           (a) $n_{\rm rms}$ and, (b) $\xi$ as a function of $\nimp$.
           (c) Area, $A_0$, over which  $|n(\rr)-\langle n\rangle|<n_{rms}/10$.
           (d) Average excess charge $\delta Q$ vs. $\nimp$.
           Adapted from \citet{kn:rossi2008}
          } 
  \label{fig:ln:nrms_xi_nimp}
 \end{center}
\end{figure}

In the presence of disorder, in order to make quantitative predictions verifiable experimentally
it is necessary to calculate disordered averaged quantities. 
Using TFD, both the disorder average $\langle X\rangle$ of a given quantity  $X$ 
and its spatial correlation function
\beq
 \delta X^2(\rr) \equiv \langle (X(\rr)-\langle X\rangle) 
                                (X(0)  -\langle X\rangle)\rangle
\enq 
can be efficiently calculated. For conditions typical in experiments
500 disorder realizations are sufficient.
From $\delta X^2(\rr)$ one can extract the
quantities:
\beq
 X_{rms}\df\sqrt{\langle(\delta X(0))^2\rangle}; 
 \hspace{0.5cm}
 \xi_X\df FWHM {\;\rm of\;} \langle(\delta X(\rr))^2\rangle;
\enq 
respectively as the root mean square (rms) and the typical spatial correlation
of the fluctuations of $X$. 
Using the TFD theory both the spatial correlation
function of the screened potential, $V_{sc}$, and carrier density 
are found to decay at long distance as $1/r^3$. This is
a consequence of the weak screening properties of graphene 
and was pointed out in \cite{kn:galitski2007, kn:adam2007a}.
From the spatial correlation functions
$\nrms$ and $\xi\equiv\xi_n$ are extracted.
Fig.~\ref{fig:ln:nrms_xi_nimp} (a), (b) show the calculated $\nrms$ and $\xi$ at the Dirac
point as a function of $\nimp$. The disorder averaged results
show the effect of the exchange-correlation potential in
suppressing the amplitude of the density inhomogeneities, $\nrms$, and
in slightly increasing their correlation length. The effect of the
exchange-correlation potential increases as $\nimp$ decreases.
At the Dirac point, 
the quantity $\xi$ can be interpreted as
the {\em effective non-linear screening length}.  Fig.~\ref{fig:ln:nrms_xi_nimp}
shows that $\xi$ depends weakly on $\nimp$.
The reason is that $\xi$ only
characterizes the spatial correlation of the regions in which the
density is relatively high.  
If a puddle is defined as a continuous region with same sign charges then
at the CNP the puddles have always a size of the same order of the system
size. 
Inside the puddles there are small areas with high density
and size $\xi$ of the order of tens of nanometers for typical
experimental conditions, much smaller than the system size, $L$.
%
%
This picture is confirmed in Fig.~\ref{fig:ln:nrms_xi_nimp}~(c) in which the
disorder averaged area fraction, $A_0$, over which $|n(\rr)-\langle
n\rangle|<n_{rms}/10$ is plotted as a function of $n_{imp}$.  As
$n_{imp}$ decreases $A_0$ increases reaching more than 1/3 at the
lowest impurity densities.  The fraction of area over which
$|n(\rr)-\langle n\rangle|$ is less than $1/5$ of $n_{rms}$ surpasses
50~\% for $n_{imp}\lesssim 10^{10}$~cm$^{-2}$.  
%
%
Fig.~\ref{fig:ln:nrms_xi_nimp}~(b) shows the average excess charge
$\delta Q \df n_{rms}\pi\xi^2$ at the Dirac point
as a function of $\nimp$. Notice that as defined $\delta Q$,
especially at low $\nimp$,
grossly underestimates the number of charges both in the 
electron-puddles and in the small regions of size $\xi$.
This is becuase in the regions of size $\xi$ the density
is much higher than $\nrms$ whereas the electron-hole
puddles have a typical size much larger than $\xi$.
Using for the small regions the estimate $|\nabla n(\rr)|/n=1/\xi$ and 
the local value of $n$ inside the regions and, for the electron-hole puddles the
estimates:
\begin{align}
 n\approx \nrms, \hspace{0.3cm}& |\nabla n(\rr)| \approx \frac{\nrms}{L}
 \label{eq:ln:estbigpud}
\end{align}
we find that 
the inequality \ceq{eq:ln:tfineq} is satisfied guaranteeing the 
validity of the TFD theory even at the Dirac point.
%
%
%

\begin{figure}[!!!b]
 \begin{center}
  \includegraphics[width=8.5cm]{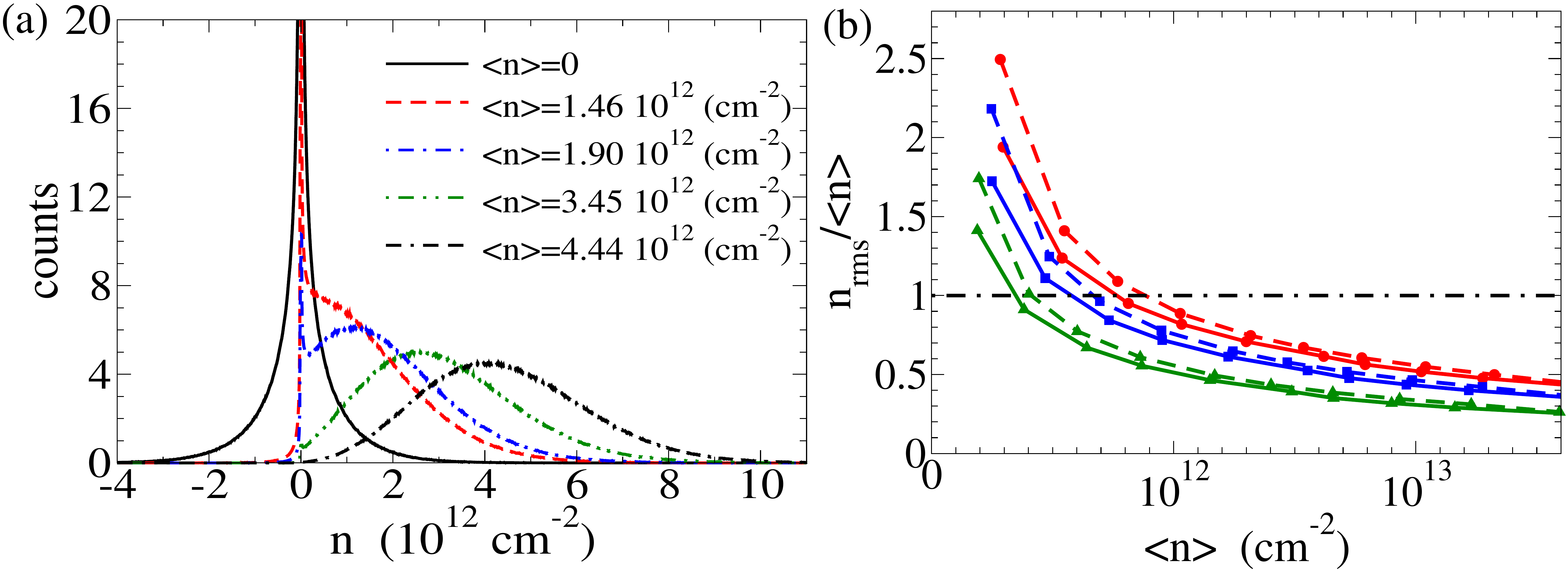}
\caption{(Color online).
          (a) Density distribution averaged over disorder
          for different values of the applied gate voltage
          assuming $\kappa = 2.5$, $d=1$~nm and $n_{imp}=10^{12}\;{\rm cm^{-2}}$.
          (b) $n_{\rm rms}/\langle n\rangle$ as a function of $\langle n\rangle$ for $d=1$~nm and different
          values of $n_{imp}$: circles, $n_{imp} = 1.5\times 10^{12}\;{\rm cm^{-2}}$;
          squares, $n_{imp} = 10^{12}\;{\rm cm^{-2}}$;
          triangles, $n_{imp} = 5\times 10^{11}\;{\rm cm^{-2}}$. Adapted from \citet{kn:rossi2008}.
         }
  \label{fig:ln:Pn}
 \end{center}
\end{figure}

As we move away from the Dirac point more of the
area is covered by electron (hole) puddles. However the
density flucutuations remain large even for relatively large values
of $V_g$. This is evident from Fig.~~\ref{fig:ln:Pn}
where the probability distribution
$P(n)$ of the density for different values of $\nav$ 
is shown in panel (a) and the
ratio $\nrms/\nav$ as a function of $\nav$ is shown in panel (b).
The probability distribution $P(n)$ is non-Gaussian \cite{kn:galitski2007,kn:adam2007a,kn:adam2009}.
For density $\nav\lesssim\nimp$, $P(n)$ does not exhibit a 
single peak around $\nav$ but rather a bimodal structure with a 
strong and narrow peak arond zero. The double peak structure for finite
$V_g$ provides direct evidence for the existance of puddles over a finite
voltage range. 
%
$\nrms/\nav$ decreases with $\nav$, a trend that is expected and
that has been observed indirectly in experiments by measuring the inhomogenous
broadening of the quasiparticle spectral function \cite{kn:hong2009}.

In the limit $r_s\ll 1$ it is possible to obtain analytic results using
the TFD approach \cite{fogler-prl-103-236801-2009}. The first step is to separate
the inhomogeneities of the carrier density and screened potential
in {\em slow}, $\bar n$, $\bar\Vsc$  and fast components, $\delta n$, $\delta\Vsc$:
\begin{equation} \label{eq:tfd_an:1}
n(\mathbf{r}) = \bar{n}(\mathbf{r}) + \delta n(\mathbf{r})\,,
\quad
V_{sc}(\mathbf{r}) = \bar{V_{sc}}(\mathbf{r}) + \delta V_{sc}(\mathbf{r})\,,
\end{equation}
where $\bar n$ and $\bar\Vsc$ contain only Fourier harmonics with $k < \Lambda$
where $1/\Lambda$ is the spatial scale below which the spatial variation
of $n$ and $\Vsc$ are irrelevant for the physical properties measured.
For imaging experiments $1/\Lambda$ is the spatial resolution
of the scanning tip and for transport experiments $1/\Lambda$ is of the
order of the mean free path. 
Let the $\limp\equiv 1/2r_s\sqrt{\nimp}$ and $R$ the non-linear screening length.
It is assumed that $l\lesssim 1/\Lambda \ll R$. With these assumptions
and neglecting exchange-correlation terms, from the TFD functional
in the limit $\Lambda\ll k_F$ and small non-linear screening terms compared
to the kinetic energy term it follows \cite{fogler-prl-103-236801-2009}:
\begin{equation} 
 \bar{n}(\bar{\Vsc}) \simeq
               -\frac{\bar{\Vsc} \left|\bar{\Vsc} \right|} {\pi (\hbar v)^2}
               - \frac{ \text{sgn}\left(\bar{\Vsc}\right)}{2 \ell^2}\,
                \ln  \frac{\left| \bar{\Vsc} \right|}{\hbar v \Lambda}
                \,,
 \quad
 \left| \bar{\Vsc} \right| \gg \hbar v \Lambda\,.
 \label{eq:ln:tfd_an:2}
\end{equation}
with $\bar\Vsc$ given, in momentum space, by the equation:
\beq
 \bar\Vsc(\kk)= V_D(\kk) + \frac{2\pi r_s\hbar v_F}{k}n(\kk)
 \label{eq:ln:tfd_an:3}
\enq
where we have assumed for simplicity $d=0$. 
Equation \ceq{eq:ln:tfd_an:3} can be approximated by the following asymptotic expressions:
\beq
  \bar\Vsc(\kk) = \left\{
  \begin{alignedat}{2}
    &V_D(\kk)\,,& &\quad kR\gg 1\,,\\
    &{V_D(\kk)\frac{kR}{1+kR}}\,,&  &\quad kR\ll 1\,.
  \end{alignedat}
  \right.
 \label{eq:ln:tfd_an:4}
\enq
Equations \ceq{eq:ln:tfd_an:2}, \ceq{eq:ln:tfd_an:3}
define a nonlinear problem that must be solved self-consistently and that in general
can only be solved numerically. However in the limit $r_s\ll 1$ an approximate
solution with logarithmic accuracy can be found. Let $K_0$ be the solution
of the equation:
\beq
 K_0 = \ln(1/(4 r_s K_0)).
\enq 
$K_0$ is the expansion parameter. 
To order $\mathcal{O}(K_0^{-1})$
$\bar{\Vsc}$ can be treated as a Gaussian random potential
whose correlator, $\delta\Vsc^2(r)$ can be calculated using \ceq{eq:ln:tfd_an:4}
to find \cite{kn:galitski2007, kn:adam2007a, fogler-prl-103-236801-2009}
\begin{equation}
 K(r)\equiv\delta\bar\Vsc^2(r) = \frac{\pi}{2} \left( \frac{\hbar v}{\ell} \right)^2 \times
 \left\{
 \begin{alignedat}{2}   
   &\ln\left(\frac{R}{r}\right)\,,  &  &\quad l \ll r \ll R\,, \\
   &2\left(\frac{R}{r}\right)^3,    &  &\quad R \ll r\,.
 \end{alignedat}
 \right.
\label{eq:ln:tfd_an:5}
\end{equation}
with $R = 1/(4 r_s K_0).$
%
Using \ceq{eq:ln:tfd_an:5} and \ceq{eq:ln:tfd_an:2} we can find the correlation
function for the carrier density \cite{fogler-prl-103-236801-2009}:
\begin{align}
 \delta \bar n^2(r) = &\frac{K_0^2}{2\pi l^4}
                 \left[3\frac{K(r)}{K(l)}\sqrt{1-\left(\frac{K(r)}{K(l)}\right)^2} + \right. \nonumber \\
                 &\left.\left(1+2\left(\frac{K(r)}{K(l)}\right)^2 \right)\arcsin\frac{K(r)}{K(l)}
                 \right].
 \label{eq:ln:tfd_an:7}
\end{align}
The correlation functions given by Eqs.~\ceq{eq:ln:tfd_an:5} and \ceq{eq:ln:tfd_an:7} are valid in the limit
$r_s\ll 1$ but are in qualitative agreement also with the numerical results obtained
for $r_s\approx 1$ \cite{kn:rossi2008}.

The location of the disorder induced PNJ is identified
by the isolines $n(\rr)=0$, or equivalently $\Vsc(\rr)=0$. The CNP corresponds to the ``percolation''
threshold in which exactly half of the sample is covered by electron puddles and half by 
hole puddles (notice that conventionally the percolation threshold is defined as the condition
in which half of the sample has non-zero charge density and half is insulating and so the term
percolation in the context of the graphene CNP has a slight different meaning and does not
imply that the transport is percolative).
At the percolation threshold all but one PNJ are closed loops. 
Over length scales $d$ such that $1/\Lambda \ll d \ll R$, $\Vsc$ is logarithmically
rough (Eq.~\ceq{eq:ln:tfd_an:2}) and so the PNJ loops of diameter $d$ have fractal dimension $D_h=3/2$~\cite{kondev2000}.
At larger $d$ the spatial correlation of $\Vsc$ decays rapidly (Eq.~\ceq{eq:ln:tfd_an:2}) so
that for $d$ larger than $R$,  $D_h$ crosses over to the standard uncorrelated percolation 
exponent of 7/4 \cite{isichenko1992}.

%

\subsubsection{Effect of ripples on carrier density distribution}

When placed on a substrate graphene has been shown \cite{kn:ishigami2007, geringer2009}
to follow with good approximation the surface profile of the substrate
and therefore to have a finite roughness. For graphene on $\siot$ the standard
deviation of the graphene height, $h$, has been measured to be $\delta h\approx 0.19$~nm
with a roughness exponent $2H\approx 1$. More recent experiments \cite{geringer2009} have found larger roughness. 
Even when suspended, graphene is never completely flat,
and it has been shown theoretically to possess intrinsic ripples \cite{kn:fasolino2007}.
A local variation of the height profile $h(\rr)$ can induce a local change of the carrier
density through different mechanisms. 
\citet{kn:juan2007} considered the change in carrier density due to a local
variation of the Fermi velocity due to the rippling and found that assuming 
 $h(\rr)=A{\rm exp}^{-|\rr|^2/b^2}$,
a variation of 1~\% and 10~\% in the carrier density was induced for ratios $A/b$ of order 0.1
and 0.3 respectively. 
\citet{kn:brey2008} observed that local
Fermi velocity changes induced by the curvature associated
with the ripples induce charge-inhomogeneities in doped
graphene but cannot explain the existence of electron-hole puddles
in undoped graphene for which the particle-hole symmetry is preserved,
and then considered the effect on the local carrier density of a
local variation of the exchange energy associated with the local
change of the density of carbon atoms due to the presence of ripples.
They found that a modulation of the out-of-plane position of the
carbon atoms of the order of $1 - 2$~nm over a distance of
$10 - 20$~nm induces a modulation in the charge density of
the order of $10^{11}{\rm cm}^{−2}$. 
\citet{kn:kim2007} considered the effect due to the 
rehybridization of the $\pi$ and $\sigma$ orbital 
between nearest neighbor sites.
For the local shift $\delta E_F$ of the Fermi level,
\citet{kn:kim2007} found:
%
%
%
\beq
 \delta E_F = -\alpha \frac{3a^2}{4}(\nabla^2 h)^2
 \label{eq:ln:n_ripples}
\enq 
where $\alpha$ is a constant estimated to be approximately equal to $9.23$~eV.
Recently \citet{kn:gibertini2010} have used the DFT-LDA to study the 
effect of ripples on the spatial carrier density fluctuations.
Transport theories in the presence of topological disorder
were considered by \citet{cortijo-prb-79-184205-2009,cortijo-npb-763-293-2007} 
and \citet{kn:herbut2008}.

%
\subsubsection{Imaging experiments at the Dirac point}
\label{subsubsec:imaging}
\begin{figure}[htb]
 \begin{center}
  \includegraphics[width=8.5cm]{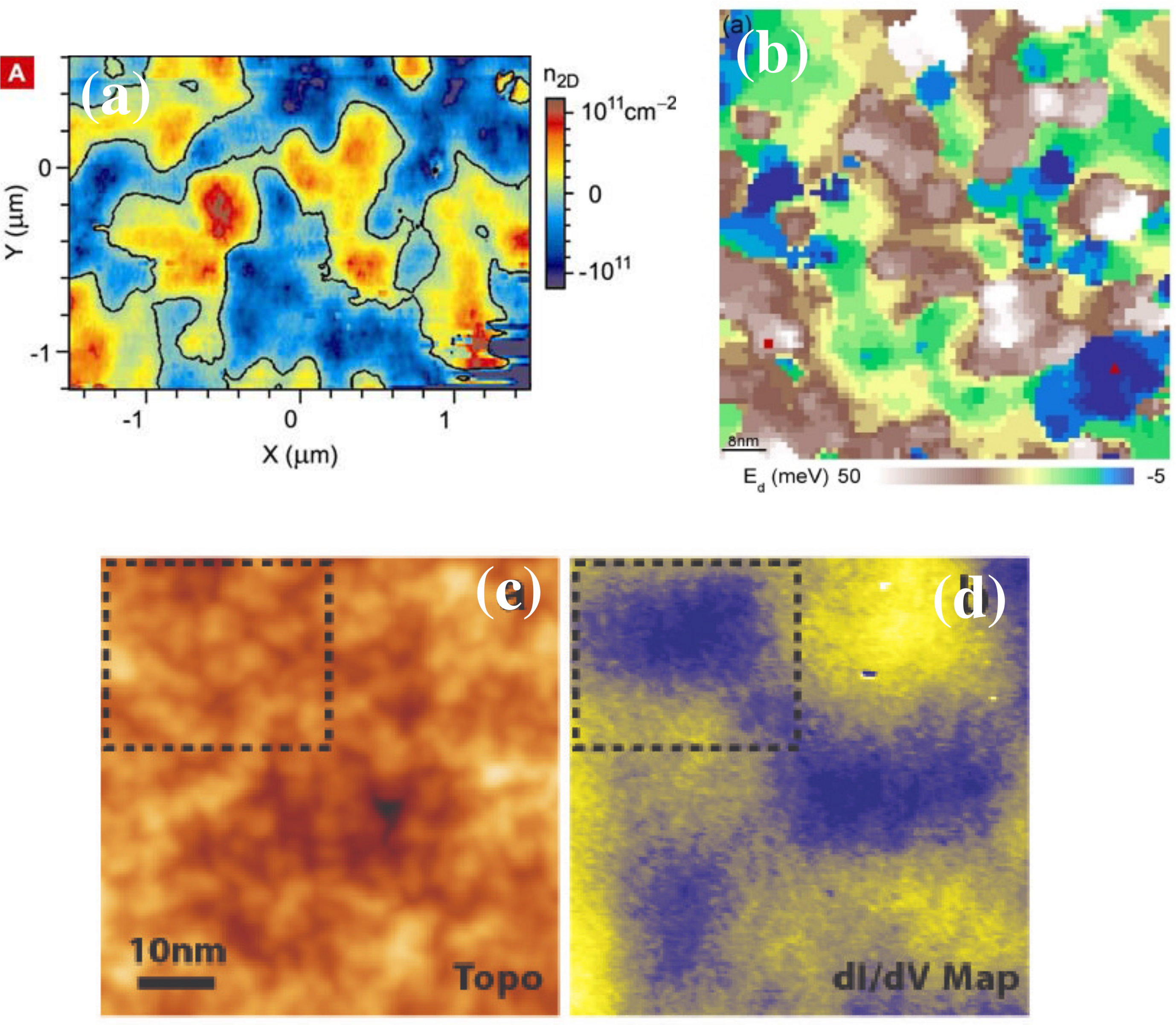}
  \caption{
           (Color online). 
           (a) Carrier density color map at the CNP measured with an SET.
               The blue regions correspond to holes and red regions to electrons. The black contour marks the
               zero density contour. 
               Adapted from \citet{kn:martin2008}.
           (b) Spatial map, on a 80~nm$\times$80~nm region, of the energy shift of the CNP in BLG
               from STM $dI/dV$ map.
               Adapted from \citet{kn:deshpande2009b}. 
               60~nm$\times$60~nm
               constant current STM topography, (c), and  simultaneous $dI/dV$ map, (d) at the CNP for SLG, 
               $(V_{bias}= -0.225\;{\rm V} I = 20\;{\rm pA})$.           
               Adapted from \citet{kn:zhang2009}.
          } 
  \label{fig:ln:set}
 \end{center}
\end{figure}

%
The first imaging experiments using scanning tunneling microscopy, STM, were
done on epitaxial graphene \cite{kn:rutter2007,kn:brar2007}.
These experiments were able to image the atomic structure of graphene
and reveal the presence of in-plane short-range defects. So far one limitation of experiments
on epitaxial graphene has been the inability to modify the graphene intrinsic doping
that is relatively high ($\gtrsim 10^{12}\;{\rm cm}^{-2}$)
in most of the samples. This fact has prevented these experiments to directly image
the electronic structure of graphene close to the Dirac point.
The first scanning probe experiment on exfoliated graphene on $\siot$ \cite{kn:ishigami2007}
revealed the atomic structure of graphene and the nanoscale morphology.
%
The first experiment that was able to directly image the electronic structure
of exfoliated graphene close to the Dirac point was performed by 
\citet{kn:martin2008} using scanning single-electron transistor, SET,
Fig.~\ref{fig:ln:set}. 
The break-up of the density landscape in electron-hole puddles as predicted
in \cite{kn:hwang2006c,kn:adam2007a} and shown by the DFT-LDA \cite{kn:polini2008}
and TFD theory \cite{kn:rossi2008} is clearly visible. 
The result shown in Fig.~\ref{fig:ln:set}~(a) however does not provide a good quantitative
characterization of the carrier density distribution due 
%
to the limited spatial resolution of the imaging technique: 
the diameter of the SET is 100~nm and the distance between the SET and
the sample is 50~nm and so the spatial resolution is approximately  150~nm.
By analyzing the width in density of the incompressible bands in the quantum 
Hall regime,  \citet{kn:martin2008} were able to extract the amplitude of the density
fluctuations in their sample. By fitting with a Gaussian
the broadened incompressible bands Martin et al. extracted the value of the
amplitude of the density fluctuations,
identical for all the incompressible bands \cite{kn:ilani2004},
and found it to be equal to $2.3\;10^{11}\;{\rm cm}^{-2}$.
Taking this value to be equal to $\nrms$ using the TFD
a corresponding value of $\nimp = 2.4\;10^{11}\;{\rm cm}^{-2}$ is found
consistent with typical values for the mobility at high density.
By calculating the ratio between the density fluctuations amplitude
extracted from the broadening of the incompressible bands in the
Quantum Hall regime and the amplitude extracted from the probability
distribution of the density extracted from the imaging results,
\citet{kn:martin2008}
obtained the upper bound of 30~nm for 
the characteristic length of the density fluctuations, consistent
with the TFD results \cite{kn:rossi2008}.

An indirect confirmation of the existence of electron-hole puddles in exfoliated graphene
close to the CNP came from the measurement of the magnetic field-dependent 
longitudinal and Hall components of the
resistivity $\rho_{xx}(H)$ and $\rho_{xy}(H)$ \cite{kn:cho2008}.
Close to the Dirac point the measurements show that 
$\rho_{xx}(H)$ is strongly enhanced and $\rho_{xy}(H)$ 
is suppressed, indicating nearly equal electron 
and hole contributions to the transport current.
In addition the experimental data were found 
inconsistent with uniformly distributed electron
and hole concentrations (two-fluid model) but in excellent agreement
with the presence of inhomogeneously distributed
electron and hole regions of equal mobility.

The first STM experiments on exfoliated graphene were
performed by \citet{kn:zhang2009}.
The STM experiments provided the most direct quantitative
characterization of the carrier density distribution of
exfoliated graphene. 
Fig.~\ref{fig:ln:set}~(c) shows the topography
of a  $60 \times 60\;{\rm nm}^2$ area of exfoliated graphene,
whereas Fig.~\ref{fig:ln:set}~(d) shows the $dI/dV$ map of the same
area. The $dI/dV$ value is directly proportional to the local density of states.
We can see that there is no correlation between topography
and $dI/dV$ map. This shows that in current exfoliated graphene samples
the rippling of graphene, either intrinsic or due to the roughness
of the substrate surface, are not the dominant cause of the 
charge density inhomogeneities.
The $dI/dV$ maps clearly reveal the presence of high density regions with
characteristic length of $\approx 20$~nm as predicted by the TFD results.
%
%
%
Recently more experiments have been performed to directly image the 
electronic structure of both exfoliated single layer graphene \cite{kn:deshpande2009} 
and bilayer graphene \cite{kn:deshpande2009b}. 
In particular \citet{kn:deshpande2009} starting from the topographic data
calculated
the carrier density fluctuations due to the local curvature of the
graphene layer using Eq.~\ceq{eq:ln:n_ripples}
and compared them to the fluctuations of the $dI/dV$ map.
The comparison shows that there is no correspondence between the density fluctuations
induced by the curvature and the ones measured directly. This leads to the conclusion 
that even though the curvature
contributes to a variation in the
electrochemical potential, it is not the main factor responsible
for the features in the dI/dV map. 
%
%
The results for BLG of
Fig.~\ref{fig:ln:set}~(b)
show that close to the CNP the density inhomogeneities are very strong
also in BLG and are in semiquantitative agreement with theoretical
predictions based on the TF theory \cite{kn:dassarma2009c}.
%
%
Using an SET \citet{martin2009} have imaged the local density of states
also in the quantum Hall regime.
The disorder carrier density landscape has also been indirectly observed
in imaging experiments of coherent transport \cite{berezovsky2009,berezovsky2010}.


\subsection{Transport in the presence of electron-hole puddles}
\label{subsec:emt}

The previous section showed, both theoretically
and experimentally, that close to the Dirac point, in the presence of long-range disorder,
the carrier density landscape
breaks up in electron hole puddles. In this situation the transport problem
becomes the problem of calculating transport properties of a system
with strong density inhomogeneities. The first step is to calculate
the conductance of the puddles $G_p$ and PNJ, $G_{\rm PNJ}$.
%

We have $G_p=\Gamma\sigma$ where $\Gamma$ is a form-factor of order one and
$\sigma$ is the puddle conductivity. 
Away from the Dirac point (see Sec.~\ref{subsec_boltzmann}) the RPA-Boltzmann transport
theory for graphene in the presence of random charge impurities is accurate.
From the RPA-Boltzmann theory we have $\sigma = e |\nav|\mu(\nav,\nimp,r_s, d, T)$.
For the purposes of this section it is convenient to explicitly write the dependence of $\sigma$ on $\nimp$ 
by introducing the function 
\beq
 F(r_s,d,T)\df \frac{h\nimp\mu}{2e}=2\pi\nimp\tau\frac{k_F}{v_F} 
\enq
so that we can write:
\begin{equation} 
 \sigma = \frac{2 e^2}{h} \frac{|\nav|}{n_{\rm imp}} {F(r_s, d,T)}.
 \label{eq:ln:sigma}
\end{equation}
Expressions for $F(r_s,d,T)$ at $T=0$ (or its inverse) were originally given in \cite{kn:adam2007a} 
(see Eq.~\ceq{eq:frs}).
%
%
We can define a local spatially varying ``puddle''
conductivity $\sigma(\rr)$ if $\sigma(\rr)$ varies on length scales that are larger
than the mean free path $l$, i.e.
\beq
 \hspace{0.2cm} \left|\frac{\nabla\sigma(\rr)}{\sigma(\rr)}\right|^{-1}\gg l.
 \label{eq:ln:sineq1}
\enq
By substituting $\nav$ with $n(\rr)$ let us use Eq.~\ceq{eq:ln:sigma} to define and calculate
the local conductivity:
\begin{equation} 
 \sigma(\rr) = \frac{2 e^2}{h} \frac{n(\rr)}{n_{\rm imp}} {F(r_s, d,T)}.
 \label{eq:ln:sigmal}
\end{equation}
Considering that $l = h\sigma/(2e^2k_F)$
%
%
and using Eq.~\ceq{eq:ln:sigmal} then inequality \ceq{eq:ln:sineq1} takes the form:
\beq
 \hspace{0.2cm} \left|\frac{\nabla n(\rr)}{n(\rr)}\right|^{-1} \gg
 \frac{F(r_s,d,T)}{\sqrt{\pi}}\frac{\sqrt{n}}{n_{imp}}.
 \label{eq:ln:sineq2}
\enq
As shown in the previous sections, at the CNP most of the graphene area
is occupied by large electron-hole puddles with size of the order of the sample size $L$
and density of the order of $\nrms\approx\nimp$.
For graphene on $\siot$ we have $r_s=0.8$ for which is $F=10$.
Using these facts we find that inequality \ceq{eq:ln:sineq2} is satisfied when
\beq
 L \gg \frac{F(r_s,d,T)}{\sqrt{\pi}}\frac{1}{\sqrt{\nimp}}
 \label{eq:ln:ineq3}
\enq 
i.e. when the sample is much larger than the typical in-plane distance between
charge impurities. Considering that in experiments on bulk graphene 
$L> 1\;\mu{\rm m}$ and $n_{imp}\in[10^{10}-10^{12}]$~cm$^{-2}$
we see that inequality \ceq{eq:ln:ineq3} is satisfied.
In this discussion we have neglected the presence of the small regions
of high density and size $\xi$. For these regions the inequality \ceq{eq:ln:ineq3}
is not satisfied. 
However these regions, because of their high carrier density,
steep carrier density gradients at the boundaries and small size $\xi< l$,
are practically transparent to the current carrying quasiparticles and
therefore, given that they occupy a small area fraction 
and are isolated (i.e. do not form a path spanning the whole sample),
give a negligible contribution to the graphene resistivity.
This fact, and the validity of inequality \ceq{eq:ln:ineq3} for the
large puddles ensure that the local conductivity $\sigma(\rr)$ 
as given by Eq.~\ceq{eq:ln:sigmal} is well defined.
In the limit $r_s\ll 1$ one can use the analytical results for the density
distribution to reach the same conclusion \cite{fogler-prl-103-236801-2009}.

To calculate the conductance across the PNJ, quantum effects must be taken
into account.
In particular, as seen in Sec.~\ref{Sec:KT_LC} 
for Dirac fermions, we have the phenomenon of Klein tunneling
\cite{kn:klein1929,dombey1999} i.e. the property of perfect transmission
through a steep potential barrier perpendicular to the direction of motion.
The PNJ conductance, $G_{\rm PNJ}$, can be estimated using the results
of \citet{kn:cheianov2006} and \citet{kn:zhang2008}. The first step is to estimate the 
steepness of the electrostatic barrier at the PNJ, i.e. the ratio
between the length scale, $D$, over which the screened potential varies
across the PNJ and the Fermi wavelength of the carriers at the side 
of the PNJ. From the TFD theory we have that at the sides of the PNJ
$n\approx\nrms$ so that $k_F=\sqrt{\pi\nrms}$, and that $D\approx 1/k_F$
so that $k_FD\approx 1$. 
In this limit, the conductance per unit length of a PNJ, $g_{\rm PNJ}$  
is given by \cite{kn:cheianov2006b, kn:zhang2008}:
\beq
 g_{\rm PNJ}=\frac{e^2}{h}k_F.
 \label{eq:gpnj}
\enq
so that the total conductance across the boundaries of the electron-hole puddles 
is $G_{\rm PNJ} = p g_{\rm PNJ}$ with $p$ the perimeter of the typical puddle \cite{kn:rossi2008b,fogler-prl-103-236801-2009}.
%
Because the puddles have size comparable to the sample
size, $p\approx L$, for typical experimental conditions 
($L\gtrsim 1\;\mu{\rm m}$ and $n_{imp}\in[10^{10}-10^{12}]$~cm$^{-2}$) using \ceq{eq:ln:sigmal} 
and \ceq{eq:gpnj} we find:
\beq
 G_{\rm PNJ}=\frac{e^2}{h}\sqrt{\pi \nrms} p\gg G_p=\Gamma F(r_s, d, T) \frac{2 e^2}{h} \frac{|\nrms|}{n_{\rm imp}}
 \label{eq:ln:gpnj1}
\enq
i.e. $G_{\rm PNJ}\gg G_p$. In the limit $r_s\ll 1$ the inequality \ceq{eq:ln:gpnj1}
is valid for any value of $\nimp$ \cite{fogler-prl-103-236801-2009}.
Inequality \ceq{eq:ln:gpnj1} shows that in exfoliated
graphene samples, transport close to the Dirac point is not percolative:
the dominant contribution to the electric resistance is due to scattering
events inside the puddles and not to the resistance of the puddle boundaries
\cite{kn:rossi2008b,fogler-prl-103-236801-2009}.  
This conclusion is consistent with the
results of~\citet{kn:adam2008d} in which the graphene
conductivity in the presence of Gaussian disorder obtained
using a full quantum mechanical calculation was
found to be in agreement with the semiclassical Boltzmann theory even at zero
doping provided the disorder is strong enough.
Given {\em (i)} the random position of the electron-hole puddles,
{\em (ii)} the fact that because of inequality \ceq{eq:ln:ineq3} the local
conductivity is well defined, and {\em (iii)}  the fact that $G_{\rm PNJ}\gg G_p$,
the effective medium theory, EMT, \cite{bruggeman1935,landauer1952,hori1975}
can be used to calculate the electrical conductivity of graphene.
The problem of the minimum conductivity at the CNP can be expressed
as the problem of correctly averaging the individual puddle conductivity.
Using Eq.~\ceq{eq:ln:sigmal}, given a carrier density distribution,
the conductivity landscape can be calculated.
%
%
In the EMT an {\em effective medium} with homogeneous transport properties equivalent
to the bulk transport properties of the inhomogeneous medium is introduced.
Starting from the local relation between current $\JJ$ and electric potential $V$:
\beq
 \JJ(\rr) = -\sigma(\rr)\nabla V(\rr)
 \label{eq:ln:trl}
\enq
the effective-medium conductivity, $\sigma_{\rm EMT}$, is defined through the equation:
\beq
 \langle\JJ(\rr)\rangle = -\sigma_{\rm EMT}\langle \nabla V(\rr)\rangle
 \label{eq:ln:semt}
\enq
where the angle bracket denotes spatial and disorder averages.
Equations \ceq{eq:ln:trl}, \ceq{eq:ln:semt} along
with the condition that
in the effective medium the electric field $-\nabla V$ is uniform
are sufficient to calculate $\semt$. The derivation of the 
relation between $\semt$ and $\sigma(\rr)$ using \ceq{eq:ln:trl}, \ceq{eq:ln:semt}, 
requires the solution of the electrostatic problem 
in which a homogeneous region of conductivity $\sigma(\rr)$ 
is embedded in an infinite medium of conductivity $\semt$.
When the shape of the homogeneous regions, puddles, in the real medium is random
the shape of the homogeneous regions used to derive the 
expression of $\semt$ is unimportant, and they can be assumed
to be spheres having the same volume as each puddle.
\cite{bruggeman1935,landauer1952}.
For a 2D system the solution of the electrostatic problem gives \cite{bruggeman1935,landauer1952}:
\beq
 \int d^2 r\frac{\sigma(\rr)-\sigma_{EMT}}{\sigma(\rr) + \sigma_{EMT}} = 0.
 \label{eq:ln:semt2}
\enq
Equation \ceq{eq:ln:semt2} can also be viewed as an approximate resummation
of the infinite diagrammatic series for the macroscopic $\sigma$ 
using the self-consistent single-site approximation \cite{hori1975}.
Disorder averaging Eq.~\ceq{eq:ln:semt2} we find:
\beq
 \left\langle\int d^2 r\frac{\sigma(\rr)-\sigma_{EMT}}{\sigma(\rr) + \sigma_{EMT}} = 0\right\rangle 
 \Leftrightarrow
 \int d \sigma\frac{\sigma-\sigma_{EMT}}{\sigma + \sigma_{EMT}}P(\sigma) = 0,
 \label{eq:ln:semt3}
\enq
where $P(\sigma)$ is the probability for the local value of $\sigma$. 
Using the relation between the local value $\sigma$ and the local
value of the carrier density $n$, Eq.~\ceq{eq:ln:sigmal}, Eq.~\ceq{eq:ln:semt3} can be
rewritten in the form:
\begin{equation}
 \int dn\frac{\sigma(n)-\sigma_{\rm EMT}}{\sigma(n)+\sigma_{\rm EMT}}P[n] = 0.
 \label{eq:ln:semt4}
\end{equation}
where $P[n]$ is the density probability distribution that can be calculated
using the TFD theory, Fig.~\ref{fig:ln:Pn}.
Using the TFD results and equation \ceq{eq:ln:semt4} the conductivity
at the Dirac point and its vicinity can be calculated.
%

\begin{figure}[ht]
 \begin{center}
  \includegraphics[width=8.00cm]{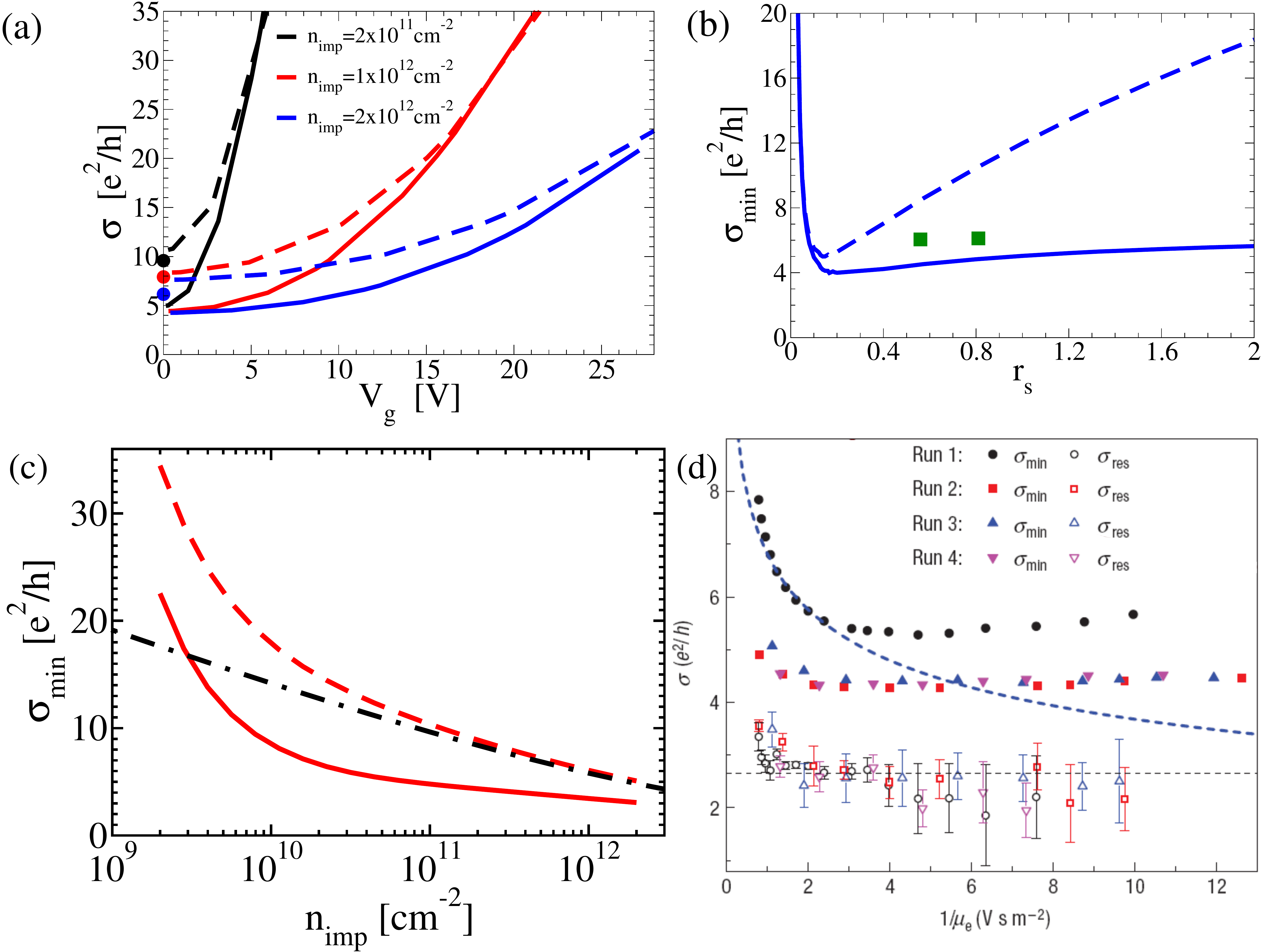}
  \caption{
           (Color online).
           Solid (dashed) lines show the EMT-TFD results with (without) exchange. 
	   (a) $\sigma$ as function of $V_g$ for three different values of $\nimp$. 
	       The dots at $V_g=0$ are the results presented in
               Ref.~\citet{kn:adam2007a} 
	       for the same values of $\nimp$.
           (b) Minimum conductivity as a function of interaction
               parameter.  Solid squares show the experimental results
               of \citet{kn:jang2008}. 
           (c) $\sigma_{min}$ as a function of $n_{imp}$,
               $r_s=0.8$ and $d=1$~nm.
               For comparison the results obtained in
               Ref.~\citet{kn:adam2007a} are also shown, dot-dashed
               line. 
           (d) $\smin$ as a function of $r_s$ for $d=1$ and
               $n_{imp}=10^{11}$~cm$^{-2}$. 
           (a), (b), (d) Adapted from \citet{kn:rossi2008b}; (c)
               adapted from \citet{kn:chen2008}. 
          }
    \label{fig:ln:s_vg}
 \end{center}
\end{figure}

Fig.~\ref{fig:ln:s_vg} shows $\sigma(V_g)$ as obtained using the TFD+EMT theory \cite{kn:rossi2008b}.
The theory correctly predicts a finite value of $\sigma$
very close to the one measured experimentally.
%
At high gate voltages, the theory predicts the linear scaling of $\sigma$
as a function of $V_g$.  
The theory correctly describes the crossover of $\sigma$
from its minimum at $V_g=0$ to its linear behavior
at high gate voltages. Fig.~\ref{fig:ln:s_vg}~(a) also shows the importance
of the exchange-correlation term at low gate voltages. 
%
The dependence of $\smin$ on $\nimp$ is shown in 
Fig.~\ref{fig:ln:s_vg}. In panel (a) the theoretical results
are shown. $\smin$ increases
as $\nimp$ decreases, the dependence of $\smin$ on $\nimp$ 
is weaker if exchange-correlation terms are taken into account.
Panel (c) of Fig.~\ref{fig:ln:s_vg} shows the dependence
of $\smin$ on the inverse mobility $1/\mu\propto \nimp$
as measured by \citet{kn:chen2008}.
In this experiment the amount of charge impurities is controlled
by potassium doping.
%
%
Fig.~\ref{fig:ln:s_vg}~(b) shows the results for $\smin$
as a function of $r_s$.
The solid (dashed) line shows the calculated values of $\smin$
including (neglecting) exchange.  $\smin$ has a
non-monotonic behavior due to the fact that $r_s$ affects
both the carrier density spatial distribution by controlling the
strength of the disorder potential, screening and exchange as well as the
scattering time $\tau$. 
The dependence of $\smin$ on $r_s$ has been measured
in two recent experiments \cite{kn:jang2008,kn:ponomarenko2009}.
In these experiments the fine structure constant of graphene, $r_s$,
is modified by placing the graphene on substrates with different
$\kappa$ and/or by using materials with $\kappa\neq 1$ as top dielectric layers.
In \citet{kn:jang2008} graphene was placed on $\siot$ and $r_s$ 
was reduced from $0.8$ (no top dielectric layer) to $0.56$ by placing ice in vacuum as a top
dielectric layer. The resulting change of $\smin$ is shown in 
Fig.~\ref{fig:ln:s_vg}~(b). As predicted
by the theory, when $V_{xc}$ is included, for this range of values of $r_s$, $\smin$ 
is unaffected by the variation of $r_s$. Overall the results presented
in \cite{kn:jang2008} are consistent with charge impurity being the
main source of scattering in graphene.
In \citet{kn:ponomarenko2009}, $r_s$ was varied by placing graphene
on substrates with different dielectric constants and  
by using glycerol, ethanol and water as a top dielectric layer.
\citet{kn:ponomarenko2009}
found very minor differences in the
transport properties of graphene with different dielectric layers, thus
concluding that 
charge impurities are not the dominant source of scattering.
At the time of this writing the reasons for the
discrepancy between the results of \citet{kn:jang2008} and
\citet{kn:ponomarenko2009} are not well understood. The experiments
are quite different. It must be noted that changing the substrate
and the top dielectric layer, in addition to
modifying $r_s$, is likely to modify the amount of disorder
seen by the carriers in the graphene layer.
By not modifying the substrate and by placing
ice in ultra-high-vacuum, \citet{kn:jang2008} minimized the change
of disorder induced by modifying the top dielectric layer.
%
%

The approach presented above based on TFD and EMT theories can be used
to calculate other transport properties of SLG and BLG close to the Dirac point.
In \citet{kn:hwang2009} the same approach was used to calculate the thermopower
of SLG with results in good agreement with experiments 
\cite{checkelsky2009, wei2009, zuev2009}.
In \citet{kn:dassarma2009c} the TFD+EMT approach was used to calculate
the electrical conductivity in BLG.
In \citet{tiwari2009} the EMT based on a simple two-fluid model was used
to calculate the magnetoresistance at low magnetic fields of SLG
close to the CNP. 

\begin{figure}[ht]
 \includegraphics[width=7.0cm]{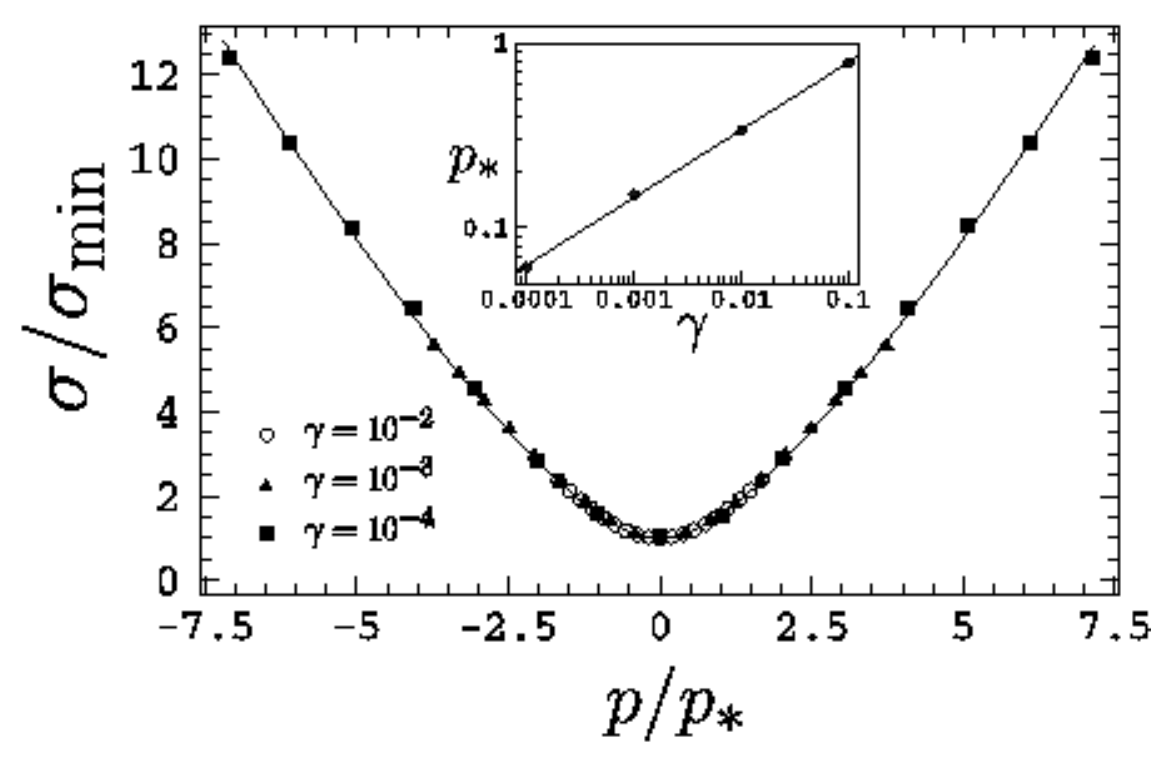}
 \caption
  {
   Collapse of the conductivity data
   obtained for RNNs with various $-1/2<p<1/2$  and values of the
   parameter $\protect\gamma $ onto a single curve.
   Points: numerical results; line: best fit obtained using 
   the equations presented in the text.
   Adapted from \citet{kn:cheianov2007}
  }
 \label{fig:ln:smin_percolation}
\end{figure}

%
%
It is interesting to consider the case when $G_p \gg G_{\rm PNJ}$. This limit
is relevant for example when a gap in the graphene spectrum is opened.
The limit $G_p \gg G_{\rm PNJ}$
was studied theoretically in \cite{kn:cheianov2007} by considering
a random resistor network (RRN) model on a square lattice in which 
only nearest neighbors and next nearest neighbor are connected directly.
Mathematically the model is expressed by the following equations
for the conductance between the sites $(i,j)$ and $(i',j')$ with 
$|i-i'|\leq 1$, $|j-j'|\leq 1$:
\begin{align}
 &G_{(i,j)}^{(i+1,j+1)} ={g}\left[1+(-1)^{i+j}\eta_{i,j}\right]/2;
 \label{eq:ln:gnn} \\
 &G_{(i,j+1)}^{(i+1,j)}={g}\left[1-(-1)^{i+j}\eta_{i,j}\right]/2;
 \label{eq:ln:gpp}\\
 &G_{(i,j)}^{(i+1,j)}=G_{(i,j)}^{(i,j+1)}=\gamma g, \quad \gamma \ll 1.
 \label{eq:ln:gnp}
\end{align}
where $\eta_{i,j}$ is a random variable, 
\beq
 \eta_{i,j}=\pm 1,\ \langle\eta_{i,j}\rangle=p,  
 \ \langle\eta_{i,j}\eta_{k,l}\rangle=\delta_{ik}\delta_{jl},
\enq
and $p$ is proportional to the doping $\nav$.
%
For $\gamma=0, $$p=0$ we have percolation. Finite $p$ and $\gamma$
are relevant pertirbations for the percolation leading to a finite
correlation length $\xi(p,\gamma)$. On scales much bigger than $\xi$
the RRN is not critical, and consists of independent regions of size $\xi$
so that $\sigma$ is well defined with scaling \cite{kn:cheianov2007}
$\sigma(p,\gamma) = [a/\xi(p,\gamma)]^xg$
%
%
with 
$\xi(p,\gamma) \sim a \gamma^{-\gamma}/F(p/p^*)$,
$ p^*=\gamma^{\mu/\nu}$
%
%
and
$\nu=4/3$, $\mu=1/(h+x)$
%
%
where $h=7/4$ and $x\approx 0.97$  are respectively the fractal dimension of the boundaries between the electron-hole
puddles and the conductance exponent, $\langle G(L)\rangle=(a/L)^xg$, at the percolation threshold
\cite{isichenko1992,kn:cheianov2007}.
Fig.~\ref{fig:ln:smin_percolation} shows the results obtained solving
numerically the RRN defined by Eq.~\ref{eq:ln:gnp}.
The numerical results are well fitted using for $F(p/p^*)$ the function
$F(z)=(1+z^2)^{(\nu/2)}$.
%
%
Estimating $g\sim\frac{e^2}{\hbar}a k_F$ and $\gamma g\sim\frac{e^2}{\hbar}(ak_F)^{1/2}$
\cite{kn:cheianov2006b}, \citet{kn:cheianov2007} estimate that 
\beq
 \sigma_{\rm min}\sim \frac{e^2}{\hbar}(a^2\delta n)^{0.41}.
 \label{eq:ln:sigma:perc2} 
\enq
From the TFD results
one gets $\delta n\sim\nimp$ and so Eq.~\ceq{eq:ln:sigma:perc2} predicts that
that $\sigma_{min}$ should increase with $\nimp$, a trend that is not observed
in experiments. The reason for the discrepnacy is due to the fact that
for current experiments the relevant regime is expected to be the one
for which $G_{\rm PNJ} \gg G_p$.
The minimum conductivity was also 
calculated for bilayer graphene~\cite{kn:adam2008a, kn:dassarma2009c}.
Other works calculated $\sigma_m$ using different models and approximations
for regimes less relevant for current experiments 
\cite{kn:cserti2007,cserti-prl-99-066802-2007,kn:katsnelson2006,kn:adam2008a,groth-prl-100-176804-2008,kn:trushin2007,trushin2010}.

Although the semiclassical approach presented in this section is
justified for most of the current experimental conditions for
exfoliated graphene its precise range of validity and level of
accuracy close to the Dirac point can only be determined by a full
quantum transport calculation that takes into account the presence of
charge impurities. This is still an active area of research and work
is in progress to obtain the transport properties of graphene using a
full quantum transport treatment \cite{kn:rossi2010}. 

\setcounter{sub3section}{0}


\section{Quantum Hall effects}
\label{sec:qhe}


\subsection{Monolayer graphene}


\subsubsection{Integer quantum Hall effect}

\begin{figure}[htb]
 \begin{center}
  \includegraphics[width=6.25cm]{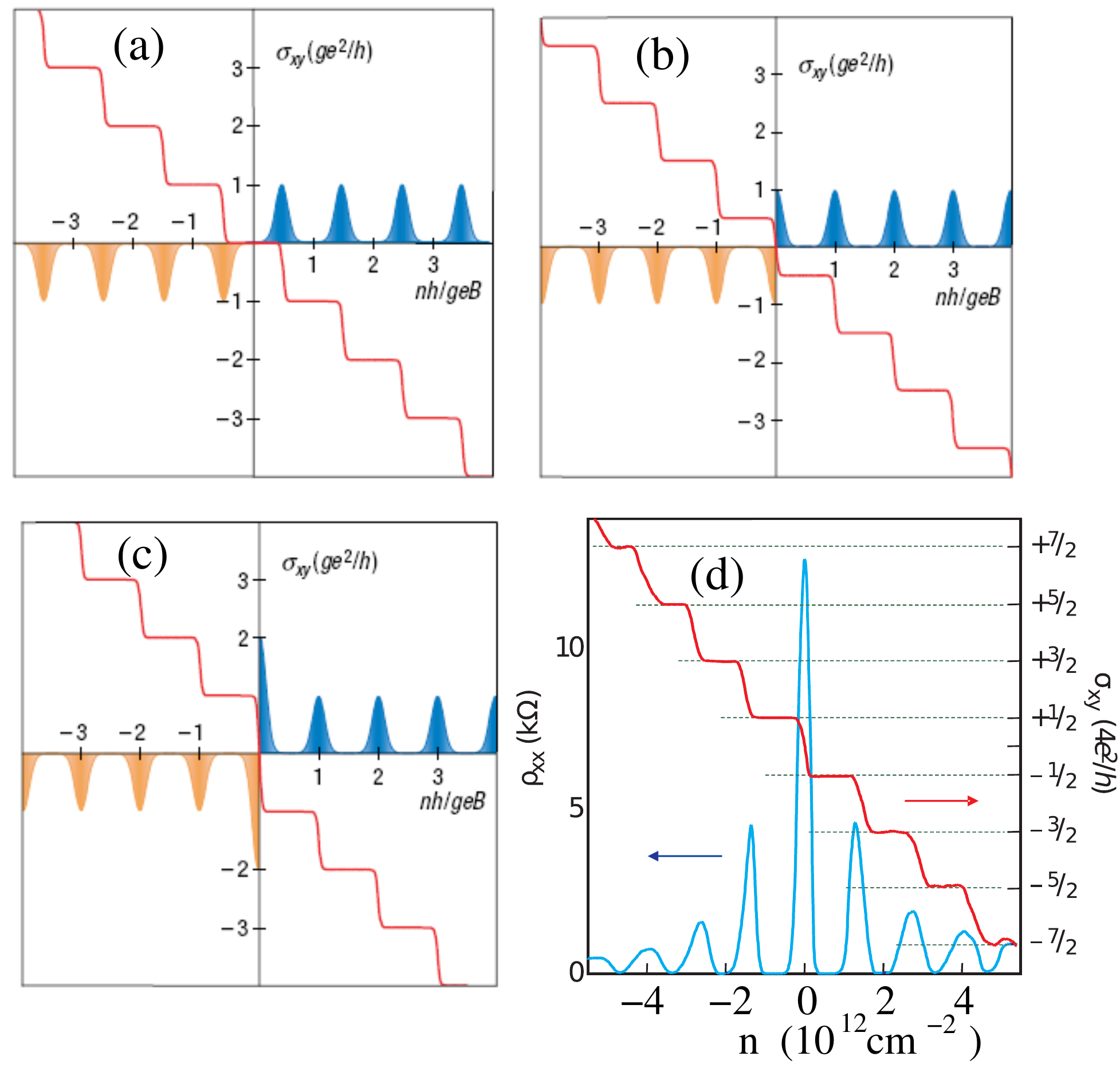}  
\caption
  {(Color online)
   Illustration of the integer QHE found in 2D semiconductor systems (a),
   incorporated from \citet{prange1990, macdonald1990}, SLG (b), BLG (c)
   The sequences of
   Landau levels as a function of carrier concentrations n are shown as dark and light peaks
   for electrons and holes, respectively.
   Adapted from \citet{kn:novoselov2006}.
   (d) $\sigma_{xy}$ (red) and $\rho_{xx}$ (blue)
   of SLG as a function of carrier density
   measured experimentally at $T=4K$ and $B=14T$.
   Adapted from  \citet{kn:novoselov2005}.
  }
  \label{fig:qh:schematic1}
 \end{center}
\end{figure}

The unique properties of the quantum Hall effect in graphene are among the most
striking consequences of the Dirac nature of the massless low energy fermionic
excitations in graphene. In the presence of a perpendicular magnetic field, $B$, electrons (holes) confined in
two dimensions are constrained to move in close cyclotron orbits
that in quantum mechanics are quantized. The quantization of the cyclotron
orbits is reflected in the quantization of the energy levels: at finite $B$ the $B=0$
dispersion is replaced by a discrete set of energy levels, the Landau levels (LL).
For any LL there are $N_\phi=BA/\phi_0$ degenerate orbital states, where $A$ is the area
of the sample and $\phi_0$ is the magnetic quantum flux. 
Quantum Hall effects \cite{dassarma1996,prange1990,macdonald1990}
appear when $N$ is comparable to the total number of quasiparticles
present in the system.
In the quantum Hall regime the Hall conductivity $\sigma_{xy}$
exhibits well developed plateaus as a function
of carrier density
(or correspondingly magnetic field) 
at which it takes quantized values.
At the same time, for the range of densities for which $\sigma_{xy}$ 
is quantized, the longitudinal conductivity $\sigma_{xx}$ is zero 
\cite{laughlin1981,halperin1982}.
For standard parabolic 2DEG, (such as the ones created in GaAs and Si quantum wells),
the LL have energies energy $\hbar\omega_c(n + 1/2)$ where $n=0,1,2,...$
and $\omega_c=eB/mc$, $m$ being the effective mass, is the cyclotron frequency.
Because the low energy fermions in graphene are massless 
it is immediately obvious that for graphene we cannot apply the results
valid for standard 2DEG ($\omega_c$ would appear to be infinite).
In order to find the energy levels, $E_n$, for the LL the 2D Dirac equation 
must be solved in the presence of a magnetic field
\cite{jackiw1984,kn:haldane1988,kn:gusynin2005,kn:peres2006}. The result 
is given by 
Eq.~\ceq{SDS_Eq_LandauLevels} (a). 
Differently
from parabolic 2DEG, in graphene we have a LL at zero energy.
In addition we have the unconventional Hall quantization rule for $\sigma_{xy}$
\cite{zheng2002,kn:gusynin2005,kn:peres2006}: 
\beq
 \sigma_{xy} = g\left(n+\frac{1}{2}\right)\frac{e^2}{h}
 \label{eq:qh:sxy_slg}
\enq
compared to the one valid for regular 2DEGs:
\beq
 \sigma_{xy} = g n\frac{e^2}{h}.
 \label{eq:qh:sxy_2deg}
\enq
shown in Fig.~\ref{fig:qh:schematic1},
where $g$ is the spin and valley degeneracy. Because in graphene
the band dispersion has two inequivalent valleys valleys $g=4$
(for GaAs quantum wells we only have the spin degeneracy so that $g=2$).
The additional $1/2$ in Eq.~\ceq{eq:qh:sxy_slg} is the hallmark of the chiral
nature of the quasiparticles in graphene. 
The factor $1/2$ in \ceq{eq:qh:sxy_slg} can be understood
as the term induced by the additional Berry phase that the electrons,
due to their chiral nature,
acquire when completing a close orbit 
\cite{mikitik1999,lukyanchuk2004}.
Another way to understand its presence is by considering the analogy to the 
relativistic Dirac equation \cite{geim2007,yang2007}. 
From this equation two
main predictions ensue: {\em (i)} the electrons have
spin $1/2$, {\em (ii)} the magnetic $g-$factor is exactly
equal to 2  for the ``spin'' in the non-relativistic limit.
As a consequence the Zeeman splitting is exactly equal to 
the orbital-splitting. In graphene the pseudospin plays the
role of the spin and instead of Zeeman splitting we have
``pseudospin-splitting'' but the same holds true: the pseudospin splitting
is exactly equal to the orbital splitting. As a consequence
the $n-th$ LL can be thought as composed of the degenerate 
pseudospin-up states of LL $n$ and the 
pseudospin-down states of LL $n-1$.
For zero mass Dirac fermions the first LL in the conduction
band and the highest LL in the valence band merge contributing
equally to the joint level at $E=0$ resulting in the half-odd-integer
quantum Hall effect described by Eq.~\ceq{eq:qh:sxy_slg}.
For the $E=0$ LL, because half of the degenerate states
are already filled by hole-like (electron-like) particles, we only need $1/2N_\phi$
electron-like (hole-like) particles to fill the level.

The remarkable quantization rule for $\sigma_{xy}$ has been observed experimentally
\cite{kn:novoselov2005, kn:zhang2005} as shown in Fig.~\ref{fig:qh:schematic1}~(d).
The experimental observation of Eq.~\ceq{eq:qh:sxy_slg} shows clearly
the chiral nature of the massless quasiparticles in graphene.
%
%
There is another important experimental consequence of the Dirac nature
of the fermions in graphene. Because in graphene $E_n$ scales as $\sqrt{nB}$
(Eq.~\ceq{SDS_Eq_LandauLevels} (a)) rather than linearly as in regular
2DEG (Eq.~\ceq{SDS_Eq_LandauLevels} (c)) at low energies ($n$) the 
energy spacing $\Delta_n\df E_{n+1}-E_n$ between LL can be rather large. Because the 
observation of the quantization of $\sigma_{xy}$ relies on the 
condition $\Delta_n\gg k_BT$ ($T$ being the temperature) it follows that
in graphene the quantization of the LL should be observable at temperatures
higher than in regular parabolic 2DEG. This fact has been confirmed
by the observation in graphene of the QH effect at room temperature \cite{kn:novoselov2007}
Graphene is the only known 
material whose quantum Hall effect has been observed at ambient temperature
(albeit at high magnetic fields).
%
%

By applying a top gate, p-n junctions, PNJ, can be created in graphene.
In the presence of strong perpendicular fields graphene PNJ exhibit
unusual fractional plateaus for the conductance that have been studied
experimentally by \citet{kn:williams2007} and \citet{kn:oezyilmaz2007}
and theoretically by \citet{kn:abanin2007}. 
Numerical studies in the presence of disorder have been performed by
\citet{long_sun2008}, \citet{li_shen2008}, and \citet{kn:low2009}.


\subsubsection{Broken-symmetry states}
\label{sec:qh:slg:brsym}

The sequence of plateaus for $\sigma_{xy}$ given by Eq.~\ceq{eq:qh:sxy_slg}
describes the QH effect due to fully occupied Landau levels including
the spin and valley degeneracy. In graphene for the fully occupied LL
we have the ``filling factors'' $\nu\df gN/N_\phi=4(n+1/2)=\pm 2,\pm 6,\pm10,...$.
In this section we study the situation in which the spin or valley, or both,
degeneracies are lifted. In this situation QH effects are observable
at intermediate filling factors $\nu=0,\pm 1$ for the lowest LL and
$\nu=\pm 3,\pm 4,\pm 5$ for $n=\pm 1$ LL. The difficulty
in observing these intermediate QH effects is the lower value of the
energy gap between successive splitted LL. If the gap
between successive Landau levels is comparable or smaller than
the disorder strength, the disorder mixes adjacent LL preventing
the formation of well defined QH plateaus for the Hall conductivity.
For the most part of this section we neglect the Zeeman coupling
that turns out to be the lowest energy scale in most of the 
experimentally relevant conditions.

\citet{koshino2007}  showed that randomness in the  bond couplings
and on-site potential can lift the valley degeneracy and 
cause the appearance of intermediate Landau Levels.
\citet{fuchs2007} considered the electron-phonon coupling
as the possible mechanism for the lifting of the degeneracy.
However in most of the theories the spin/valley degeneracy
is lifted due to interaction effects 
\cite{kn:nomura2006a,goerbig2006,alicea2006,yang2006,abanin2007z,gusynin2006,herbut2007,ezawa2007,ezawa2008},
in particular, electron-electron interactions.
When electron-electron interactions are taken into account, the
quasiparticles filling a LL can polarize in order to minimize
the exchange energy (maximize it in absolute value). 
In this case, given the $SU(4)$ invariance of the Hamiltonian,
the states
\beq
 |\Psi_0\rangle =\prod_{1\le i\le M}\prod_{k}c_{k,\sigma}^\dagger|0\rangle. 
 \label{eq:qh:ground}
\enq
where $i$ is the index of the internal states that runs from 1 to $M=\nu-4(n-1/2)\leq 4$, 
and $|0\rangle$ is the vacuum, are exact eigenstates of the Hamiltonian.
For a broad class of repulsive interaction $|\Psi_0\rangle$ is expected to be
the exact ground state \cite{yang2006,yang2007}.
The state described $|\Psi_0\rangle$ is a ``ferromagnet'', sometimes
called a QH ferromagnet, in which 
either the real spin or the pseudospin associated with the valley
degree of freedom is polarized.
The problem of broken symmetry states in the QH regime of graphene is 
analogous to the problem of ``quantum Hall ferromagnetism'' 
studied in regular 2DEG in which, however, normally only the $SU(2)$
symmetry associated with the spin can be spontaneously broken
(notice however that for silicon quantum wells the 
valley degeneracy is also present so that in this case the Hamiltonian is 
$SU(N)$ $(N>2)$ symmetric).
Because in
the QH regime the kinetic energy is completely quenched, the formation
of polarized states depends on the relative strength of 
interaction and disorder. For graphene, \citet{kn:nomura2006a},
using the Hartree-Fock approximation derived a ``Stoner criterion''
for the existence of polarized states, i.e. QH ferromagnetism, 
for a given strength of the disorder. 
\citet{chakraborty2007}  have numerically verified that QH ferromagnetic
states with large gaps are realized in graphene.
\citet{sheng2007} using exact diagonalization 
studied the interplay of long-range Coulomb interaction
and lattice effects in determining the robustness of the 
$\nu=\pm 1$ and $\nu=\pm 3$ states with respect to disorder.
\citet{kn:nomura2008b} studied the effect of strong long-range disorder.
\citet{wang2008} have performed numerical studies that show that various CDW phases
can be realized in the partially-filled $\nu=\pm 3$ LL.

\begin{figure}[htb]
 \begin{center}
  \includegraphics[width=8.5cm]{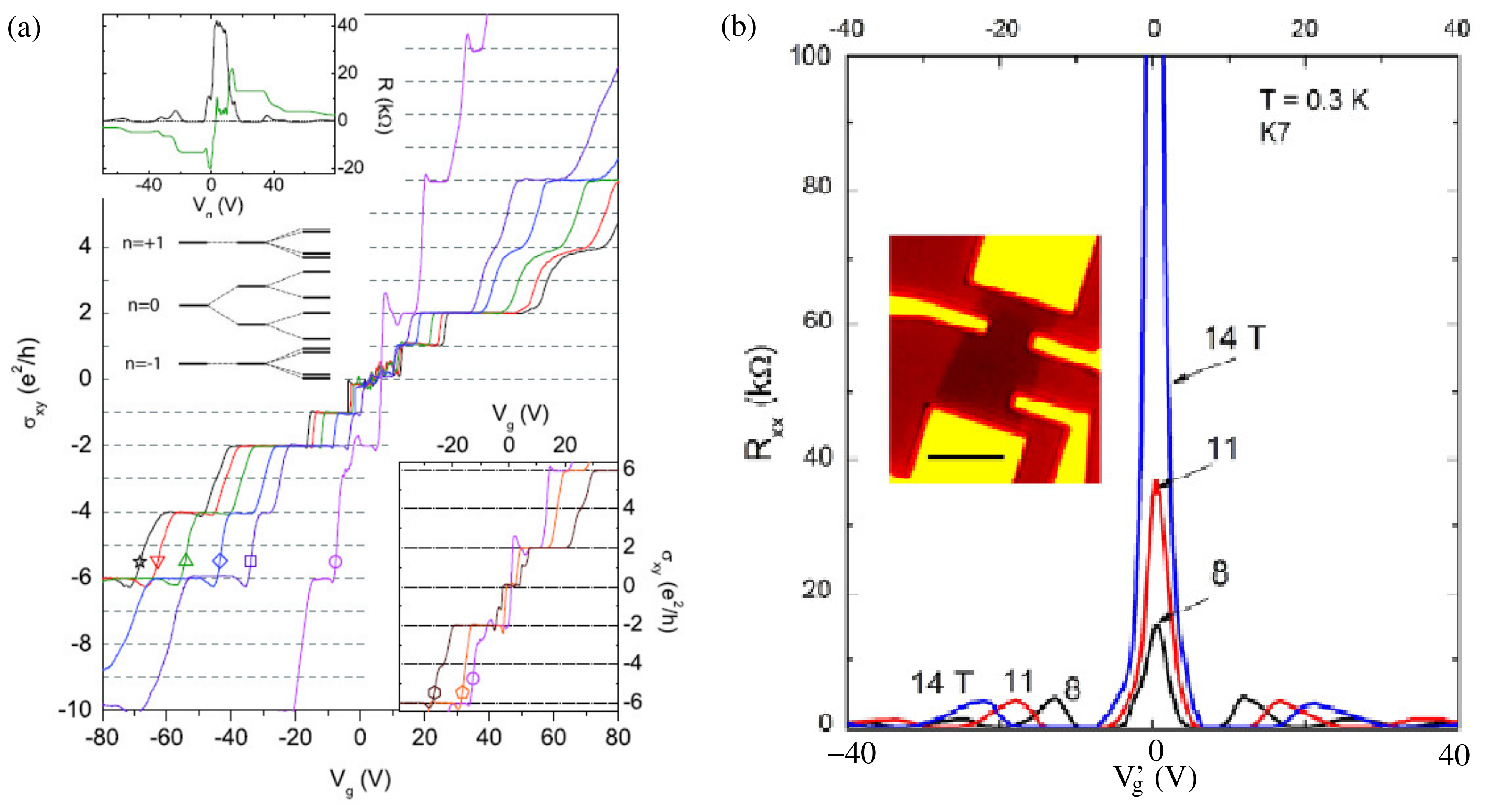}
  \caption
   {
    (Color online)
    (a) $\sigma_{xy}$, as a function of gate voltage at different
    magnetic fields: 9 T (circle), 25 T (square), 30 T (diamond), 37 T
    (up triangle), 42 T (down triangle), and 45 T (star). All the data
    sets are taken at T=1.4 K, except for the B=9 T curve, which
    is taken at T=30 mK. Left upper inset: $R_{xx}$ and $R_{xy}$ for the
    same device measured at B=25 T. 
    Right inset: detailed  $\sigma_{xy}$ data near the Dirac point for 
    B=9 T (circle), 11.5 T (pentagon), and 17.5 T (hexagon) at T=30 mK.
    Adapted from \citet{kn:zhang2006}.
    (b) Longitudinal resistance $R_{xx}$ as a function of gate voltage $V_g'= V_g-V_0$
    at 0.3 K and several values of the magnetic field: 8, 11 and 14 T.  
    The inset shows in false color a graphene crystal (dark red) with Au leads 
    deposited (yellow regions). The bar indicates 5 $\mu$m. 
    At $V_g' = 0$, the peak in $R_{xx}$ grows to 190 k$\Omega$ at 14 T.    
    Adapted from \citet{kn:checkelsky2008}.
   }
  \label{fig:qh:qhfm_slg}
 \end{center}
\end{figure}

Experimentally the existence of broken-symmetry states has been
verified by \citet{kn:zhang2006}, Fig.~\ref{fig:qh:qhfm_slg}~(a),
that showed the existence of the intermediate Landau levels with
$\nu=0,\pm 1$ for the $n=0$ LL and the intermediate level $\nu=\pm 4$
for the $n=1$ LL that is therefore only partially resolved.
Given that the magnetic field by itself does not lift the valley degeneracy,
interaction effects are likely the cause for the full resolution
of the $n=0$ LL. On the other hand a careful analysis of the data
as a function of the tilting angle of the magnetic field suggests
that the partial resolution of the $n=1$ LL is due to Zeeman splitting
\cite{kn:zhang2006}.
%


\subsubsection{The $\nu=0$ state}

In the previous section we have seen that in strong magnetic fields the lowest
LL can be completely resolved and the spin and valley degeneracies may be lifted.
In particular, an approximate plateau for $\sigma_{xy}$ appear for $\nu =0$.
This state has been experimentally studied in a series of works by
\citet{kn:checkelsky2008,checkelsky2009,kn:jiang2007,giesbers2009},
see Fig.~\ref{fig:qh:qhfm_slg}~(b).
The state is unique in that the plateau of $\sigma_{xy}$ corresponds
to a maximum of the longitudinal resistivity $\rho_{xx}$ in contrast
to what happens for $\nu\neq 0$ where a plateau of $\sigma_{xy}$
corresponds to zero longitudinal resistivity.  
In addition, the $\nu=0$ edge states are supposed to not carry
any charge current, but only spin currents \cite{abanin2006,abanin2007x,abanin-ssc-143-77-2007}.
As pointed out by \citet{dassarma2009nu0} however the situation
is not surprising if we recall the relations between the 
resistivity tensor and the conductivity tensor:
\beq
 \rho_{xx} =\frac{\sigma_{xx}}{\sigma_{xx}^2+\sigma_{xy}^2};\quad
 \rho_{xy} =\frac{\sigma_{xy}}{\sigma_{xx}^2+\sigma_{xy}^2}.
 \label{eq:qh:rho_sigma}
\enq
and the fact that the quantization of $\sigma_{xy}$ is associated
with the vanishing of $\sigma_{xx}$. This can be seen from
Laughlin's gauge argument \cite{laughlin1981,halperin1982}.
Using Eq.~\ceq{eq:qh:rho_sigma}, a possible resolution of the $\nu=0$
``anomaly'' is obvious: for any finite $\sigma_{xy}$ 
the vanishing of $\sigma_{xx}$ corresponds to the vanishing of $\rho_{xx}$,
however for $\sigma_{xy}=0$ we have $\rho_{xx}=1/\sigma_{xx}$ so that
$\sigma_{xx}\to 0$ implies $\rho_{xx}\to\infty$.
This is very similar to the Hall insulator phase in ordinary 2D
parabolic band electron gases.
This simple argument shows that the fact that $\rho_{xx}$ seem to diverge
for $T\to 0$ for the $\nu=0$ state is not surprising.
However this argument may not be enough to explain the details of the dependence
of $\rho_{xx}(\nu =0)$ on temperature and magnetic field.
In particular \citet{checkelsky2009} found evidence  for a
field-induced transition to a strongly insulating state at a finite
value of $B$.
These observations suggest that the $\nu=0$ ground state might differ
from the $SU(4)$ eigenstates \ceq{eq:qh:ground} and theoretical
calculations proposed that it could be a spin-density-wave or charge-density-wave \cite{jung2009,herbut2007}.
It has also been argued that the divergence of $\rho_{xx}(\nu=0)$ might be
the signature of Kekule instability \cite{hou2010,nomura2009}

\citet{giesbers2009} have interpreted their experimental data 
using a simple model involving the opening of a field dependent spin gap.
\citet{zhang2009} have observed a cusp in the longitudinal
resistance $\rho_{xx}$ for $\nu\approx 1/2$ and have interpreted
this as the signature of a transition from a Hall insulating
state for $\nu>1/2$ to a collective insulator, like a Wigner crystal \cite{zhang-joglekar2007},
for $\nu<1/2$. 
No consensus has been reached so far, and much more work 
is needed to understand the $\nu=0$ state in graphene.
%
%


\subsubsection{Fractional quantum Hall effect}

\begin{figure}
 \begin{center}
\includegraphics[width=\columnwidth]{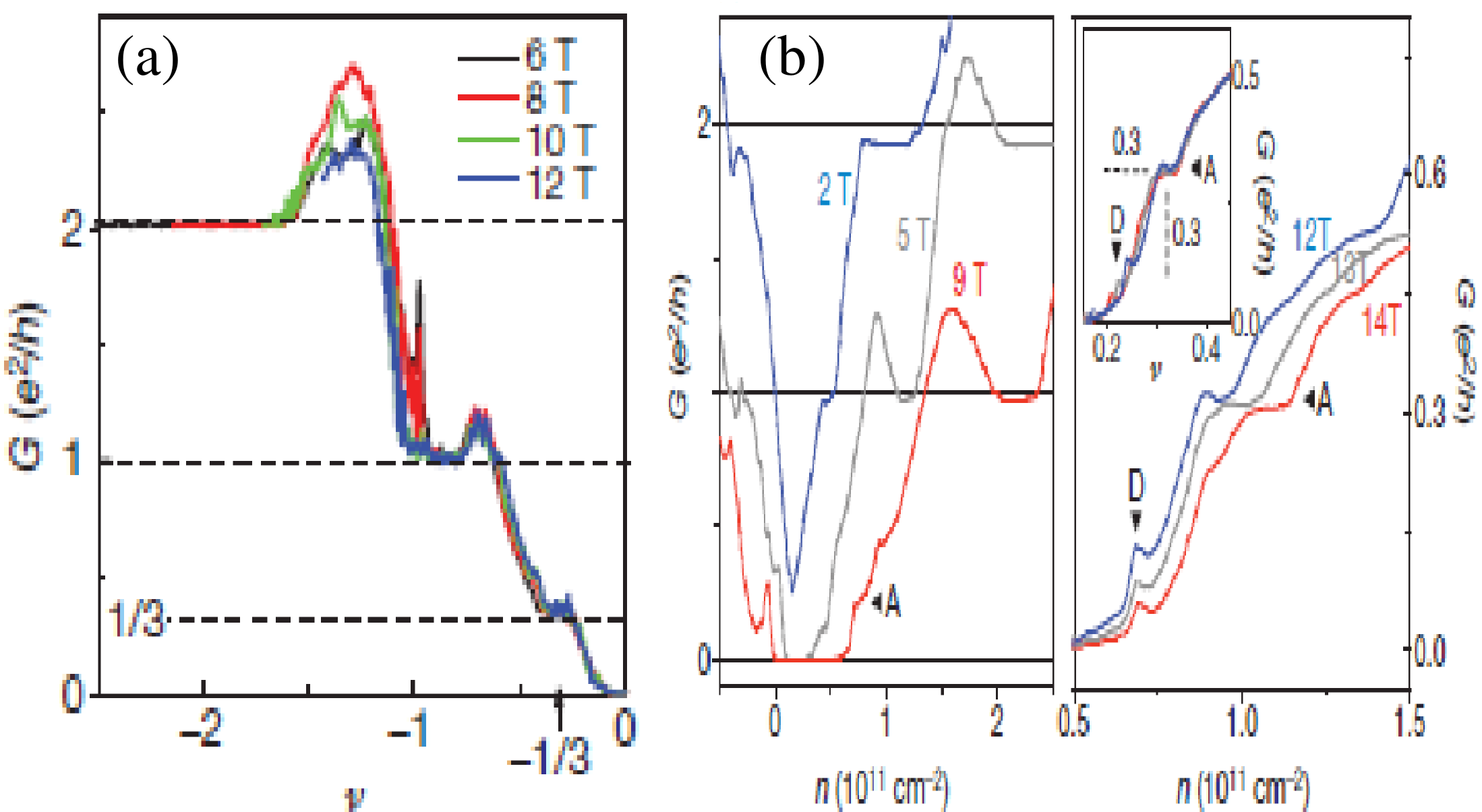}
 \end{center}
 \caption{
          (Color online)
          Graphene fractional
          quantum Hall data taken from (a) \citet{kn:du2009} and (b)
          \citet{kn:bolotin2009}, observed on two probe suspended
          graphene samples.
         }
 \label{fig:qh:fqhe}
\end{figure}

In addition to QH ferromagnetism, the electron-electron interaction
is responsible for the fractional quantum Hall effect, FQHE.
For the FQHE the energy gaps are even smaller than for the QH
ferromagnetic states. For graphene the FQHE gaps have
been calculated by \citet{kn:toke2006,apalkov2006}.
For the $\nu=1/3$ the gap has been estimated to be of the order
of $0.05e^2/\kappa l_B$, where $l_B\df (\hbar c/eB)^{1/2}$ is the magnetic length.
Because of the small size of the gaps 
the experimental observation of
FQHE requires high quality samples.
For graphene very low amount of disorder
can be achieved in suspended samples and in these suspended samples
two groups \cite{kn:du2009,kn:bolotin2009} have recently
observed signatures of the $\nu=1/3$ fractional quantum Hall
state in two-terminal measurements, see Fig.~\ref{fig:qh:fqhe}.
A great deal of work remains to be done in graphene FQHE.

\subsection{Bilayer graphene}
%


\subsubsection{Integer quantum Hall effect}

\begin{figure}[htb]
 \begin{center}
\includegraphics[width=8.5cm]{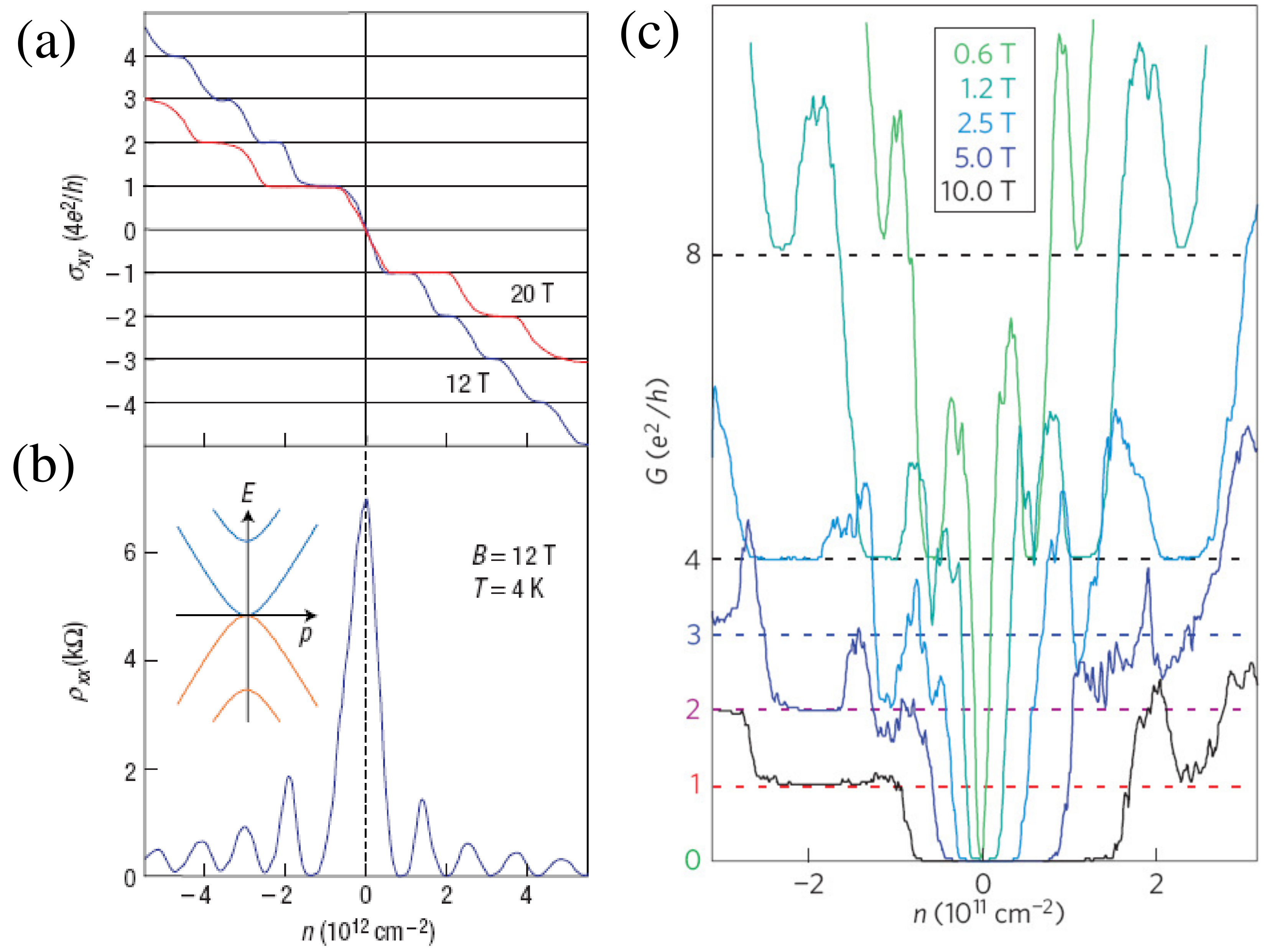}
  \caption
   {
    (Color online)
    (a) Measured Hall conductivity $\sigma_{xy}$ in BLG 
     as a function of carrier density
    for $B=12$~T, in blue, and $B=20$~T, in red, at $T$=4~K.
    (b) Measured longitudinal resistivity in BLG
    at $T$=4~K and $B=12$~T. The inset shows the calculated
    BLG bands close to the CNP.
    Adapted from \citet{kn:novoselov2006}.
    (c) Two terminal conductance, $G$, as a function of carrier density
    at $T$=100~mK for different values of the magnetic field in suspended BLG.
    Broken symmetry states in BLG. 
    Adapted from \citet{kn:feldman2009}.    
   }
  \label{fig:qh:blg}
 \end{center}
\end{figure}

In bilayer graphene the low energy fermionic excitations are massive, i.e. with good approximation
the bands are parabolic. This fact would suggest that the bilayer QH effect in graphene might
be similar to the one observed in regular parabolic 2DEG. There are however two
important differences: the band structure of bilayer graphene is gapless and 
the fermions in BLG, as in SLG, are also chiral but with a Berry phase equal to $2\pi$
instead of $\pi$ \cite{kn:mccann2006b}. As a consequence, as shown in Eq.~\ceq{SDS_Eq_LandauLevels} (c),
the energy levels have a different sequence from both regular 2DEGs and SLG.
In particular BLG also has a LL at zero energy, however, because the Berry phase
associated with the chiral nature of the quasiparticles in BLG is $2\pi$,
the step between the plateaus of $\sigma_{xy}$ across the 
CNP is twice as big
as in SLG (as shown schematically in Fig.~\ref{fig:qh:schematic1})~(c).
One way to understand the step across the CNP is to consider that in BLG 
the $n=0$ and $n=1$ orbital LL are degenerate.
%
%
The spin and valley degeneracy factor, $g$, in BLG is equal to $4$ as in MLG.
In BLG the valley degree of freedom can also be regarded
as a layer degree of freedom considering that without loss of generality
we can use a pseudospin representation in which the $K$ valley states
are localized in the top layer and the $K'$ states in the bottom layer.
The QH effect has been measured experimentally. Fig.~\ref{fig:qh:blg}~(a)
shows the original data obtained by \citet{kn:novoselov2006}.
In agreement with the theory the data show a double size step, compared to SLG, for $\sigma_{xy}$ across the CNP.
%


\subsubsection{Broken-symmetry states}

As discussed in the previous section, the $E_n=0$ LL in BLG has an 8-fold degeneracy
due to spin degeneracy, valley (layer) degeneracy and $n=0$, $n=1$
orbital LL degeneracy. The $E_n\neq 0$ LL have only a 4-fold degeneracy
due to spin and valley degeneracy.
As discussed for SLG, it is natural to expect that the degeneracy of the full $LL$
will be lifted by external perturbations and/or interactions.
Similar considerations to the ones made in section \ref{sec:qh:slg:brsym}
for SLG apply here: the splitting can be due to Zeeman effect \cite{giesbers2009},
strain-induced lifting of valley degeneracy \cite{abanin2007z}
or Coulomb interactions \cite{barlas2008,ezawa2007}.
\citet{barlas2008} considered the splitting of the $E_n=0$ 
LL in BLG due to electron-electron interactions and calculated the
corresponding charge-gaps and filling sequence.
As in SLG, the charge gaps of the splitted LLs will be smaller than the 
charge gap, $\hbar\omega_c$, for the fully occupied LLs and so the observation
of QH plateaus due to the resolution of the LL requires higher quality
samples. This has recently been achieved in suspended BLG samples
\cite{kn:feldman2009,kn:zhao2009} in which 
the full resolution of the eightfold degeneracy of the zero energy LL
has been observed, Fig.~\ref{fig:qh:blg}~(b). By analyzing the dependence of the maximum resistance
at the CNP on $B$ and $T$ \citet{kn:feldman2009} concluded that the
the observed splitting of the $E_n=0$ LL cannot be attributed to 
Zeeman effect. Moreover the order in magnetic fields in which
the broken-symmetry states appear is consistent with the theoretical predictions
of \cite{barlas2008}. These facts suggest that in BLG the resolution
of the octet zero energy LL is due to electron-electron interactions.
%
%


\setcounter{sub3section}{0}


\section{Conclusion and summary}

In roughly five years, research in graphene physics has made
spectacular advance starting from the fabrication of gated
variable-density 2D graphene monolayers to the
observations of fractional quantum Hall effect and Klein
tunneling.  The
massless chiral Dirac spectrum leads to novel integer quantum Hall
effect in graphene with the existence of a $n=0$ quantized Landau level
shared equally between electrons and holes. The non-existence
of a gap in the graphene carrier dispersion leads to a direct transition
between electron-like metallic transport to hole-like metallic
transport as the gate voltage is tuned through the charge neutral
Dirac point. By contrast, 2D semiconductors invariably become
insulating at low enough carrier densities. In MLG nanoribbons and in BLG structures 
in the presence of an
electric field, graphene carrier transport manifests a transport gap
because there is an intrinsic spectral gap induced by the confinement
and the bias field, respectively. The precise relationship between the
transport and the spectral gap is, however, not well understood at
this stage and is a subject of much current activity.
Since back-scattering processes are suppressed,
graphene exhibits weak anti-localization behavior in contrast to the
weak localization behavior of ordinary 2D systems. The presence of any
short-range scattering, however, introduces inter-valley coupling,
which leads to the eventual restoration of weak localization. Since
short-range scattering, arising from lattice point defects, is weak in
graphene, the weak
anti-localization behavior is expected to cross over to weak
localization behavior only at very low temperatures although a direct
experimental observation of such a localization crossover is still lacking and
may be difficult.

The
observed sequence of graphene integer quantized Hall conductance follows the
expected formula, $\sigma_{xy}=(4e^2/h) (n+1/2)$, indicating the
Berry phase contribution and the $n=0$ Landau level shared between
electrons and holes. 
For example, the complete
lifting of spin and valley splitting leads to the observation of the
following quantized Hall conductance sequence $\nu=0,\pm1,\pm2, ...$
with $\sigma_{xy}=\nu e^2/h$ whereas in the presence of spin and
valley degeneracy (i.e., with the factor of 4 in the front) one gets
the sequence $\nu=\pm2,\pm6,...$
The precise nature of the $\nu=0$ IQHE, which
seems to manifest a highly resistive ($\rho_{xx}\rightarrow \infty$)
state in some experiments but not in others, is still an open question
as is the issue of the physical mechanism or the quantum phase
transition associated with the possible spontaneous symmetry breaking
that leads to the lifting of the degeneracy. 
We do mention, however, that similar, but
not identical, physics arises in the context of ordinary IQHE in 2D
semiconductor structures. For example, the 4-fold spin and valley
degeneracy, partially lifted by the applied magnetic field, occurs in
2D Si-(100) based QHE, as was already apparent in the original
discovery of IQHE by Klaus von Klitzing \cite{vonklitzing1980}. The issue of spin and valley
degeneracy lifting in the QHE phenomena is thus generic to both
graphene and 2D semiconductor systems although the origin of valley
degeneracy is qualitatively different~\cite{mcfarland2009,eng2007}
in the two cases.  The other similarity
between graphene and 2DEG QHE is that both systems tend to manifest 
strongly insulating phases at very high magnetic field when $\nu
\ll 1$. In semiconductor based high mobility 2DEG, typically such a
strongly insulating phase occurs \cite{jiang1990,jiang1991} for $\nu < 1/5-1/7$ whereas in
graphene the effect manifest near the charge neutral Dirac point
around $\nu\approx 0$. Whether the same physics controls both
insulating phenomena or not is an open question.

Very recent experimental observations of  $\nu=1/3$ FQHE in graphene 
has created a great deal of excitement. These
preliminary experiments involve 2-probe measurements on suspended
graphene samples where no distinction between $\rho_{xx}$ and
$\rho_{xy}$ can really be made. Further advance in the field would
necessitate the observation of quantized plateaus in $\rho_{xy}$ with
$\rho_{xx}\approx 0$. Since FQHE involves electron-electron interaction
effects, with the non-interacting part of the Hamiltonian playing a
rather minor role, we should not perhaps expect any dramatic
difference between 2DEG and graphene FQHE since both systems manifest
the standard $1/r$ Coulomb repulsion between electrons. Two possible
quantitative effects distinguishing FQHE in graphene and 2DEG, which
should be studied theoretically and numerically, are the different
Coulomb pseudopotentials and Landau level coupling in the two
systems. Since the stability of various FQH states depends crucially
on the minute details of Coulomb pseudopontentials and inter-LL
coupling, it is conceivable that graphene may manifest novel FQHE not
feasible in 2D semiconductors.

We also note that many properties reviewed in this article should also
apply to topological insulators \citep{kn:hasan2010}, which have only
single Dirac 
cones on their surfaces.
Although we now have a reasonable theoretical understanding of the broad
aspects of transport in monolayer graphene, much work remains to be done
in bilayer and nanoribbons graphene systems.

\section*{Aknowledgments}
This work is supported by US-ONR and NSF-NRI.


\end{document}